\documentclass[twoside,12pt]{article}
\usepackage{epsfig,graphicx}
\usepackage{axodraw2}

\usepackage{amssymb}
\usepackage{amsmath}
\usepackage[colorlinks=true,linkcolor=black]{hyperref}

\def\Journal#1#2#3#4{{#1} {#2} (#4) #3 }

\def\PHYS{{\em Physica}}
\def\NPA{{\em Nucl. Phys.} A}

\def\PRO{{\em Prog. Theor. Phys.}}
\def\NPB{{\em Nucl. Phys.} B}

\def\PLB{{\em Phys. Lett.} B}

\def\PL{{\em Phys. Lett.}}
\def\PRL{\em Phys. Rev. Lett.}
\def\PREV{\em Phys. Rev.}
\def\PREP{\em Phys. Rep.}

\def\PRD{{\em Phys. Rev.} D}
\def\PRC{{\em Phys. Rev.} C}

\def\ZP{\em Z. Phys.}
\def\ZPC{{\em Z. Phys.} C}

\def\EPJC{{\em Eur. Phys. J.} C}
\def\EPJA{{\em Eur. Phys. J.} A}
\def\ANNP{\em Ann. Phys. (N.Y.)}
\def\RMP{{\em Rev. Mod. Phys.}}

\def\INT{{\em Int. J. Mod. Phys.} E}
\def\PPNP{{\em Prog. Part. Nucl. Phys}}
\def\RPP{{\em  Rep. Prog. Phys.}}

\newcommand{\vh}{\varphi}

\newcommand{\vs}{\vskip 5pt}

\newcommand{\vp}{\mathbf{p}}
\newcommand{\vq}{\mathbf{q}}
\newcommand{\vr}{\mathbf{r}}
\newcommand{\vx}{\mathbf{r}}
\newcommand{\vz}{\mathbf{z}}
\newcommand{\vk}{\mathbf{k}}
\newcommand{\up}{{\hat{\mathbf{p}}}}

\newcommand{\ep}{\epsilon}
\newcommand{\vep}{\varepsilon}
\newcommand{\ve}{\varepsilon}
\newcommand{\De}{\Delta}

\newcommand{\cN}{{\cal N}}
\newcommand{\cD}{{\cal D}}
\newcommand{\cS}{{\cal S}}
\newcommand{\cI}{{\cal I}}

\newcommand{\cdd}{M_{\rm CDD}}

\def\XXint#1#2#3{{\setbox0=\hbox{$#1{#2#3}{\int}$}
		\vcenter{\hbox{$#2#3$}}\kern-.5\wd0}}

\newcommand{\be}{\begin{equation}}
\newcommand{\ee}{\end{equation}}
\newcommand{\bea}{\begin{eqnarray}}
\newcommand{\ea}{\end{eqnarray}}
\newcommand{\nn}{\nonumber}

\begin{document}

\title{ \vspace{1cm} Coupled-channel approach in hadron-hadron scattering}
\author{J.\ A.\ Oller\\
\\
 Departamento de F\'{\i}sica, Universidad de Murcia, E-30071 Murcia, Spain}
\maketitle
\begin{abstract}
  Coupled-channel dynamics for scattering and
  production processes in partial-wave amplitudes is discussed from a perspective that emphasizes unitarity and analyticity.
  We elaborate on several methods that have driven to important results in hadron physics, either by themselves or in
  conjunction with effective field theory. 
  We also develop and compare with the use of the Lippmann-Schwinger equation in near-threshold scattering.
  The final(initial)-state interactions are discussed in detail for the elastic and coupled-channel case.
  Emphasis has been put in the derivation and discussion of the methods presented, with some  applications
  examined as important examples of their usage.  
\end{abstract}
%\eject
\newpage
\tableofcontents

\def\theequation{\arabic{section}.\arabic{equation}}

\newpage
%%%%%%%%%%%%%%%%%%%%%%%%%%%%%%%%%%%%%%%%%%%%%%%%%%%%
\section{Introduction}
\setcounter{equation}{0}   
\label{sec.190414.1}

We would like to discuss along this review a series of methods used in coupled-channel dynamics.
The list of methods offered here is not really exhaustive but rather shows the idiosyncrasy of the author.
As a result, we would like to apologize to all of those who consider that their work should have been quoted in a
review on coupled-channel dynamics but it is not the case. 
The choice offered along this manuscript is based on the work that we have deployed along the years in the study
of hadron-hadron interactions, both directly in scattering reactions or in production processes, many of them involving actual
external probes.
In addition, this has forged a personal preference towards some other works in the literature, even if we have not been directly
involved on them. Nonetheless, they have been reviewed in detail, like in the cases of
Secs.~\ref{sec.190224.1} and \ref{sec.1903291.1}, where we have also developed a more compact and general notation.
A decision on writing this review has also been made on discussing only the methods valid in the continuum,
without any reference to lattice theories. 

The emphasis is always put in the method rather than in the particular applications.
In this way, the latter are quoted either because they have been crucial for the development of the approach
and/or they are good examples of its application. But again, we would like to emphasize that no particular interest has been
put in offering an exhaustive list of references, even at the level of a particular example.

An important basis for the methods here considered is the unitarity and analytical properties of the scattering amplitudes,
in particular of the partial-wave amplitudes  for which the unitarity requirements can be expressed in a simpler way.
On the contrary, crossing symmetry is much more involved at the level of partial-wave amplitudes.
An introduction to some basic aspects on these regards are discussed in Sec.~\ref{181024.1}.

These analytical properties and unitarity are employed in the next section to settle the mathematical framework of dispersion
relations  and obtain general results for partial-wave amplitudes.
These are based on methods that allow one to write the partial-wave amplitudes
in terms of two functions, each of which has only a unitarity cut or crossed-channel cuts, like e.g. the $N/D$ method.
We also present another approach whose exact solution requires to solve a non-linear integral equation, but it is
 specially suitable for a perturbative treatment of the contributions from crossed-channel cuts. 
The problem gets simplified if the latter cuts are neglected in a first approximation, and this simplicity
is exploited in Sec.~\ref{sec.181105.1} to offer a deep picture of the lightest resonances in Quantum Chromodynamics.
 The important Castillejo-Dalitz-Dyson poles are introduced as well, rescued from the oblivion in Ref.~\cite{oller.181105.1},
as  a source of dynamics even though no crossed channels are considered. 

There has also been a  revival in the interest of non-relativistic techniques, in particular of the
Lippmann-Schwinger equation, for the study of meson-meson scattering
since the discovery of the well-known $X(3872)$ \cite{Choi:2003ue}.
This discovery has been followed by many others in which  heavy-quark hadrons in the near-threshold region
of open heavy-flavor channels  do not fit well within quark model
expectations or non-relativistic Quantum Chromodynamics, so that a full implementation of coupled-channel dynamics
is required for an understanding of their properties.
The reflection of this interest on this review prompted the emergence of Sec.~\ref{sec.181202.1}, in which the uncoupled case
is treated in the first two subsections and the latter ones also deal with coupled channels, including elastic and inelastic ones.
Here we have made use of the general results that follow by applying the $N/D$ method without crossed-channel dynamics or
by combining meson-meson direct scattering plus the exchange of bare states in  a Lippmann-Schwinger equation.

The next step undertaken in Sec.~\ref{sec.181111.1} is implementing crossed-channel dynamics,
which in a relativistic theory is a requirement due to crossing
symmetry, while it stems from the potential in a non-relativistic treatment.
The matching with an underlying effective field theory allows a perturbative solution on the contributions of the crossed-channel
cuts. As a particular realization of this scheme we have the so-called inverse-amplitude method.
We also discuss the perturbative solution of the $N/D$ by iterating on the discontinuity along the crossed-channel cuts.
 % up to different orders.

An important source for the study of strong interactions has been the use of external sources so as to trigger a given reaction,
which is afterwards strongly modulated by the final-state interactions. Of course, the process can be reversed in time and we have
initial-state interactions or both simultaneously.
This is the object of the last section~\ref{sec.181117.1}, where we start first discussing the uncoupled case and
afterwards proceed with the coupled-channel one, with the prototypical Omn\`es solution discussed with special detail.
The possible presence of crossed-channel dynamics is taken into account from the start, and the methods are developed
accordingly. The final section is dedicated to discuss the Khuri-Treiman formalism, exemplified in the eta to three-pion
decays. It is an interesting and rather subtle approach that requires for its development many 
relevant techniques discussed along the review. We end with some remarks in Sec.~\ref{sec.190516.1}.

%%%%%%%%%%%%%%%%%%%%%%%%%%%%%%%%%%%%%%%%%%%%%%%%%%%%
\section{$S$ and $T$ matrices. Unitarity, crossing and partial-wave amplitudes}
\setcounter{equation}{0}   
\label{181024.1}

We take for granted in the following that the interactions involved in the scattering process are of finite range, so that 
the states at asymptotic times are made up of free particles.\footnote{This does not mean that we cannot discuss 
processes involving electromagnetic interactions. However, these interactions are treated at the first non-vanishing order and 
are not further iterated.} 
The free multiparticle states are given by the direct product of free monoparticle states. 
Every of them is characterized  by its three-momentum $\vp$, spin $s$, third-component of spin $\sigma$,
mass $m$, and other quantum numbers (like the charges) are denoted by $\lambda$.
A monoparticle free state is written as
\be
\label{181031.1}
|\vp,\sigma,m,s,\lambda\rangle~,  
\ee
with $ \sigma=-s,-s+1,\ldots,s-1,s$. 
These states are normalized according to the Lorentz invariant normalization
\be
\label{181031.2}
\langle \vp',\sigma',m,s',\lambda'|\vp,\sigma,m,s,\lambda\rangle =
\delta_{s's} \delta_{\sigma'\sigma}\delta_{\lambda'\lambda} (2\pi)^3 2p^0 \delta(\vp'-\vp)~,
\ee
with $p^0=\sqrt{m^2+\vp^2}$.

The probability amplitudes for the transition of a  free initial state $|i\rangle$, at asymptotic time $t\to -\infty$, 
to a free final state  $|f\rangle$,  at asymptotic time $t\to +\infty$, 
are the matrix elements of  the $S$ matrix \cite{heisenberg.190106.1}, denoted by $S_{fi}$, with  
\be
\label{181101.2}
S_{fi}=\langle f|S|i\rangle~.
\ee
The $S$ matrix is a unitary operator $S$, 
\be
\label{181101.1}
S S^\dagger=S^\dagger S=I~.
\ee
Because of the space-time homogeneity the matrix elements $S_{fi}$ always comprise  a total
energy-momentum Dirac delta function, $(2\pi)^4 \delta^{(4)}(p_f-p_i)$, with $p_i$ and $p_f$ the
initial an final four-momenta, respectively. 
In the unitarity relation obeyed by the $S$ matrix one can consistently factor out a Dirac delta function
of total energy and momentum conservation. 
We do not always show explicitly such simplification, though it should be clear from the context.

We next introduce the $T$ matrix, defined by its relation with the $S$ matrix as
\be
\label{181101.4}
S=I+i T~,
\ee
so that the $T$ matrix requires at least of one interaction to take place. Importantly, the 
unitarity relation  of Eq.~\eqref{181101.1} in terms of the $T$ matrix reads
\bea
\label{181101.5}
T-T^\dagger &=& i T T^\dagger\\
\label{181101.5b}
&=&i T^\dagger T~.
\ea
The first (second) line follows from the first (second) term  in Eq.~\eqref{181101.1}.
We include a resolution of the identity by free states between the product of the operators $T$ and $T^\dagger$, which implies 
at the level of the matrix elements the following non-linear unitarity relation
\bea
\label{181101.6}
\langle f|T|i\rangle-\langle f|T^\dagger |i\rangle& = &i \sum \int
\left[(2\pi)^4\delta^{(4)}(p_f-\sum_{j=1}^n q_j) \prod_{j=1}^n \frac{d^3q_j}{(2\pi)^3 2 q_j^0}\right]\\
& \times& \langle f|T^\dagger|\vq_1,\sigma_1,m_1,s_1,\lambda_1;\ldots;\vq_n,\sigma_n,m_n,\lambda_n\rangle
\langle \vq_1,\sigma_1,m_1,\lambda_1;\ldots;\vq_n,\sigma_n,m_n,\lambda_n|T|i\rangle~,\nn
\ea
where the total energy-momentum conservation, $p_f=p_i$, should be understood.
The sum extends over all the possible intermediate states allowed by the appropriate quantum numbers and with thresholds below
the total center-of-mass (CM) energy $\sqrt{p_f^2}$ (otherwise the intermediate Dirac delta function would vanish).
An analogous relation also holds with $T\leftrightarrow T^\dagger$ exchanged in the right-hand side (rhs) 
of the previous equation.

The basic content of Hermitian analyticity (section 4.6 of Ref.~\cite{olive.181102.1})
is precisely to show that the matrix elements of $T^\dagger$  are also given by the same analytical function as those of $T$ itself, 
but with a slightly negative imaginary part in the total energy (or partial ones
for subprocesses) along the real axis, instead of the slightly positive imaginary part employed for the matrix elements
of $T$ in Eq.~\eqref{181101.6}.
As a result, the unitarity relation of Eq.~\eqref{181101.6} gives rise to the presence  in the scattering amplitudes 
of the right-hand cut (RHC), or unitarity cut, when the  energy involved is real
and larger than the smallest threshold, typically  a two-body state. 
It could also comprise other singularities like the pole ones, 
while its iteration from the simplest singularities (pole and normal thresholds) could generate more complicated 
ones, such as the anomalous thresholds (sections 4.10 and 4.11 of Ref.~\cite{olive.181102.1}), cf.
the discussion in the first full paragraph after Eq.~\eqref{190403.10}. 

The factor between square brackets in Eq.~\eqref{181101.6} is the differential
phase space, $dQ$, which could be partial or totally integrated,
\be
\label{181101.7}
\int  d Q=\int
(2\pi)^4\delta^{(4)}(p_f-\sum_{i=1}^n q_i) \prod_{i=1}^n \frac{d^3q_i}{(2\pi)^3 2 q_i^0}~.
\ee
An important property of phase space is that it  is Lorentz invariant.

Given an initial two-particle state with four-momenta $p_1$ and $p_2$, its cross section to a generic final state 
$|f\rangle$, denoted by $\hat{\sigma}_{fi}$, is defined as the number of particles scattered per unit time 
 divided by the incident flux $\phi_0$.  
The latter  is necessary because the number of collisions rises  in a given experiment 
as the number of incident particles does. 
In our normalization, Eq.~\eqref{181031.2}, it follows that $\hat{\sigma}_{fi}$ in the CM frame is  
\be
\label{181101.8}
\hat{\sigma}_{fi} = \frac{1}{4|\vp_1|\sqrt{s}}\int d Q_f \left|\langle 
f|T|\vp_1,\sigma_1,m_1,s_1,\lambda_1;\vp_2,\sigma_2,m_2,s_2,\lambda_2\rangle\right|^2~,
\ee
with $s=(p_1+p_2)^2$.  
 The total cross section, $\hat{\sigma}_i$,  is given by the sum over all the possible
final sates and from Eq.~\eqref{181101.8} we then have that 
\be
\label{181101.10}
\hat{\sigma}_{i} =\frac{1}{4|\vp_1|\sqrt{s}}\sum_f \int d Q_f \left|\langle f|T|\vp_1,\sigma_1,m_1,s_1,\lambda_1;\vp_2,\sigma_2,m_2,s_2,\lambda_2\rangle\right|^2~.
\ee

In the following, for brevity in the notation, we denote the monoparticle states  by
$|\vp_1\hat{\sigma}_1\lambda_1\rangle$, omitting some labels that can be inferred from the information given.
 We can relate the total cross section $\hat{\sigma}_i$ with the imaginary part of the forward $T$-matrix element
$T_{ii}$ by taking $|f\rangle=|i\rangle$ in the unitarity relation of Eq.~\eqref{181101.6}, which then implies that 
\be
\label{181101.11}
\Im T_{ii} =\frac{1}{2}\sum_f\int d Q_f |T_{fi}|^2=2 |\vp_1|  \sqrt{s}\,\hat{\sigma}_i~.
\ee
This result is usually referred as the optical theorem.  For the other  order  
$T T^\dagger$ in the unitarity relation Eq.~\eqref{181101.5}, it results 
\be
\label{181101.12}
\Im T_{ii}=\frac{1}{2}\sum_f\int d Q_f |T_{if}|^2~.
\ee
The comparison with Eq.~\eqref{181101.11} implies the equality 
\be
\label{181101.13}
\sum_f\int d Q_f |T_{fi}|^2=\sum_f\int d Q_f |T_{if}|^2~,
\ee
which entails the important Boltzmann $H$-theorem in statistical mechanics
(as discussed in section 3.6. of Ref.~\cite{weinberg.181101.1} and in chapter 1 of Ref.~\cite{oller.190503.1}),
\be
\label{181101.19}
\frac{d \mathfrak{S}}{dt}\geq 0~,
\ee
where $\mathfrak{S}$ is the entropy.

%%%%%%%%%%%%%%%%%%%%%%%%%%%%%%%%%%%%%%%%%%%%%%%%%%%%%%%%%%%%%%%%%%%%%
%\subsection{Two-body scattering. Partial-wave expansion}
%\label{sec.181101.1}
%\setcounter{equation}{0}   

The basic process under consideration is the scattering of two particles of four-momenta $p_1$ and $p_2$  into  other 
two  of momenta $p_3$ and $p_4$. For a two-body state the Lorentz-invariant differential phase-space factor of Eq.~\eqref{181101.7}, expressed in the CM variables, is
\be
\label{181101.20}
dQ=\frac{|\vp_1|d\Omega}{16\pi^2\sqrt{s}}~,
\ee
where $d\Omega$ is the differential of solid angle of $\vp_1$ in the CM. 
We have introduced $s$ which is one of the standard Mandelstam variables $s$, $t$ and $u$, defined as
\bea
\label{181102.2}
s&=&(p_1+p_2)^2=(p_3+p_4)^2~,\\
t&=&(p_1-p_3)^2=(p_2-p_4)^2~,\nn\\
u&=&(p_1-p_4)^2=(p_2-p_3)^2~.\nn
\ea
From Eq.~\eqref{181102.2} it follows the equality
\bea
\label{181102.2b}
s+t+u =\sum p_i^2~. 
\ea
Thus, for on-shell scattering ($p_i^2=m_i^2$), $s+t+u$ is equal to the sum of the masses squared.

Making use of $dQ$ given in Eq.~\eqref{181101.20},  the two-body unitarity relation, which is
valid below the threshold of the intermediate states with three or more particles, reads
\bea
\label{181101.6a}
\langle f|T|i\rangle-\langle f|T^\dagger |i\rangle&=&i \sum \int
\frac{|\vq_1|d\Omega}{16\pi^2\sqrt{s}}
\langle f|T^\dagger|\vq_1,\sigma_1,\lambda_1;\vq_2,\sigma_2,\lambda_2\rangle\nn\\
&\times& \langle \vq_1,\sigma_1,\lambda_1;\vq_2,\sigma_2,\lambda_2|T|i\rangle~.
%\langle f|T^\dagger|\vq_1,\sigma_1,m_1,s_1,\lambda_1;\vq_2,\sigma_2,m_2,\lambda_2\rangle\nn\\
%&\times& \langle \vq_1,\sigma_1,m_1,\lambda_1;\vq_2,\sigma_2,m_2,\lambda_2|T|i\rangle~.
\ea

The differential cross section, Eq.~\eqref{181101.8}, between two-body states is 
\bea
\label{181101.21}
\frac{d\hat{\sigma}}{d\Omega}=\frac{|\vp'_1|}{|\vp_1|}
\frac{|\langle \vp'_1,\sigma'_1,\lambda'_1;\vp'_2,\sigma'_2,\lambda'_2
	|T|\vp_1,\sigma_1,\lambda_1;\vp_2,\sigma_2,\lambda_2\rangle|^2}{64\pi^2 s}~.
\ea

It is convenient to perform a partial-wave expansion of the scattering amplitudes employing  states with well-defined total angular momentum $J$, total spin $S$ and orbital angular momentum $\ell$, the so-called $ LSJ$ basis.
A main reason is because unitarity adopts a simpler form in terms of partial-wave amplitudes (PWAs). 
Of course, a PWA expansion stems from the invariance under rotations of the $S$ and $T$ operators,  
\bea
\label{181101.22}
R S R^\dagger&=&S~,\\
R T R^\dagger&=&T~,
\ea
where $R$ is a generic rotation. 
In the manipulations that follow we only show the active variables under rotations and suppress any other labels that would play a parametric role. The relations between the scattering amplitudes in momentum space and the PWAs in the $LSJ$ basis that 
we obtain here have evolved from the corresponding ones in Refs.~\cite{lacour.181101.1,gulmez.181101.2}.  
Let us consider a two-body state of two particles with spins $s_1$ and $s_2$, which is further  characterized 
by its CM three-momentum $\vp$,  and the third components of spin  $\sigma_1$ and $\sigma_2$ in the rest frame of every particle. 
This state is denoted by $|\vp,\sigma_1\sigma_2\rangle$, and we also introduce the state 
$|\ell m,\sigma_1\sigma_2\rangle$, with well-defined orbital angular-momentum $\ell$ and third component $m$, which is 
given by 
\be
\label{280916.6}
|\ell m,\sigma_1\sigma_2\rangle=\frac{1}{\sqrt{4\pi}}\int d\up\,Y_\ell^m(\hat{\vp})|\vp,\sigma_1\sigma_2\rangle~.
\ee 
It is shown in the chapter 2 of Ref.~\cite{oller.190503.1} that the state $|\ell m,\sigma_1\sigma_2\rangle$ transforms 
under the action of  a rotation $R$  as the direct product of the irreducible representations associated
with the orbital angular momentum $\ell$ and the spins $s_1$ and $s_2$  of the two particles.  
Therefore, we can combine the orbital angular momentum and the spins 
in a state of the $LSJ$ basis having well-defined total angular momentum $J$ with third component $\mu$. 
These states are denoted by $|J\mu,\ell S\rangle$ and can be calculated from the composition of the angular momenta as
\be
\label{290916.1}
|J\mu,\ell S \rangle =\sum_{\sigma_1,\sigma_2,m,M} (\sigma_1\sigma_2 M|s_1s_2S)(m M \mu|\ell S J) |\ell m,\sigma_1\sigma_2\rangle~,
\ee
where the Clebsch-Gordan coefficient $(m_1 m_2 m_3|j_1j_2j_3)$ corresponds to the  composition for
$\mathbf{j}_1+\mathbf{j}_2=\mathbf{j}_3$, with $m_i$ referring to the third components of the spins. 
We also consider in the following isospin indices  $\alpha_1$ and $\alpha_2$ for the third components of the isospins of the 
two particles, $\tau_1$ and $\tau_2$, respectively.  

By inverting Eq.~\eqref{280916.6} one can express the states with definite three momentum in terms of the $LSJ$ basis as, 
\bea
\label{290916.3}
|\vp,\sigma_1\sigma_2,\alpha_1\alpha_2\rangle &=&\sqrt{4\pi} 
\sum_{\ell,m} Y_\ell^m(\hat{\vp})^* |\ell m,\sigma_1\sigma_2,\alpha_1\alpha_2\rangle\\
&=&\sqrt{4\pi} \!\! \sum_{\scriptsize{
		\begin{array}{c}
		J,\mu,\ell,m\\S,M,I,t_3
		\end{array}}}  \!\! 
Y_\ell^m(\hat{\vp})^* (\sigma_1\sigma_2 M|s_1s_2S) (m M \mu|\ell S J) 
(\alpha_1 \alpha_2 t_3|\tau_1\tau_2 I)
|J\mu,\ell S,I t_3\rangle~,\nn
\ea
with $I$ the total isospin of the particle pair and $t_3$ its the third component.
By a straightforward calculation, taking into account the orthogonality properties of the spherical harmonics and the 
Clebsch-Gordan coefficients \cite{rose.021016.2}, Eq.~\eqref{290916.3} allows to express the states $|J\mu,\ell S,I t_3\rangle$ 
as the following linear superposition of the momentum-defined states, 
\be
\label{061016.1}
|J\mu,\ell S, I t_3\rangle = 
\frac{1}{\sqrt{4\pi}}\!\!
\sum_{\scriptsize{
		\begin{array}{c}
		\sigma_1,\sigma_2\\
		M,m\\
		\alpha_1,\alpha_2 
		\end{array}}  }\!\!
\int d\hat{\vp} \,Y_\ell^m(\hat{\vp}) (\sigma_1\sigma_2 M|s_1s_2S)
(m M \mu|\ell S J)  (\alpha_1 \alpha_2 t_3|\tau_1\tau_2 I)
|\vp,\sigma_1\sigma_2,\alpha_1\alpha_2\rangle~.
\ee

The two-body CM state $|\vp,\sigma_1\sigma_2,\alpha_1\alpha_2\rangle$ is the direct product of the states
$|\vp,\sigma_1,\alpha_1\rangle$ and  $|\!-\!\vp,\sigma_2,\alpha_2\rangle$. 
From the normalization of the monoparticle states, Eq.~\eqref{181031.2}, this two-body state satisfies the normalization
\be
\label{061016.2}
\langle \vp',\sigma'_1\sigma'_2,\alpha'_1\alpha'_2|\vp,\sigma_1\sigma_2,\alpha_1\alpha_2\rangle
=\frac{16 \pi^2 \sqrt{s}}{|\vp|} \delta(\up'-\up)\delta_{\sigma'_1\sigma_1}\delta_{\sigma'_2\sigma_2}\delta_{\alpha'_1\alpha_1}\delta_{\alpha'_2\alpha_2}~,
\ee
where a  factor $(2\pi)^4\delta^{(4)}(p'-p)$ has been factorized out. 
The total energy conservation implies the conservation of the moduli of the final and initial three-momenta, 
that we denote by $|\vp|$.

In terms of Eqs.~\eqref{061016.1}, \eqref{061016.2} and the orthogonality properties of the spherical harmonics and the 
Clebsch-Gordan coefficients, one has that the $|J\mu,\ell S\rangle$  states are normalized as 
\be
\label{061016.3}
\langle  J' \mu' , \ell' S' , I' t_3' |  J  \mu , \ell S , I t_3 \rangle = \frac{ 4 \pi \sqrt{s} }{|\vp|}
\delta_{J'J}\delta_{\mu'\mu}\delta_{\ell'\ell}\delta_{S'S}\delta_{I'I}\delta_{t'_3t_3} ~.
\ee

A PWA is the transition between states in the $LSJ$ basis, which then corresponds to the matrix element
\be
\label{011016.2}
T^{(IJ)}_{\ell S;\bar{\ell}\bar{S}}=
\langle J\mu,\ell S,I t_3|T|J\mu,\bar{\ell} \bar{S}, I t_3\rangle~,
\ee
where the quantum numbers that refer to the initial state are barred. 
Let us notice that because of rotational invariance in ordinary and isospin spaces, the Wigner-Eckart theorem 
implies that a PWA is independent of $\mu$ and $t_3$. Expressing  the $LSJ$ states in Eq.~\eqref{011016.2} 
in terms of the ones with well-defined three-momentum, we obtain in a first step that 
\bea
\label{021016.1}
T^{(IJ)}_{\ell S;\bar{\ell}\bar{S}}&=&
\frac{1}{4\pi}\sum\int d\hat{\vp}\int d\hat{\vp}' 
\,Y_\ell^m(\hat{\vp}')^* Y_{\bar\ell}^{\bar{m}}(\hat{\vp})
(\sigma_1\sigma_2 M|s_1s_2 S)(m M \mu|\ell S J)(\alpha_1\alpha_2 t_3|\tau_1\tau_2I)\nn\\
&\times& (\bar{\sigma}_1\bar{\sigma}_2 \bar{M}|\bar{s}_1\bar{s}_2 \bar{S})(\bar{m} \bar{M} \mu|\bar{\ell} \bar{S} J)
(\bar{\alpha}_1\bar{\alpha}_2 t_3|\bar{\tau}_1\bar{\tau}_2I)
_S\langle \vp',\sigma_1\sigma_2,\alpha_1\alpha_2|T|\vp,\bar{\sigma}_1\bar{\sigma}_2,\bar{\alpha}_1\bar{\alpha}_2\rangle_S~.
\ea
Here, the explicit indices on which the sum is done are not displayed so as not to overload the notation. Nonetheless, they 
correspond to those  indicated under the summation symbol in Eq.~\eqref{061016.1}, both for the initial and final states.
 One can also use the rotational invariance of the $T$-matrix operator  to simplify the previous equation,
  so that only one integral over the solid angle of the final moment has to be performed. 
 The basic point is to make a rotation that takes the 
 initial three-momentum $\vp$ to $\hat{\mathbf{z}}$ and proceed consistently with rotational invariance.
 These needed steps are  
 discussed in detail in the chapter 2 of Ref.~\cite{oller.190503.1} or in the appendix A of Ref.~\cite{gulmez.181101.2}.  
 The final resulting expression for $T^{(IJ)}_{\ell S;\bar{\ell}\bar{S}}$ is,
\bea
\label{051016.6}
T^{(IJ)}_{\ell S;\bar{\ell}\bar{S}}&=&\frac{Y_{\bar{\ell}}^{0}(\hat{\mathbf{z}})}{2J+1} \!\!
\sum_{\scriptsize{\begin{array}{l} 
		\sigma_1,\sigma_2,\bar{\sigma}_1\\
		\bar{\sigma}_2,\alpha_1,\alpha_2\\
		\bar{\alpha}_1,\bar{\alpha}_2,m
		\end{array}}}  \!\!
\int d\hat{\vp}'' \, \langle \vp'',\sigma_1\sigma_2,\alpha_1\alpha_2|T|\,
|\vp|\hat{\vz},\bar{\sigma}_1\bar{\sigma}_2,\bar{\alpha}_1\bar{\alpha}_2\rangle Y_\ell^{m}(\hat{\vp}'')^* 
(\sigma_1\sigma_2M|s_1s_2S)\nn \\
&\times & 
(m M \bar{M}|\ell S J)(\bar{\sigma}_1\bar{\sigma}_2\bar{M}| \bar{s}_1\bar{s}_2\bar{S}) (0\bar{M}\bar{M}|\bar{\ell}\bar{S}J)  (\alpha_1\alpha_2t_3|\tau_1\tau_2I)
(\bar{\alpha}_1\bar{\alpha}_2t_3|\bar{\tau}_1\bar{\tau}_2I)~,
\ea
where it should be understood that  
 $M=\sigma_1+ \sigma_2$ and $\bar{M}=\bar{\sigma}_1+\bar{\sigma}_2$.

The unitarity in PWAs can be obtained by first considering the unitarity requirements of the $T$- matrix operator, 
cf. Eqs.~\eqref{181101.5} and \eqref{181101.5b}. 
We also assume that time-reversal invariance is a good symmetry, which implies that the PWAs are symmetric, e.g. 
this is demonstrated in the footnote~9 of \cite{oller.181101.1}, as well as in the chapters 3 and 5 of Ref.~\cite{martin.290916.1}. 
 Taking the matrix element of Eq.~\eqref{181101.5} between states in the $LSJ$ basis one has that 
\be
\label{051016.10}
2 \Im T^{(IJ)}_{\ell S;\bar{\ell}\bar{S}}= \langle J\mu,\ell S,T t_3|T T^\dagger |J\mu,\bar{\ell}\bar{S},T t_3\rangle~.
\ee

From the normalization of a $ LSJ$ state, cf. Eq.~\eqref{061016.3}, one has from Eq.~\eqref{051016.10} the following 
expression for two-body unitarity in  PWAs,
\be
\label{051016.11}
\Im T^{(IJ)}_{\ell S;\bar{\ell}\bar{S}}=\sum_{\ell'',S''} \frac{|\vp''|}{8\pi\sqrt{s}}T^{(IJ)}_{\ell,S;\ell'',S''}{T^{(IJ)}}^*_{\ell'',S'';\bar{\ell}\bar{S}}~.
\ee
On the rhs we have inserted between $T$ and $T^\dagger$  a resolution of the 
identity in terms of the states $|J\mu,\ell S,I t_3\rangle$, which is strictly valid below the threshold of multiparticle production.  
Nonetheless, in some cases the restriction to only two-body intermediate states could be a good phenomenological 
approximation if the multiparticle-state contributions were suppressed for some reason. 
E.g. this is the case for  the $4\pi$ state in meson-meson  scattering when $\sqrt{s}\lesssim 1$~GeV \cite{roy.190114.1,royndu.190114.1,alba.190114.1}. 

The phase-space factor in Eq.~\eqref{051016.11} [which is actually the true phase-space factor divided by 2, which stems from the left-hand side (lhs) of Eq.~\eqref{051016.10}] is included in the diagonal matrix $\rho(s)$. 
Its matrix elements are 
\be
\label{051016.12}
\rho_{ij}(s)=\frac{|\vp|_i}{8\pi\sqrt{s}}\theta(s-s_{{\rm th};i})\delta_{ij}~,
\ee
where $s_{{\rm th};i}$ is the Mandelstam variable $s$ at the threshold of the $i_{\rm th}$ state. 
We typically denote the different partial waves that are coupled by a Latin index that runs from 1 to $n$, being $n$ the 
number of them.  
In this regard, the coupled matrix elements between PWAs are gathered together as the matrix elements 
of the matrix $T^{(IJ)}(s)$. With this matrix notation, Eq.~\eqref{051016.11} then becomes
\be
\label{051016.12a}
\Im T^{(IJ)}
%-{T^\dagger}^{(IJ)}
= T^{(IJ)} \rho {T}^{(IJ)\dagger} ~.
\ee
We have another interesting form for this unitarity relation that results by rewriting the lhs of the previous equation as 
$(T^{(IJ)}-{T^{(IJ)\dagger}})/2i$ and multiplying it to the left by ${T^{(IJ)}}^{-1}$ 
 and to the right by ${T^{\dagger (JI)}}^{-1}$. It results,  
\be
\label{051016.12b}
\Im {T^{(IJ)}}^{-1}(s)=- \rho(s)~,
\ee
for $s$ above the thresholds of the channels involved. 
Along this work, we often denote by channel any of the states interacting in a given process.

In order to introduce the $S$ matrix in partial waves, we take the matrix elements of the $S$-matrix operator  
between $LSJ$ states, cf. Eq.~\eqref{181101.4} for the relation between the $S$- and $T$-matrix operators. 
However, a more standard definition of the $S$-matrix for partial waves, $S^{(IJ)}$, is 
\be
\label{051016.13}
S^{(IJ)}=I+2i\rho^{\frac{1}{2}}T^{(IJ)}\rho^{\frac{1}{2}}~.
\ee
This redefinition amounts to evaluate the $S$ matrix between the re-normalized partial-wave projected states
\be
\label{181117.1}
\sqrt{\frac{|\vp|_i}{4\pi}}|J\mu,\ell S;i\rangle~.
\ee
These states are  normalized to the product of Kronecker deltas with unit coefficient, instead of the
original normalization in Eq.~\eqref{061016.3}.
As a result, the diagonal matrix elements of the identity operator $I$ and $S^{(IJ)}$ are  1 and 
\be
\label{181109.4}
S^{(IJ)}_{ii}=\eta_i e^{2i\delta_i}~,
\ee
respectively, 
where $0\leq \eta_i\leq 1$ is the elasticity parameter for channel $i$  and $\delta_i$ its phase shift.

Another matrix that we introduce for PWAs and that coincides with $S^{(IJ)}(s)$ for the uncoupled case is the matrix 
$\cS^{(IJ)}(s)$, which is defined as
\begin{align}
\label{181125.13}
{\cal S}^{(IJ)}(s)&=I+T^{(IJ)}(s) 2i\rho(s)~.
\end{align}
Note that even though $T^{(IJ)}(s)$ is symmetric this is not the case in general for ${\cS}^{(IJ)}(s)$. 
 It is straightforward to prove  the property that  for $s$ along the RHC,
\begin{align}
\label{181125.14}
\cS^{(IJ)}(s)\cS^{(IJ)}(s)^*&=\cS^{(IJ)}(s)^*\cS^{(IJ)}(s)=I, 
\end{align}
as it follows by taking into account  Eq.~\eqref{181125.13} and the unitarity relation of Eq.~\eqref{051016.12a}.
 To avoid confusion let us remark that the asterisk corresponds to complex conjugation and not to the
Hermitian conjugate of  ${\cS}^{(IJ)}(s)$.
While the $S$ matrix in partial waves is unitary and symmetric, 
neither of these properties applies in general to ${\cS}^{(IJ)}(s)$ for $n>1$.

A slight change in the formalism for projecting into PWAs is needed when the two-body
states are composed of two identical particles, or when the two particles are treated as indistinguishable within the
isospin formalism. 
This has been discussed for fermions and bosons
in Refs.~\cite{lacour.181101.1} and \cite{gulmez.181101.2}, respectively.
These (anti)symmetric states are 
\be
\label{290916.2}
|\vp,\sigma_1\sigma_2, \alpha_1\alpha_2\rangle_S=\frac{1}{\sqrt{2}}\left(
|\vp,\sigma_1\sigma_2,\alpha_1\alpha_2\rangle \pm  |-\vp,\sigma_2\sigma_1,\alpha_2\alpha_1\rangle
\right)~,
\ee
with the subscript $S$ referring the (anti)symmetrized nature of the state under the exchange of the two particles,
the sign $+$ is for bosons and the $-$  for fermions. 
 One can write the (anti)symmetric states in terms of the partial-wave projected 
ones by invoking Eq.~\eqref{290916.3}. Thus,  instead of Eq.~\eqref{290916.3}, we have now
\bea
\label{290916.4}
|\vp,\sigma_1\sigma_2,\alpha_1\alpha_2\rangle_S=\sqrt{4\pi}\!\! 
\sum_{\scriptsize{ 
		\begin{array}{c}
		J,\mu,\ell,m\\S,M,I,t_3
		\end{array}}} \!\! 
%\frac{1\pm (-1)^{\ell+S+I}}{\sqrt{2}}
\chi(\ell S I)
(\sigma_1\sigma_2 M|s_1 s_2 S) (m M \mu|\ell S J) (\alpha_1 \alpha_2 t_3|\tau_1\tau_2 I) Y_\ell^m(\hat{\vp})^* |J\mu, \ell S, I t_3 \rangle~,\nn
\ea
where
\be
\label{190114.1}
\chi(\ell S I)=\frac{1\pm (-1)^{\ell+S+I}}{\sqrt{2}}~.
\ee

For deducing Eq.~\eqref{290916.4} we have used the well-known symmetry properties of the Clebsch-Gordan 
coefficients \cite{rose.021016.2}
\bea
\label{290916.5}
(\sigma_2\sigma_1 M|s_2 s_1 S)&=& (-1)^{S-s_1-s_2}(\sigma_1\sigma_2 M|s_1 s_2 S)~,\nn\\
(\alpha_2\alpha_1 t_3|t_2 t_1 I)&=&(-1)^{I-t_1-t_2}(\alpha_1\alpha_2 t_3|\tau_1\tau_2 I)~,
\ea
and also the parity property of spherical harmonics, 
\begin{align}
\label{190122.1}
Y_\ell^m(-\hat{\vp})=(-1)^\ell Y_\ell^m(\hat{\vp})~.
\end{align} 
Of course, in the present case $s_1=s_2$, $\tau_1=\tau_2$.
It is clear from Eq.~\eqref{190114.1} that only the states with (odd)even $\ell+S+I$ are allowed for (fermions)bosons. 
The  inversion of Eq.~\eqref{290916.4} gives,
\begin{align}
\label{011016.1}
|J\mu,\ell S, I t_3\rangle & = 
\frac{1}{\sqrt{8\pi}}\!\!\!\!
\sum_{\scriptsize{
		\begin{array}{c}
		\sigma_1,\sigma_2\\
		M,m\\
		\alpha_1,\alpha_2 
		\end{array}}  }
\!\!\!\!
\int d\hat{\vp} Y_\ell^m(\hat{\vp}) (\sigma_1\sigma_2 M|s_1s_2S)
(m M \mu|\ell S J)
(\alpha_1 \alpha_2 t_3|\tau_1\tau_2 I) |\vp,\sigma_1\sigma_2,\alpha_1\alpha_2\rangle_S~,
\end{align}
where we have assumed that $\ell+S+I$=even(odd) for bosons(fermions),  so that $\chi(\ell S I)=\sqrt{2}$. 
The comparison of Eqs.~\eqref{011016.1} and \eqref{061016.1} leads us to conclude that the only difference in writing 
the states $|J\mu,\ell S,I t_3\rangle$ involving the (anti)symmetric ones is a factor $1/\sqrt{2}$. 
Thus, we can still use  Eq.~\eqref{051016.6} for determining the PWAs but including an extra factor $1/\sqrt{2}$ for
every state involved that obeys (anti)symmetric properties under the exchange of its two particles.
This is the so-called unitary normalization introduced in Ref.~\cite{oller.181101.2}.

%______ Crossing ____________________

For making the exposition more self contained, let us make a few basic and general remarks on crossing symmetry in 
relativistic scattering, first introduced by Gell-Mann and Goldberger \cite{gell.mann.190503.1}.
These authors noticed that the scattering amplitude $T_{\nu\mu}(q',q)$ for $\gamma N\to \gamma N$,
where $q$ and $q'$ are the four-momenta of the initial and finial photons and
$\mu$, $\nu$ are the space-time components of the corresponding polarization vectors, in the same order,
satisfies the symmetry property
\begin{align}
T_{\nu\mu}(q',q)=T_{\mu\nu}(-q,-q')~.
\end{align}
By exchanging the photons between them from the initial to the final state and vice versa,
one has to change the order of the
discrete labels and of the momenta, multiplying the latter ones by a minus sign.
This is because for each direct Feynman diagram one has the corresponding crossed one, with the exchange of roles of the
two photons. 

Devoted discussions on crossing symmetry can be found e.g. in the books~\cite{martin.290916.1,chew.181102.1}.
In perturbative QFT  a given field $\phi_i(x)$ contains both the annihilation operators of a particle and the 
 creation ones of the antiparticle \cite{weinberg.181101.1}. 
 To get the basic idea involved in crossing let us take the simplest case of a zero-spin field. 
While the annihilation operators are multiplied by the space-time factor $\exp(-ipx)$, 
the creation ones are multiplied by $\exp(ipx)$.   
Therefore, any vertex in a scattering process can be associated with a
particle of four-momentum $p$ or with its antiparticle of four-momentum $-p$.
 As a result, given the scattering
\begin{align}
\label{181102.3}
a_1(p_1)+a_2(p_2)+\ldots \to b_1(p'_1)+b_2(p'_2)+\ldots
\end{align}
the same scattering amplitude gives any other reaction in which one or several particles
are changed from the initial/final state to the final/initial one and, at the same time, the four-momenta of the 
interchanged particles flip sign.  
 In this way, the reaction in Eq.~\eqref{181102.3} is related to many others by crossing, for instance to 
\begin{align}
\label{181102.3b}
a_1(p_1)+a_2(p_2)+\ldots+\bar{b}(-p'_1) \to b_2(p'_2)+\ldots
\end{align}
where the bar indicates the corresponding antiparticle. This is the basic content of crossing.

For the two-body reaction $a+b\to c+d$, we have the following processes related by crossing: 
\begin{align}
\label{181102.4}
a(p_1)+b(p_2)&\to c(p_3)+d(p_4)~,\\
\label{181102.4b}
a(p_1)+\bar{c}(-p_3)&\to \bar{b}(-p_2)+d(p_4)~,\\
\label{181102.4c}
a(p_1)+\bar{d}(-p_4)&\to c(p_3)+\bar{b}(-p_2)~.
\end{align}
From top to bottom, they are referred as $s$-, $t$- and $u$-channel scattering, respectively.
It is also common to denote the $s$-channel as the direct one while the $t$- and $u$-channels are the crossed ones.
Apart from the reactions indicated in Eqs.~\eqref{181102.4}-\eqref{181102.4c},
there are other three in which instead of the exchange $b(p_2)\to\bar{b}(-p_2)$, 
 there is the exchange $a(p_1)\to \bar{a}(-p_1)$ from the initial to the final state. 
These  processes are also related by CPT invariance to the ones shown in Eqs.~\eqref{181102.4}-\eqref{181102.4c}.

Because of the change of sign in the four-momentum of every of the particles involved 
in the crossing process, the $s$, $t$ and $u$ variables for each channel are related.
Let us add a subscript $t$ and $u$ to the Mandelstam variables for the $t$- and $u$-channels, respectively.
As a result,  for the $t$-channel one has that
\begin{align}
\label{181102.5}
s_t&=(p_1-p_3)^2=t~,\\
t_t&=(p_1+p_2)^2=s~,\nn\\
u_t&=(p_1-p_4)^2=u~,
\end{align}
and for the $u$-channel the relations are,
\begin{align}
\label{181102.6}
s_u&=(p_1-p_4)^2=u~,\\
t_u&=(p_1-p_3)^2=t~,\nn\\
u_u&=(p_1+p_2)^2=s~.\nn
\end{align}
The physical regions for these processes do not overlap, cf. Fig.~\ref{fig.190330.1} for an explicit example showing the
different physical regions in the Mandelstam plane. 
 Analyticity assumes that the scattering amplitudes in the three disjoint physical regions for
the $s$-, $t$- and $u$-channels correspond to the same analytical function $A(s,t,u)$ of complex $s$ and $t$ variables 
[$u$ can be thought to be given by the constraint in Eq.~\eqref{181102.2b}].
The physical values of the scattering amplitude for the different channels are boundary values of this analytic function.

Let us take a constant $t$, and focus on the unitarity cut in the $u$-channel
that stems from the associated normal threshold. 
Because of the constraint in Eq.~\eqref{181102.2b}, with $t$ fixed, we can take the scattering amplitude 
as an analytical function in the complex $s$ plane. 
Therefore, the unitarity cut in the $u$ variable gives rise to a crossed cut in the complex $s$ plane, given by 
the set values
\begin{align}
\label{181102.12}
s&=4m^2-t-u\leq -t~,
\end{align}
where the equal mass case has been assumed for simplicity.  
Of course, in addition we also have for $s\geq 4m^2$ the $s$-channel unitary cut.

For particles with spin the analytical continuation of the scattering amplitude in the complex $s$ and $t$
variables is more involved because of the kinematical singularities.
 They arise from the solution of the wave equations for the particles with spins, like the spinors for spin 1/2. 
 For a general account on kinematical singularities we refer to \cite{chew.181103.1,chew.181102.1,lutz.190509.1}.
A possible way to deal with the kinematics singularities is to isolate Lorentz invariant functions out of
the scattering amplitudes \cite{chew.181102.1}.
Other more modern techniques to accomplish this is to employ the spinor helicity formalism
for scattering amplitudes, see e.g. Ref.~\cite{arkani.190509.1,elvang.190509.1} and references therein. 

As an example, for pion-nucleon scattering, $\pi^a(q)N(p,\sigma;\alpha)\to \pi^{a'}(q')N(p',\sigma';\alpha')$,
where $a$ and $a'$ denote the Cartesian coordinates of the pions in the isospin space, we can write the 
scattering amplitude as \cite{martin.181103.1,chew.181103.1}
\begin{align}
\label{181103.4}
T^{\pm}=\bar{u}(p',\sigma')\left[A^\pm(s,t,u)+\frac{1}{2}(q\!\!\! /+q'\!\!\!\! /)B^\pm(s,t,u)\right]u(p,\sigma)~.
\end{align}
with 
\begin{align}
\label{181103.2}
T_{aa'}&=\delta_{a'a} T^+ + \frac{1}{2}[\tau_a,\tau_{a'}]T^-~.
\end{align}
And the functions $A^\pm(s,t,u)$ and $B^\pm(s,t,u)$ are analytical functions in the complex $s$ and $t$ variables, and 
are amenable to the same analytical continuation in its argument as a scattering amplitude in the zero-spin scattering case.
  The other factors in Eq.~\eqref{181103.4} are of course important to  establishing the link between analyticity and experimental results.

The presence of poles in the scattering amplitude for real values of any of the Mandelstam variables correspond to the 
exchange of a bound state in the corresponding direct or crossed channels. 
We denote bound state to any pole in the physical Riemann sheet of the scattering amplitudes. 
We should remark that this notation does not actually imply any preconception about the nature of this exchange particle. 
In this regard, from our present knowledge based on Quantum Chromodynamics (QCD) \cite{qcd.190122.1,qcd.190122.2,gross.190122.1}, 
we would not consider  the proton as a state  predominantly composed by  
a $n$ and a $\pi^+$ in $P$ wave \cite{weise.190122.1,brodsky.190122.1}. 
We refer to Ref.~\cite{oller.190122.1} for a general discussion on the problem of compositeness of a pole in the scattering 
amplitudes in QFT. A point worth stressing is that a pole in the crossed channels, $t$ or $u$, gives also rise to 
 crossed cuts in a PWA as a function of complex $s$. 
For instance, coming  back to $\pi^- n\to \pi^- n$ scattering and its $u$-channel proton pole, we have that 
 in the CM frame the $u$ variable reads
\begin{align}
\label{181103.5}
u&=m_N^2+m_\pi^2-2\omega E-2\vp^2\cos\theta~,
\end{align}
with $m_N$ and $m_\pi$ the nucleon and pion masses (we assume the validity of the isospin symmetry).
We also have denoted by $E$ and $\omega$ the nucleon and pion CM energies, in this order. 
When performing the partial-wave projection one has to integrate over the scattering angle, 
cf. Eq.~\eqref{051016.6}, with $\cos\theta\in[-1,1]$. The proton poles occurs for  
$u=m_N^2$ in Eq.~\eqref{181103.5}. Once the kinetic variables $\omega$, $E$ and $\vp^2$ are expressed in terms of $s$,
 the solutions in $s$ to the equation $u(s,x)=m_N^2$ are ($x=\cos\theta$),
\begin{align}
\label{181103.7}
s_1(x)=\frac{m_N^2 x + m_\pi^2(1+x)-\sqrt{m_N^4+2m_\pi^4(1+x)+2m_N^2m_\pi^2(-1+x+2x^2)}}{1+x}~,\\
s_2(x)=\frac{m_N^2 x + m_\pi^2(1+x)+\sqrt{m_N^4+2m_\pi^4(1+x)+2m_N^2m_\pi^2(-1+x+2x^2)}}{1+x}~.\nn
\end{align}

The first solution $s_1(x)$ implies a cut along the negative real $s$ axis, 
as the  radicand is larger than the square of the term before it.
 This cuts extends from  $-\infty$ ($x=-1$) to zero ($x=1$) and it is a clear example of a left-hand cut (LHC).
The other solution $s_2(x)$ implies a finite cut along the positive real $s$ axis 
from $(m_N^2-m_\pi^2)^2/m_N^2$ up to $m_N^2+2m_\pi^2$. 

The analysis for $\pi\pi$ scattering is simpler.
There are no  bound-state poles due to the strong interactions and the two-pion unitarity cuts 
in the $t$- and $u$-channel imply a LHC in the complex $s$ plane.
In this case, the initial and final energy of the pions in the CM frame is the same
because all of them have the same mass. As a result, $t=(p_1-p_3)^2=-(\vp-\vp')^2=-2\vp^2(1-x)$ and
$u=(p_1-p_4)^2=-(\vp+\vp')^2=-2\vp^2(1+x)$, where $\vp$ and $\vp'$ are the initial and final
CM three-momenta, respectively.
Taking into account the kinematical identity, $s=4(m_\pi^2+\vp^2)$,
we have that  $\vp^2=s/4-m_\pi^2$, which is used to express $t$ and
$u$ as a function of $s$ and $x$. 
The starting of the two-pion unitarity cut in the $t$-channel requires that 
$t=-2(s/4-m_\pi^2)(1-x)\geq 4m_\pi^2$. Similarly, the on-set of the two-pion intermediate
states in the $u$-channel implies that $u=-2(s/4-m_\pi^2)(1+x)\geq 4m_\pi^2$.
Solving for $s$ it follows that
\begin{align}
\label{190812.1}
s\leq -4m_\pi^2\frac{1+x}{1\mp x}~,
\end{align}
where the minus sign applies to the crossed cut due to the $t$-channel unitarity and
the plus sign to the $u$-channel one. 
We have then that for both cases the resulting LHC extends 
for $s\in(-\infty,0]$, as $x$ varies within the interval $[-1,1]$.

Let us now consider two-body scattering with different masses. The simplest case 
is the one involving two different  particles $a$ and $b$, namely, $a+b\to a+b$, of masses $m_a$ and $m_b$,
respectively. 
Needless to say that here as well the $t-$ and $u$-channel unitarity cuts, the former extends for $t>4m_a^2$
and the latter for $u>(m_a+m_b)^2$, imply crossed cuts for the PWAs in the $s$ channel. 
We now have the kinematical relations $t=-2\vp^2(1-x)$
and $u=(w_a-w_b)^2-2\vp^2(1+x)$, with the CM energies of the particles $a$ and $b$
denoted by $w_a$ and $w_b$, in this order.
%Since $w_{a}=(s+m_a^2-m_b^2)/2\sqrt{s}$ and
%$w_b=(s+m_b^2-m_a^2)/2\sqrt{s}$, the term $(w_a-w_b)^2$ is given by $(m_a^2-m_b)^2/s$.
However, due to the more involved relation between three-momentum and the $s$
variable the geometry of the cuts in the complex  $s$ plane is richer and one also has a circular cut. 
A detailed discussion of the kinematics involved can be found in Ref.~\cite{martin.290916.1}.
As it is also pointed out in this book, the cuts are only linear in the $\vp^2$ complex plane.
This is trivially seen for the $t$-channel unitarity cut because the $t$ variable in terms
of $\vp^2$ is given by the same expression as in the equal-mass scattering.
For the $u$-channel case, using that $s=(\sqrt{\vp^2+m_a^2}+\sqrt{\vp^2+m_b^2})^2$, one finds that
$u=m_a^2+m_b^2-2(\sqrt{(m_a^2+\vp^2)(m_b^2+\vp^2)}+\vp^2 x)$. The requirement for the
unitarity $u$-channel cut, $u>(m_a+m_b)^2$, implies a
quadric algebraic equation for $\vp^2$ as a function of $x\in [-1,1]$. Nonetheless, it can be
easily proved that the radicand of the solution is always positive, so that the cut remains linear
in the $\vp^2$ complex plane. 
  
In this manuscript the analyticity of the $S$ matrix is taken as a basic property to be satisfied,
so that the physical transition amplitudes are the boundary values of some analytic function,
as mentioned above in connection with two-body scattering after Eq.~\eqref{181102.6}.
It is usually believed that causality results in analyticity properties of the scattering amplitudes.
However, the exact manner in which this presumably occurs is not clear (Sec.~4.1 of Ref.~\cite{olive.181102.1}).

Historically, the first statement in $S$-matrix theory in connection with causality was performed by
Kronig \cite{kronig.190726.1,cushing.190728.1} in analogy with the scattering of light by atoms.
For the latter case, the Kramers-Kronig dispersion relation for the index of refraction
for the propagation of light in a material resulted.
The basic idea is that the effect cannot precede the cause.
Thus, if the incident wave is zero before $t=0$ around the center of force where the interactions take place,
there cannot be scattered waves for $t<0$. 
The connection of this statement with analyticity can be easily illustrated by
considering the Fourier transform of a function $f(t)$ which is zero for $t<0$.
E.g. this can represent the reaction of a physical system to an incident wave that reaches it at $t=0$,
and because of the causality statement it must vanish before the incident wave has arrived.
Its Fourier transform, denoted by $f(\omega)$, is
\begin{align}
\label{190727.1}
f(\omega)&=\int_{-\infty}^{+\infty} dt e^{i\omega t} f(t)~.  
\end{align}
It is clear from the previous equation hat if $f(t)=0$ for $t<0$ then the Fourier transform $f(\omega)$ can
be extended to the upper complex $\omega$ plane with $\Im \omega>0$, because of the exponential dumping in the
integration for $t\to +\infty$. Nonetheless, we have to take into account that this is a  qualitative
formulation of causality.
As discussed in Refs.~\cite{vanKampen.190727.1,vanKampen.190727.2}  there are no
in-going or out-going wave packets that are exactly zero up a certain time if one allows only positive
energies in the Fourier decomposition of the wave function.
The argument is rather straightforward \cite{vanKampen.190727.2} so that we reproduce its main 
reasoning thread.  
If we write the (spherically symmetrical) reduced wave function as
\begin{align}
\label{190727.2}
r\psi_{{\rm in}}(r,t)&=\int_0^\infty dp A(p)e^{-ipr-iEt}~,
\end{align}
where  $E=p^2/2m$ and $A(p)$ is a square integrable function, it follows that $\psi_{{\rm in}}(r_1,t)$,
 with $r=r_1$, cannot be zero in any open interval of time, e.g. for $t<0$.
This is so because $E\geq 0$ in Eq.~\eqref{190727.1}, so that the rhs is an analytic function of $t$
with $\Im t<0$.
Then, if $\psi_{{\rm in}}(r_1,t)$ were zero for an open domain along the real $t$ axis then one can provide the analytical extrapolation into the upper half plane of the $t$ complex plane by applying
the Schwarz reflection principle, which is introduced in Sec.~\ref{sec.190124.1}.
But since the function vanishes in an open domain it is then zero in its whole analytic domain
in the complex $t$ plane, which is an absurd result. 
Nonetheless, such exactly vanishing wave packets at a time are possible but with the caveat 
 of including negative energies in the linear superposition of the wave function, as earlier employed
by Ref.~\cite{Tiomno.190725.1}.
In this regard,  the latter reference also shows that in non-relativistic potential scattering the 
poles of the $S$ matrix in the $k$ complex plane can only lie on the positive imaginary axis or
in the half-lower plane of $k$. The former poles correspond to bound states and the latter to
resonances or virtual states (for which $k=i \Im k$). If there were a pole with $\Im k\geq 0$, and having also
a non-zero real part of $k$, then the  Hamiltonian would have complex eigenvalues, which is not possible because
it is a self-adjoint operator \cite{Gottfried.190728.1} [section~8.2(c)]. 
However, we do not know of any general proof of an analogous result (which is typically
accepted as a constraint) for relativistic scattering theory.

Finally, let us also mention that in standard QFT one introduces the notion of microcausality which, in
simple terms, is the requirement that two quantized (fermionic) bosonic field operators, $\phi_i(t,\vx)$,
(anticommute) commute at spatial separations,\footnote{For fermionic quantum fields one imposes that they anticommute for 
a spatial space-time separation between them.}
\begin{align}
\label{190727.3}
[\phi_{i'}(t',\vx'),\phi_i(t,\vx)]=0~,
\end{align}
for $(t'-t)^2-c^2(\vx'-\vx)^2<0$. In this way, one can guarantee the commutation at spatial
space-time separation of observables constructed out of the quantum fields $\phi_i(t,\vx)$.
This together with the reduction formalism originally developed by  Goldberger \cite{Goldberger.190727.1}
allows one to express the scattering amplitudes in terms of the commutator of current operators.
The vanishing of these commutators in certain regions of the integration allows to derive the analyticity
properties of the $T$ matrix, similarly as indicated above for $f(\omega)$, cf. Eq.~\eqref{190727.1}.
The reference~\cite{weinberg.181101.1} qualifies as causal the quantum fields that satisfy Eq.~\eqref{190727.3}
because  the  Hamiltonian density ${\cal H}(t,\vx)$,  written in terms of them, commutes with itself at a 
spatial space-time separation. The same Ref.~\cite{weinberg.181101.1} (section 3.5) shows that
this property then guarantees the Lorentz invariance of the $T$ matrix by making use of perturbation theory.

%%%%%%%%%%%%%%%%%%%%%%%%%%%%%%%%%%%%%%%%%%%%%%%%%%%%%%%%%%%%%%%%%%
%\section[General results for two-meson scattering in partial waves]{General results for two-meson scattering in partial
%	waves.  Introduction to DRs, the $N/D$ method and the CDD poles}
\section{Introduction to DRs, the $N/D$ method and the CDD poles}
\label{sec.181104.1}
\setcounter{equation}{0}   

Many approaches discussed in this work are based on the use of unitarity and analyticity. 
These important properties of the scattering amplitudes can be combined to provide integral representations 
of them on the basis of dispersion relations (DRs). 
To begin with, we first introduce some important mathematical results of use when implementing dispersion relations. 
Afterwards, we elaborate on the $N/D$ method and settle it in terms of the appropriate DRs.

%%%%%%%%%%%%%%%%%%%____________________%%%%%%%%%%%%%%%%%%
\subsection{Some mathematical preliminaries}
\label{sec.190124.1}

Let us discuss briefly the Schwarz reflection principle. It states that given a function of a complex 
variable $z$,  $f(z)$, which is real along some interval of the real $z$ axis contained in its domain of  analyticity, then 
\begin{align}
f(z) &= f(z^*)^*~,
\label{180929.1}
\end{align}
in this domain. The Schwarz reflection principle is often used when writing down DRs. 
This is based on the fact that for a given PWA $T(s)$ there is typically a separation between the RHC and LHC, 
so that $T(s)$ is real in the interval between the cuts and the Schwarz reflection principle is applicable. 
 It follows then that the discontinuity of $T(s)$ along any of these cuts is 
\begin{align}
\label{181019.8}
f(x+i\ve)-f(x-i\ve)=f(x+i\ve)-f(x+i\ve)^*=2i \Im f(x+i\ve)~.
\end{align}
A region between cuts along the real $s$ axis is also expected to be found where an invariant part
 of a scattering amplitude, $A(s,t,u)$, is real for fixed and real $t$ (such that this value of $t$ is not on top of intermediate 
states in the $t$ channel). Therefore, $A(s,t,u)$ would satisfy the Schwarz 
reflection principle as function of $s$ in the complex $s$ plane.

\begin{figure}
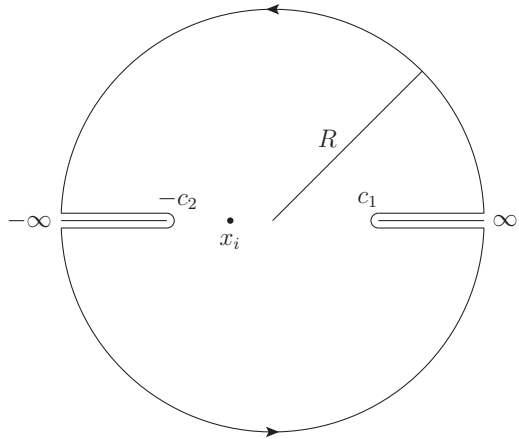

	\begin{center}
\scalebox{0.8}{		\begin{tabular}{ll}
			\begin{axopicture}(200,200)
				\Line(0,100)(50,100) %LHC
				\Line(150,100)(200,100) %RHC
				\Arc[arrow,arrowpos=0.5](100,100)(100,2,178) %upper arc at infinity
				\Arc[arrow,arrowpos=0.5](100,100)(100,182,358) %lower arc at infinity
				\Text(55,110){$-c_2$} %-c_2
				\Text(145,110){$c_1$} %c_1
				\Vertex(80,100){1.5} %x_i
				\Text(80,90){$x_i$}  %x_i
				\Text(210,100){$\infty$} 
				\Text(-15,100){$-\infty$}
	\Line(100,100)(170.72,170.72)
	\Text(126,139){$R$} %Indicating the radius $R$
				\Line(0,103.5)(50,103.5) %up encircling LHC
				\Line(0,96.5)(50,96.5)   %down encircling LHC
				\Line(150,103.5)(200,103.5) %up encircling RHC
				\Line(150,96.5)(200,96.5)   %down encircling RHC
				\Arc(50,100)(3.5,270,90) %arc surrounding LHC branch-point 
				\Arc(150,100)(3.5,90,270) %arc surrounding RHC branch-point
			\end{axopicture}
		\end{tabular}
	}
		\caption{{\small The integration contour ${\cal C}$ used to formulate  the DR of Eq.~\eqref{181011.2}.}
			\label{fig.181202.1}}
	\end{center}
\end{figure}

Another important mathematical result for DRs is the Sugawara-Kanazawa theorem \cite{sugawara.180929.1}, which is 
indicated for determining the number of subtractions that are needed in a DR. It can be stated as follows:
 Let  $f(z)$ be a function analytic in the complex $z$ plane except for two non-overlapping cuts along the real axis and
poles between them, as represented schematically in Fig.~\ref{fig.181202.1}. It is also assumed that 
\begin{itemize}
\item[i)] The limits $f(x\pm i\ve)$ with  $x\to \infty$  and $\ve\to 0^+$ along the $c_1$ cut or RHC are finite. These limits are 
denoted by $f(\infty\pm i\ve)$.
	
\item[ii)] The function $f(z)$ for $z\to \infty$ diverges less strongly than $z^N$ with $N\geq 1$.
	
\item[iii)] The limits $f(x\pm i \ve)$ with $x\to -\infty$  
 along the $c_2$ cut or LHC are definite (not necessarily finite).
\end{itemize}

In such a circumstances,  $f(z)$ has  the limits
\begin{align}
\label{181011.1}
\lim_{z\to \infty} f(z)&= f(\infty+i\ve)~,~{\Im z>0}~, \\
\lim_{z\to \infty} f(z)&= f(\infty-i\ve)~,~{\Im z<0}~, \nn
\end{align}
and it can be represented by the DR
\begin{align}
\label{181011.2}
f(z)=\sum_i\frac{R_i}{z-x_i}+\frac{1}{\pi}\left(\int_{c_1}^\infty+\int_{-\infty}^{-c_2}\right)
\frac{\Delta f(x)}{x-z}dx+\bar f(\infty)~,
\end{align}
with
\begin{align}
\label{181011.3}
\Delta f(x)&=\frac{1}{2i}\left[ f(x+i\ve) - f(x-i\ve) \right]~,\\
\bar{f}(x)&=\frac{1}{2}\left[ f(x+i\ve) + f(x-i\ve) \right] ~.\nn
\end{align}
This is the end of the statement of the theorem. 

Let us remark that only one subtraction constant, namely $\bar f(\infty)$, is required in Eq.~\eqref{181011.2}, despite the fact 
that the limit of $f(z)$ for $z\to\infty$ could be in principle divergent as any finite power of $z$. 
The number of subtraction constants, as required by the Sugawara-Kanazawa theorem, is then determined by the finiteness 
of the limits $f(\infty\pm i\ve)$.  
This theorem reflects a typical situation found in many applications, as already remarked above in connection with the
Schwarz reflection principle, together with the possible presence of poles in the first Riemann sheet, which would correspond
 to bound states.

 One step in the demonstration of this theorem, for which we refer to Refs.~\cite{sugawara.180929.1,oller.190503.1},
is of particular interest since it shows how one can determine the 
asymptotic behavior (for infinite argument) of some typical integrals appearing in DRs.
In this way, let us consider the integral
\begin{align}
\label{181012.5}
\frac{z}{\pi}\int_{c_1}^\infty\frac{\Delta f(x)}{x(x-z)}dx
&=\Delta f(\infty)\frac{z}{\pi}\int_{c_1}^\infty\frac{1}{x(x-z)}dx
+\frac{z}{\pi}\int_{c_1}^\infty\frac{\Delta f(x)-\Delta f(\infty)}{x(x-z)}dx~,
\end{align}
where $\De f(\pm \infty)$ is the limit
\begin{align}
\label{181012.6}
\De f(\pm \infty)&=\lim_{x\to \pm \infty} \De f(x)~.
\end{align}
The first integral on the rhs of Eq.~\eqref{181012.5} diverges logarithmically for $z\to \infty$ unless $\Delta f(\infty)=0$,
while the last integral has the limit
\begin{align}
\label{181012.7}
\lim_{z\to \infty} \frac{z}{\pi}\int_{c_1}^\infty\frac{\Delta f(x)-\Delta f(\infty)}{x(x-z)}dx
&=-\frac{1}{\pi}\int_{c_1}^\infty\frac{\Delta f(x)-\Delta f(\infty)}{x}dx~.
\end{align}
For more mathematical rigor in the discussion the interested reader is referred to the original Ref.~\cite{sugawara.180929.1}.\footnote{In the manipulations for the demonstration of this theorem we take that the 
	vanishing limits for $x\to\infty$ tend to zero at least as $x^{-\gamma}$ with $\gamma>0$.} 
From Eq.~\eqref{181012.5} one cannot really conclude that $\Delta f(\infty)$ is zero because there are other integral
contributions in the DR of $f(z)$ along the LHC that could cancel such logarithmic divergent, cf. Eq.~\eqref{181011.2}.
% This way of determining the asymptotic behavior of a dispersive integral is of interest in actual applications of DRs.

We can also deduce the following corollaries from the  Sugawara-Kanazawa theorem:

\begin{enumerate}
\item If the RHC and LHC are of finite extent, the results of the theorem are then trivial.
	\item If $f(z)$ has only one finite cut then it is also the case that $\De f(\infty)=0$, 
because the function is continuous at the opposite end of the real axis (since there is not cut there), and then
	$\bar{f}(\infty)=f(\infty\pm i\ve)$.	
\item If the Schwarz theorem holds, cf. Eq.~\eqref{180929.1}, then $\Delta f(x)=\Im f(x)$ and $\bar{f}=\Re f(\infty)$.
	\item In the point i) of the formulation of the theorem we have assumed that the two limits  
$f(\infty\pm i\ve)$ are finite. Let us first notice that this is necessarily the case if the Schwarz reflection principle is 
fulfilled and one of them is known to be finite.
	Even in the case in which only one of these limits were finite, let us say $f(\infty+i\infty)$, and the other 
 infinite, the first line of Eq.~\eqref{181011.1} would still be valid. 

\item If $f(z)$ tends to zero when $z\to \infty\pm i\ve$, it follows from the theorem that $f(z)$ approaches zero in any
	other direction and the unsubtracted dispersion relation of Eq.~\eqref{181011.2} with $\bar{f}(\infty)=0$ is valid.
	\item If $f(z)$  diverges in either one or both of the limits $\infty\pm i\ve$, we can not apply directly the theorem 
	because i) does not hold. However, we can then consider an intermediate function $F(z)$,  
which diverges at least as strongly as $f(z)$ in the limit $z\to\infty$, 
and then apply the results of the theorem for the ratio  $f(z)/F(z)$, as long as 
this new function fulfills the points i)-iii).  
Since  $F(z)$ is supposed to be known, the residues and discontinuity of $f(z)/F(z)$ in terms of those of $f(z)$ can be calculated.
	For this case,  $f(z)$ diverges at infinity as $F(z)$ times constants, which are the limits of $f(z)/F(z)$ when $z\to \infty\pm i\ve$.
\end{enumerate}

%%%%%%%_____End Kanazawa_________________________

 The Herglotz theorem \cite{moments.181017.1} states that given an analytic function 
 $g(\xi)$ of the complex variable $\xi$  and such that $\Im g(\xi)\geq 0$ for  with $|\xi|<1$,  
 the function $g(\xi)$ fulfills the integral representation 
\begin{align}
\label{181019.1}
g(\xi)&=i\int_0^{2\pi} \frac{e^{i\theta}+\xi}{e^{i\theta}-\xi} d\beta(\theta)+c~,
\end{align}
where $c$ is a real constant and $\beta(\theta)$ an increasing bounded function.

The change of variables 
\begin{align}
\label{181019.2}
\xi&=\frac{z-i}{z+i}~,~z=i\frac{1+\xi}{1-\xi}~,
\end{align}
maps the region $|\xi|<1$ with the upper-half plane $\Im z>0$. 
In turn, the transformation
\begin{align}
\label{181019.3}
x&=i\frac{1+e^{i\theta}}{1-e^{i\theta}}=-\cot \frac{\theta}{2}~,
\end{align}
maps the circle $|\xi|=1$ with the real $z$ axis. 
In terms of the new complex variable $z$ we can rewrite the integral in Eq.~\eqref{181019.1} as \cite{moments.181017.1}
\begin{align}
\label{181017.1}
f(z)&=c+A z+\int_{-\infty}^{\infty}\frac{1+x z}{x-z}d\alpha(x)~,~\Im z>0~.
\end{align}
Here $c$ is the same constant as in Eq.~\eqref{181019.1}, $\alpha(x)=\beta(\theta(x))$
(and then it is an increasing bounded function in $x$) 
and $A$ is another real constant, such that \cite{moments.181017.1} 
\begin{align}
\label{181019.4}
\lim_{z\to\infty}\frac{f(z)}{z}=A~,~\Im z>0~.
\end{align}

Let us notice that Eq.~\eqref{181017.1} also implies the analytical representation
of $f(z)$ in the lower half plane, $\Im z<0$, because it is obvious that it satisfies the Schwarz reflection principle,
as  $c$, $A$ and $\alpha(x)$ are real. Thus, $f(z^*)=f(z)^*$.
The integral representation in Eq.~\eqref{181017.1} can also be written as a twice-subtracted DR in the form 
\begin{align}
\label{181019.6}
f(z)=f(z_0)+f'(z_0)(z-z_0)+\frac{(z-z_0)^2}{\pi} \int_{-\infty}^\infty
\frac{\sigma(x)}{(x-z_0)^2(x-z)}dx~,
%\frac{\pi (1+t^2)d\alpha(t)}{(t-z_0)^2(t-z)}~.
\end{align}
where $f'(z)=df(z)/dz$ and 
\begin{align}
\label{181019.7}
\sigma(x)  &=\Im f(x+ i\ve)=  \pi (1+x^2)\frac{d\alpha(x)}{dx}\geq 0~.
\end{align}
 The Herglotz theorem is applied in Ref.~\cite{weinberg.181208.1} to provide the DR satisfied by the 
eigenvalues of the kernel of the LS equation.

%%%_ End Herglotz

%%%%%%%%%_________________________________________%%%%%%%%%%%%%%%%
%\subsection{The $N/D$ method I. Structure of a PWA without explicit LHC}
\subsection{The $N/D$ method. Structure of a PWA without explicit LHC}
\label{sec.190203.1}

The Ref.~\cite{oller.181105.1} studies the $S$- and $P$-wave 
two-body scattering between the lightest pseudoscalars ($\pi$, $K$ and $\eta$),  
as well as the related spectroscopy. This study is based on the application of the $N/D$ method in coupled channels, 
which is then matched with an underlying chiral effective field theory description of the  two-meson scattering. 
As an interesting result of this study one has the derivation of the general structure of a PWA when the LHC 
is not explicitly realized. This could be an appropriate approximation for many reactions of interest.

 For equal-mass scattering $a+a\rightarrow a+a$, with $m_i=m_a$, there is only a 
LHC for $s<s_{\rm Left}$ due to the crossed-channel cuts. However, this is not the only case and 
e.g. for the scattering process $a+b\rightarrow a+b$, with $m_1=m_3=m_a$ and $m_2=m_4=m_b$, apart from a LHC  
there is also a circular cut in the complex $s$ plane for $|s|=m_2^2-m_1^2$ \cite{martin.290916.1}, 
where we have taken for definiteness that $m_2>m_1$. 
We typically refer in the following, for brevity, to the LHC as if it comprises all the crossed cuts. 
 Nonetheless, it is worth keeping in mind that if we have worked in the  complex $p^2$ plane 
 all the cuts would be linear and only a LHC would be then present [let us recall the discussion in the long paragraph
after Eq.~\eqref{190812.1}]. 
One could apply in such a situation an analogous analysis to the one developed here in the complex $s$ plane.

Following Ref.~\cite{oller.181105.1} we denote by $T_L(s)$ a two-meson PWA with angular 
momentum $L$.\footnote{If the mesons had non-zero spins we could proceed similarly as develop here by applying first Eq.~\eqref{051016.6} to calculate the given set of PWAs under interest.}
We first discuss the uncoupled case and then we generalize the formalism for coupled-channel scattering.
The imaginary part of $T^{-1}_L(s)$ along the  RHC  is given by unitarity, 
cf. Eq.~\eqref{051016.12b}, as
\begin{align}
\label{181105.2a}
%\Im T^{-1}_L(s)&=-\frac{\varkappa(s)}{8\pi}~,~s\geq s_{\rm th}~.
\Im T^{-1}_L(s)&=-\frac{p}{8\pi\sqrt{s}}~,~s\geq s_{\rm th}~.
\end{align}
In this equation,  $s_{\rm th}=(m_1+m_2)^2$, the CM three-momentum of the two-meson system is denoted by $p$,  
\begin{align}
\label{181105.2}
%\varkappa(s)&=\sqrt{\frac{(s-(m_1+m_2)^2)(s-(m_1-m_2)^2)}{4s^2}}
%\equiv \sqrt{\frac{\lambda(s,m_1^2,m_2^2)}{4 s^2}}
p=|\vp|&=\frac{\sqrt{(s-(m_1+m_2)^2)(s-(m_1-m_2)^2)}}{2 \sqrt{s}}
\equiv\frac{\lambda^{1/2}(s,m_1^2,m_2^2)}{2 \sqrt{s}}
\end{align}
and  $\lambda(s,m_1^2,m_2^2)$  is the K\"allen triangle
function.\footnote{The context makes clear when $p$ is a four-momentum or the modulus of $\vp$.} 
The discontinuity of $T_L(s)$ along the LHC, $s<s_L$, reads
\begin{align}
\label{lhc}
T_L(s+i\epsilon)-T_L(s-i\epsilon)=2 i \Im T_L(s)~.
\end{align}

We now apply the $N/D$  method \cite{Chew1} in order to calculate a $T_L(s)$ which fulfills 
Eqs. (\ref{181105.2a}) and (\ref{lhc}). In this method the PWA  $T_L(s)$  is expressed as the quotient of two functions,
\begin{align}
\label{n/d}
T_L(s)=\frac{N_L(s)}{D_L(s)}~,
\end{align}
where the numerator and denominator functions, $N_L(s)$ and $D_L(s)$, in this order, only have LHC  and RHC, respectively.
 
To enforce the right behavior of a PWA near threshold, which vanishes like $p^{2L}$,  
we divide $T_L(s)$ by $p^{2L}$. The new function that results is denoted by $T'_L(s)$, 
\begin{align}
\label{T'}
T'_L(s)=\frac{T_L(s)}{p^{2L}}~.
\end{align} 
We apply the $N/D $ method to this function and then we write 
\begin{align}
\label{n/d'}
T'_L(s)=\frac{N'_L(s)}{{D}'_L(s)}~.
\end{align}
It follows from the Eqs. (\ref{181105.2a}), (\ref{lhc}) and (\ref{T'}),  that the discontinuities of ${N}'_L(s)$ and 
${D}'_L(s)$ along the LHC and RHC, respectively, are
\begin{align}
\label{eqs1} 
{\Im  D}'_L&=\Im T'^{-1}_L \; {N}'_L=-\rho(s) {N}'_L p^{2L}~,  &s>s_{\rm th} \\
{\Im D}'_L&=0~,   &s<s_{\rm th}\nn\\
\label{eqs2}
{\Im  N}'_L&=\Im T'_L \; {D}'_L=\frac{\Im T_L}{p^{2L}}D'_L~,  &s<s_{\rm Left}  \\
{\Im N}'_L&=0~.  &s>s_{\rm Left}\nn
\end{align}

Next, we divide ${N}'_L$ and ${D}'_L$ by the polynomial %aqui made out of
whose roots are the possible poles of $N'_L(s)$.
Notice that this operation does not change the ratio of these functions, which is ${T}'_L$, 
nor Eqs.~(\ref{eqs1}) and (\ref{eqs2}). 
It follows then that we can take in the subsequent that  ${N}'_L(s)$ is free of poles. 
Thus, the poles of a PWA would correspond to the zeros of ${D}'_L(s)$.

We can write DRs for  ${D}'_L(s)$ and ${N}'_L(s)$, taking into account their discontinuities given by 
 Eqs.~(\ref{eqs1}) and (\ref{eqs2}), as

\begin{align}
\label{d'}
{D}'_L(s)=-\frac{(s-s_0)^n}{\pi}\int^\infty_{s_{\rm th}} ds' 
\frac{p(s')^{2L} \rho(s') {N}'_L(s')}{(s'-s)(s'-s_0)^n}+\sum_{m=0}^{n-1}
\overline a _m s^m~.
\end{align}
Here $n$ is the number of subtractions needed to guarantee that  
\begin{align}
\label{n}
\displaystyle \lim_{s \to \infty} \frac{{N}'_L(s)}{s^{n-L}}= 0~.
\end{align}
To settle this equation notice that  from Eq.~(\ref{181105.2}) it results that $\displaystyle{\lim_{s \to \infty} \frac{p^{2L} \rho(s)}{s^L}=\frac{1}{4^{L+2} \pi}~.}$

Consistently with  Eq.~(\ref{n}), we write for $N'_L(s)$ the DR                                      
\begin{align}
\label{n2}
{N}'_L(s)=\frac{(s-s_0)^{n-L}}{\pi}\int_{-\infty}^{s_{\rm Left}} ds' 
\frac{{\Im  T}_L(s') {D}'_L(s')}{p(s')^{2L} (s'-s_0)^{n-L} (s'-s)}+
\sum_{m=0}^{n-L-1} \overline a '_m s^m~.
\end{align}                         

The Eqs.~(\ref{d'}) and (\ref{n2}) constitute a system of coupled linear integral equations (IEs) for the functions 
$N'_L(s)$ and $D'_L(s)$, which input  is ${\Im  T}_L(s)$ along the LHC. 
 However, there could be other solutions apart from Eqs.~(\ref{d'}) and (\ref{n2}) 
 because of the possible zeros of ${T}_L$ which do not arise from the solution of these equations. 
 Therefore, these zeros have to be accounted for explicitly and we include them as poles in the function  
${D}'_L$ (typically known as  Castillejo-Dalitz-Dyson (CDD) poles after Ref.~\cite{Castillejo}). 
Let us notice that in a zero the function $T_L(s)^{-1}$ does not exist and Eq.~\eqref{181105.2a} is not defined at that point. 
Following Ref.~\cite{Castillejo}, we introduce the auxiliary function $\lambda(s)$ such that 
\begin{align}
\label{dl1}
{\Im D}'_L(s)=\frac{d \lambda (s)}{ds}~,
\end{align} 
and rewrite Eq.~(\ref{eqs1}) as, 
\begin{eqnarray}
\label{dl2}
{\displaystyle \frac{d\lambda}{ds}}&=-\rho(s) p^{2L}{N}'_L~, & s>s_{\rm th} \\
{\displaystyle \frac{d\lambda}{ds}}&=0~.~~~~~~~~~~~~~~ & s<s_{\rm th} \nn
\end{eqnarray}

Denoting by  $s_i$ the zeros of $T'_L(s)$ along the real axis,  ${T}'_L(s_i)=0$, 
we can solve Eq.~\eqref{dl2} for $\lambda(s)$ with the result 
\begin{align}
\label{l2}
\lambda(s)=-\int_{s_{\rm th}}^s p(s')^{2L} \rho(s') {N}'_L(s') ds'+
\sum_i \lambda(s_i)\theta(s-s_i)~,
\end{align}
where the $\lambda(s_i)$ are unknown, because the inverse of $T'_L(s'_i)$ is not defined at this point, and 
$\theta(s)$ is the usual Heaviside or step function. Therefore, it follows from Eqs.~(\ref{dl1}) and (\ref{l2}) that
\begin{align}
\label{tower1}
{D}'_L(s) &= \sum_{m=0}^{n-1} \overline a_m s^m + 
\frac{(s-s_0)^n}{\pi}\int_{s_{\rm th}}^\infty \frac{
	{\Im  D}'_L(s') ds'}{(s'-s)(s'-s_0)^n} \\ 
& =\sum_{m=0}^{n-1} \overline a_m
s^m -\frac{(s-s_0)^n}{\pi}\int_{s_{\rm th}}^\infty \frac{p(s')^{2L} \rho(s') 
	{N}'_L(s')}{(s'-s)(s'-s_0)^n}ds' +
\frac{(s-s_0)^n}{\pi} \int_{s_{\rm th}}^\infty \frac{\sum_i \lambda(s_i) 
	\delta(s'-s_i)}{(s'-s)(s'-s_0)^n}ds'\nonumber\\
&=\sum_{m=0}^{n-1} \overline a_m s^m
-\frac{(s-s_0)^n}{\pi}\int_{s_{\rm th}}^\infty \frac{p(s')^{2L} \rho(s') 
	{N}'_L(s')}{(s'-s)(s'-s_0)^n}ds'+\sum_i \frac{\lambda(s_i)}{\pi(s_i-s) }\frac{(s-s_0)^n}{(s_i-s_0)^n}~.\nonumber
\end{align}

This equation also results  from Eq.~(\ref{eqs1}) and the  
 the Cauchy theorem for complex integration by taking into account the CDD 
poles of ${D}'_L$ (zeros of ${T}'_L$) that could appear inside and along 
the integration contour, which consists of a circle at infinity that engulfs the RHC. 
As a result one can also account for higher-order zeroes and that some of the $s_i$ could indeed be complex 
which, because of the Schwarz reflection theorem, would be accompanied by another zero at $s_i^*$. 
 However, Ref.~\cite{oller.181105.1} shows that for meson-meson scattering with  
 $L\leq 1$, after matching with the low-energy chiral effective field theory employed, the only 
 CDD poles that appear are along the real axis and correspond to simple zeroes. 
% By taking ${T}'_L$ one avoids having to consider a pole of $D_L$ at threshold of order $L$  in the DR of Eq.~(\ref{tower1}).

It is convenient to rewrite the last term in Eq.~(\ref{tower1}) as
\begin{align}
\label{cano}
\frac{(s-s_0)^n}{s-s_i}&=\sum_{i=0}^{n-1}(s-s_0)^{n-1-i}(s_i-s_0)^i+\frac{(s_i-s_0)^n}{s-s_i}~.
\end{align}
The contributions $\displaystyle{\sum_{i=0}^{n-1}(s-s_0)^{n-1-i}(s_i-s_0)^i}$ can be reabsorbed in 
 $\displaystyle{\sum_{m=0}^{n-1} \overline a _m s^m}$.  Thus, Eq.~\eqref{tower1} can be simplified as  
\begin{align}
\label{d'2}
{D}'_L(s)=-\frac{(s-s_0)^n}{\pi}\int_{s_{\rm th}}^\infty 
\frac{p(s')^{2L} \rho(s') {N}'_L(s')}{(s'-s)(s'-s_0)^n}ds' + 
\sum_{m=0}^{n-1} \widetilde{a}_m s^m + 
\sum_i \frac{\widetilde{\gamma}_i}{s-s_i}~,
\end{align} 
where $\widetilde{a}_m$,  $\widetilde{\gamma}_i$ and $s_i$  are arbitrary parameters. 
However, for any complex  $s_i$ there is another $s_j$ such that $s_j$=$s_i^*$ and $\widetilde{\gamma}_j=
\widetilde{\gamma}_i^*$, as indicated above. 
The last sum in  Eq.~(\ref{d'2}) comprises the CDD poles \cite{Castillejo}. 
The Eqs.~(\ref{d'2}) and (\ref{n2}) correspond to a way of presenting the IEs for ${D}'_L$ and ${N}'_L$ 
which result from the $N/D$ method. 

Next, we proceed under the approximation of neglecting the LHC,  ${\Im  T}_L(s)=0$ in Eq.~(\ref{n2}), which then becomes
\begin{align}
\label{naprox}
{N}'_L(s)=\sum_{m=0}^{n-L-1} \widetilde a '_m s^m={\mathcal{C}} \prod_{j=1}^{n-L-1}(s-s_j)~,
\end{align} 
and $N'_L(s)$ is just a polynomial. The constant ${\cal C}$ must be real so that the Schwarz reflection theorem holds. 
By dividing simultaneously ${N}'_L$ and ${D}'_L$ by $ N'_L(s)$ itself the PWA $T'_L(s)$ remains invariant and we 
can make that $N'_L(s)\to 1$, generating all the zeroes contained originally in $N'_L(s)$ as CDD poles in $D'_L(s)$. 
In this way, we can write 
\begin{eqnarray}
\label{fin/d}
{T}'_L(s)&=&\frac{1}{{D}'_L(s)}~,\\
{N}'_L(s)&=&1~,\nn\\
{D}'_L(s)&=&-\frac{(s-s_0)^{L+1}}{\pi}\int_{s_{\rm th}}^\infty  
\frac{p(s')^{2L} \rho (s')}{(s'-s)(s'-s_0)^{L+1}}ds'+\sum_{m=0}^L a_m s^m+
\sum_i^{M_L} \frac{R_i}{s-s_i}~.\nn
\end{eqnarray}
The number of real free parameters in the previous equation  is $L+1+2M_L$.  
The Ref.~\cite{Chew2} links the presence of CDD poles with 
 the appearance of ``elementary" particles, that is, particles which do not stem from a given 
 `potential' or the exchange forces between the scattering states. 
Notice that given a ${D}'_L(s)$ we could adjust the position and residue of a CDD pole so that 
the real part of $D'_L$ vanishes  at the desired position. Typically, this situation corresponds to the presence 
of a bound state or a nearby (narrow) resonance.  
This is why the parameters of a CDD pole are usually related  with the coupling constant and mass 
of a pole in the $S$ matrix. 
In other instances, the need to include a CDD pole might be motivated by 
the presence of zeroes in the PWA as required by the underlying theory, e.g. because of the Adler zeros \cite{adler.181115.1} for the S-wave meson-meson interaction in QCD \cite{oller.181101.2,oller.181105.1}.  
The location of the zero is the same as that of the CDD pole, while the derivative of the PWA at the zero corresponds to 
the inverse of the residue of the CDD pole, $\widetilde{\gamma}_i$. 
The other $L+1$ parameters emerge  for having enforced the behavior of a PWA near to threshold, so that it vanishes as $p^{2L}$. 

Let us remark then that Eq.~(\ref{fin/d}) is the structure of an uncoupled PWA  when the explicit appearance LHC 
is neglected. This could be an interesting approximation if the LHC is far away and/or it is weak for some reason. 
 The free parameters in Eq.~\eqref{fin/d} could be fitted to experiment or  calculated from the basic underlying theory, 
 e.g. by reproducing lattice QCD (LQCD) results \cite{guo.190126.1,guo.190126.2,guo.190126.3,albaladejo.190126.1}.
The Ref.~\cite{oller.181105.1}  deals with strong interactions, which basic dynamics should stem from QCD.  
Of course, Eq.~(\ref{fin/d}) could also be applied to other interactions, as the Electroweak Symmetry Breaking 
Sector  \cite{oller.181105.2}, which also shares the needed symmetries \cite{ESBS} and the basic analytical properties 
employed in the derivation of Eq.~(\ref{fin/d}), cf. Sec.~\ref{sec.181111.1}.

%%%%___
Let us now proceed to the generalization of Eq.~\eqref{fin/d} to coupled channels employing a  matrix notation.
From the beginning the unphysical cuts are neglected (as in the previous equation). 
Therefore, ${T}_L(s)_{ij}$ is proportional to $p_i^L p_j^L$, which makes that this PWA has, apart of the RHC, 
another cut for odd $L$ between $s_{{\rm th}; i}$ and $s_{{\rm th}; j}$ because of the square roots in $p_i$ and $p_j$.
We can avoid this cut by employing the generalization to coupled channels of Eq.~\eqref{T'}, so that the  matrix ${T}'_L$ 
is defined as 
\begin{align}
\label{A.1}
{T}'_L(s)=p^{-L} {T}_L(s) p^{-L}~.
\end{align}
Here $p$ is a diagonal matrix which elements are $p_{ij}=p_i \delta_{ij}$, with 
$p_i$ the modulus of the CM three-momentum for the channel $i$,
$\displaystyle{p_i=\frac{\lambda^{1/2}(s,m_{1i}^2,m_{2i}^2)}{2\sqrt{s}}}$; 
$m_{1i}$ and $m_{2i}$ are the masses of the two mesons involved.\footnote{If needed, the 
formalism can be generalized for different $L$s corresponding to the initial and final states.}

The unitarity relation along the RHC reads 
\begin{align}
\label{A.2}
{\Im T}'^{ -1}_L(s)=-p^L \rho(s) p^L=-\rho(s) p^{2L}~,
\end{align} 
where the diagonal matrix $\rho(s)$ is already defined in Eq.~\eqref{051016.12}. 

 Now, we express ${T}'_L$ as 
\begin{align}
\label{A.4}
{T}'_L={D}'^{ -1}_L {N}'_L~,
\end{align}
with  ${N}'_L$ and ${D}'_L$ two matrices having only LHC and RHC, respectively. 
This is the straightforward generalization to coupled channel of Eq.~\eqref{n/d}
by  making use of the coupled-channel version of the $N/D$ method in matrix notation, 
as introduced originally in Ref.~\cite{Bjorken}. 
By multiplying simultaneously $N'_L$ and $D'_L$ by an adequate common matrix, we can always 
take ${N}'_L=I$.
In such a case all the zeroes of the determinant of ${T}'_L$ correspond to CDD poles in the
determinant of $D'_L(s)$. This is the generalization to the coupled-channel case of the
CDD poles introduced above for the uncoupled PWAs.
As a result,  we can write under the present circumstances that 
\begin{eqnarray}
\label{A.8}
{T}'_L&=&\widetilde{{D}}'^{-1}_L \\
\widetilde{{N}}'_L&=&1~, \nonumber\\
\widetilde{{D}}_L'&=&-\frac{(s-s_0)^{L+1}}{\pi}\int_0^\infty ds'
\frac{\rho(s')p^{2L}(s')}{(s'-s)(s'-s_0)^{L+1}}+{R}(s)~,\nonumber
\end{eqnarray}
where ${R}(s)$ is a matrix of rational functions which poles give rise to the CDD poles
in $D'_L$.

%%%%%%%%%%%%%%%%%%%%%%%%%%%%%%%%%%%%%%%%%%%%%%%%%%%%%%%%%%%%%%%%
%%%%%%%%%%%%%%%%%%%%%%%%%%%%%%%%%%%%%%%%%%%%%%%%%%%%%%%%%%%%%%%%%
%\subsection[Another DR formula for the PWAs. Reaching the unphysical Riemann sheets]{Another DR representation for  PWAs. Reaching the unphysical Riemann sheets}
\subsection{Another DR representation for PWAs.}% Reaching the unphysical Riemann sheets}
\label{sec.181109.1}

An integral representation for a PWA can be obtained by performing a DR of its inverse, which allows one 
to isolate explicitly the RHC by using  Eq.~\eqref{051016.12b}. Then we can write for $T_L(s)^{-1}$ the following expression
\begin{align}
\label{181109.6}
T_L(s)^{-1}&={\cal N}_L(s)^{-1}+a(s_0)-\frac{s-s_0}{\pi}\int_{s_{\rm th}}\frac{\rho(s')ds'}{(s'-s_0)(s'-s)}~,\\
\label{181109.6b}
T_L(s)&=\left[\cN_L(s)^{-1}+g(s)\right]^{-1}~,
\end{align}
where a subtraction at $s_0$ has been taken because the phase-space function $\rho(s)$ tends to constant as $s\to \infty$. 
By construction  the matrix ${\cal N}_L(s)$ has only crossed-channel cuts (although its inverse could generate  CDD poles).
In the limit situation in which the LHC is neglected,  $\cN_L(s)$ and the function $N_L(s)$ in the $N/D$ method can be matched. 
We also introduce the diagonal matrix $g(s)$, defined as the sum of the dispersive integral in Eq.~\eqref{181109.6} 
plus the  subtraction constant $a(s_0)$  (which is necessary to ensure that this sum is independent of the subtraction point $s_0$). 
Explicitly, the diagonal matrices elements of $g(s)$ are
\begin{align}
\label{181110.1a}
g_i(s)&=a_i(s_0)-\frac{s-s_0}{\pi}\int_{s_{{\rm th};i}}^\infty \frac{\rho_i(s')ds'}{(s'-s_0)(s'-s)}~.
\end{align}

We could apply the Sugawara-Kanazawa theorem, Sec.~\ref{sec.190124.1}, for the DR of a PWA 
 if the only singularities of $T_L(s)_{ij}$ were a RHC, a LHC, possible poles
in between the two cuts, and  if it were  bounded by some power of $s$ for $s\to \infty$ in the complex $s$ plane. 
We can guarantee  that  $T(\infty\pm i\ve)$ is finite because of unitarity and the 
fulfillment of the Schwarz reflection principle. Notice, that $\displaystyle{T_{ij}=(S_{ij}-1)/\big(2i \rho_i^{1/2}{\rho_j^{1/2}}\big)}$,  and $|S_{ij}|\leq 1$. 
Furthermore,  we would expect for the case of finite-range interactions that 
$S_{ij}$ tends to a definite limit for $s\to\infty+i\ve$ (which would be also applicable to the limit $s\to\infty-i\ve$ by 
applying the Schwarz reflection theorem), at least if the interactions become trivial in this limit. 
Similarly, we are tempted to admit on the same grounds  that the limit of the PWA for $-\infty\pm i\ve$ is definite.  
Thus, accepting as plausible the application of the Sugawara-Kanazawa theorem, it implies that $T_{ij}(s)$ 
tends to constant for $s\to \infty$ in the complex $s$ plane, like $(S_{ij}(\infty+i\ve)-1)/(2i \rho_i(\infty+i\ve)^{1/2}\rho_j(\infty+i\ve)^{1/2})$  for $\Im s>0$, and like its complex conjugate for $\Im s<0$. 
Nonetheless, in practical applications (typically at the effective level) one has to handle with singular interactions \cite{case.180502.1,plesset.180501.1,frank.180502.1,meetz.180503.1,arriola.180502.1}, so that  it is expected that  
the PWAs are not bounded in the complex $s$ plane in the limit $s\to \infty$. 
In the case of potential scattering examples of this situation are provided in Ref.~\cite{oller.181101.1}, 
where a master formula is deduced that allows one to calculate the exact discontinuity of a PWA along the LHC 
for a given potential. This formula is applicable to both regular and singular potentials. 
It happens for the latter ones that  the modulus of the discontinuity along the LHC is not bounded and diverges 
in the limit $p^2\to-\infty$ more strongly than any polynomial. 
As a result, the conditions for the applicability of the  Sugawara-Kanazawa theorem are not met and 
$T_{ij}(s)$ is actually divergent for $p^2\to -\infty$, as the explicit calculation of the discontinuity along the LHC shows 
\cite{oller.181101.1}. 

\begin{figure}
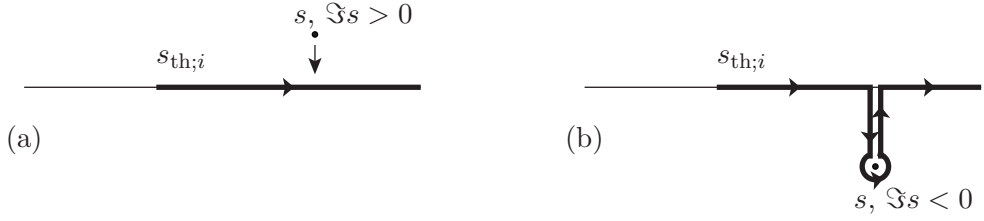

	\begin{center}
		\begin{tabular}{ll}
			\begin{axopicture}(200,100)
				\Line(0,50)(150,50) %x-axis
				\GCirc(110,70){1}{.1} %circle for s
				\Text(102,77)[l]{{\small $s$, $\Im s>0$}}
				\Text(0,30)[c]{{\small (a)}}
				\Text(50,60)[l]{{\small $s_{{\rm th};i}$}}
				\LongArrow(110,66)(110,58)
				\SetWidth{2}
				\SetArrowAspect{0.6}
				\Line[arrow,arrowpos=0.5](50,50)(150,50)  %cut
			\end{axopicture}
			&
			\begin{axopicture}(200,100)
				\SetWidth{0.5}
				\Line(0,50)(150,50) %x-axis
				\GCirc(110,20){1}{.1} %circle for s
				\Text(102,7)[l]{{\small $s$, $\Im s<0$}}
				\Text(0,30)[c]{{\small (b)}}
				\Text(50,60)[l]{{\small $s_{{\rm th};i}$}}
				\SetWidth{2}
				\SetArrowAspect{0.6}
				\Line[arrow,arrowpos=0.5](50,50)(109,50) % cut-1
				\Line[arrow,arrowpos=0.7](108,50)(108,23) % cut-2
				\Line[arrow,arrowpos=0.65](112,23)(112,50) % cut-3
				\Line[arrow,arrowpos=0.5](111,50)(150,50) % cut-4
				\Arc[arrow,arrowpos=.5](110,20)(5,120,80) %circle deformed contour
			\end{axopicture}
		\end{tabular}
		\caption{{\small Contour deformation (thick solid line) to proceed with the analytical extrapolation of $g_i(s)$ from the 
1st (a) to the 2nd (b) RS after crossing the RHC. 
The deformation of the integration contour is needed so as to avoid the pole singularity  at $s'=s$ of the integrand in Eq.~\eqref{181110.1a}.}
			\label{fig.181109.1}}
	\end{center}
\end{figure}

Now, let us discuss how to proceed to calculate the $T$ matrix of PWAs in an unphysical Riemann sheet (RS).
This is accomplished by performing the analytical continuation of $g_i(s)$, which has a branch-point singularity at 
$s_{{\rm th};i}$ and a cut which, as usual, is taken  to run along the real $s$ axis for $s>s_{{\rm th};i}$. 
The second Riemann sheet of $g_i(s)$ is reached by crossing the RHC but at the same time the integration 
contour in Eq.~\eqref{181110.1a} has to be deformed so as to guarantee the smooth process of analytical continuation. 
The actual deformation needed to cross the RHC from the upper (1st RS) to the lower (2nd RS) half complex $s$ planes 
is depicted in Fig.~\ref{fig.181109.1}. 
The only change is to add to $g_i(s)$ the result of the integral along the closed contour around $s$. 
Thus, with $g_{II;i}(s)$ the unitary loop function in the 2nd RS, we deduce the result
\begin{align}
\label{181110.1}
g_{II;i}(s)&=g_i(s)-2i \rho_{II;i}(s)=g_i(s)+2i\rho_{I;i}(s)~.
\end{align}
Here  the function $\rho_{I;i}(s)$ in the complex $s$ plane is
\begin{align}
\label{181110.2}
\rho_{I;i}(s)=\frac{1}{16\pi}\sqrt{\frac{\lambda(s,m_1^2,m_2^2)}{s^2}}~,
\end{align}
where $\sqrt{z}$ is taken in its 1st RS with a RHC for $s>0$, such that ${\rm arg}z\in[0,2\pi[$. 
%The minus sign in front of $\rho_{II;i}(s)$ in Eq.~\eqref{181110.1} is introduced because this is the analytical continuation of $\rho_{I;i}(s)$ in its 2nd RS. 
Notice that for continuing analytically  an integral by the deformation of its integration contour requires to consider appropriately 
a multivalued integrand as defined in its multiple Riemann sheets. 

The Eq.~\eqref{181110.1} implies that the RHC in $g_i(s)$ is a two-sheet one, since by crossing it again 
and going into the upper half plane, one has to add  $+2i\rho_{I;i}(s)$, but this time added to
$g_{II;i}(s)=g_i(s)-2i\rho_{II;i}(s)$ (which becomes $g_i(s)-2i\rho_{I;i}(s)$ when crossing the RHC upwards). 
%(the square-root function in $\rho_{I;i}(s)$, Eq.~\eqref{181110.1} has to be analytically continued to  its 2nd RS).
As a result, the extra terms cancel and one ends again with $g_i(s)$ in the 1st or physical RS. 
This analysis also allows us to conclude that the RHC of a PWA is a two-sheet one too. 
Because of this reason  the different RSs can be characterized similarly as the RSs of the square root present in 
the definition of the CM three-momentum $p_i(s)$,
\begin{align}
\label{181110.3}
p_i(s)&=\pm \sqrt{\frac{\lambda(s,m_{1i}^2,m_{2i}^2)}{4s}}~.
\end{align}
We denote the $2^n$ different RSs as follows.
The physical or first Riemann sheet (RS) corresponds to take the plus sign in the definition of $p_i(s)$ for all the channels, $(+,+,\ldots)$. The second RS implies to calculate $p_1(s)$ in its 2nd RS, that is, we include a minus sign for the first channel, $(-,+,\ldots)$. The third RS is then represented by $(+,-,+,\ldots)$, the fourth RS by $(-,-,+,\ldots)$, and so on. 
Thus, there are $2^{m-1}$ RSs before the sign of $p_m$ for the $m_{\rm th}$ channel is reversed.

Following the derivation of Ref.~\cite{lacour.181101.1}, let us present a DR representation of $\cN_L(s)$, which also 
gives rise to a non-linear IE for this function. 
To avoid unessential complications we study  an uncoupled PWA as a function of the CM three-momentum squared, $p^2$. 
This variable is chosen so as to evade the circular cuts for unequal mass scattering when taking the variable $s$, as referred above, see e.g. \S1.1 of chapter 8 of Ref.~\cite{martin.290916.1}. 
In this way, $T_L(p^2)$ has only a LHC and a RHC. 
The procedure discussed can be easily generalized to coupled channels.

It follows from Eq.~\eqref{181109.6b}  that $\Im \cN(p^2)$  along the LHC obeys (we omit the subscript $L$ to shorten 
the writing),
\begin{align}
\label{181110.4}
\Im T(p^2)&=\Im \frac{1}{\cN(p^2)^{-1}+g(p^2)}=-\frac{\Im \cN(p^2)^{-1}}{|\cN(p^2)^{-1}+g(p^2)|^2}
=\Im \cN(p^2)\frac{|T(p^2)|^2}{|\cN(p^2)|^2}~.
\end{align}
Therefore,
\begin{align}
\label{181110.5}
\Im \cN(p^2)%&=\frac{|\cN(p^2)|^2}{|T(p^2)|^2}\Im T(p^2)\\
&=\left|1+g(p^2)\cN(p^2)\right|^2\Delta(p^2)~,~p^2<p^2_{\rm Left}~.
\end{align}
Where the function $\Delta(p^2)$ is defined as
\begin{align}
\label{181110.6}
\Delta(p^2)&=\Im T(p^2)~,~p^2<p^2_{\rm Left}~,
\end{align}
and $p^2_{\rm Left}$ is the upper bound of the LHC.
 If $\cN(p^2)/p^{2n}$ vanishes for $p^2\to\infty$ we can write the following DR,
\begin{align}
\label{181110.7}
\cN(p^2)&=\sum_{m=0}^{n-1}a_m {p}^{2m}
+\frac{{p}^{2n}}{\pi}\int_{-\infty}^{p^2_{\rm Left}}\frac{\left|1+g(q^2)\cN(q^2)\right|^2\Delta(q^2)dq^2}{{q}^{2n}(q^2-p^2)}~.
\end{align}
In turn, this expression is a non-linear IE, whose input is $\Delta(p^2)$ along the LHC. 
Implementing this result into Eq.~\eqref{181109.6b}, we can write $T(p^2)$ as
\begin{align}
\label{181110.8}
T(p^2)&=\left[\left( \sum_{m=0}^{n-1}a_m {p}^{2m}
+\frac{{p}^{2n}}{\pi}\int_{-\infty}^{p^2_{\rm Left}}\frac{\left|1+g(q^2)\cN(q^2)\right|^2\Delta(q^2)dq^2}{{q}^{2n}(q^2-p^2)}
\right)^{-1}
+g(p^2)\right]^{-1}~.
\end{align}
The subtraction constants $a_m$ can be fixed by reproducing physical observables, e.g.
by fitting phase shifts, reproducing the effective range expansion (ERE) shape parameters, etc.

We can also show that  $T(p^2)$ is independent of the subtraction constant in $g(p^2)$ by following the argument 
given in Ref.~\cite{lacour.181101.1}. Let us take only one subtraction constant in $\cN(p^2)$, as this is  enough for illustrating the point.
A DR for $T^{-1}(p^2)$ is performed directly by taking into account Eq.~\eqref{181110.6} and 
that $\Im T(p^2)^{-1}$ along the RHC is $-\rho(p^2)$,
Eq.~\eqref{051016.12b}. One can then write 
\begin{align}
\label{181110.9}
T^{-1}(p^2)=\beta-\frac{p^2}{\pi}\int_{0}^\infty \frac{\rho(q^2)dq^2}{q^2(q^2-p^2)}
-\frac{p^2}{\pi}\int_{-\infty}^{p^2_{\rm Left}}\frac{\Delta(q^2)dq^2}{|T(q^2)|^2q^2(q^2-p^2)}+R(p^2)~.
\end{align}
 In this equation $R(p^2)$ is a rational function to account for the possible zeroes of $T(p^2)$ and which does not play
an active role here. There is a free parameter (subtraction constant) to be determined, $\beta$, 
even though in Eq.~\eqref{181110.8} it is split in two contributions. Precisely, one of them is added to the integral along the RHC giving 
rise to the unitary loop function $g(p^2)$. 
Thus, the inclusion of a subtraction constant in this function is just a  matter of convenience.

%%%%%%%%%%%%%%%%%%__________________________________%%%%%%%%%%%%
%\subsection{The subtraction constants of $g_i(s)$ in the $SU(3)$ limit}
%\label{sec.181110.1}

Let us consider the flavor $SU(3)$  limit in which the lightest quarks ($u$, $d$ and $s$) have equal mass.  
The $T$ matrix is an $SU(3)$ singlet and then it is diagonal in a basis of states with well-defined transformation properties under $SU(3)$. 
These states are denoted by $|R,\lambda\rangle $, where $R$ indicates the $SU(3)$ irreducible representation and $\lambda$ 
takes into account all the other quantum numbers that are necessary to distinguish among the states in $R$ \cite{lichtenberg.181110.1}.
 In this  notation the matrix elements of the $T$ operator for two-body scattering are then written as 
\begin{align}
\label{181110.12}
\langle R',\lambda'|T|R,\lambda\rangle&=T_R \delta_{RR'}=\frac{1}{\cN_{R}^{-1}+g_R(s)}\delta_{RR'}~, 
\end{align}
where we have used  Eq.~\eqref{181109.6b}. Let us also denote by $a_R$ the subtraction constant in the unitary loop function 
$g_R(s)$.
The scattering PWAs $T_{ij}(s)$ in the physical basis are  given by Eq.~\eqref{181109.6b}, in terms of 
$\cN_{ij}(s)$, calculated in the charge basis, and $g_i(s)$, involving the subtraction constants $a_i$.

An interesting result originally deduced in Ref.~\cite{oller.181111.1}, and revisited in Ref.~\cite{oller.190503.1},
is that all the subtraction constants $a_i$  are equal in the $SU(3)$ limit.
In order to be more specific, this result holds for all the channels $i$ that are connected to each other when proceeding to
their decompositions in irreducible representations of the $SU(3)$ group.
 A straightforward corollary from this result is that all the subtraction constants $a_R$ in the previous decomposition are also the same.

%%%%%%%%%%%%%%%%%%%%%%%%%%%%%%%%%%%%%%%%%%%%%%%%%%%%%%
%%%%%%%%%%%%%%%%%%%%%%%%%%%%%%%%%%%%%%%%%%%%%%%%%%%%%%
%\section[Meson-meson scattering: the lightest scalar resonances \& $\rho(770)$]{Meson-meson scattering I in the light-quark sector. Results without including an explicit LHC}
\subsection{Meson-meson scattering in the light-quark sector.}
\label{sec.181105.1}

Let us first apply the formalism of Sec.~\ref{sec.181104.1} for the study of the main features of  
the low-energy $I=0$  $S$-wave and $I=1$ $P$-wave $\pi\pi$  amplitudes. 
The phase shifts of these PWAs are characterized by the presence of a broad shoulder  ($\sigma$) for the
former PWA and by a sharp rise ($\rho$) for the later one. 

As primary source of the dynamics among the pions we consider the lowest or leading order (LO) Chiral Perturbation 
Theory scattering amplitudes.
 Chiral Perturbation Theory (ChPT) is a low-energy effective field theory (EFT) of QCD that is written
by implementing the chiral symmetry of QCD as well as its spontaneous breaking.
The degrees of freedom of ChPT are the lightest pseudoscalars, corresponding to the
pseudo-Goldstone bosons associated with the spontaneous  chiral symmetry breaking of strong interactions. 
 The pseudo-Goldstone bosons finally acquire a small mass because of the explicit breaking  of the chiral symmetry due to the quark
  masses. 
  Another consequence of the Goldstone theorem is that the interactions in which the Goldstone bosons participate 
 are of derivative nature and become zero in the limit $p_i^2\to 0$ (with $p_i$ being a Goldstone-boson four-momentum). 
This makes that  there are near-threshold zeroes even for the $S$ waves, the so-called Adler zeroes, 
while for the $P$ and higher waves these are just the zero at threshold.
For detailed accounts on ChPT  the interested reader is referred to the chapter 19
of Ref.~\cite{weinberg.181105.1}, Ref.~\cite{leutwyler.181105.1} 
or the reviews \cite{pich.181105.1,meissner.181105.1,ecker.181105.1}, just to quote a few.

The LO ChPT amplitudes for the $I=L=0$ and $I=L=1$ $\pi\pi$ PWAs are 
\begin{align}
\label{181106.1}  
V_0(s)&=\frac{s-m_\pi^2/2}{f_\pi^2}~,\\
V_1(s)&=\frac{s-4m_\pi^2}{6f_\pi^2}~,\nn
\end{align}
 respectively \cite{oller.181105.1}.  
In the previous equation, we denote the weak pion decay constant by $f_\pi\simeq 92.4~$MeV. 
The amplitudes in Eq.~\eqref{181106.1} are  first order polynomials that can be accounted for by 
including  a CDD pole in the $D(s)$ function,  cf. Eq.~\eqref{fin/d}. The position and the residue of this 
pole is adjusted by matching with Eq.~\eqref{181106.1}. We then have the following expression for the 
non-perturbative PWAs by applying Eq.~\eqref{fin/d} in the way described:   
\begin{align}
\label{181106.2}
T_0(s)&=\left(\frac{f_\pi^2}{s-m_\pi^2/2}+\widetilde{a}_0-\frac{s}{\pi}\int_{s_{\rm th}}^\infty ds'\frac{\rho(s')}{s'(s'-s)}\right)^{-1}~,\\ 
\label{181106.2b}
T_1(s)&=\left(\frac{6f_\pi^2}{s-4m_\pi^2}+\widetilde{a}_1-\frac{s}{\pi}\int_{s_{\rm th}}^\infty ds'\frac{\rho(s')}{s'(s'-s)}\right)^{-1}~. 
\end{align}

We have to discuss about the physical meaning of the subtraction constants $\widetilde{a}_I$ in Eq.~\eqref{181106.2} 
in order to realize from the hadronic point of view (by taking the pions as the explicit degrees of freedom in the theory) the sharp difference between the nature of the resonances $f_0(500)$ (or $\sigma$) and the $\rho(770)$. 
One can perform explicitly the integration involved in the definition of the unitary loop function $g(s)$ in Eq.~\eqref{181110.1a}, 
where the two particles in the intermediate state have masses $m_1$ and $m_2$, with the result
\begin{align}
\label{181106.3}
g(s)&=\widetilde{a}-\frac{s}{\pi}\int_{s_{\rm th}}^\infty ds'\frac{\rho(s')}{s'(s'-s)}\\
&=\frac{1}{16\pi^2}\left[
a(\mu)+\log\frac{m_1^2}{\mu^2}-x_+\log\frac{x_+-1}{x_+}-x_-\log\frac{x_--1}{x_-}\right]~,\nn\\
x_\pm&=\frac{s+m_2^2-m_1^2}{2s}\pm\frac{1}{2s}\sqrt{(s+m_2^2-m_1^2)^2-4s(m_2^2-i0^+)}~.\nn
\end{align}
The parameter $\mu$ is a renormalization scale introduced for dimensional reasons, so as to 
end with a dimensionless argument of the first logarithm. A change in the value of $\mu$ can always be 
reabsorbed in a corresponding variation of $a(\mu)$, so that the combination $a(\mu)-2\log\mu$ is independent of $\mu$. 
The value of $g(s)$ at threshold, $g(s_{\rm th})$, can be written in terms of  $a(\mu)$ as 
\begin{align}
\label{181106.4}
%\widetilde{a}
g(s_{\rm th})&=\frac{a(\mu)}{16\pi^2}+\frac{1}{8\pi^2(m_1+m_2)}(m_1\log\frac{m_1}{\mu}+m_2\log\frac{m_2}{\mu})~.
\end{align}
The unitarity loop function $g(s)$ corresponds to the one-loop two-point function
\begin{align}
\label{181106.5}
g(s)&=i\int\frac{d^4p}{(2\pi)^4}\frac{1}{[(P/2-p)^2-m_1^2+i\ve][(P/2+p)^2-m_2^2+i\ve]}\\
&=\int_0^\infty \frac{p^2 dp}{(2\pi)^2}\frac{\omega_1+\omega_2}{\omega_1\omega_2[s-(\omega_1+\omega_2)^2+i\ve]}~,\nn
\end{align}
with $P=p_1+p_2$  the total four-momentum and $\omega_i=\sqrt{m_i^2+\vp^2}$. 
This loop integral diverges logarithmically and requires regularization, which is the reason why a  
subtraction constant is included Eq.~\eqref{181106.3}. 
This last equation can also be obtained by employing dimensional regularization  in $d$ dimensions   
and then reabsorbing  the diverging term $1/(d-4)$ for $d\to 4$ in the subtraction constant $a(\mu)$. 
In turn, a three-momentum cutoff regularization could be also used, so that the integration in the modulus of the 
three-momentum $p$ in Eq.~\eqref{181106.5} is done up to a maximum value $\Lambda$. For this case, the
resulting expression for the function $g(s)$, and  denoted by $g_\Lambda(s)$,  
can be found in Ref.~\cite{oller.181115.1b} and it reads
\begin{align}
\label{181106.6}
&  g_\Lambda(s)=\frac{1}{32\pi^2}\Bigg(
-\frac{\Delta}{s}\log\frac{m_1^2}{m_2^2}
+2\frac{\Delta}{s}\log\frac{1+\sqrt{1+m_1^2/\Lambda^2}}{1+\sqrt{1+m_2^2/\Lambda^2}}
+\log\frac{m_1^2m_2^2}{\Lambda^4} \\
& +\frac{\nu}{s}\left\{
\log\frac{s-\Delta+\nu\sqrt{1+m_1^2/\Lambda^2}}{-s+\Delta+\nu\sqrt{1+m_1^2/\Lambda^2}}
+\log \frac{s+\Delta+\nu\sqrt{1+m_2^2/\Lambda^2}}{-s-\Delta+\nu\sqrt{1+m_2^2/\Lambda^2}}\right\} \nn\\
& -2\log\left[\left(1+\sqrt{1+m_1^2/\Lambda^2}\right)\left(1+\sqrt{1+m_2^2/\Lambda^2}\right)\right] \Bigg)~,\nn
\end{align}
where $\nu=\lambda(s,m_1^2,m_2^2)^{1/2}$ and $\Delta=m_2^2-m_1^2$.

The natural size for a three-momentum cutoff in hadron physics  is the typical size of a
hadron, as resulting from the strong dynamics binding the quarks and gluons together. 
Thus, we expect a value of $\Lambda$  around 1~GeV, 
because de Broglie wave length associated to the higher three-momenta would reveal 
elementary degrees of freedom not accounted explicitly in the EFT for low energies.  
In the non-relativistic limit the functions $g(s)$ and $g_\Lambda(s)$ ($|\vp|\ll m_1$, $m_2$) are a constant (corresponding 
to the value at threshold of every function) plus $-ip/(8\pi(m_1+m_2))+{\cal O}(p^2)$. 
For $g(s)$ the referred constant is given in Eq.~\eqref{181106.4}, and  $g_\Lambda(s_{\rm th})$ can be worked out 
explicitly from Eq.~\eqref{181106.6} with the result \cite{guo.190126.1}
\begin{align}
\label{181106.7}
g_\Lambda(s_{\rm th})&=-\frac{1}{8\pi^2(m_1+m_2)}\left[
m_1\log\left(1+\sqrt{1+m_1^2/\Lambda^2}\right)\right.\\
&\left. +m_2\log\left(1+\sqrt{1+m_2^2/\Lambda^2}\right)
-m_1\log\frac{m_1}{\Lambda}-m_2\log\frac{m_2}{\Lambda}
\right]~.\nn
\end{align}     
By equating Eqs.~\eqref{181106.4} and \eqref{181106.7} we find this expression for $a(\mu)$ as a function $\Lambda$, 
\begin{align}
\label{181106.8}
a(\mu)&=-\frac{2}{m_1+m_2}\left[m_1\log\left(1+\sqrt{1+m_1^2/\Lambda^2}\right)
+m_2\log\left(1+\sqrt{1+m_2^2/\Lambda^2}\right)    \right]-\log\frac{\mu^2}{\Lambda^2}~. %$\mu\leftrightarrow \Lambda$ in the log
\end{align}
For example,  in the case of $\pi\pi$ scattering and $\Lambda=1$~GeV we have
\begin{align}
\label{181106.9}
a(\mu)&=-1.40-\log\frac{\mu^2}{\Lambda^2}~,~\Lambda=1~{\rm GeV}~.
\end{align}
The natural value for a subtraction constant results by employing $\mu=\Lambda\simeq 1$~GeV in Eq.~\eqref{181106.8}, 
 as originally introduced in Ref.~\cite{meissner.181106.1}. 
So that both the renormalization scale and the cut off take a value corresponding to the transition region from the 
low-energy EFT to the underlying QCD dynamics.
 
There is an interesting twist given in Ref.~\cite{hyodo.190128.1} concerning the interpretation of the departure  of 
the subtraction constants with respect to the natural value. 
This reference shows that this can be associated to the exchange of bare particles in the 
equivalent interaction kernel  that one should employ if keeping the natural value for the subtraction constants. 
This is quite clear from Eq.~\eqref{181109.6}, because a variation in the subtraction constant, $\Delta a(\mu)$, can be reabsorbed 
by a change in $\cN(s)$. If we write this equation as 
\begin{align}
\label{190129.1}
T_L(s)^{-1}&={\cal N}_L(s)^{-1}+a(s_0)-\frac{s-s_0}{\pi}\int_{s_{\rm th}}^\infty \frac{\rho(s')ds'}{(s'-s_0)(s'-s)}\\
&=	{\cal N}_L(s)^{-1}+a(s_0)-a'(s_0)+a'(s_0)-\frac{s-s_0}{\pi}\int_{s_{\rm th}}^\infty \frac{\rho(s')ds'}{(s'-s_0)(s'-s)}~,\nn
\end{align}
then the corresponding $\cN_L(s)$ when $a'(s_0)$ is used as a new value for the subtraction constant is 
\begin{align}
\label{190129.2}
\bar{\cN}_L(s)&=\left[I+\cN_L(s)(a(s_0)-a'(s_0))\right]^{-1}\cN_L(s)\\
&=\left[I+\cN_L(s)\frac{a(\mu)-a'(\mu)}{(4\pi)^2}\right]^{-1}\cN_L(s)~,\nn
\end{align}
%where the same expression is obtained either by using the subtraction constants within a DR or after performing algebraically 
%the integration for $g(s)$ in the form of Eq.~\eqref{181106.3}, as stressed in Eq.~\eqref{190129.2} by employing either $s_0$ or $\mu$ as renormalization scales. 
 E.g. if we use  for $\cN(s)$ in Eq.~\eqref{190129.2} any of the LO ChPT amplitudes in Eq.~\eqref{181106.1}, which are 
first order polynomial in $s$, it is then clear that $\bar{\cN}_L(s)$ exhibits a denominator which resembles the propagator of a  
bare ``elementary" particle. This is also the case when employing the LO ChPT amplitudes for meson-baryon 
scattering, the system on which the discussions of Ref.~\cite{hyodo.190128.1} are focused.

\begin{figure}[tb]
\begin{center}
\begin{minipage}[t]{8 cm}
\epsfig{file=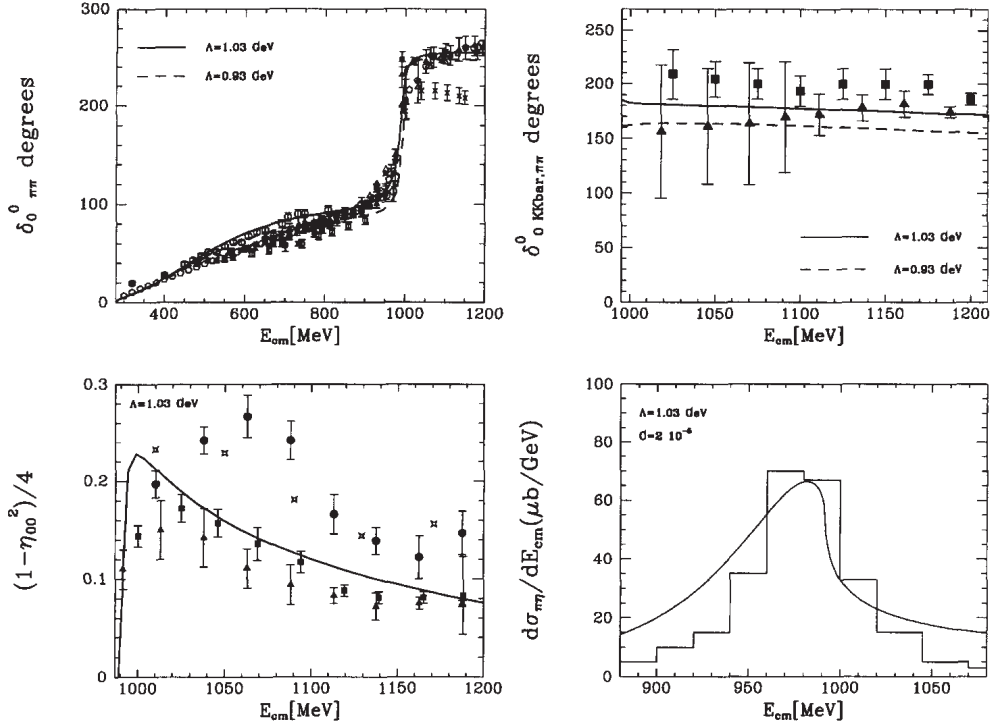,scale=0.5,angle=-90}
\end{minipage}
\begin{minipage}[t]{16.5 cm}
\caption{{\small The reproduction of several scattering observables by Ref.~\cite{oller.181101.2}.  
From top to bottom and left to right, the first three panels correspond to the isoscalar scalar  $\pi\pi\to \pi\pi$ and  the $\pi\pi\to K\bar{K}$ phase shifts and to the inelasticity $(1-\eta_{00}^2)/4$, with $\eta_{00}$ the elasticity parameter, respectively. 
The references to the experimental data (points with errorbars) can be found in the original work \cite{oller.181101.2}. } 
\label{fig.190130.1}}
\end{minipage}
\end{center}
\end{figure}

By  taking the natural value for  $a(\mu)$ in the isoscalar scalar $\pi\pi$ partial-wave amplitude, so that 
$\mu=\Lambda=1$~GeV in Eq.~\eqref{181106.9}, we find a pole for the resonance $\sigma$ or $f_0(500)$ 
in the 2nd RS at
\begin{align}
\label{181106.9b}
s_\sigma&=(0.47-i\,0.20)^2~{\rm GeV}^2.
\end{align}
Despite the simplicity of the model based on $T_0(s)$ given in Eq.~\eqref{181106.2} (which incorporates a CDD pole, 
whose properties are fixed by the Adler zero in the LO  ChPT amplitude $V_0(s)$, 
and a subtraction constant required to have natural value)  the 
$f_0(500)$ pole in Eq.~\eqref{181106.9b} is already  compatible with the value given in Particle Data Group (PDG) \cite{pdg.181106.1}, 
 \begin{align}
\label{181106.9c}
s_\sigma&=((0.4-0.5)-i\,(0.20-0.35))^2~{\rm GeV}^2. 
\end{align}
In addition, the phase shifts for this PWA are also compatible with data from threshold up to the 
rise associated with the appearance of the resonance  $f_0(980)$ for $\sqrt{s}\gtrsim 0.9$~GeV, see 
the first panel in Fig.~\ref{fig.190130.1}. 

Then, the resonance $f_0(500)$ has a quite simple explanation as a dynamically generated resonance 
in terms of pions as effective degrees of freedom. 
It emerges from an interplay between the Adler zero (as required by chiral symmetry), located at 
around $m_\pi^2/2$ (small higher-order corrections occur), and the rescattering of the two pions when propagating. 
The latter effect is accounted for by using the non-perturbative expression of Eq.~\eqref{181106.2} for $T_0(s)$. 
A recent and comprehensive review devoted to the $\sigma$ resonance is given in Ref.~\cite{pelaez.181109.1}.

However, the same idea if applied to the isovector vector $\pi\pi$ PWA does not give rise to an acceptable reproduction 
of the corresponding phase shifts nor of the pole of the $\rho(770)$ resonance. 
In order to give account of the latter a much bigger absolute value  of the subtraction constant $a(\mu)$ 
is required.  In this respect, for $a(\mu)=-14$, $\mu=1$~GeV, we find then a good pole position for the $\rho(770)$ 
resonance \cite{pdg.181106.1} at 
\begin{align}
\label{181106.10}
s_\rho&=(0.777-i\,0.072)^2~{\rm GeV}^2~,~a(1~{\rm GeV})=-14~.
\end{align}
This value of $a(\mu)$, if reproduced by employing Eq.~\eqref{181106.8}, would require a three-momentum 
cutoff of  around $\Lambda=600$~GeV, near the TeV region, which is completely an unreasonable scale for the 
QCD dynamics. 
It is then clear that the $\rho(770)$ cannot be explained as dynamically generated from the pion interactions, being a 
resonance of very different nature as compared with the $f_0(500)$ or $\sigma$.\footnote{In hadron physics a more 
natural	justification for the appearance of the $\rho(770)$ in the spectrum is obtained  by gauging 
the chiral symmetry in the non-linear chiral Lagrangians \cite{bando.181110.1,bando.181110.2,ulf.181110.1}.}
 Continuing with the discussion on the $\rho(770)$, Ref.~\cite{oller.181105.1} includes one more CDD pole in 
 the function $D(s)$ and matches it to the tree-level PWA that results by adding to $V_1(s)$ in Eq.~\eqref{181106.1} the contribution 
 of the explicit exchange of a bare $\rho$ resonance, calculated by employing the chiral Lagrangians of Ref.~\cite{leutwyler.190130.1,EP}.
  In this way, the large non-natural  value of the subtraction constant $a(\mu)$ needed to give rise to the $\rho(770)$ resonance pole 
 is a manifestation of the ``elementary" character of this resonance with respect to $\pi\pi$ rescattering, as 
 it is clear from quark models \cite{lichtenberg.181110.1}, LQCD \cite{hsc.190130.1,molina.190130.1}, evolution of the $\rho$ pole with the number of colors of QCD ($N_c$) \cite{pelaez.190130.1,pelaez.190130.2,guo.181123.1,elvira.190130.1}, 
its variation under changes in the quark (pion) masses  \cite{hanhart.190130.1,nebreda.181116.1}, etc. 

The tree-level LO ChPT amplitude plus the bare exchange of a $\rho$ resonance
employing the Lagrangians of Ref.~\cite{EP} is
\cite{oller.181105.1} %aqui
\begin{align}
\label{t1n2}
t_1(s)=\frac{2}{3}\frac{p^2}{f_\pi^2}\left[1
+g_v^2 \frac{s}{M_\rho^2-s}\right]~,
\end{align} 
where $M_\rho$ is the (bare) mass of the $ \rho(770)$ resonance and $g_v^2$ measures the deviation 
of the coupling of the $\rho(770)$ to $\pi\pi$ with respect to the prediction given by the KSFR relation \cite{KSFR}, which 
 implies $g_v=1$. 
One can match this expression with two CDD poles. 
 One of them accounts for  the zero at threshold, which is the main structure of $T_1(s)$ in Eq.~\eqref{181106.2b}. 
 The new CDD pole location and its residue, needed to reproduce the new zero of $t_1(s)$ in Eq.~\eqref{t1n2}, are
\begin{align}
\label{181108.1}
s_2&=\frac{M_\rho^2}{1-g_v^2}~,\\
\gamma_2&=\frac{6f_\pi^2}{1-g_v^2}\frac{g_v^2M_\rho^2}{M_\rho^2-4(1-g_v^2)m_\pi^2}~.\nn
\end{align}
%The position of this zero coincides with the one of the CDD pole and the derivative of
%$t_1(s)$ at this zero is the inverse of the residue of the CDD pole in $D(s)$. 
In the limit $g_v^2\to 1$ the zero $s_2$ tends to infinity and this is the reason why it was possible before 
to give rise to the pole of the $\rho(770)$ in terms of a large and negative subtraction constant  added to $g(s)$
in Eq.~\eqref{181106.2b}. In relation with this, we have that 
\begin{align}
\label{181108.2}
\lim_{s_2\to \infty} \frac{\gamma_2}{s-s_2}=-\frac{6f_\pi^2}{M_\rho^2}~.
\end{align}
This result times $16\pi^2$ gives 
\begin{align}
\label{181108.3}
-\frac{96\pi^2f_\pi^2}{M_\rho^2}=-13.6~,
\end{align}
which is the value required above for $a(1~{\rm GeV})$ in the $I=L=1$  $\pi\pi$ scattering to generate an adequate 
$\rho(770)$ pole. 
 This analysis clearly shows that this number reflects the ``elementary" nature of the $\rho(770)$ when taking the 
 pions as effective low-energy degrees of freedom of strong interactions.
Thus, the final isovector vector $\pi\pi$ PWA that results is
\begin{align}
\label{181108.4}
T_1(s)&=\left[\frac{6f_\pi^2}{s-4m_\pi^2}-\frac{6 f_\pi^2}{M_\rho^2}
+\frac{1}{16\pi^2}\left(\log\frac{m_\pi^2}{\mu^2}-x_+\log\frac{x_+-1}{x_+}-x_-\log\frac{x_--1}{x_-}\right)\right]^{-1}~,
\end{align}
with $\mu\simeq 1$~GeV. An analogous analysis holds  for the $I=1/2$ vector $K\pi$ scattering
and the $K^*(892)$ resonance, as explicitly worked out in Ref.~\cite{oller.181105.1}. 

In the same reference, higher energies are reached for the scalar sector by employing the matrix notation of Eq.~\eqref{A.8} that allows 
to take into account the coupling among several two-body channels and also by including  explicit bare resonance fields for a singlet 
and an octet with bare masses around 1 and 1.3~GeV, respectively. The phase shifts and elasticity parameters 
are studied for the different isospins and the channels involved, given between brackets, are 
$I=0$ ($\pi\pi$, $K\bar{K}$, $\eta_8\eta_8$), 1 ($\pi\eta_8$, $K\bar{K}$)  and 1/2 ($K\pi$, $K\eta_8$). 
The unitarized amplitude for every isospin is given by 
\begin{align}
\label{181108.5}
T_I(s)&=\left[t_I(s)^{-1}+g(s)\right]^{-1}~,
\end{align}
where the $t_I(s)$ are the tree-level amplitudes obtained from LO ChPT and
the chiral Lagrangians of Ref.~\cite{EP}, which are employed for evaluating the exchange of bare-resonance fields. 
As an example, the different matrix elements of  $t_1(s)$   read 
\begin{align}
\label{181108.6}
t_{1;11}(s)&=\frac{m_\pi^2}{3f_\pi^2}+\frac{\beta_1^2}{M_S^2-s}~,\\
t_{1;12}(s)&=-\frac{\sqrt{3/2}}{12f_\pi^2}(6s-8m_K^2)+\frac{\beta_1\beta_2}{M_S^2-s}~,\nn\\
t_{1;22}(s)&=\frac{s}{4f_\pi^2}+\frac{\beta_2^2}{M_S^2-s}~,\nn
\end{align}
with the channels  $\pi\eta_8$ and $K\bar{K}$  labeled as 1 and 2, respectively. 
 Next, the bare-resonance couplings $\beta_i$ are
\begin{align}
\label{181108.7}
\beta_1&=\frac{\sqrt{2}}{\sqrt{3}f_\pi^2}\left(c_d(s-m_\pi^2-m_{\eta_8}^2)+2c_m m_\pi^2\right)~,\\
\beta_2&=-\frac{2}{f_\pi^2}\left(c_d\frac{s}{2}+(c_m-c_d)m_K^2\right)~,\nn
\end{align}
where $m_K$ is the kaon mass and $m_{\eta_8}^2=4 m_K^2/3-m_\pi^2/3$, which is the Gell-Mann \cite{mann.190130.1}-Okubo \cite{okubo.190130.1} mass relation. This relation provides an $\eta_8$ mass which is only around a 3\% higher than the physical 
mass of the $\eta$ pseudoscalar, $m_\eta$.  
The tree-level amplitudes in Eq.~\eqref{181108.6}, as well as the ones for the other channels, 
are explicitly given in Ref.~\cite{oller.181105.1}.

The unitary loop function $g(s)$ is given in Eq.~\eqref{181106.3} and Ref.~\cite{oller.181105.1} takes 
the subtraction constant $a(\mu)$ to be the same for all channels (which is justified in the $SU(3)$ limit 
in virtue of the discussion at the end of Sec.~\ref{sec.181109.1}). 
The final value for $a(\mu)$ results from a fit to data and Ref.~\cite{oller.181105.1} obtains  $a(M_\rho)=-0.7$, with $\mu=M_\rho$. 
%This value is around a factor of 2 smaller in absolute value compared with the prediction in Eq.~\eqref{181106.8} by taking $\mu\simeq \Lambda=1$~GeV. 
As a result,  this study obtains poles corresponding to the whole set of 
light scalar resonances, $f_0(500)$, $f_0(980)$ ($I=0$), $a_0(980)$ $(I=1)$, $K^*_0(700)$ or $\kappa$ ($I=1/2$),
and to an octet of scalar resonances with masses around 1.4~GeV with $I=0$, $1/2$, and $1$.
This mass indeed is very close to the bare mass of the $SU(3)$ octet of bare explicit scalar resonances introduced
in the calculation of the tree-level amplitudes $t_I(s)$. 
In addition, Ref.~\cite{oller.181105.1} also includes a singlet bare resonance with a mass around 1~GeV, which mainly  
gives a contribution to the physical $f_0(980)$. This latter contribution is necessary in order to reproduce 
the elasticity parameter $\eta_{00}$ associated with the $\pi\pi$ channel, 
once the $\eta_8\eta_8$ also enters in the formalism as a one more channel. 
 The inclusion of this singlet scalar resonance also solves the objection  of Ref.~\cite{kaiser.181109.1} 
 with regards the unsatisfactory reproduction  of the elasticity parameter $\eta_{00}$ by the unitarized $I=L=0$  LO 
ChPT amplitudes once  the $\eta_8\eta_8$ state is included in addition to $\pi\pi$ and $K\bar{K}$. Additionally,  Ref.~\cite{oller.181105.1} provides  too a fine reproduction of the experimental phase shifts and inelasticities, 
cf. Figs.~2--8 in this reference.

All these results are obtained without including explicitly the LHC in the formalism, cf. Eq.~\eqref{fin/d}. 
Its contributions are estimated in Ref.~\cite{oller.181105.1} and, remarkably, they are small for the resonant
scalar  two-meson channels for  $\sqrt{s}\lesssim 1$~GeV.
This estimation results by evaluating: i) the crossed loop diagrams at next-to-leading order (NLO) or ${\cal O}(p^4)$ in the ChPT meson-meson amplitudes,  and ii) the $t$- and $u$-channel exchanges of explicit resonances with spin$\leq 1$.
The contributions i) and ii), whose sum is  denoted by $T_{I;{\rm Left}}(s)$, can be taken from the ChPT calculation 
of the meson-meson scattering amplitudes undertaken in  Ref.~\cite{bernard.181109.1} with some bare resonances included. 
In more detail, Ref.~\cite{oller.181105.1} matches the calculated PWA $A(s)$
by Ref.~\cite{bernard.181109.1} with the  expansion of Eq.~\eqref{181108.5} up to one-loop,
\begin{align}
\label{181109.1}
t_I(s)-t_I(s) g(s) t_I(s)~.
\end{align}
Then,  $T_{I;{\rm Left}}(s)$ is defined as 
\begin{align}
\label{181109.2}
T_{I;{\rm Left}}(s)&=A(s)-t_I(s)+t_I(s)g(s)t_I(s)~,
\end{align}
and we remove from $A(s)$ the tree-level amplitude and the once-iterated LO ChPT amplitudes because they are already 
accounted for   by  $T_I(s)$.\footnote{The Ref.~\cite{oller.181105.1} does not include  the tadpole contributions for calculating $T_{I;{\rm Left}}(s)$	in $A(s)$ and it keeps  the ones that involve explicit LHC, while  
	tadpoles are contact terms. 
Note that the tadpole	contributions could also be removed by employing a different regularization scheme in the 
calculations \cite{mandl.190130.1}.} 
 The  Ref.~\cite{oller.181105.1} concludes that for  the elastic scattering amplitudes of 
 $L=0$, $I=0$ $\pi\pi$ and $L=0$, $I=1/2$ $K\pi$, the absolute value of the ratio $T_{I;{\rm Left}}(s)/t_I(s)$ is
 $\lesssim 5$\% for CM energies up to around $\sqrt{s}=1$~GeV.
This smallness of the LHC contribution is important and stems from a  largely cancellation  between
the crossed exchange of resonances and the crossed loops. Indeed, every of these contributions separately 
is around a 15\% of $t_I(s)$ in the same region.

This cancellation between crossed resonance exchanges and crossed  loops is a manifestation of a fact worth stressing in the meson-meson scalar sector, which is the violation of large $N_c$ QCD expectations. 
The point is that  meson-meson loops are subleading compared with resonance exchanges according to the large 
$N_c$ counting \cite{manohar.181109.1}. 
Indeed,  Ref.~\cite{oller.181105.1} was the first study in the literature to conclude that
the mass of the $f_0(500)$ resonance does not follow the expected behavior  for a $q\bar{q}$ resonance 
according to large $N_c$ QCD. 
However, it is found that the mass of this resonance increases with $N_c$ instead of being ${\cal O}(N_c^0)$.
 This result can be easily grasped  from the expressions for $T_0(s)$, Eq.~\eqref{181106.2}, and $a(\mu)$, Eq.~\eqref{181106.8}. 
 The latter shows that $a(\mu)$ is ${\cal O}(N_c^0)$, as the rest of terms in $g(s)$.
Therefore, when looking for a pole in Eq.~\eqref{181106.2} we have an equation for $s$ as
\begin{align}
s_\sigma\to &  -f_\pi^2/ g(s_\sigma)={\cal O}(N_c)~,
\end{align}
where use has been made that   $f_\pi^2$ varies as $N_c$ in the large $N_c$ QCD counting \cite{manohar.181109.1}.

For the case of the $\rho(770)$ resonance the situation is entirely different as follows by considering the equations 
for $T_1(s)$, Eq.~\eqref{181106.2b}, and for $a(\mu)$, Eq.~\eqref{181108.3} (let us notice that 
$\widetilde{a}_1=a(\mu^2)/16\pi^2$). The resulting equation for the pole position of $T_1(s)$ is now 
\begin{align}
s_\rho&\to M_\rho^2+{\cal O}(N_c^{-1})~,
\end{align}
which runs as ${\cal O}(N_c^0)$ in the large $N_c$ counting. 
These derivations also reflect the so much different nature of the $\rho(770)$ and the $\sigma$ or $f_0(500)$  
 regarding their origin.

Two more interesting facets of the meson-meson scalar spectroscopy  
are also concluded in a novel way in Ref.~\cite{oller.181105.1}.
The first point already obtained in Ref.~\cite{oller.181105.1} is the strong sensitivity of the 
masses of the pole positions of the lightest scalar resonances [$f_0(500)$, $K^*_0(700)$, $f_0(980)$ and $a_0(980)$] 
 with the pseudoscalar masses. 
The second finding is that these resonances form an octet of degenerate scalar resonances plus a singlet in the 
flavor $SU(3)$ limit. 
 In particular, in the chiral limit $(m_\pi=m_K=0)$ Ref.~\cite{oller.181105.1}
obtains that the pole position of the $SU(3)$ octet is around $500-i\,350$~MeV and the one of the singlet  is lighter, at
around $400-i\,250$~MeV. 
Let us note that these pole positions are very different from those of the physical pseudoscalar masses, whose values found in 
Ref.~\cite{oller.181105.1} are collected Table~\ref{table.181109.1}. 
The strong dependence of the $f_0(500)$ pole position with the pion mas has been confirmed more recently by studies of
LQCD results \cite{hanhart.190130.1,hsc.181109.1}. 

  The coupling constants $\xi_i$ of every resonance to the different channels calculated in 
 Ref.~\cite{oller.181105.1} are also given in Table \ref{table.181109.1}. 
 These couplings are defined by the residues of the  $T$ matrix of PWAs at the pole position, 
\begin{align}
\label{181109.3}
\xi_i\xi_j&=\lim_{s\to s_R}(s-s_R)T_{ij}(s)~,  
\end{align}
with $i$ and $j$ referring to the different  channels and $s_R$ is the resonance pole. 
\begin{table}
	\begin{center}
		\begin{tabular}{|l|l|l|l|}
			\hline
			$\sigma$ & $f_0(980)$ & $a_0(980)$ & $\kappa$ \\
     			\hline
			$\sqrt{s_\sigma}=450-i\,221$ &  $\sqrt{s_{f_0}}=987-i\,14$ & $\sqrt{s_{a_0}}=1053-i\,24$ &  $\sqrt{s_\kappa}=779-i\,330$ \\ 
			$|\xi_{\pi\pi}|=4.25$ & $|\xi_{K\bar{K}}|=3.63$ & $|\xi_{K\bar{K}}|=5.50$ & $|\xi_{K\pi}|=5.00$ \\
			$\left|\frac{\xi_{K\bar{K}}}{\xi_{\pi\pi}}\right|=0.25$ &
			$\left|\frac{\xi_{\pi\pi}}{\xi_{K\bar{K}}}\right|=0.51$ &
			$\left|\frac{\xi_{\pi\eta_8}}{\xi_{K\bar{K}}}\right|=0.70$ &
			$\left|\frac{\xi_{K\eta_8}}{\xi_{K\pi}}\right|=0.62$ \\
			$\left|\frac{\xi_{\eta_8\eta_8}}{\xi_{\pi\pi}}\right|=0.04$ &
			$\left|\frac{\xi_{\eta_8\eta_8}}{\xi_{K\bar{K}}}\right|=1.11$ & & \\
			\hline
		\end{tabular}
		\caption{{\small The results of  Ref.~\cite{oller.181105.1} for the pole positions [MeV] and moduli of the
				couplings [GeV] of the lightest scalar resonances.
				The couplings to the $\pi\pi$ and $\eta_8\eta_8$
				channels have been multiplied by $\sqrt{2}$ to account for the
				unitary normalization.}
			\label{table.181109.1}}	
	\end{center}
\end{table}

The resulting scalar meson-meson PWAs from Ref.~\cite{oller.181105.1} were employed also in Ref.~\cite{oller.190506.1} to
determine the mixing angle of the aforementioned nonet made by the lightest scalar resonances, because of the mixing between the $I=0$ 
$f_0(500)$ and the $f_0(980)$ resonances. Firstly, this reference studied the continuous movement of the resonance poles by
varying a parameter $\lambda\in[0,1]$ from the physical point ($\lambda=0$) to a $SU(3)$ limit ($\lambda=1$),
with a common pseudoscalar mass of 350~MeV.
For this value Ref.~\cite{oller.190506.1} finds a pole position for the degenerate octet
of scalar resonance at around $600-i\,175$~MeV, and a pole position of the lightest singlet scalar resonances at around
$430-i\,225$~MeV. Further support on the membership of the $a_0(980)$ and $K^*_0(700)$ (or $\kappa$)
to the same octet is given in Ref.~\cite{oller.190506.1} by studying the couplings $\xi(a_0(980)\to \pi\eta_8)$, $\xi(a_0(980)\to K\bar{K})$ and
$\xi(\kappa\to K\pi)$, given in Table~\ref{table.181109.1} (together with other cases obtained from some variance in the modelling of the interactions
\cite{oller.190506.1}). Taking into account the $SU(3)$ Clebsch-Gordan coefficients \cite{oller.190506.1} one has
\begin{align}
\label{190506.1}  
\xi(a_0(980)\to \pi\eta_8)&=\frac{1}{\sqrt{5}}\xi_8~,\\
\xi(a_0(980)\to K\bar{K})&=-\sqrt{\frac{3}{10}}\xi_8~,\nn\\
\xi(\kappa\to K\pi)&=\frac{3}{\sqrt{20}}\xi_8~,\nn
\end{align}
where $\xi_8$ is the coupling to the states of the $SU(3)$ basis. From this analysis Ref.~\cite{oller.190506.1}
finds
\begin{align}
\label{190506.2}
|\xi_8|&=8.7\pm 1.3~{\rm GeV}~,
\end{align}
with a relative uncertainty of a 15\%, well within the expected accuracy of an $SU(3)$ analysis of around a 20\%.

The next step undertaken in Ref.~\cite{oller.190506.1} is to take into account the mixing between the $f_0(500)$ and
the $f_0(980)$, which are written in terms of the singlet and octet states,  $|S_1\rangle$ and $|S_8\rangle$,
respectively, as
\begin{align}
\label{190506.3}
|f_0(500)\rangle&=\cos\theta_S |S_1\rangle + \sin\theta_S|S_8\rangle~,\\
|f_0(980)\rangle&=-\sin\theta_S|S_1\rangle + \cos\theta_S |S_8\rangle~,\nn
\end{align}
with $\theta_S$ the mixing angle of the lightest scalar nonet.
The couplings constants $\xi_8$, $\xi_1$ and the mixing angle $\theta_S$ are fitted to the
set of couplings of the scalar resonances $f_0(500)$, $f_0(980)$, $K^*_0(700)$ and
$a_0(980)$ to the meson-meson channels,
once they are related by applying the Wigner-Eckart theorem
to $SU(3)$ taking into account the $SU(3)$ Clebsch-Gordan coefficients (also provided in Ref.~\cite{oller.190506.1}).
The result of the fit is
\begin{align}
\label{190506.4}
|\xi_8|&=8.2\pm 0.8~{\rm GeV}~,\\
|\xi_1|&=3.9\pm 0.8~{\rm GeV}~,\nn\\
|\theta_S|&=19^{\rm o}\pm 5^{\rm o}~,\nn
\end{align}
with the relative signs fulfilling that ${\rm sign}(\xi_1\xi_8\theta_S)=+1$.
It is noticeable the  agreement within errors between this new determination of $\xi_8$ and
the previous one in Eq.~\eqref{190506.2} by only taking into account the resonances
$a_0(980)$ and $K^*_0(700)$. 

Another interesting evaluation is performed in Ref.~\cite{oller.190506.1} that allows one 
to determine the sign of $\theta_S$ as well as the relative sign between $\xi_8$ and $\xi_1$. This is accomplished by considering
the  coupling of the $f_0(980)$ to the $q\bar{q}$ scalar sources,
$n\bar{n}=(u\bar{u}+d\bar{d})/2$ and $s\bar{s}$, by making use of the
results of Ref.~\cite{oller.180804.1}. The $SU(3)$ decomposition of these
sources reads \cite{oller.190506.1},
\begin{align}
\label{190506.5}
\bar{s}s&=\frac{1}{\sqrt{3}} s_1-\sqrt{\frac{2}{3}} s_8~,\\
\bar{n}n &=\sqrt{\frac{2}{3}} s_1+\frac{1}{\sqrt{3}} s_8~,\nn
  \end{align}
where in an obvious notation, $s_1$ and $s_8$ are the singlet and octet scalar sources.
Therefore, taking into account the mixing in Eq.~\eqref{190506.3} one also has
\begin{align}
\label{190506.6}
\langle 0|\bar{s}s |f_0(980)\rangle&=-\sin \theta_S\frac{1}{\sqrt{3}}
\langle 0|s_1|S_1\rangle-\cos\theta_S \sqrt{\frac{2}{3}}\langle 0|s_8|S_8\rangle~,\nn\\
\langle 0|\bar{n}n |f_0(980)\rangle&=-\sin \theta_S \sqrt{\frac{2}{3}}\langle
0|s_1|S_1 \rangle+\cos\theta_S \frac{1}{\sqrt{3}}\langle 0|s_8|S_8\rangle~.
\end{align}

Using the value of $|\theta_S|$ given in Eq.~\eqref{190506.4} and
the approximate equality $\langle 0|s_1|S_1\rangle \approx \langle 0|s_8|S_8\rangle$, which is expected to be
enough for a semiquantitative estimate  in virtue of the $U(3)$ symmetry, Ref.~\cite{oller.190506.1} finds that
depending on the sign of $\theta_S$ one has that 
\begin{align}
\label{190506.7}
\theta_S=+19^{\rm o}&\rightarrow \left|\frac{f_s}{f_n}\right|=3.5~,\\
\theta_S=-19^{\rm o}&\rightarrow \left|\frac{f_s}{f_n}\right|=0.7~.\nn
\end{align}
where $f_s=\langle 0|\bar{s}s |f_0(980)\rangle$ and $f_n=\langle 0|\bar{n}n |f_0(980)\rangle$.
This exercise clearly favors the positive sign for $\theta_S$ because the $f_0(980)$ couples much more strongly
to the strange source than to the $n\bar{n}$, so that
$|f_s|/|f_n|\simeq 5.5$
\cite{oller.180804.1,oller.190506.1}. Thus, by taking a positive sign for $\xi_8$ the final result of
Ref.~\cite{oller.190506.1} is 
\begin{align}
\label{190506.8}
\xi_8&= 8.2 \pm 0.8 \, \hbox{ GeV}~,\\
\xi_1&= 3.9 \pm 0.8\, \hbox{ GeV} ~,\nn\\
\theta_S&=+19^{\rm o}\pm 5^{\rm o}~.\nn
  \end{align}
This value for lightest scalar mixing angle implies that the $f_0(500)$ is mainly a singlet and the
$f_0(980)$ an octet. It is in agreement with the determination
of Ref.~\cite{napsuciale.190506.1}, and a factor of 2 smaller than the ideal mixing angle.
For other approaches studying the mixing of the light scalar nonet of
resonances see e.g. Refs.~\cite{morgan,jaffe,scad,putative}.

It is of interest to connect the  formalisms developed in Secs.~\ref{sec.181104.1}
and \ref{sec.181109.1}, based on the unitarity and the analytical properties of the PWAs, with the so-called on-shell factorization
of the vertices in a Bethe-Salpeter equation, as developed in Ref.~\cite{oller.181101.2}.\footnote{Our 
	convention for the scattering amplitudes differs in a minus sign with respect to the one used in Ref.~\cite{oller.181101.2}. 
Furthermore, we indicate the different coupled channels with a label that increases as the thresholds associated do.}
In this reference the $L=0$ LO  ChPT amplitudes with $I=0$ and 1 are unitarized by first settling  a Bethe-Salpeter equation 
for the PWA under study. Namely, 
\begin{align}
\label{181125.1}
T(k,p)=V(k,p)-i\int\frac{d^4q}{(2\pi)^4}\frac{V(k,q)T(q,p)}{[q^2-m_1^2][(P-q)^2-m_2^2]}~,
\end{align}
where we indicate a four-momentum (e.g.  of the first particle)
for the initial (right) and final (left argument) states. Any other momentum can be inferred from the conservation 
of the total momentum $P$. In Eq.~\eqref{181125.1}  $V(k,q)$ is the matrix of $S$-wave projected off-shell LO ChPT amplitudes. 
 To illustrate the argument let us take the off-shell LO  ChPT isoscalar $\pi\pi$ scattering amplitude that reads \cite{oller.181101.2} 
\begin{align}
\label{181125.3}
V_{11}(s)&=%\frac{1}{2}\left[3A(s,t,u)+A(t,s,u)+A(u,t,s)\right]\\&=
\frac{1}{9f^2}\left(9s+\frac{15}{2}m_\pi^2-3\sum_{i=1}^4p_i^2\right)~,
\end{align}
which is  already in $S$ wave. 
In Eq.~\eqref{181125.3}   $p_i$ is the four-momentum of any pion and $f$ is the pion decay constant in the chiral limit. 
The main point for the factorization is to realize that the off-shell part in the
LO ChPT meson-meson scattering amplitudes considered, cf. Eq.~\eqref{181125.3}, is of 
 the form $\sum_i c_i (p_i^2-m_i^2)$, where the $c_i$ are constants.
Therefore, the off-shell part cancels one or two of the propagators in the Bethe-Salpeter equation, giving rise to 
 terms  that only contribute to the renormalization of the bare parameters which appear 
at LO \cite{oller.181101.2,nieves.181125.1},
and having values  rather constraint by phenomenology.
In fact, at LO in the chiral expansion these parameters can be taken with their physical values. 
This is why Ref.~\cite{oller.181101.2} gave the matrix of coupled PWAs, $T(s)$, in the form,
\begin{align}
T(s)&=\left[\cN(s)^{-1}+g(s)\right]^{-1} ~,
\label{181125.4}
\end{align}
with $\cN(s)$  the  on-shell LO PWAs in ChPT ($p_i^2=m_i^2$). E.g. for
 $I=L=0$ meson-meson scattering these on-shell PWAs are 
\begin{align}
\label{181125.5}
\cN_{11}(s)&=\frac{1}{f_\pi^2}(s-\frac{m_\pi^2}{2})~,\\
\cN_{12}(s)&=\frac{\sqrt{3}s}{4f_\pi^2}~,\nn\\
\cN_{22}(s)&=\frac{3s}{4f_\pi^2}~,\nn
\end{align}
and $\cN_{21}(s)=\cN_{12}(s)$.

 %%%%%%%%%%%%%%%%%%%%%%%%%%%%%%%%%%%%%%%%%%%%%%%%%%%%%%%%%%%%%%%%%%%%%%%%%%%%%%%
 \section{Near-threshold scattering}
 \label{sec.181202.1}
 \setcounter{equation}{0}   

 Let us consider the validity of non-relativistic scattering to describe the dynamics in a near-threshold energy interval. 
We also assume that  the LHC is  far away and/or weak.
 In the former case the LHC  admits a smooth Taylor expansion in the energy region under study 
 and its contributions can be taken into account without including it explicitly.
 For the later case we assume that the LHC is weak enough so that it can be neglected in good approximation.
For instance, a relevant example is that of the scattering of two heavy-flavor hadrons,
where it is plausible that the exchanges of pions
can be considered in a  perturbative manner~\cite{Fleming:2007rp,pavon.181208.1,Wu:2010jy,Wu:2010vk,Wu:2012md}.
For exchanges of heavier vector resonances, given the fact that their masses are much larger than the typical 
three-momenta involved, their contributions can be effectively included via %0
contact interactions. 
 Within this scenario we can apply the general results deduced in Sec.~\ref{sec.181104.1}, cf. Eq.~\eqref{fin/d}, 
which are derived under the assumption that  the LHC is not realized explicitly.
 
 We  consider the $S$-wave scattering, which for low energies is of special relevance. 
Under such circumstances the general structure of a PWA is given by Eq.~\eqref{fin/d} applied with $L=0$.
% The relativistic phase space in the integral along the RHC is $p(s)/8\pi\sqrt{s}$. 
For non-relativistic scattering it is convenient to introduce the
 kinetic energy $E$ given by its non-relativistic expression, 
 \begin{align}
 \label{181203.1}
 E=\frac{p^2}{2\mu}~,
 \end{align}
such that $\sqrt{s}=m_1+m_2+E+{\cal O}(p^4)$.  
The unitary loop function, corresponding to the RHC integral in Eq.~\eqref{fin/d}, is given algebraically in  Eq.~\eqref{181106.3}. 
 Its power series expansion in $p$  can be easily worked out, and its first terms are
 \begin{align}
 \label{181204.1}
 g(p)&=\frac{a}{16\pi^2}+\frac{1}{8\pi^2(m_1+m_2)}(m_1\log\frac{m_1}{\mu}+m_2\log\frac{m_2}{\mu})
 -i\frac{p}{8\pi(m_1+m_2)}
 +{\cal O}(\frac{p^2}{\Sigma^2})~.
 \end{align}
Here,  $\Sigma$ has dimension of mass and it is made out of $m_1$ and $m_2$. 
In this expansion there are no enhanced terms that could result by the appearance of powers of $m_1/m_2$ with 
$m_1\gg m_2$. 
This is clear  from  the non-relativistic reduction of  the  integral in Eq.~\eqref{181106.3},
 \begin{align}
 \label{181204.2}
 g(p)&=\widetilde{a}+\frac{s}{4\pi^2}\int_{\sqrt{s_{\rm th}}}^\infty\frac{q\,dW}{W^2(s-W^2)}
 =\widetilde{a}+\frac{1}{8\pi(m_1+m_2)}\int_0^\infty\frac{q dq^2}{p^2-q^2}+{\cal O}(\frac{p^2}{\Sigma^2})~,
 \end{align}
 where $\omega_i(q)$ is defined in Eq.~\eqref{181106.5} and $W=\omega_1(q)+\omega_2(q)$. 
 
 Let us adopt the more standard non-relativistic normalization of a PWA $t(E)$, such that
 \begin{align}
 \label{190212.1}
 t(E)&=\frac{1}{p\cot\delta-i p}~.
 \end{align}
 We also denote by $\beta$ the constant term on the rhs of Eq.~\eqref{181204.1}
 times $8\pi(m_1+m_2)$, because of the change in normalization, so that
 \begin{align}
 \label{181204.3}
 \beta&=\frac{a(m_1+m_2)}{2\pi}+\frac{1}{\pi}(m_1\log\frac{m_1}{\mu}+m_2\log\frac{m_2}{\mu})~.
 \end{align}
 Considering the master Eq.~\eqref{fin/d} with $L=0$,  other relevant 
 structures in the near-threshold region, apart from the threshold branch-point singularity, 
 result from the CDD pole contributions. 
 Following Ref.~\cite{kang.181206.1}, let us explore the addition of one of them, so that Eq.~\eqref{fin/d} is recast as
 \begin{align}
 \label{181204.4}
 t(E)&=\left(\frac{\gamma}{E-M_{\rm CDD}}+\beta-ip(E)\right)^{-1}~.
 \end{align}
The CDD pole is characterized by its residue $\gamma$ and its position in energy $M_{\rm CDD}$. 
 
The Eq.~\eqref{181204.4} is derived from two  basic analytical properties of the PWAs, 
namely, the  RHC and a pole in the inverse of the PWA.
This equation goes beyond an ERE, up to and including ${\cal O}(p^4)$, cf. Eq.~\eqref{181113.4}. 
In this regard, let us notice first that any values for $a$, $r_2$ and $v_2$ in the ERE can be 
reproduced by choosing adequately $\gamma$, $\cdd$ and $\beta$, with the explicit expressions
 \begin{align}
 \label{181204.5}
 \frac{1}{a}&=\frac{\gamma}{\cdd}-\beta~,\\
 r&=-\frac{\gamma}{\mu\cdd^2}~,\nn\\
 v_2&=-\frac{\gamma}{4\mu^2\cdd^3}~.\nn
 \end{align}
However, the presence of a CDD pole implies that the ERE has a radius of convergence in $p^2$ less than 
$2\mu |\cdd|$,  because $p\cot\delta$ becomes infinite at its position.
 Thus, if the zero in the PWA generated by the CDD pole occurs very close to the threshold 
 the region of validity of the ERE is reduced accordingly.
This fact typically invalidates the ERE  as an adequate tool to study the corresponding near-threshold scattering. 

 An important output of Eq.~\eqref{181204.5} is that a near-threshold CDD pole,
 $\cdd\to 0$, gives rise to an $r$ of large absolute value  ($\gamma$ can take any sign). 
 This is also the case for the other higher-order shape parameters in the ERE $v_i$ with $\geq 2$, e.g. for $v_2$ 
as explicitly shown in Eq.~\eqref{181204.5}.  However, the  scattering length $a$ tends to zero as $\cdd/\gamma$
  in the same limit. Of course, the larger $ \gamma $, the sooner this scenario takes place. 
 
 Another parameterization  used in the literature to describe near-threshold resonances
 is the Flatt\'e parameterization \cite{flatte.181204.1},  $t_F(E)$. It is defined as
 \begin{align}
 \label{181206.1}
 t_F(E)&=\frac{g^2/2}{M_F-E-i\frac{1}{2}\Gamma(E)}~,\\
 \Gamma(E)&=g^2 p(E)~,\hspace{0.4cm} E>0~,\nn\\
 \Gamma(E)&=ig^2 |p(E)|~,E<0~,\nn
 \end{align}
 and the coupling squared  $g^2\geq 0$, so that the width $\Gamma(E)\geq 0$ for $E>0$. 
 The so-called Flatt\'e mass $M_F$ is the energy at which the real part of $t_F(E)^{-1}$ becomes zero.
 The energy dependence of the width is driven by its linear dependence on the three-momentum. 
 
 It is worth noticing that $t_F(E)$  is a particular case of an ERE up to and including ${\cal O}(p^2)$. 
 The relationship between $a$, $r$ and the mass and coupling of a Flatt\'e parameterization is
 \begin{align}
 \label{181206.2}
 a&=-\frac{g^2}{2M_F}~,\\
 r&=-\frac{2}{g^2\mu}~.\nn
 \end{align}
Importantly, a restriction on the Flatt\'e parameterization is that it can only reproduce  negative values for the effective range, $r<0$.
 The scattering length is proportional to $1/M_F$ and it becomes infinity when $M_F$ is zero, in which case  $t_F(0)$ is  also 
 infinity. In this respect, the  pole content of $t_F(E)$ exhibits a qualitatively different behavior depending on the sign of $M_F$. 
 Solving for the poles of $t_F(E)$, we can write the latter as
 \begin{align}
 \label{181206.3}
 t_F(E)&=\frac{-\mu g^2}{(p(E)-p_1)(p(E)-p_2)}~,\\
 \label{181206.3b}
 p_{1,2}&=-i\frac{g^2\mu}{2}\left(1\pm\sqrt{1-\frac{8M_F}{g^4\mu}}\right)~,
 \end{align}
 where 1(2) corresponds to the $+(-)$ in front of the square root, respectively.
 In the range $M_F< 0$ both momenta are purely imaginary but having opposite sign so that 
 $p_1$ corresponds to a virtual state and $p_2$ to a bound state. Furthermore, $|p_1|>|p_2|$ and 
 then the former state is deeper than the later. There is a transition when $M_F=0$, so that 
 the bound state is at threshold, and for  $0<M_F<g^4\mu/8$ it becomes into another virtual state closer to threshold 
 than the one corresponding to $p_2$. 
 
 For $M_F\to g^4\mu/8$ another transition occurs, as the radicand in Eq.~\eqref{181206.3b} tends to vanish, and the two poles  
 merge giving rise to a double virtual-state pole \cite{kang.181206.1}. 
 For $M_F>g^4\mu/8$ the pole positions have the same negative imaginary part but they acquire a real part, which 
 is the same in absolute value but with opposite sign. 
 This is a consequence of the existence of two poles in complex conjugate positions in the complex $E$ plane, as 
 required by the  Schwarz theorem. 
 Another  limitation of the  Flatt\'e model is that it cannot give rise to higher than double poles. 
We also conclude from Eq.~\eqref{181206.1} that in the limit $\Gamma(M_F)\ll M_F$ the nearest pole position to the physical axis ($s+i\ve$ with $\sqrt{s}$ above threshold)  is 
 \begin{align}
 \label{181206.4}
 E_F&\simeq M_F-i\frac{\Gamma(M_F)}{2}~,
 \end{align}
that corresponds to the narrow-resonance case.

In the complex $E$ plane the pole positions  are  $E_{1,2}=p_{1,2}^2/2\mu$,
 with the $p_{1,2}$  given in Eq.~\eqref{181206.3b}. The expressions for $E_{1,2}$ are then 
 \begin{align}
 \label{181206.5}
 E_{1,2}&=M_F-\frac{g^4 \mu}{4}\mp i\frac{g^4\mu}{4}\sqrt{\frac{8M_F}{g^4\mu}-1}~,~M_F>\frac{g^4\mu}{8}~, 
 \end{align}
  Identifying the resonance mass, $M_R$,  as $\Re E_{1,2}$, it results that 
 \begin{align}
 \label{190109.1b}
 M_R&=M_F-\frac{g^4 \mu}{4}~,
 \end{align}
being the variation of $M_R$ with respect to $M_F$  due to the self-energy, $-g^4\mu/4$. 
 On the other hand, defining the width of the resonance, $\Gamma$, as twice the modulus of the imaginary part of $E_{1,2}$, 
 one has
 \begin{align}
 \label{181206.6}
 \Gamma&=\frac{g^4\mu}{2}\sqrt{\frac{8M_F}{g^4\mu}-1}~.
 \end{align}
 The previous expression for $\Gamma$ is equal to $g^2\sqrt{2\mu M_R}$, 
 cf. Eq.~\eqref{181206.1}, only for $M_F\gg g^4\mu$. 
 In this situation,  $-2\Im E_F$ in Eq.~\eqref{181206.4} is also close to $\Gamma$ in good approximation. 
As a result, the  narrow-resonance limit requires that  $M_F\gg g^4\mu$. 
 
The couplings of the poles are defined as their residue, either in the complex momentum or energy spaces, as
 \begin{align}
 \label{181206.7}
 \gamma_k^2&=-\lim_{p\to p_i}(p-p_i)t_F(p^2/2\mu)~,\\
 \label{181206.7b}
 \gamma_E^2&=-\lim_{E\to E_P}(E-E_i)t_F(E)~,
 \end{align}
 respectively. They are related by 
 \begin{align}
 \label{181206.8}
 \gamma_E^2&=\gamma_k^2\left.\frac{dE}{dp}\right|_{p_{1,2}}=g_k^2\frac{p_{1,2}}{\mu}~.
 \end{align}
The residue $\gamma_k^2$ from Eq.~\eqref{181206.3} is 
 \begin{align}
 \label{181206.9}
 \gamma_k^2&=\pm \frac{\mu g^2}{p_1-p_2}=\pm\frac{1}{\sqrt{\frac{8M_F}{g^4\mu}-1}}~,
 \end{align}
 where the sign $+(-)$ applies to  the pole at $p_{1}(p_2)$. 
In the narrow-resonance case, $M_F\gg g^4\mu/8$,  $\gamma_k^2\to 0$, while it diverges 
in the limit  $M_F\to g^4\mu/8$, because there is then a double virtual-state pole \cite{kang.181206.1}. 
 
Let us notice that for $\Re E_{1,2}\geq 0$ one has that  $1\geq |\gamma_k|^2\geq 0$. 
Indeed, Refs.~\cite{guo.181206.1,kang.181205.1} develop a probabilistic interpretation for $|\gamma_k|^2$ 
applicable in this situation, such that $|\gamma_k|^2$ is called the compositeness of the resonance. 
It would correspond to the  weight in the composition of the resonance of the two-body states in the continuum. 
 
% As noted above, the effective range $r$ for a Flatt\'e parameterization must be negative, cf. Eq.~\eqref{181206.2}. 
The ERE for a PWA $t(p^2)$ up to and including $p^2$,  $t_r(E)$, is given by  
 \begin{align}
 \label{181206.10}
 t_{r}(E)&=\frac{1}{-\frac{1}{a}+\frac{1}{2}rp(E)^2-ip(E)}~.
 \end{align}
Because of  the quadratic dependence on $p$ of the denominator in the previous equation, 
 we have two poles located at
 \begin{align}
 \label{181206.11}
 p_{1,2}=\frac{1}{r}\left(i\mp\sqrt{\frac{2r}{a}-1}\right)~.
 \end{align}
Resonance poles emerge when
 \begin{align}
 \label{181206.12}
 r/a>1/2~ \text{and}~ r<0~.
 \end{align}
If these requirements are applied to the expressions of $a$ and $r$ in Eq.~\eqref{181206.2} for a Flatt\'e parameterization,  
we re-derive that  $M_F>g^2\mu/8$ in order to end with resonance poles. 
The expression for the residues  of $t_r(E)$ at the poles in Eq.~\eqref{181206.11} is
 \begin{align}
 \label{181206.13}
 \gamma_k^2&=\frac{1}{r p_{1,2}-i}=\mp\frac{1}{\sqrt{\frac{2r}{a}-1}}~.
 \end{align}
The pole positions in energy are
 \begin{align}
 \label{181206.15}
 E_{1,2}&=\frac{p_{1,2}^2}{2\mu}=\frac{1}{ar\mu}\left(1-\frac{a}{r}\right)\mp i \frac{1}{r^2\mu}
 \sqrt{\frac{2r}{a}-1}~.
 \end{align}
 So that $\Re E_{1,2}\geq 0$ for 
 \begin{align}
 \label{181206.14}
 \frac{r}{a}\geq 1~.
 \end{align}
 This is also the constraint for the probabilistic interpretation of
 $|\gamma_k|^2$ \cite{guo.181206.1,kang.181205.1},  $\gamma_k^2\in[0,1]$ as follows from Eq.~\eqref{181206.13}.  
Of course, this constraint in the case of a Flatt\'e parameterization translates into the requirement that $M_F\geq g^2\mu/4$, 
which was already discussed in this regard. 
We can also read the width of the resonance from the imaginary part of $E_{1,2}$ given in \eqref{181206.15}, 
 \begin{align}
 \label{181206.16}
 \Gamma&=\frac{2}{r^2\mu}\sqrt{\frac{2r}{a}-1}~.
 \end{align}

The values of $a$ and $r$ for one CDD pole included are given in Eq.~\eqref{181204.5}. It follows that as 
$\cdd\to 0$ the ratio $|r/a|$  increases and $|\gamma_k|^2$ decreases, cf. Eq.~\eqref{181206.13}.
In virtue of the probabilistic interpretation of $\gamma_k^2$ for a resonance, one interprets that  the scenario 
$\cdd\to 0$ corresponds to a purely `elementary' state (the weight in the resonance state of the 
asymptotic two-body states tends to vanish).\footnote{It is necessary to have a negative $\cdd$ and a positive $\gamma$ in the limit $\cdd\to 0$ so as to fulfill the requirements in Eq.~\eqref{181206.12}.}
 According to Eq.~\eqref{181206.15} when $\cdd\to 0$  the resonance poles tend to 
 \begin{align}
 \label{181206.17}
 E_{1,2}&\xrightarrow[\cdd\to 0]{} -\frac{\cdd^3}{\lambda^2}\mp i\frac{(-\cdd)^{7/2} \sqrt{2\mu}}{\lambda^2}~.  
 \end{align}
As a result,  the width vanishes more rapidly than the mass by an extra  factor $(-\cdd)^{1/2}$, and 
the narrow-resonance limit holds.  
 One can think that the decoupling limit of a bare `elementary' resonance from the two-body continuum 
requires a zero in the PWA in order to guarantee the removal of the bare resonance propagator from $t(E)$. 
This offers an intuitive picture relating a weakly coupled resonance and the occurrence of a nearby CDD pole.
 Of course, this is connected with the common association of a CDD pole explicitly introduced 
 in the equations, and the exchange of an `elementary' bare resonance.
See e.g. Ref.~\cite{dyson.190510.1} where Dyson constructs 
a model in which each one of the extra solutions to the Low's scattering equation, and associated with CDD poles, 
may be the correct physical one. 
 
For $r/a\to 1$ we have that $\gamma_k^2\to 1$, and it follows from Eq.~\eqref{181206.15} 
that the difference between the mass of the resonance and the threshold tends to vanish.
This indicates that a resonance which is purely composite of the 
asymptotic two-body states has a width which is much larger than this difference \cite{kang.181205.1}.
 
Let us stress an important point.  
The Eq.~\eqref{181204.4} is more general than an ERE up to and including
 ${\cal O}(p^4)$, because the latter cannot be applied beyond the position of a near-threshold zero, 
 and this fact may limit severely its application. 
It follows then also that Eq.~\eqref{181204.4} is more general than a Flatt\'e parameterization, 
because the latter is a particular case of an ERE up to ${\cal O}(p^2)$ and, furthermore, it can only 
give rise to a negative effective range.  

%%%%%%%%%%%%%%%%%%%%%%%%%%%%%%%%%%%%%%%%%%%%%%%%%%%%%%
\subsection{Potential scattering}
\label{190121.2}

%Let us show explicitly how  unitarity results in potential scattering.
For more details on the solution of a Lippmann-Schwinger equation given a potential
the reader can consult section 2 of Ref.~\cite{oller.181101.1} or the monograph \cite{faddeev.170929.1}.
We split the Hamiltonian $H$ in a free, $H_0$, and an interacting part, $v$, such that 
$H=H_0-v$. The minus sign in front of $v$  is introduced so as to conform with the sign convention 
employed for the relation between the $T$ and $S$ matrices, cf.  Eq.~\eqref{181101.4}. 
%When the interacting part is denoted with a capital letter,  $V$,  then it corresponds to the ordinary splitting $H=H_0+V$. 

Let us denote  by $r_0(z)$ and $r(z)$ the resolvents of $H_0$ and $H$, 
\bea
\label{180803.1}
r_0(z)&=&\left(H_0-z\right)^{-1}~,\nn\\
r(z)&=&\left(H-z\right)^{-1}~,
\ea
in order, with $\Im z\neq 0$. 
The following equations for $r(z)$ can be obtained rather straightforwardly \cite{oller.181101.1,faddeev.170929.1}
\bea
\label{180803.2}
r(z)&=&r_0(z)+r_0(z)v r(z)\\
\label{180803.3}
&=&r_0(z)+r(z) v r_0(z)~. 
\ea

The $T$-matrix operator $T(z)$ is introduced via its relation with the resolvent $r(z)$, and it is defined by 
\be
\label{180803.4}
T(z)r_0(z)=v r(z)~.
\ee
As a consequence of this definition and of Eqs.~\eqref{180803.2} and \eqref{180803.3} one also has that
\be
\label{180803.6}
r_0(z) T (z)=r(z) v~.
\ee
 The Lippmann-Schwinger (LS) equation for $T(z)$  follows from Eqs.~\eqref{180803.4} and \eqref{180803.2}, 
with the latter one multiplied to the left by $v$, so that
\be
\label{180803.8}
T(z)=v+v r_0(z) T(z)~.
\ee
This equation can also be written as
\be
\label{180803.8b}
T(z)=v+T(z) r_0(z) v~.
\ee
as follows if we had used instead Eqs.~\eqref{180803.3} and \eqref{180803.6}.
 An interesting property of $T(z)$ is that 
\be
\label{180803.9}
T(z)^\dagger=T(z^*)
\ee
which follows from the fact that $r(z)^\dagger=r(z^*)$.\footnote{The potential is also required to fulfill this property, $v(z)^\dagger=v(z^*)$.} The Eq.~\eqref{180803.9} also illustrates the Hermitian analyticity \cite{olive.181102.1} in 
potential scattering, which was already introduced in Sec.~\ref{181024.1}.

There is an interesting relation between two resolvent operators calculated with different 
values of  $z$, known as the Hilbert identity, which reads
\be
\label{180803.10}
r(z_1)-r(z_2)=(z_1-z_2)r(z_1)r(z_2)~.
\ee
Its demonstration readily follows from difference
$r(z_2)^{-1}-r(z_1)^{-1}=z_1-z_2$, which is then 
multiplied  to the left by $r(z_1)$ and to the right by $r(z_2)$. 
The Hilbert identity becomes in terms of the $T$ matrix 
\be
\label{180803.12}
T(z_1)-T(z_2)=(z_1-z_2)T(z_1)r_0(z_1)r_0(z_2)T(z_2)~,
\ee
as it turns out by the multiplication of Eq.~\eqref{180803.10} on the left and right by $v$, and subsequent use of the LS equation.
This relation is very useful to derive the unitarity properties of the $T$ matrix in PWAs.

Given the three-dimensional nature of the LS equation it is convenient  to keep the Dirac delta function 
of conservation of energy within the normalization of the two-body states projected in a  partial wave, 
$|k,\ell S,J\mu \rangle$, with $k$ the modulus of the three-momentum.\footnote{This is why in standard text books on Quantum Mechanics a $\delta(E'-E)$ is typically 
introduced in the relationship between the $S$ and $T$ matrices, cf. Eq.~\eqref{181101.4}.}
 For brevity in the exposition we typically  denote the discrete indexes simply by $\lambda$, so that $|k,\lambda\rangle$ 
stands for the same state. These states are normalized as
\be
\label{180804.1}
\langle k,\lambda|k',\lambda'\rangle=2\pi^2\frac{\delta(k-k')}{k^2}\delta_{\lambda\lambda'}~.
\ee
The normalization in Eq.~\eqref{180804.1} is consistent with the non-relativistic limit of Eq.~\eqref{061016.3}, 
with $\sqrt{s}=p_1^0+p_2^0$ replaced by $m_1+m_2$. 
The latter equation should also be  multiplied by $2\pi\delta({k}^2/2\mu-{k'}^2/2\mu)=2\pi\mu\delta(k-k')/k$ and divided by  $4m_1m_2$. 
This last factor arises  because in a non-relativistic theory a plane wave is normalized 
as $(2\pi)^3\delta(\vk-\vk')$, without the factor $2 p_i^0$ present in Eq.~\eqref{181031.2} (which in the non-relativistic limit just becomes $2m_i$). Notice that now we have indicated by $\mu$ the reduced mass of the two particles,
\begin{align}
\label{190115.1}
\mu&=\frac{m_1m_2}{m_1+m_2}~.
\end{align}

The matrix elements of the scattering operator $T$ between partial-wave projected states are the PWAs, 
which are denoted by $T_{ij}(k,k';z)$ and correspond to  
\be
\label{180804.4}
T_{ij}(k,k';z)=\langle k,\lambda_i|t(z)|k',\lambda_j\rangle~.
\ee
Similarly, we also introduce $v_{ij}(k,k')$ defined as  
\be
\label{180804.6}
v_{ij}(k,k')=\langle k,\lambda_i|v|k',\lambda_j\rangle~.
\ee

The two forms of the LS equation in partial waves are obtained by taking the matrix elements 
of Eqs.~\eqref{180803.8} and \eqref{180803.8b} between partial-wave states, 
\bea
\label{180804.5}
T_{ij}(k,k';z)&=&v_{ij}(k,k')+\frac{\mu}{\pi^2}\sum_{n}  \int_0^\infty \frac{dq q^2}{q^2-2\mu z}v_{in}(k,q)T_{nj}(q,k';z)\nn\\
&=&v_{ij}(k,k')+\frac{\mu}{\pi^2}\sum_{n} \int_0^\infty \frac{dq q^2}{q^2-2\mu z}T_{in}(k,q;z)v_{nj}(q,k')~,
\ea
in order. In these equations we have included between
the operators $T(z)$ and $v$ an intermediate set of two-body partial-wave states $|q,\lambda_j\rangle$.

For the Hilbert identity in partial waves we have from Eq.~\eqref{180803.12} that 
\bea
\label{180804.7}
T_{ij}(k,k';z_1)-T_{ij}(k,k';z_2)&=&(z_1-z_2) \frac{2\mu^2}{\pi^2} 
\sum_n\int_0^\infty \frac{dq q^2}{(q^2-2\mu z_1)(q^2-2\mu z_2)} \nn\\ 
&\times& T_{in}(k,q;z_1)T_{nj}(q,k';z_2)~.
\ea

We refer to Refs.~\cite{oller.181101.1,oller.190503.1} for a detailed derivation
of the off-shell unitarity relation of PWAs  from the Hilbert identity, which can be written as
\be
\label{180804.14}
\Im T_{ij}(k,k';z)=\theta(E) \frac{\mu \kappa}{2\pi} \sum_n T_{in}(k,\kappa;z)T_{jn}(k',\kappa;z)^*~,
\ee
with $\kappa$ the modulus of the on-shell three-momentum momentum, 
\be
\label{180804.13}
\kappa=\sqrt{2\mu \Re z}~.
\ee
Two important particular cases of Eq.~\eqref{180804.14} are the half-off-shell and on-shell unitarity relations.
For the former, one takes  $E={k'}^2/2\mu$ (so that $\kappa=k'$) in Eq.~\eqref{180804.14}, which then reads
\be
\label{180804.15}
\Im T_{ij}(k,\kappa;E+i\vep)=\theta(E) \frac{\mu \kappa}{2\pi} \sum_n  
T_{in}(k,\kappa;E+i\vep)T_{jn}(\kappa,\kappa;E+i\vep)^*~.
\ee
This is of the same type as the unitarity relation for form factors \cite{oller.180804.1,basdevant.181119.1,martin.290916.1},
cf. Secs.~\ref{sec.181117.2} and \ref{sec.181117.3}.

The on-shell unitarity relation, or simply unitarity, arises by taking  furthermore that $k=\kappa$ . 
The Eq.~\eqref{180804.15} then becomes
\be
\label{180804.16}
\Im T_{ij}(\kappa,\kappa;E+i\vep)=\theta(E)\frac{\mu \kappa}{2\pi}\sum_n  
T_{in}(\kappa,\kappa;E+i\vep)T_{jn}(\kappa,\kappa;E+i\vep)^*~.
\ee
This imaginary part implies the right-hand cut (RHC) or unitarity cut in the PWAs for positive real values of the 
kinetic energy.  Because of the Hermitian analyticity, Eq.~\eqref{180803.9}, and the symmetric character of  the 
PWAs due to time reversal, it follows then that $T_{ij}(\kappa,\kappa,E+i\ve)-T_{ij}(\kappa,\kappa,E-i\ve)=2i\Im T_{ij}(\kappa,\kappa,E+i\ve)$, with the latter given by $2i$ times the rhs of Eq.~\eqref{180804.16}.

% Let us note that in order to deduce the presence of the RHC in the PWA one does not really need to invoke time-reversal 
% invariance, since we can apply directly Eq.~\eqref{180803.9} with $z_1=z_2^*=E+i\ve$  and, proceeding 
% analogously, we would have that 
%\be
%\label{180804.16n}
%T_{ij}(\kappa,\kappa,E+i\vep)-T_{ij}(\kappa,\kappa,E-i\vep)=2i\theta(E)\frac{\mu \kappa}{2\pi}\sum_n  
%T_{in}(\kappa,\kappa;E+i\vep)T_{jn}(\kappa,\kappa;E+i\vep)^*~.
%\ee

The partial-wave decomposition of the $S$ matrix, cf. Eq.~\eqref{051016.13}, is
\be
S_{ij}(E)=\delta_{ij}+i\frac{\mu \kappa}{\pi}T_{ij}(\kappa,\kappa;E+i\vep)~,
\label{130116.1} 
\ee
which is symmetric if $T$ is so. 
The unitarity in partial waves, Eq.~\eqref{180804.16}, implies that $S$ is a unitary operator for $E\geq 0$, so that  
\be
S(E) S(E)^\dagger=S(E)^\dagger S(E)=I~,~E\geq 0~.
\label{130116.3}
\ee

%______ Crossing _____________________________
In a non-relativistic theory, the quantum fields only contain annihilation operators (or creation ones
for the Hermitian conjugate field, chapter 5 of Ref.~\cite{thirring.181101.1}) and there is no crossing symmetry.
Nevertheless, the scattering amplitudes still could have  a LHC  due to the particles exchanged which make up the
potential.  For instance, let us take a  Yukawa potential
\begin{align}
\label{181104.1}
V(r)&=\alpha\frac{ e^{-r m_\pi}}{r}~.  
\end{align}
Its Fourier transform is
\begin{align}
\label{181104.2}
V(\vq^2)&=\alpha\int d^3 r e^{-i\vq \vr}\frac{e^{-r m_\pi}}{r}=\frac{4\pi\alpha}{\vq^2+m_\pi^2}~,
\end{align}
with $\vq=\vp'-\vp$,  the momentum transfer. Its partial-wave projection for particles
with zero spin is 
\begin{align}
\label{181104.3}
V_J(p,p')&=\frac{1}{2}\int_{-1}^{+1}d\cos\theta \,V(\vq^2)\,P_J(\cos\theta)\\
&=-\frac{\pi \alpha}{pp'}\int_{-1}^{+1}d\cos\theta \frac{P_J(\cos\theta)}{\cos\theta-(p^2+{p'}^2+m_\pi^2)/(2pp')}~. \nn
\end{align}
In the following we denote by $\xi$ the combination
\begin{align}
\xi&=\frac{p^2+{p'}^2+m_\pi^2}{2pp'}~.
\end{align}
The LHC results from the zeroes in the denominator of Eq.~\eqref{181104.3}. 
In order to make more explicit its appearance we rewrite Eq.~\eqref{181104.3} as 
\begin{align}
\label{181104.4}
V_J(p,p')&%=-\frac{\pi \alpha}{pp'}\int_{-1}^{+1}d\cos\theta \frac{P_J(\cos\theta)}{\cos\theta-\xi}
=-\frac{\pi \alpha}{pp'}\int_{-1}^{+1}d\cos\theta \frac{P_J(\cos\theta)-P_J(\xi)}{\cos\theta-\xi}
-\frac{\pi \alpha}{pp'}P_J(\xi)\int_{-1}^{+1} \frac{d\cos\theta}{\cos\theta-\xi}~.
\end{align}
It is clear that the LHC stems from the last term in the previous equation, since the term before it has a finite 
limit $\xi\to \cos\theta$. 
The integration over $\cos\theta$ giving rise to the LHC can be done explicitly with the result
\begin{align}
\label{181104.5}
-\frac{\pi \alpha}{pp'}P_J(\xi)\int_{-1}^{+1} \frac{d\cos\theta}{\cos\theta-\xi}
&=-\frac{\pi \alpha}{pp'}P_J(\xi)\left[
\log(1-\xi)-\log(-1-\xi)
\right]~.
\end{align}
For physical values of $p$ and $p'$ the difference of logarithms can be written 
as
\begin{align}
\label{181104.6}
\frac{\pi \alpha}{pp'}P_J(\xi)\left[
\log((p+p')^2+m_\pi^2)-\log((p-p')^2+m_\pi^2) \right]~.
\end{align}
This expression allows us to perform the analytical continuation to complex values of $p$ and $p'$ in 
much simpler terms than its original integral representation. %, cf. Eq.~\eqref{181104.5}. 
This idea has been introduced and fully exploited in Ref.~\cite{oller.181101.1}.

It is then clear from Eq.~\eqref{181104.6} that the cuts in the $p$ variable for a given $p'$
happens when $(p+p')^2+m_\pi^2<0$, first logarithm, or when $(p-p')^2+m_\pi^2<0$, second logarithm. 
This translates into the vertical cuts in the complex $p$ plane
\begin{align}
\label{181104.7}
p=(\pm)p'\pm i \sqrt{m_\pi^2+x^2}~,~x\in \mathbb{R}~,
\end{align}
where the first and second $\pm$ symbols are uncorrelated with each other. 
An analogous relation can be also derived for the cuts in the variable variable $p'$ in its complex plane as a function of $p$

The Ref.~\cite{oller.181101.1} is able to demonstrate that the cuts in the complex momenta of a half-off shell PWA lie along
the vertical cuts of Eq.~\eqref{181104.7} too, as in the much simpler case of the potential.
It also derives a linear integral equation that allows one to calculate exactly the discontinuity across those cuts of an on-shell PWA,
both for regular and singular potentials. When this exact discontinuity along the LHC is implemented within the
$N/D$ method, the later turns out to be exact, providing then solutions that agree with those obtained
with the LS equation when the minimum
number of subtractions is taken in the $N/D$ method.
However, it also has the capability of obtaining new solutions that do not stem from a LS equation, when the latter  
implements short-range dynamics through contact interactions in the potential \cite{entem.190510.1}.\footnote{We refer here to 
regulator independent solutions.}

For on-shell scattering, $p=p'$, a cut stems when taking the minus sign in the first $\pm$  in Eq.~\eqref{181104.7}. 
 One then obtains that the cut corresponds to the values of $p$ given by 
\begin{align}
\label{181104.9}
p=\pm \frac{i}{2}\sqrt{m_\pi^2+x^2}~.
\end{align}
In the complex variable $p^2$ we have a LHC running through the values
\begin{align}
\label{181104.10}
-\infty<p^2\leq -\frac{m_\pi^2}{4}~.
\end{align}
E.g. this is the LHC that occurs when studying non-relativistic $NN$ scattering \cite{oller.181101.1}.
 The $LSJ$ partial waves are function of the variable $p^2$, as follows by imposing parity invariance
of the $T$ matrix,
\begin{align}
\label{181104.11}
P T P&=T~.
\end{align}
Taking this relation into Eqs.~\eqref{051016.6} and \eqref{190122.1}, one concludes that  
\begin{align}
\label{180811.1}
T_{ij}(-p)=\tilde{\eta}_i\tilde{\eta}_j(-1)^{\ell_i+\ell_j} T_{ij}(p)=T_{ij}(p)~,
\end{align}
with  $\tilde{\eta}_i\tilde{\eta}_j (-1)^{\ell_i+\ell_j}=+1$, being $\tilde{\eta}_i$ the product of the intrinsic
parities of the particles making the $i_{{\rm th}}$ partial wave.

 %%%%%%%%%%%%%%%%%%%%%%%%%%%%%%%%%%%%%%%%%%%%%%%%%%%%%%
 \subsection{LS-based parameterizations in non-relativistic scattering}
 \label{sec.190210.1}
 
\begin{figure}
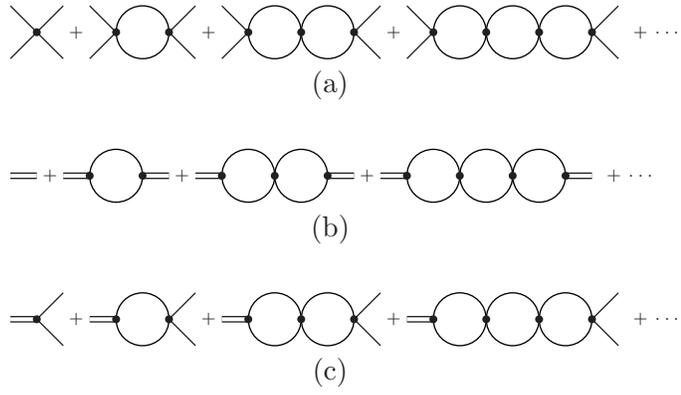

 	\begin{center}
 		\begin{tabular}{c}
 			\begin{axopicture}(200,50)(0,-30)
 				\Vertex(10,20){1.5} %Single Vertex
 				\Line(0,30)(10,20)
 				\Line(0,10)(10,20)
 				\Line(10,20)(20,30)
 				\Line(10,20)(20,10)
 				\Text(25,20)[c]{{\tiny $+$}} %Once-iterated vertex
 				\GCirc(50,20){10}{1} 
 				\Vertex(40,20){1.5}
 				\Vertex(60,20){1.5}
 				\Line(30,30)(40,20)
 				\Line(30,10)(40,20)
 				\Line(60,20)(70,30)
 				\Line(60,20)(70,10)
 				\Text(75,20)[c]{{\tiny $+$}} %Twice-iterated vertex
 				\GCirc(100,20){10}{1}
 				\GCirc(120,20){10}{1}
 				\Vertex(90,20){1.5}
 				\Vertex(110,20){1.5}
 				\Vertex(130,20){1.5}
 				\Line(80,30)(90,20)
 				\Line(80,10)(90,20)
 				\Line(130,20)(140,30)
 				\Line(130,20)(140,10)
 				\Text(145,20)[c]{{\tiny $+$}} %Three-times iterated vertex
 				\GCirc(170,20){10}{1}
 				\GCirc(190,20){10}{1}
 				\GCirc(210,20){10}{1}
 				\Vertex(160,20){1.5}
 				\Vertex(180,20){1.5}
 				\Vertex(200,20){1.5}
 				\Vertex(220,20){1.5}
 				\Line(150,30)(160,20)
 				\Line(150,10)(160,20)
 				\Line(220,20)(230,30)
 				\Line(220,20)(230,10)
 				\Text(245,20)[c]{{\tiny $+\,\cdots$}}
 				\Text(121,-0)[c]{{\small (a)}}
 			\end{axopicture}
 			\\
 			\begin{axopicture}(200,50)(0,-30)
 				\Line[double](0,20)(10,20) %single propagator
 				\Text(15,20)[c]{{\tiny $+$}}
 				\Line[double](20,20)(30,20) %double propagator
 				\Line[double](50,20)(60,20)
 				\GCirc(40,20){10}{1}
 				\Vertex(30,20){1.5}
 				\Vertex(50,20){1.5}
 				\Text(65,20)[c]{{\tiny $+$}}  %triple propagator
 				\GCirc(90,20){10}{1}
 				\GCirc(110,20){10}{1}
 				\Line[double](70,20)(80,20)
 				\Line[double](120,20)(130,20)
 				\Vertex(80,20){1.5}
 				\Vertex(100,20){1.5}
 				\Vertex(120,20){1.5}
 				\Text(135,20)[c]{{\tiny $+$}}  %Four propagators
 				\GCirc(160,20){10}{1}
 				\GCirc(180,20){10}{1}
 				\GCirc(200,20){10}{1}
 				\Line[double](140,20)(150,20)
 				\Line[double](210,20)(220,20)
 				\Vertex(150,20){1.5}
 				\Vertex(170,20){1.5}
 				\Vertex(190,20){1.5}
 				\Vertex(210,20){1.5}
 				\Text(235,20)[c]{{\tiny $+\,\cdots$}}
 				\Text(121,-0.)[c]{{\small (b)}}
 			\end{axopicture}
 			\\
 			\begin{axopicture}(200,50)(0,-30)
 				\Line[double](0,20)(10,20) %single propagator
 				\Vertex(10,20){1.5}
 				\Line(10,20)(20,10)
 				\Line(10,20)(20,30)
 				\Text(25,20)[c]{{\tiny $+$}}
 				\Line[double](30,20)(40,20) %Two propagators
 				\GCirc(50,20){10}{1}
 				\Vertex(40,20){1.5}
 				\Vertex(60,20){1.5}
 				\Line(60,20)(70,10)
 				\Line(60,20)(70,30)
 				\Text(75,20)[c]{{\tiny $+$}}
 				\Line[double](80,20)(90,20) %Three propagators
 				\GCirc(100,20){10}{1}
 				\GCirc(120,20){10}{1}
 				\Vertex(90,20){1.5}
 				\Vertex(110,20){1.5}
 				\Vertex(130,20){1.5}
 				\Line(130,20)(140,10)
 				\Line(130,20)(140,30)
 				\Text(145,20)[c]{{\tiny $+$}}
 				\Line[double](150,20)(160,20) %Four propagators
 				\GCirc(170,20){10}{1}
 				\GCirc(190,20){10}{1}
 				\GCirc(210,20){10}{1}
 				\Vertex(160,20){1.5}
 				\Vertex(180,20){1.5}
 				\Vertex(200,20){1.5}
 				\Vertex(220,20){1.5}
 				\Line(220,20)(230,10)
 				\Line(220,20)(230,30)
 				\Text(245,20)[c]{{\tiny $+\,\cdots$}}
 				\Text(121,-0.)[c]{{\small (c)}}
 			\end{axopicture}
 		\end{tabular}
 		\caption{{\small Graphic representation of the solution of the LS in Eq.~\eqref{181207.2}  for $T(\vp,\vp',E)$. 
 The diagrams in (a) correspond to the solution of the LS for the direct-scattering amplitude $T_V(\vp,\vp',E)$, Eq.~\eqref{181207.3}.
The row of diagrams (b) represents the rising of the self-energy for the bare propagator. 
The diagrams in (c) produce the dressing of the bare coupling because of the FSI due to the direct scattering  between the two particles.}
 			\label{fig.181207.1}}
 	\end{center}
 \end{figure}

 A common strategy in the literature for describing near-threshold
 scattering consists of solving a LS equation with a potential that 
includes the exchange of an explicit bare resonance.
The total potential, $V_T(\vp,\vp',E)$, consists of  an energy-independent potential $V(\vp,\vp')$, that gives rise to the so-called 
direct scattering between the two-body states in the continuum \cite{hanhart.190123.1}, plus the exchange of a bare state,
 \begin{align}
 \label{181207.1}
 V_T(\vp,\vp',E)&=V(\vp,\vp')+\frac{f(\vp)f(\vp')}{E-E_0}~.  
 \end{align}
 Where $E_0$ is the bare mass and the real function $f(\vp)$ is the bare coupling to the two-body states.
 The scattering amplitude results by solving the LS equation in momentum space,  cf. Eq.~\eqref{180803.8},
 \begin{align}
 \label{181207.2}
 T(\vp,\vp',E)&=V_T(\vp,\vp',E)+\int\frac{d^3q}{(2\pi)^3}\frac{V_T(\vp,\vq,E)T(\vq,\vp',E)}{q^2/(2\mu)-E-i\ve}~.
 \end{align}
Let us first discuss graphically the solution of this IE. 
The set of  diagrams without the exchange of any bare-state propagator is 
depicted in Fig.~\ref{fig.181207.1}(a). 
This row represents  the iteration of $V(\vp,\vp')$ which produces the 
direct-scattering amplitude, $T_V(\vp,\vp',E)$, 
 \begin{align}
 \label{181207.3}
 T_V(\vp,\vp',E)&=V(\vp,\vp')+\int\frac{d^3q}{(2\pi)^3}\frac{V(\vp,\vq)T_V(\vq,\vp',E)}{q^2/(2\mu)-E-i\ve}~.
 \end{align}
 In Fig.~\ref{fig.181207.1}(a)  an insertion of $V(\vq,\vq')$ is represented by a vertex
  with four lines attached, and the circles joining 
vertices represent the loop for the propagation of the two-body intermediate states in the continuum.
 
 Next, we move on to those contributions involving at least the exchange of one bare state, represented   by a double line 
in the second and third rows of Fig.~\ref{fig.181207.1}. 
In the iteration process for calculating the scattering amplitude there are intermediate states comprising both bare and 
 two-particle  states in the continuum. 
As a result, once the self-energy of the bare state is calculated,  one has the standard 
 Dyson resummation for calculating the dressed propagator, Fig.~\ref{fig.181207.1}(b).
 In addition, there is also the dressing of the bare coupling of the exchanged state to the continuum because  the 
 direct scattering of the latter, which is depicted in Fig.~\ref{fig.181207.1}(c). 
Therefore, the diagrams in the rows (b) and (c) of  Fig.~\ref{fig.181207.1}  build the exchange of a  particle 
 with dressed propagator and couplings, which can be written as
 \begin{align}
 \label{181207.4}
 R(\vp,\vp',E)&=\frac{\Theta(\vp,E)\Theta(\vp',E)}{E-E_0+G(E)}~.
 \end{align}
The notation is quite obvious, so that $\Theta(\vp,E)$ is the dressed coupling, $1/[E-E_0+G(E)]$ is the dressed propagator, 
and $G(E)$ is the self-energy. 
Diagrammatically we then see that $T(\vp,\vp',E)$ must be given by the sum of $T_V$ and $R$,
 \begin{align}
 \label{181207.5}
 T(\vp,\vp',E)&=T_V(\vp,\vp',E)+\frac{\Theta(\vp,E)\Theta(\vp',E)}{E-E_0+G(E)}~.
 \end{align}
 Let us show then that the previous equation is a solution of  the LS equation in
 Eq.~\eqref{181207.2} for appropriately built functions $\Theta(\vp,E)$ and $G(E)$. 
 The tentative solution of Eq.~\eqref{181207.5} is inserted in Eq.~\eqref{181207.2}, 
so that we end with the following equation for $R(\vp,\vp',E)$ [use is also made %that $T_V(\vp,\vp',E)$ satisfies
of Eq.~\eqref{181207.3}], 
 \begin{align}
 \label{181207.6}
 \frac{\Theta(\vp,E)\Theta(\vp',E)}{E-E_0+G(E)}&=
 \frac{f(\vp)f(\vp')}{E-E_0}
 +\int\frac{d^3q}{(2\pi)^3}\frac{1}{q^2/(2\mu)-E-i\ve}\Big[
 V(\vp,\vq)\frac{\Theta(\vq,E)\Theta(\vp',E)}{E-E_0+G(E)}\nn\\
 &+\frac{f(\vp)f(\vq)}{E-E_0}T_V(\vq,\vp',E)
 +\frac{f(\vp)f(\vq)}{E-E_0}\frac{\Theta(\vq,E)\Theta(\vp',E)}{E-E_0+G(E)} 
 \Big]~.
 \end{align}
Taking  $E\to E_0$ and  $E\to E_0-G(E)$  we derive the equations for $G(E)$ and $\Theta(\vp,E)$ (notice that $E_0$ itself can 
take any value),
 \begin{align}
 \label{181207.7}
 &\Theta(\vp',E)\frac{-1}{G(E)}\int\frac{d^3q}{(2\pi)^3}\frac{f(\vq)\Theta(\vq,E)}{q^2/(2\mu)-E-i\ve}
 =f(\vp')+\int\frac{d^3q}{(2\pi)^3}\frac{f(\vq)T_V(\vq,\vp',E)}{q^2/(2\mu)-E-i\ve}~.\\
 &\Theta(\vp,E)=
 -\frac{f(\vp)}{G(E)}\int\frac{d^3q}{(2\pi)^3}\frac{f(\vq)\Theta(\vq,E)}{q^2/(2\mu)-E-i\ve}
 +\int\frac{d^3q}{(2\pi)^3}\frac{V(\vp,\vq)\Theta(\vq,E)}{q^2/(2\pi)-E-i\ve}~.\nn
 \end{align}
 These two equations are satisfied by identifying
 \begin{align}
 \label{181207.8}
 G(E)&=-\int\frac{d^3q}{(2\pi)^3}\frac{f(\vq)\Theta(\vq,E)}{q^2/(2\mu)-E-i\ve}~,
 \end{align}
 and
 \begin{align}
 \label{181207.9}
 \Theta(\vp',E)&=f(\vp')+\int\frac{d^3q}{(2\pi)^3}
 \frac{f(\vq)T_V(\vq,\vp',E)}{q^2/(2\mu)-E-i\ve}~.
 \end{align}
 Let us notice that this IE can also be rewritten as
 \begin{align}
 \label{181207.9b}
 \Theta(\vp',E)&=f(\vp')+\int\frac{d^3q}{(2\pi)^3}\frac{T_V(\vp',\vq,E)f(\vq)}{q^2/(2\mu)-E-i\ve}\\
 &=f(\vp')+\int\frac{d^3q}{(2\pi)^3}\frac{V(\vp',\vq)\Theta(\vq,E)}{q^2/(2\mu)-E-i\ve}~.\nn
 \end{align}
 The three IEs in Eqs.~\eqref{181207.9} and \eqref{181207.9b} are equivalent, e.g. this can be seen by the Neumann series expansion
 of $T_V(\vq,\vp',E)$, Eq.~\eqref{181207.3}, and similarly for  $\Theta(\vp',E)$, Eq.~\eqref{181207.9b}.
 As a check, it is straightforward to see
 that Eq.~\eqref{181207.6} is fulfilled once  Eqs.~\eqref{181207.8} and \eqref{181207.9} are satisfied.

A more general derivation can be elaborated for the representation of $T(\vp,\vp',E)$ in Eq.~\eqref{181207.5}.  
The  Hamiltonian $H$ is split in the free part $H_0$ and the potential $V$, $H=H_0-V$, and 
$|0\rangle$ is an eigenstate of $H_0$, $H_0|0\rangle=E_0|0\rangle$. 
Next, we define a new $T$ matrix $T_1(E)$ that results by removing the state $|0\rangle $ among 
the intermediate states in the LS equation, and obeys the equation 
 \begin{align}
 \label{181207.11}
 T_1(E)&=V+ V(H_0-E)^{-1} \theta T_1(E)~,
  \end{align}
 where $\theta=I-|0\rangle\langle 0|$, so that $\theta|0\rangle=0$ ($\langle 0|0\rangle=1$). 
  This equation can be transformed into the IE for the resolvent of the kernel of a linear IE by 
 multiplying it  to the right by $(H_0-E)^{-1}\theta$, with the result
 \begin{align}
 \label{181207.12}
 T_1(E)(H_0-E)^{-1}\theta=V(H_0-E)^{-1}\theta+V(H_0-E)^{-1}\theta T_1(E) (H_0-E)^{-1}\theta~.
 \end{align}
Its  kernel is $ V(H_0-E)^{-1}\theta$ and its resolvent, denoted by $K_1(E)$,  is then \cite{tricomi.181021.1}
 \begin{align}
 \label{181207.13}
 K_1(E)&=T_1(E)(H_0-E)^{-1}\theta~.
 \end{align}
 
For the LS equation satisfied by the full $T$ matrix $T(E)$ one has
 \begin{align}
 \label{181208.1}
 T(E)&=V+V|0\rangle (E_0-E)^{-1}\langle 0|T(E)+V(H_0-E)^{-1}\theta T(E)~. 
 \end{align}
In this equation we can identify the same kernel as in Eq.~\eqref{181207.11}, but now with a different independent term
 $V+V|0\rangle (E_0-E)^{-1}\langle 0|T(E)$. Thus,  we can write for the solution of $T(E)$ the expression  
 \begin{align}
 \label{181208.2}
 T(E)&=V+V|0\rangle(E_0-E)^{-1}\langle 0|T(E)+
 K_1(E)\Big[V+V|0\rangle(E_0-E)^{-1}\langle 0|T(E)\Big]\nn\\
 &=T_1(E)+T_1(E)|0\rangle(E_0-E)^{-1}\langle 0|T(E)~.
 \end{align}
Here we have taken into account that $T_1(E)=V+K_1(E) V$, as a consequence of $K_1(E)$ being the resolvent 
of the IE of Eq.~\eqref{181207.11}.\footnote{The reader less familiar with the basic mathematics of IEs \cite{tricomi.181021.1} can   
convince himself of this expression for $T_1(E)$ by performing the Neumann series of $K_1(E)$,  given by the iterative solution  
of Eq.~\eqref{181207.12}.}
The previous IE is multiplied to the left by $\langle 0|$, which allows us to give  
$\langle 0|T(E)$ in terms of known matrix elements as
 \begin{align}
 \label{181208.4}
 \langle 0| T(E)&=\Big[1-\langle 0|T_1(E)|0\rangle(E_0-E)^{-1}\Big]^{-1}
 \langle 0|T_1(E)~.
 \end{align}
We can express $T(E)$ in terms of $T_1(E)$ by implementing Eq.~\eqref{181208.4} in Eq.~\eqref{181208.2}. 
The final expression is
 \begin{align}
 \label{181208.5}
 T(E)&=T_1(E)+ \frac{T_1(E)|0\rangle \langle 0|T_1(E)}{E-E_0-\langle 0|T_1(E)|0\rangle } ~.
 \end{align}
It is clear from here that the dressed propagator is 
 \begin{align}
 \label{181208.5b}
 \Delta(E)&=\frac{1}{E-E_0-\langle 0|T_1(E)|0\rangle }~,
 \end{align}
with  $\langle 0|T_1(E)$ corresponding to the coupling operator. E.g. when acting 
on the states in the continuum it gives rise to the coupling of the exchanged state with them, 
 \begin{align}
 \label{181209.1}
 \Theta(\vp_n,E)&=\langle \vp_n|T_1(E)|0\rangle~,   
 \end{align}
so that  the bare coupling is then dressed because of the  final-state interactions (FSI) due to the direct coupling of the the states in the continuum. 
Let us also note that Eq.~\eqref{181207.5} is a particular case of the operational Eq.~\eqref{181208.5},  
as it can be seen by taking its matrix elements between the  states in the continuum. 
The Eq.~\eqref{181208.5} is introduced in Ref.~\cite{weinberg.181208.1},  though its derivation is not given. 
  
Following Ref.~\cite{kang.181206.1},  let us compare Eq.~\eqref{181204.4} with the
scattering model of Ref.~\cite{han.181206.1}. This is obtained by employing Eq.~\eqref{181207.5} 
with $T_V(\vp,\vp',E)$ given by an ERE including only the scattering length approximation, 
 \begin{align}
 \label{181208.6}
 T_V(\vp,\vp',E)&=\frac{2\pi}{\mu}\frac{1}{-\frac{1}{a_V}-ik(E)}~,\\
 k(E)&=\sqrt{2\mu E}~.\nn
 \end{align}
Since in this model $T_V(\vp,\vp',E)$ only depends on the energy we denote it as $T_V(E)$. 
The calculation of  $G(E)$ and  $\Theta(\vp,E)$, cf. Eqs.~\eqref{181207.8} and \eqref{181207.9}, respectively, 
is direct once $f(\vp)$ is known. 
However, Ref.~\cite{han.181206.1} considers that because  $k\alpha\ll 1$, 
where $\alpha$ is the range of the interactions, one could approximate  $f(\vp)$ by 
$f_0=f(0)/(2\pi)$. Then, the resulting diverging integrals from
 Eqs.~\eqref{181207.8} and \eqref{181207.9} are regularized by naive dimensional analysis, with the expressions
 \begin{align}
 \label{181208.7}
 \widetilde{g}_1(E)&=\int\frac{d^3q}{(2\pi)^3}\frac{f(\vq)^2}{q^2/(2\mu)-E-i\ve}
 =(R+\mu ik)f_0^2~,\\
 \widetilde{g}_2(E)&=\int\frac{d^3q}{(2\pi)^3}\frac{f(\vq)}{q^2/(2\mu)-E-i\ve}
 =(R'+\mu ik)f_0~,\nn
 \end{align}
 where  the constants $R$ and $R'$ are expected to be of ${\cal O}(\mu/\alpha)$. 
 These expressions also follow by calculating them with a sharp cutoff and then keeping the leading
linear divergence.  
Taking the results of these integrals to  Eqs.~\eqref{181207.5}, \eqref{181207.8}  and
 \eqref{181207.9}, it follows the expression for the on-shell 
$T$ matrix \cite{han.181206.1}
 \begin{align}
 \label{181208.8}
 t(E)&=-\frac{2\pi}{\mu}\frac{E-E_f+\frac{1}{2}g_f \gamma_V}{(E-E_f)(\gamma_V+ik)+i\frac{1}{2}g_f\gamma_V k}~,
 %\frac{1}{R_V+\frac{f_0^2(E_C-E_0)(RR_V-{R'}^2)}{E-E_C}-i\mu k/2\pi}~,\\
 %R_V&=\frac{\mu}{a_V}~,\nn\\
 %E_C&=E_0-f_0^2(R+R_V-2R')~.
 \end{align}
with $\gamma_V=1/a_V$,  $g_f$ and $E_f$ are functions of  $R$, $R'$ and $E_0$ (explicit expressions for this 
functional dependence can be found in Ref.~\cite{han.181206.1}).
Multiplying $t(E)$ by $\mu/(2\pi)$, we have a particular case of
 Eq.~\eqref{181204.4} with its parameters corresponding to
 \begin{align}
 \label{181208.9}
 \beta&=-\gamma_V~,\\
 \gamma&=\frac{1}{2}g_f \gamma_V^2~,\nn\\
 \cdd&=E_f-\frac{1}{2}g_f\gamma_V~.\nn
 \end{align}
 However, it is important to emphasize that Eq.~\eqref{181204.4} is not a particular
 case of Eq.~\eqref{181208.8}.
In order to illustrate this statement,  let us notice that the effective range $r$ resulting from Eq.~\eqref{181208.8} must be  negative
 \cite{kang.181206.1}, being given by the expression 
 \begin{align}
 \label{181208.10}
 r&=-\frac{g_f \gamma_V^2}{2\mu(E_f-g_f\gamma_V/2)^2}\leq 0~.
 \end{align}
The last inequality follows because $g_f=2\mu f_0^2 (R-R_V)^2/R_V^2>0$ \cite{han.181206.1}.

 The non-relativistic near-threshold scattering is of particular interest in nowadays heavy-quark hadron spectroscopy 
  near a relevant threshold. A vivid interest is on going because 
  states with unexpected properties from quarkonium spectroscopy have been detected under such circumstances
  \cite{pdg.181106.1,Chen:2016qju,fkguo.181209.1}. 
In this regard, there is nowadays great excitement because of the latest results of the LHCb
which have driven to the discovery of the pentaquark states $P_c(4312)$, $P_c(4440)$ and
$P_c(4457)$ \cite{aaij.190511.1}, which have superseded the original
finding of the $P_c(4380)$ and $P_c(4450)$ resonances \cite{aaij.190511.2}.
These results have triggered great theoretical interest, see e.g.
Ref.~\cite{guo.190511.1,meissner.190511.1,fkguo.181209.1,chen.190511.1} and references therein. 
An important point is that the coupling of the pion with these heavy-quark mesons is suppressed as compared 
with the  light-quark sector.
As an example, let us mention that  the $P^*P$ potential derived in  Ref.~\cite{pavon.181208.1} (here $P^*$ is a heavy-quark ($c$ or $b$) vector-meson and $P$  a heavy-quark pseudoscalar),
has a central and tensor components of the one-pion mediated interaction which is
weaker by around a factor $g^2/(2g_A^2)\approx 0.07$ compared to the corresponding ones for the $NN$ interactions.
The parameter $g$ is the  $P^*P\pi$ coupling, which is around 0.5 \cite{pavon.181208.1}, and $g_A$ is the axial-vector coupling
of the nucleon. We have taken $g_A=1.32$ for this estimate based on the $\pi N$ coupling constant $g_{\pi N}=13.4$ and the
Goldberger-Treiman relation $g_A=g_{\pi N}f_\pi/m_N$ \cite{kaiser.181113.1}.   
 This makes that a perturbative treatment of the LHC due to pion exchanges seems adequate, and that 
 one could even neglect it as a first approximation. In such a case,  we may apply the results presented in this section  \cite{kang.181205.1,kang.181206.1}.
 
 We have referred briefly to some interesting results for  quantifying the compositeness and
 elementariness of a pole in the $S$ matrix. For a through approach to the problem the interested reader
 can consult Ref.~\cite{oller.181104.1}.
There are many other valuable references on this topic, among which one has  \cite{weinberg.181208.1,weinberg.181208.2,baru.181208.1,jido.181208.1,aceti.190111.1,kang.181205.1,guo.181206.1,kang.181206.1,sekihara.181208.1,hernandez.190215.1,hyodo.190511.1,sekihara.190511.1,oller.190511.2}.
 
%%%%%%%%%%%%%%%%%%%%%%%%%%%%%%%%%%%%%%%%%%%%%%%%%%%%%%
 %%%%%%%%%%%%%%%%%%%%%%%%%%%%%%%%%%%%%%%%%%%%%%%%%%%%%%
 \subsection[Near-threshold parameterization: bare, elastic and inelastic contributions]{Parameterization for near-threshold states accounting for open inelastic channels and bare states}
 \label{sec.190224.1}

 Let us now  consider the approach settled recently in Ref.~\cite{hanhart.190224.1}.
 The prototypical example considered by the authors of this reference  concerns the existence of states around 
 an $S$-wave threshold with exotic properties in the quarkonium sector (charmonium or bottomonium ones). 
 These states are manifest in reactions involving the open-flavor and hidden-flavor channels. 
 A clear example is the case of the $X(3872)$ near the $D^0\bar{D}^{* 0}$ threshold, at which energy the channels 
 $J/\psi \rho$, $J/\psi \omega$, $\ldots$, are open. Other important examples are the $Z_b(10610)$ and $Z_b(10650)$ near 
 the $B^{(*)}\bar{B}^{*}$ thresholds, as well as the $Z_c(3900)$ and $Z_c(4020)/Z_c(4025)$ close to the 
 $D^{(*)}\bar{D}^{*}$ thresholds \cite{pdg.181106.1}, etc. 
The approach of Refs.~\cite{hanhart.190224.1,guo.190224.1} is a further generalization of the one 
previously developed in Refs.~\cite{baru.190303.1,hanhart.190303.1}, or that derived 
in Refs.~\cite{nieves.190304.1,braaten.190304.1}.  
Since it is based on solving the LS
equation, the  unitarity in PWAs is fulfilled, which is certainly an important constraint,
mixing simultaneously the elastic and inelastic channels.
It furthermore relates scattering and production processes with their respective line shapes.
The reproduction of the latter ones is important to match with experiment. 

Let us proceed with the formalism of Ref.~\cite{hanhart.190224.1},  which is settled  here in a more compact and general manner. 
Our subsequent derivations are based on applying the operational method used in Sec.~\ref{sec.190210.1} to obtain Eq.~\eqref{181208.5}. 
In this way, the  solution of the full LS equation,
\begin{align}
\label{190224.1}
T(E)&=V+ V(H_0-E)^{-1}T(E)~,
\end{align}
can be expressed in terms of the $T$ matrix obtained without the explicit exchange of  the bare state $|0\rangle$.
 The point now is to iterate further this method by splitting explicitly  the coupled channels in two sets. 
 One of them (referred by $L$) comprises the so-called elastic channels, as denoted in  Ref.~\cite{hanhart.190224.1},
 whose thresholds are close to each other.
 The other set (referred by $I$) includes the open channels, with typically much lower thresholds, and are also qualified as
 inelastic in Ref.~\cite{hanhart.190224.1}. In addition we also have  the bare state $|0\rangle$.  
 As a result, the full $T$ matrix is written in terms of the one involving only the exchanges of the channels in the set $L$. 
 
 The Hamiltonian $H$ is split in the usual manner as $H=H_0-V$, 
and the potential $V$ couples among themselves the states of the type $L$, $I$ and 
the bare state $|0\rangle$ ($H_0|0\rangle=E_0|0\rangle$).
The potential $V$ is written in a synthetic notation as 
 \begin{align}
 \label{190225.1}
 V&=\left(
 \begin{matrix}
 V^{00} & V^{0L} &  V^{0I}  \\
V^{L0} & V^{LL} & V^{LI} \\ 
V^{I0} & V^{IL}  & V^{II} 
 \end{matrix}
 \right)~,
 \end{align}
 where the superscripts refer to the subsets that are coupled. For the practical example developed below the potential 
 is symmetric, as in Ref.~\cite{hanhart.190224.1}. 
 Because of time-reversal invariance this is the case for a partial-wave projected potential.
 
  We further define the projector $\theta_1$ such that it excludes the 
intermediate states $|0\rangle$ and the ones in $I$, while $\theta_1 L=L$. 
Below we use again the projector $\theta$ to exclude the state $|0\rangle$ among the intermediate states, cf. Eq.~\eqref{181207.11}.  
 The $T$ matrix $T_1$ is the solution of the LS equation without summing over $|0\rangle$ and $I$ in the intermediate states, 
 \begin{align}
 \label{190225.2}
 T_1&=V+ V(H_0-E)^{-1}\theta_1 T_1~.
 \end{align}
Its formal solution is 
\begin{align}
 \label{190225.3}
T_1=\left[I-V J^L\right]^{-1}V~,
\end{align}
where $J^L$ is the unitarity loop operator comprising only the intermediate states in the set $L$, $J^L=(H_0-E)^{-1}\theta_1$.
We can rewrite this solution 
so that only the states in $L$ appear in the matrix to be inverted, which also reduces the numerical burden.
For that, let us notice that
\begin{align}
 \label{190225.4}
\left[I-V J^L\right]^{-1}-I+I=VJ^L \left[I-V J^L\right]^{-1}+I=VJ^L \left[I-V^{LL} J^L\right]^{-1}+I~.
\end{align}
  Thus, we can then write $T_1$ conveniently  as 
\begin{align}
\label{180225.4}
T_1^{AB}&=
\left\{
\begin{array}{rl}
  \left[I-V^{LL} J^L\right]^{-1}V^{LB} &, A= L~,\\
  V +  V J^L \left(I-V^{LL} J^L\right)^{-1}V &, A \neq L~.
\end{array}
\right.
\end{align}
The next step is to consider the inclusion of the intermediate states of $I$ and the resulting $T$ matrix is denoted by $T_2(E)$, 
which therefore satisfies the equation,
\begin{align}
\label{180225.5}
T_2&=V+V(H_0-E)^{-1}\theta T_2~.
\end{align}
Let us connect it with $T_1$ by rewriting the previous equation as
\begin{align}
\label{180225.6}
T_2&=V+V J^I T_2
+V(H_0-E)^{-1}\theta_1 T_2~,
\end{align}
where $J^I$ is the unitarity loop operator with only intermediate states belonging to $I$, $J^I=(H_0-E)^{-1}(\theta-\theta_1)$. 
The Eq.~\eqref{180225.6} is similar to Eq.~\eqref{181208.1}, but with a  continuum of intermediate states $I$ instead of just $|0\rangle$. 
We then adopt similar steps as above in oder to end with the solution of $T_2$ written in terms of $T_1$.
First, we multiply to the right Eq.~\eqref{190225.2} by $(H_0-E)^{-1}\theta_1$ which then results into
\begin{align}
\label{180225.7}
T_1(H_0-E)^{-1}\theta_1&=V(H_0-E)^{-1}\theta_1+ V(H_0-E)^{-1}\theta_1 T_1(H_0-E)^{-1}\theta_1~,
\end{align}
which is analogous to Eq.~\eqref{181207.12}. The kernel of this IE is $V(H_0-E)^{-1}\theta_1$ and its  resolvent is 
$T_1(H_0-E)^{-1}\theta_1$. The kernel is the same as in Eq.~\eqref{190225.2} and then we can write the solution of Eq.~\eqref{180225.6} as
\begin{align}
\label{180225.8}
T_2&=V+V J^I T_2+T_1 (H_0-E)^{-1}\theta_1 (V+V J^I T_2)\\
&=T_1+T_1 J^I T_2~.\nn
\end{align}
We multiply this equation to the left by $J^I$, which allows us to express $J^I T_2$ in terms of $T_1$, and write for $T_2$ the expression 
\begin{align}
\label{180225.9}
T_2&= \left(I-T_1 J^I\right)^{-1} T_1~.
\end{align}
In applications it is also interesting to express $T_2$ so that one has to invert only a matrix restricted to the subspace $I$.
For that we can apply a similar procedure as in Eq.~\eqref{190225.4},
\begin{align}
 \label{190225.10}
\left[I-T_1 J^I\right]^{-1}-I+I=T_1J^I \left[I-T_1 J^I\right]^{-1}+I=T_1J^I \left[I-T_1^{II} J^I\right]^{-1}+I~.
\end{align}
Thus, Eq.~\eqref{180225.9} can be recast as
\begin{align}
\label{190225.11}
 T_2^{AB}&=\left\{
 \begin{array}{rr}
\left(I-T_1^{II} J^I\right)^{-1} T_1^{IB} &, A=I~,\\ 
T_1 + T_1J^I \left(I-T_1^{II} J^I\right)^{-1}T_1 &, A \neq I~.
\end{array}
\right.
\end{align}
The full $T$ matrix, solution of Eq.~\eqref{190224.1}, can be expressed in terms of $T_2$, by the same process as used in Sec.~\ref{sec.190210.1} to
derive Eq.~\eqref{181208.5}. Namely, we rewrite Eq.~\eqref{190224.1} as
\begin{align}
 \label{190225.12}
T&=V+V|0\rangle (E_0-E)^{-1}\langle 0|T+V(H_0-E)^{-1}\theta T~,
\end{align}
which can be seen as the same IE as Eq.~\eqref{180225.5} but with a different independent term.
The resolvent of the latter IE is $T_2(H-E_0)^{-1}\theta$ and, by proceeding analogously as above to obtain
Eq.~\eqref{181208.5} from Eq.~\eqref{181208.1}, we have now
\begin{align}
\label{190225.13}
T&=T_2+\frac{T_2|0\rangle \langle 0|T_2}{E-E_0-\langle 0|T_2|0\rangle}~.
\end{align}

As a result,  a given matrix element of the full $T$ matrix can be calculated in terms of the matrix
elements of $T_2$, Eq.~\eqref{190225.11}.
In turn, $T_2$ is determined by the knowledge of $T_1$, which is given by Eq.~\eqref{180225.4}.
For obtaining the latter one has to invert the matrix $\langle \vp'\,\alpha|I-V^{LL} J^{L}|\vp \, \beta\rangle$,
and for obtaining the former
the matrix that has to be inverted is $\langle \vp'\,j|I-T_1^{II} J^I|\vp\, i\rangle$.
 We use the Greek letters to denote the states in $L$, and the Latin letters
for the states in $I$.

\begin{figure}
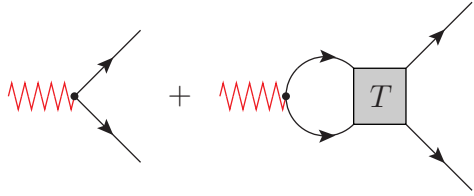

  \begin{center}
    \begin{axopicture}(200,100)
\ZigZag[color=Red](0,50)(25,50){5}{5} %1st zigzag
\Vertex(25,50){1.5}                   %1st Vertex
\Line[arrow](25,50)(50,75)            %Outgoing line
\Line[arrow](25,50)(50,25)            %Outgoing line
\Text(65,50)[c]{+}                    %Plus sign
\ZigZag[color=Red](80,50)(105,50){5}{5}%2nd zigzag
\Vertex(105,50){1.5}                  %2nd vertex
\Arc[arrow,arrowpos=0.61](120,50)(15,135,0)  %Internal arc
\Arc[arrow,arrowpos=.36,clockwise](120,50)(15,135,0)  %Internal arc
\GBox(130.607,39.3934)(150.393,60.6066){0.8}
\Line[arrow](150.393,60.6066)(175.393,85.6066)            %Outgoing line
\Line[arrow](150.393,39.3934)(175.393,14.3934)            %Outgoing line
\Text(141,50){$T$} % T matrix symbol
    \end{axopicture}
    \caption{{\small Production process accounting for the final-state interactions, which  correspond to the gray box to the right of the figure. 
In turn, the left vertex in each diagram is the ``elementary'' production amplitude or Born term, which is either parameterized ad-hoc or  calculated typically from a Lagrangian. 
The external probe is indicated by the red wiggly line.}
  \label{fig.190225.1}}
  \end{center}
  \end{figure}

Regarding the FSI in the production amplitudes, although they are treated in more depth in Sec.~\ref{sec.181117.1},
we introduce them here already because they are the main object of desire in Refs.~\cite{hanhart.190224.1,au.190224.1}.
 The calculation of the production amplitudes, gathered in the column vector $F$,
follows by considering the diagrams in Fig.~\ref{fig.190225.1}. The first diagram on
the left corresponds to the ``elementary'' production amplitudes or Born terms, ${\cal F}$, and
the rightmost digram represents the FSI driven by the full $T$ matrix, cf. Eq.~\eqref{190225.13}.
Then, the column vector of production amplitudes $F$ can be expressed as
\begin{align}
  \label{190226.1}
F&={\cal F}+{\cal F}(H_0-E)^{-1} T  
\end{align}
or, equivalently
\begin{align}
  \label{190226.2}
F&={\cal F}(I-(H_0-E)^{-1}V)^{-1}~.
\end{align}
A simplifying assumption of Refs.~\cite{hanhart.190224.1} is to consider that in the FSI only the states in $L$ contribute in the
sum over the intermediate states. In such a case Eqs.~\eqref{190226.1}  simplifies as
\begin{align}
\label{190226.3}
F&={\cal F} + {\cal F}^L J^L T~,
\end{align}
where as usual the superscript $L$ indicates the restriction to the subspace of states in the set $L$.

The Ref.~\cite{hanhart.190224.1} proposes a handy parameterization for practical applications for studying the spectrum of near-threshold
states in an energy window around threshold.
This reference takes a potential of the form
\begin{align}
  \label{190227.1}
  V^{00}&=0~,\\
  V^{0L}_\alpha(\vp)=V^{L0}_\alpha(\vp)&=f_\alpha(\vp)~,\nn\\
  V^{0I}_i(\vp)=V^{I0}_i(\vp)&=f_i(\vp)~,\nn\\
  V^{LL}_{\alpha\beta}(\vp',\vp)&=v_{\alpha \beta}(\vp',\vp)~,\nn\\
  V^{LI}_{\alpha \,i}(\vp',\vp)=V^{IL}_{i\,\alpha}(\vp,\vp')&=v_{\alpha \, i}(\vp',\vp)~,\nn\\
  V^{II}_{ij}(\vp',\vp)&=v_{ij}(\vp',\vp)~,\nn
\end{align}
and proceeds with a set of extra simplifying assumptions on the final functions involved in the potential given.
Namely,
\begin{align}
  \label{190227.2}
  f_\alpha(\vp)&=f_\alpha~,\\
  f_i(\vp)&=f_i|\vp|^{\ell_i}~,\nn \\
  v_{\alpha\beta}(\vp',\vp)&=v_{\alpha\beta}~,\nn\\
  v_{\alpha i}(\vp',\vp)&=g_{i\alpha}|\vp|^{\ell_i}~,\nn\\
  v_{ij}(\vp',\vp)&=0~,\nn
\end{align}
where $f_\alpha$, $v_{\alpha\beta}$, $f_i$ and $g_{\alpha i}$ are constants.
 Furthermore, the matrices $v_{\alpha\beta}$ and $g_{\alpha i}$ are symmetric.
%\begin{align}
%  \label{190227.3}
%  v_{\alpha\beta}&=v_{\beta\alpha}~,\\
%  g_{\alpha i}&=g_{i\alpha}~.\nn
%\end{align}
The Ref.~\cite{hanhart.190224.1} puts forward several arguments to justify the  of Eq.~\eqref{190227.2}.
The vanishing of $v_{ij}(\vp',\vp)$ is based on the fact that ``there are good reasons to neglect the direct interactions'' between
the states in the inelastic channels. Some concrete examples are  the
transitions between the $\rho J/\psi$ and $\omega J/\psi$ channels that are forbidden in the limit
of isospin conservation, and these channels are of relevance for the $X(3872)$ resonance.
In addition, because of the absence of light quarks
in the $J/\psi$ the direct $\rho J/\psi \to \rho J/\psi$ interactions are expected to be suppressed.
A similar argument is also invoked in Ref.~\cite{hanhart.190224.1}
to expect the weakness of the pion interactions with the
quarkonia $\Upsilon(nS)$ and $h_b(mP)$, which is important for the study of the
$Z_b^{(}{'}^{)}$ (as emphasized e.g. in Ref.~\cite{kang.181205.1}).
In this respect, the EFT evaluations \cite{ref8.190301.1} and calculations in
LQCD \cite{ref9.190301.1} obtain very small values for the scattering length of a pion with a $c\bar{c}$ or
$b\bar{b}$.\footnote{More results on the weak scattering of pions on heavy-quark hadrons was also commented above
in regards with the work of Ref.~\cite{pavon.181208.1}, due to a large extent to the relatively small value of the 
coupling of a pion to a vector and a pseudoscalar heavy-quark mesons.  Pion exchanges could be part of 
$v_{\alpha\beta}$, whose energy and momentum exchanged dependence is neglected. One could then advocate the
results of \cite{pavon.181208.1}.}
Regarding $v_{\alpha i}(\vp',\vp)$ its final form in Eq.~\eqref{190227.2} implies that it is  only sensitive to the
momentum dependence of the inelastic channel, which is furthermore assumed to interact with a dominant orbital
angular momentum $\ell_i$. Given the fact that the thresholds for the inelastic channels are supposed to be relatively far away
from the elastic ones,  the energy dependence of the direct transition amplitudes involving these channels is 
supposed to be weak. In other terms, the main features of the scattering amplitudes are then supposed to arise from
to the inclusion of the explicit bare state and the
threshold branch points of the elastic channels. Although appealing, the picture stemming from the potential in Eq.~\eqref{190227.2}
could fail to reproduce interesting dynamical situations,
as already discussed in Sec.~\ref{sec.190210.1} and at the beginning of Sec.~\ref{sec.181202.1}. 

Since the potential of Eq.~\eqref{190227.2} fulfills that $V^T=V$ (exchanging both discrete and continuum indices),
 all the $T$ matrices introduced above, $T_1$, $T_2$ and $T$ are also symmetrical. 
 The simple form of the potential $V$ in Eq.~\eqref{190227.2} allows one to obtain algebraic matrix expressions for
 the $T$ matrices.

 Let us now discuss the unitarity loop functions that appear in the final expressions of Eqs.~\eqref{180225.4} and
 \eqref{190225.11} for $T_1$ and $T_2$, respectively. We show next that  $I-V^{LL}J^L$ and $I-T_1^{II}J^I$ become
 algebraic matrices (depending only on the total energy),
 so that their inversion and the evaluation of $T_1$ and $T_2$ is straightforward. 
 For the channel $\alpha$ in $L$ we have
 \begin{align}
\label{190302.1}
J_\alpha(E)&=\int\frac{d^3q }{(2\pi)^3}\frac{1}{q^2/2\mu_\alpha-p_\alpha^2/2\mu_\alpha-i\ve}
=\frac{1}{\pi}\int_0^\infty dq^2 \frac{\mu_\alpha q /2\pi}{q^2-p^2_\alpha-i\ve}~,
 \end{align}
 where
\begin{align}
\label{190302.1b}
p_\alpha=\sqrt{2\mu_\alpha(E-m_{{\rm th};\alpha})}~,
\end{align}
with $m_{{\rm th};\alpha}$ the threshold for the channel $\alpha$. 
 The integral in Eq.~\eqref{190302.1} is divergent, so that we take a subtraction, e.g. at $p=0$,
 \begin{align}
 \label{190302.2}
J_\alpha(E)&=J_\alpha(0)+\frac{p^2_\alpha}{\pi}\int_0^\infty dq^2\frac{\mu_\alpha q /2\pi}{q^2(q^2-p^2_\alpha-i\ve)}~.
 \end{align}
 The latter integral can be done explicitly, cf. Eq.~(18) of Ref.~\cite{oller.181112.1}, and $J_\alpha(E)$ then reads
 \begin{align}
\label{190302.3}
J_\alpha(E)&=J_\alpha(0)+i\frac{\mu_\alpha p_\alpha}{2\pi}~.
 \end{align}
 As a result the matrix elements of $I-V^{LL}J^L$ are
\begin{align}
\label{190302.4}
\delta_{\alpha\beta}-v_{\alpha\beta}J_\beta(E)~,
\end{align}
and the ones of its inverse matrix are denoted by $(I-V^{LL}J^L)^{-1}_{\alpha\beta}$, in terms of
which all the matrix elements of $T_1(E)$ can be evaluated 
from Eq.~\eqref{180225.4} by performing matrix multiplications.
E.g. $T_1^{II}(E)$ in Eq.~\eqref{190225.11}  is given by
\begin{align}
\label{190302.5}
T_1^{ij}&=p_i^{\ell_i}p_j^{\ell_j} \sum_\alpha g_{i\alpha}J_\alpha\,(I-V^{LL} J^L)^{-1}_{\alpha\beta}\,g_{j\beta}~.
\end{align}

We are now ready to discuss the practical calculation of $T_2(E)$ from Eq.~\eqref{190225.11}.
We take advantage of the simplified form of the potential proposed in Ref.~\cite{hanhart.190224.1},
Eq.~\eqref{190227.2}, to express $T_2$ as the inverse of a finite order matrix, instead of
$\langle \vp \,j|I-T_1^{II}J^I|\vp'\,i\rangle$, which is of infinite order.
 There are several ways to accomplish this aim. The most straightforward one is to perform the Neumann series of
$(I-T_1^{II}J^I)^{-1}T_1^{IB}$ as follows
\begin{align}
\label{190302.6}
(I-T_1^{II}J^I)^{-1}T_1^{IB}=T_1^{IB}+T_1^{II}J^I T_1^{IB}+T_1^{II} J^I T_1^{II} J^I T_1^{IB}+\ldots
\end{align}
Now, the matrix elements of $T_1^{IB}$ have the form  $p^{\ell_I}\widetilde{T}_1^{IB}$, where $p^{\ell_I}$ is a diagonal matrix with its matrix
elements given by $p_i^{\ell_i}$. In turn, $\widetilde{T}_1^{IB}$ factorizes out when acting to the right in a loop integral.
The particular $T$-matrix $T_1^{II}$ is explicitly given in Eq.~\eqref{190302.5}, we write it as
$T_1^{II}=p^{\ell_I}\Theta_1^{II}p^{\ell_I}$, and then $\widetilde{T}_1^{II}=\Theta_1^{II}p^{\ell_I}$. 
In this way we can rewrite Eq.~\eqref{190302.6} as
\begin{align}
\label{190302.7}
(I-T_1^{II}J^I)^{-1}T_1^{IB}&=
p^{\ell_I}\sum_j \left( I + \Theta_1^{II}[q^{\ell_I}J^Iq^{\ell_I}]
  +\Theta_1^{II}[q^{\ell_I}J^I q^{\ell_I}]\Theta_1^{II}[q^{\ell_I}J^Iq^{\ell_I}]+\ldots\right)\widetilde{T}_1^{IB}\\
&=p^{\ell_I}\left(I-\Theta_1^{II}[q^{\ell_I}J^Iq^{\ell_I}]\right)^{-1}\widetilde{T}_1^{IB}~,\nn
\end{align}
where $[q^{\ell_I}J^Iq^{\ell_I}]$ and $I-\Theta_1^{II}[q^{\ell_I}J^Iq^{\ell_I}]$ are $n_I\times n_I$ order matrices,
with $n_I$ the number of inelastic channels.
We denote in the following the matrix $[q^{\ell_I}J^Iq^{\ell_I}]$ by $\widetilde{J}^I$,
\begin{align}
\label{190302.7b}
\widetilde{J}^I&=[q^{\ell_I}J^Iq^{\ell_I}]~.
\end{align}
It is a diagonal matrix whose matrix elements,
in the non-relativistic limit, are given formally by the integral
\begin{align}
\label{190302.8}
J_i(E)\equiv %[q^{\ell_I}J^I q^{\ell_I}]_{ii}
\widetilde{J}^I_{ii}(E)&=\frac{1}{\pi}\int_0^\infty dq^2\frac{q^{2\ell_i} \, \mu_i q/2\pi}{q^2-p_i^2-i\ve}~.
\end{align}
This integral requires to take $\ell_i+1$ subtractions to obtain a finite and subtraction-point independent result.
The Ref.~\cite{hanhart.190224.1} advocates to approach it simply by its imaginary part.
The plausibility of this assumption rests on the fact that the inelastic channels typically
have far away thresholds from the elastic ones, so that the energy dependence of the integral in Eq.~\eqref{190302.8} in
the physical region around the elastic threshold is expected to be smooth. As a result, Ref.~\cite{hanhart.190224.1}
argues that this energy dependence from the real part of $J_i(E)$ could be reabsorbed by
refitting the parameters of the matrix elements of the potential $V$ involving the inelastic channels, cf. Eq.~\eqref{190227.1}.
The final expression that follows  according to the dynamical model of Ref.~\cite{hanhart.190224.1} is 
\begin{align}
\label{190302.9}
J_i(E)\to i\frac{\mu_i p_i^{\ell_i+1}}{2\pi}~,
\end{align}
with $p_i=\lambda(s,m_{i1}^2,m_{i2}^2)/2\sqrt{s}$. 
Nonetheless, Ref.~\cite{hanhart.190224.1} includes some relativistic corrections to the phase space factor that determines the
imaginary part of the previous equation. A quick way to see it is to recall the relativistic expression
for the phase space factor in Eq.~\eqref{051016.12}, which is next multiplied by $4 m_{1i} m_{2i}$ to accommodate with the
non-relativistic normalization, cf. the discussion after Eq.~\eqref{180804.1}. Thus,  
\begin{align}
\label{190302.10}
 4 m_{i1}m_{i2}\times \frac{p_i}{8\pi\sqrt{s}}&= \frac{m_{{\rm th};i}\mu_i p_i}{2\pi\sqrt{s}}~,
\end{align}
and Eq.~\eqref{190302.9} becomes
\begin{align}
\label{190302.10b}
J_i(E)\to i\frac{m_{{\rm th};i}\mu_i p_i^{\ell_i+1}}{2\pi\sqrt{s}}~.
\end{align}
Notice that  Ref.~\cite{hanhart.190224.1} introduces an extra factor $(2\pi)^3$ multiplying the unitarity loop functions 
$J_\alpha(E)$ and $J_i(E)$ here introduced.

Another way to end with the result of Eq.~\eqref{190302.7} is by writing directly that
$T_2^{IB}=p^{\ell_I}\widetilde{T}_2^{IB}(E)$ and $T_1(E)=p^{\ell_I}\widetilde{T}_1^{IB}$,  and then solving
 %the IE of 
Eq.~\eqref{180225.9} for $\widetilde{T}_2^{IB}(E)$. 

Since  $T_2$ is a symmetric matrix, $T_2^{AB}(\vp',\vp)=T_2^{BA}(\vp,\vp')$, we can evaluate it using only the
first line of Eq.~\eqref{190225.11}, except for $T_2^{AB}(\vp',\vp)$ with $A$ and $B$ not in $I$.
These  matrix elements can be evaluated straightforwardly from the second line in the same equation as,
\begin{align}
\label{190302.11}
T_2^{\alpha\beta}&=T_1^{\alpha\beta}
+\sum_i \widetilde{T}_1^{i\alpha}(E)J_i(E) \widetilde{T}_2^{i\beta}(E)~,\\
\label{190302.12}
T_2^{\alpha 0}&=T_1^{\alpha 0}+\sum_i\widetilde{T}_1^{i\alpha}J_i(E)\widetilde{T}_2^{i0}~,\\
\label{190302.13}
T_2^{00}&=T_1^{00}+\sum_i \widetilde{T}_1^{i0}J_i(E)\widetilde{T}_2^{i0}~,
\end{align}
which are particular cases of the  general formula
\begin{align}
\label{190302.14}
T_2^{AB}&=T_1^{AB} + \widetilde{T}_1^{IA}(E) \widetilde{J}^I %[q^{\ell_I}J^I(E) q^{\ell_I}]
\widetilde{T}_2^{IB}(E)~, A,B\neq I~.
\end{align}
Let us also notice that because of the symmetry property of $T_k^{AB}(\vp',\vp)$ (with $k=1,2$) it also follows that
$\widetilde{T}_k^{IA}=\widetilde{T}_k^{AI}$ for any $A$ (which also includes $\Theta_1^{I_1I_2}(E)=\Theta_1^{I_2I_1}(E)$).
This is clear because
\begin{align}
\label{190204.1}
  T_k^{ir}(\vp,\vq)&=p^{\ell_i}\widetilde{T}_k^{ir}(\vq)~,\\
  T_k^{ri}(\vq,\vp)&=p^{\ell_i}\widetilde{T}_k^{ri}(\vq)~,\nn
\end{align}
where $r$ is a label that can refer to any channel in the sets $I$, $L$ or $|0\rangle$.

We then have all the necessary matrix elements of $T_2$ for evaluating the total amplitude $T$ as given in
Eq.~\eqref{190225.13}. The latter can then be applied to evaluate the production amplitudes $F$ in terms of the
Born terms ${\cal F}$ by making use of Eq.~\eqref{190226.1}.
The Ref.~\cite{hanhart.190224.1} makes the assumption, based on the naturalness of the values of the ${\cal F}$ involved,
that the states in $L$ dominate the intermediate ones, as long as the energy considered is near the elastic thresholds
and  the resonance poles couple strongly to (some of) the channels in $L$. This is the type of contribution represented 
in the loop at the right diagram of Fig.~\ref{fig.190225.1}.
 The Ref.~\cite{hanhart.190224.1} also neglects the Born terms in front of the term with the FSI,
because the $T$ matrix contains the near-thresholds poles.
Nonetheless, this assumption could be violated if there were CDD poles also in the
near-threshold region \cite{kang.181206.1,oller.190303.1,oller.190304.1,bugg.190304.1,Oller:1999ag}.
For the elastic channel this hypothesis is relaxed in the related and 
extended study of Ref.~\cite{guo.190224.1}. 
As a result, the expression for the production amplitudes $F$, according to assumptions put
forward in Ref.~\cite{hanhart.190224.1}, simplifies to
\begin{align}
\label{190204.2}
F_{x}&=\sum_\alpha {\cal F}_\alpha J_\alpha T_{\alpha x }~,
\end{align}
and $x\in I$ or $L$. 
Indeed, Ref.~\cite{hanhart.190224.1} considers the standard situation for
the line shapes in the study of quarkonium near-threshold states so that, together with the subsystem
comprised in the state under study, there is a third particle. For instance, one can think of
$B\to K X(3872)$, $\Upsilon(5S)\to \pi Z_b^{(}{'}^{)}$, etc,
with the resonance then decaying into the final state of interest whose line shape has to be fitted.
This reference  advocates to use the spectator assumption, so that the FSI
with the particle 3 are neglected and expected to be somewhat accounted for by the
Born amplitudes ${\cal F}$ and the free parameters involved. In particular Ref.~\cite{hanhart.190224.1} analyzes in detail the experimental
data for of the bottomonium states $Z_b$ and $Z_b'$, which can be reproduced very well and correspond to
two poles with $J^{PC}=1^{+-}$ and $I=1$. The $Z_b(10610)$ appears as a virtual state and the $Z_b'(10650)$ as a
resonance near the $B^*\bar{B}^*$ threshold. 

It is also straightforward to generalize Eq.~\eqref{190225.13} by including more than one bare state, as also
considered in Ref.~\cite{guo.190224.1}.
The  $n_0$ bare states are denoted by $|a\rangle$, with $a=1,\ldots,n_0$, and they
span the vector subspace $\Pi_0$, which is then annihilated by the projectors $\theta_1$ and $\theta$.
Thus, instead of Eq.~\eqref{190225.12} we have in these more general terms
\begin{align}
\label{190204.3}
T&=V+\sum_a V|a\rangle (E_a-E)^{-1}\langle a|T+V(H_0-E)^{-1}\theta T~.
\end{align}
Proceeding along the same lines as after Eq.~\eqref{190225.12} we have that its solution can be written as
\begin{align}
\label{190204.4}
T=T_2+\sum_a T_2|a\rangle(E_a-E)^{-1}\langle a|T~.
\end{align}
Acting on the left with $\langle b|\in \Pi_0$, the previous equation can be given as
\begin{align}
\label{190204.5}
T=T_2+T_2 J^0(I-T_2^{00}J^0)^{-1}T_2~,
\end{align}
where $J^0$ and $T_2^{00}$ are the restriction of the operators $(H_0-E)^{-1}$ and $T_2$ in the subspace $\Pi_0$.
Of course, $\langle a|(H_0-E)^{-1}|b\rangle=(E_a-E)^{-1}\delta_{ab}$.
The Eq.~\eqref{190204.5} is the sought generalization of Eq.~\eqref{190225.13} for more than one bare state considered. 
 Within the scattering model corresponding to the potential in Eq.~\eqref{190227.2} the needed matrix elements $T_2^{0B}$ can be
calculated by applying Eq.~\eqref{190302.14}.

%%%%%%%%%%%%%%%%%%%%%%%%%%%%%%%%%%%%%%%%%%%%%%%%%%%%%%%%%%%%%%%%%%%%%%%%%%%%%%%%%%%%%%%%
\subsection{Contact interactions}
\label{sec.190728.1}

When the wave length in the CM frame of the two colliding particles is much longer than
the actual range of the interactions is pertinent to take the limit situation in which the potential
in configuration space is approximated by a sum over
Dirac delta functions and their derivatives \cite{phillips.180319.1,kolck.171107.1}.
We review here the derivation of the solution for the
LS equation in these circumstances as given in Ref.~\cite{oller.190122.1}. 

Let us express the potential which couples the channels $\alpha$ and $\beta$ in a partial-wave expansion as  
 \begin{align}
\label{180319.1}
v_{\alpha\beta}(k_\alpha,p_\beta)&=k_\alpha^{\ell_\alpha}p_\beta^{\ell_\beta} \sum_{i,j}^N v_{\alpha\beta;ij}k_\alpha^{2i}p_\beta^{2j}~,
 \end{align}
with the channels expressed by the Greek subscripts running from 1 to $n$. 
Notice that we also comprise in this manner the different PWAs that couple since the formalism is the same as
for coupled-channel scattering. 
We have explicitly kept in Eq.~\eqref{180319.1} the right threshold behavior by the factor $k_\alpha^{\ell_\alpha}p_\beta^{\ell_\beta}$,
where $k_\alpha$ and $p_\beta$ are the CM three-momenta of the final and initial states for the channels indicated, respectively.  
The coefficients $v_{\alpha\beta;ij}$  in the sum could be energy dependent too, though they are not functions
of the (off-shell) three-momenta.

A convenient compact way to express Eq.~\eqref{180319.1} is to use a matrix notation so that the
potential is written as
 \begin{align}
   \label{180319.4}
   v_{\alpha\beta}(k_\alpha,p_\beta)&=[k_\alpha]^T\cdot [v]\cdot [p_\beta]~.
 \end{align}
 Here, $[v]$ is an $N n \times N n$ matrix defined as 
 \begin{align}
   \label{180319.2}
   [v]&=\left(\begin{matrix}
     [v_{11}] & [v_{12}] & \ldots & [v_{1n}]\\
     [v_{21}] & [v_{22}] & \ldots & [v_{2n}]\\
     \ldots  & \ldots  &\ldots & \ldots \\
     [v_{n1}] & [v_{n2}] & \ldots & [v_{nn}]\\
     \end{matrix}
   \right)~,
 \end{align}
where the  $[v_{\alpha\beta}]$ is the $N\times N$ matrix made up by the coefficients $v_{\alpha\beta;ij}$ with $\alpha$ and
$\beta$ fixed. 
In turn, these matrices  are the block entries of the potential matrix  $[v]$ as given in
Eq.~\eqref{180319.2}. 
Furthermore, in Eq.~\eqref{180319.4} the  $N  n$ column vectors $[k_\alpha]$ are  defined as
\begin{align}
\label{180319.3}
[k_\alpha]^T&=(
\underbrace{0,\ldots,0}_{N(\alpha-1) \text{~places}},k^{\ell_\alpha},k^{\ell_\alpha+2},\ldots,k^{\ell_\alpha+2N},0,\ldots,0)~.
\end{align}
Following the same matrix notation as for the potential, 
we write the  solution $t_{\alpha\beta}(k,p;E)$ for the LS equation, cf. Eq.~\eqref{190224.1},  as
\begin{align}
\label{180319.5}
t_{\alpha\beta}(k_\alpha,p_\beta;E)&=[k_\alpha]^T\cdot [t(E)]\cdot [p_\beta]~,
\end{align}
with $[t(E)]$ a squared $N n\times N n$ matrix, analogous to $[v]$ in Eq.~\eqref{180319.4}.
The fulfillment of the LS equation implies that $[t(E)]$ must satisfy the algebraic equation
\begin{align}
\label{180319.6}
[t(E)]&=[v(E)]-[v(E)]\cdot [G(E)] \cdot [t(E)]~.
\end{align}
Here  $[G(E)]$ is a  block-diagonal matrix that comprises the unitarity one-loop functions
\begin{align}
\label{180319.6b}
[G(E)]&=\sum_\alpha [G_\alpha(E)]~,
\end{align}
and  the  $N n\times N n$ matrix  $[G_\alpha(E)]$ is defined by  
\begin{align}
\label{180319.7}
[G_\alpha(E)]&=-\frac{m_\alpha}{\pi^2}
\int_0^\infty dq \frac{q^2}{q^2-2m_\alpha E} [q_\alpha] \cdot [q_\alpha]^T~,
\end{align}
being $m_\alpha$  the reduced mass of the $\alpha_{\rm th}$ channel.
From Eq.~\eqref{180319.6} one can deduce the sought solution 
\begin{align}
\label{180319.8}
[t(E)]&=[D(E)]^{-1}~,\\
[D(E)]&=[v(E)]^{-1}+[G(E)]~.\nn
\end{align}

%%%%%%%%%%%%%%%%%%%%%%%%%%%%%%%%%%%%%%%%%%%%%%%%%%%%%%%%%%%%%%%%%%%%%%%%%%%%%%%
\section{Perturbative treatment of crossed-channel cuts within several approaches}
\label{sec.181111.1}
\setcounter{equation}{0}   

We present in this section several methods for including perturbatively the  crossed-channel cuts when unitarizing PWAs obtained from some EFT. As indicated above, for simplicity, the crossed channels cuts are denoted generically as LHC.  
Four methods are discussed in the following. 
The first one employs Eq.~\eqref{181109.6} in terms of the function $\cN(s)$ and $g(s)$. 
From here another approach is introduced, that is referred in the literature as the Inverse Amplitude Method (IAM).
Next, we also employ the $N/D$ method when $\Delta(p^2)$ is calculated perturbatively in some effective QFT, in such a way 
that either the resulting $N/D$ IEs are solved fully or perturbatively at the level of its first iteration.
This last two ways of treating perturbatively the LHC is the object of  Secs.~\ref{sec.181112.1} and \ref{sec.181116.1}, 
respectively.

Let  $T(s)$ be calculated by employing Eq.~\eqref{181109.6b} and let us take the subtraction constant in the $g(s)$ 
function such that this function vanishes for some $s$ in the low-energy LHC region. 
This is motivated by Eq.~\eqref{181110.7}, because it is a way at our disposal to diminish the influence of the 
iteration of the LHC contributions  along the physical region (at least for not too large  values of $s$). 
 The iterative contributions referred  arise  from the product of $\cN(s)$ with $g(s)$ in Eq.~\eqref{181110.7}.  
If we impose that  $g(s_0)=0$, with $s_0$ along the LHC, and take the subtraction point also at $s_0$,  then  
$a(s_0)=0$, as it is clear from Eq.~\eqref{181110.1a}. 

This way of proceeding by choosing appropriately the subtraction constant, as introduced in  Ref.~\cite{lacour.181101.1},
 is convenient when looking for a perturbative solution of  Eq.~\eqref{181110.7} for $\cN(s)$. The latter can be settled by performing the 
geometric series of Eq.~\eqref{181109.6b}, 
\begin{align}
\label{181110.11}
T&=\cN-\cN g \cN-\cN g\cN g \cN+\ldots
\end{align}
and matching it  with a perturbative loop expansion  of $T$. 
As a result  $\cN$ could be determined order by order \cite{lacour.181101.1,ollerww.181110.1}.
To illustrate this point let us discuss the study of Ref.~\cite{ollerww.181110.1} on the 
scattering of the longitudinal components of the $W$ and $Z$ bosons, 
or simply the  $W_L W_L$  scattering. This reference employs the ChPT amplitudes for massless pions in $SU(2)$ up to NLO, 
which can be applied for energies much higher than the $W$ and $Z$ masses because of the equivalence theorem \cite{equivalence.181111.1}.  In this EFT the momentum and loop expansion go hand in hand, so that the chiral dimension $D$ of a perturbative
Feynman graph with $L$ loops is $D=2L+2+\sum_d N_d(d-2)$. Here,  $d$ is the chiral order of a given monomial in the ChPT
Lagrangian and $N_d$ is the number of such vertices \cite{weinberg.181111.1}.
The LO and  NLO isoscalar scalar $W_LW_L$ PWAs  in ChPT are \cite{gasser.181111.1,ollerww.181110.1}
\begin{align}
\label{181111.1}
T_2(s)&=\frac{s}{v^2}~,\\
\label{181111.1b} %181111.3
T_4(s)&=b\frac{s^2}{v^4}-\frac{s^2}{1728\pi^2 v^4}\left[
108\log\frac{-s}{m_H^2}+42\log\frac{s}{m_H^2}  \right]~,
\end{align}
where $v=(\sqrt{2}G_F)^{-1/2}\simeq 1/4~{\rm TeV}$ is analogous to $f_\pi$ for the pion case,  
$G_F$ is the Fermi coupling constant,  and $b$ is a NLO counterterm in the chiral series.
%aqui 
E.g. for the exchange of a scalar particle with the couplings  of a Standard Model Higgs boson
this latter constant is given by \cite{ollerww.181110.1}
\begin{align}
\label{181111.2}
b&=\frac{11 v^2}{6m_H^2}-\frac{1673-297\pi\sqrt{3}}{1728 \pi^2}~.
\end{align}
%At NLO all of these alternative theories would give rise to $T_4(s)$ as 
%written above in terms of $v$ and $b$, although with the latter having different values. 
The scale $m_H^2$ introduced in Eq.~\eqref{181111.1b} refers to a higher-energy scale
in which bare resonances could appear. In this equation the first logarithm generates the RHC and the second one the LHC.

In order to proceed with the unitarization of $T_2(s)+T_4(s)$ we employ Eq.~\eqref{181109.6b}, so that
\begin{align}
\label{181111.4}
T(s)&=\frac{\cN(s)}{1+g(s)\cN(s)}~,
\end{align}
and fix $\cN(s)$ by matching with the chiral expansion of $T(s)$. %$T_2(s)+T_4(s)$. 
For that we also make a chiral expansion of  $\cN(s)$ as 
\begin{align}
\label{181111.5}
\cN(s)&=\cN_2(s)+\cN_4(s)+{\cal O}(p^6)~,
\end{align}
with the subscript indicating the chiral order. The function $g(s)$ in the massless case reads 
\begin{align}
\label{181111.6}
g(s)&=\frac{1}{16\pi^2}\left(a+\log\frac{-s}{m_H^2}\right)~,
\end{align}
and it counts  as ${\cal O}(p^0)$, as follows by changing  the renormalization scale $m_H$ to any other value. 
This counting is also clear from  the loop expression of $g(s)$ in Eq.~\eqref{181106.5}. 
 By doing this chiral matching up to ${\cal O}(p^4)$ or NLO between $T(s)$, Eq.~\eqref{181111.4}, and 
 the ChPT expression $T_2(s)+T_4(s)$, one has
\begin{align}
\label{181111.7}
\cN_2(s)&=T_2(s)~,\\
\label{181111.7b}
\cN_4(s)&=T_4(s)+T_2(s)^2 g(s)=
\frac{s^2}{288\pi^2 v^4}\left(18(a + 16 b \pi^2 )-7\log\frac{s}{m_H^2}\right)~.
\end{align}
For $m_H\to \infty$ (which actually means that $m_H\gg 4\pi v$), 
while $|a|$ and $|b|$ are kept of ${\cal O}(1)$ [e.g. from Eq.~\eqref{181111.2} one obtains that 
$b\simeq 0.1$ for a scalar particle  of mass around 1~TeV with the same coupling structure as a 
Standard Model Higgs], a resonance pole with vanishing mass and width 
is dynamically generated \cite{ollerww.181110.1,dobado.181111.1,pelaez.181111.1}.

In the opposite situation of a light Higgs, $m_H\ll  4\pi v$, let us consider the explicit tree-level 
exchange of a Standard Model Higgs instead of including the low-energy counterterm $b$.\footnote{As stated above,
underlying this analysis is the applicability of the equivalence theorem so that 
it is not justified to apply it to a Higgs of mass 125~GeV, as actually observed \cite{pdg.181106.1}.
However, the derivations that follow up to Eq.~\eqref{181112.3} emphasize the formal aspects
of the non-perturbative method for unitarizing PWAs.}  
Then $T_4(s)$ reads 
\begin{align}
\label{181111.3} 
T_4(s)&=\frac{3s^2}{2v^2(m_H^2-s)}+\frac{m_H^4}{v^2s}\left[
\log\left(1+\frac{s}{m_H^2}\right)-\frac{s}{m_H^2}+\frac{s^2}{2m_H^4}\right]\\
&-\frac{s^2}{1728 \pi^2v^4}\left[1673-297\sqrt{3}\,\pi+108\log\frac{-s}{m_H^2}+42\log\frac{s}{m_H^2}\right]~,\nn
\end{align}
and $T_2(s)$ does not change because the Higgs exchanges only contribute at NLO and higher orders. 
 Now, in $T_2(s)+T_4(s)$  we ignore the polynomial terms divided by $4\pi v$ in
comparison with the ones  divided by $m_H\ll 4\pi v$, and take advantage of the fact that  
the direct Higgs boson exchange near the bare pole, $s\simeq m_H^2$, dominates over the other 
terms. After these simplifications, one has  
\begin{align}
\label{181112.1}
\cN(s)&\approx \frac{3s^2}{2v^2(m_H^2-s)}~.
\end{align}
The expression for $T(s)$ from Eq.~\eqref{181111.4} then becomes
\begin{align}
\label{181112.2}
T(s)&\approx \frac{3m_H^2/(2v^2)}{m_H-\sqrt{s}-i\frac{3m_H^3}{64\pi v^2}}~,
\end{align}
which is a fixed-width Breit-Wigner (BW) parameterization for the exchange of a resonance, with mass
$m_H$ and width $\Gamma_H$, 
\begin{align}
\label{181112.3}
\Gamma_H&=\frac{3m_H^3}{32\pi v^2}~.
\end{align}
It is noticeable that this expression for the width coincides with the QFT result for the width of a light Higgs boson
from the electroweak Lagrangian. This is a narrow resonance with $\Gamma_H\ll m_H$ because $m_H\ll 4\pi v$. 

Let us see how the basic expansion in Eq.~\eqref{181110.11} also gives rise to the 
Inverse Amplitude Method (IAM). This is another method that provides unitarized PWAs from its perturbative 
calculation \cite{truong.181115.1,dobado.181115.1,herrero.181115.1,pelaez.181115.1,oller.181115.1,oller.181115.1b}.
We come back again to Eq.~\eqref{181109.6b} and express $\cN=\cN_2+\cN_4+{\cal O}(p^6)$, which are determined 
by matching with $T=T_2+T_4+{\cal O}(p^6)$ as explained above. Then, 
\begin{align}
\label{181116.1}
T(s)&=\left(\left[T_2(s)+T_4(s)+T_2(s) g(s) T_2(s)+{\cal O}(p^6)\right]^{-1}+g(s)\right)^{-1}~.
\end{align}
The next step is to perform the chiral expansion of the terms inside the round brackets in the previous equation,  
\begin{align}
\label{181116.2}
T(s)&=\left(T_2(s)^{-1}\left[ I+T_4(s)T_2(s)^{-1} + T_2(s) g(s) + {\cal O}(p^4) \right]^{-1} + g(s) \right)^{-1} \\
&=\Big(T_2(s)^{-1} \left[ I-T_4(s)T_2(s)^{-1}-T_2(s)g(s)+{\cal O}(p^4) \right]+g(s) \Big)^{-1}\\
&=\left[ I-T_4(s)T_2(s)^{-1}+{\cal O}(p^4)\right]^{-1}T_2(s)\nn\\
&=T_2(s)\left[T_2-T_4+{\cal O}(p^6)\right]^{-1}T_2(s)~.\nn
\end{align}
The last expression is the NLO IAM \cite{truong.181115.1,dobado.181115.1,herrero.181115.1,pelaez.181115.1,oller.181115.1,oller.181115.1b}
\begin{align}
\label{181116.3}
T(s)&=T_2(s)\left[T_2(s)-T_4(s)\right]^{-1}T_2(s)~.
\end{align}
 Despite it is based on a perturbative solution of Eq.~\eqref{181110.7},
the IAM has the property that does not depend on the function $g(s)$ and, therefore, on its subtraction constant.
This can also be seen to be the case because  Eq.~\eqref{181116.3} can be obtained by the expansion of the  inverse of the PWA, 
$T(s)^{-1}=(T_2+T_4)^{-1}=T_2^{-1}(T_2-T_4+{\cal O}(p^6))T_2^{-1}$, 
and then taking the inverse of this expansion. However, the IAM gives rise to a pathological behavior of the PWAs for $s$ around a  
$S$-wave Adler zero. The problem originates because then $T_2(s)$ is zero around this value so that there is a typically narrow 
energy region in which $T_4(s)$ becomes indeed larger than $T_2(s)$. It is then not adequate to perform the expansion of the inverse 
in powers of $T_4(s)T_2(s)^{-1}$ in that region. Modified versions of the IAM  are derived in
Refs.~\cite{Hannah:1997sm,gomez.181116.1} to cure this deficiency. However, let us notice that this problem is absent by employing Eq.~\eqref{181116.1},  
$T=1/([\sum_n \cN_n(s)]^{-1}+g)$, without expanding $\cN(s)^{-1}$. 
This is the method to include perturbatively the LHC contributions explained above in this section.

There is an alternative dispersive derivation of the uncoupled IAM by performing a DR  
of the auxiliary function $G(s)=T_2(s)^2/T(s)$ \cite{truong.181115.1,dobado.181115.1,pelaez.181115.1}. 
Its  imaginary part is from Eq.~\eqref{051016.12b} 
\begin{align}
\label{181116.4}
\Im G&=-T_2(s)^2\rho(s)~.
\end{align}
A three-times subtracted DR for $G(s)$ is written because $T_2(s)^2$ diverges  at most like $s^2$ for $s\to\infty$, 
\begin{align}
\label{181116.5}
G(s)&=G(0)+G'(0)s+\frac{1}{2}G''(0)s^2
-\frac{s^3}{\pi}\int_{s_{\rm th}}^\infty ds'\frac{\rho(s')T_2(s')^2}{(s')^3(s'-s)}
-LC(G)+PC(s)~.
\end{align}
In this equation  $-LC(G)$ stems from the DR engulfing the LHC  in $G(s)$ and $PC(s)$ 
stems from possible zeroes of $T(s)$. In the original derivation this contribution is neglected
 because the zeroes in the denominator are expected to be canceled by $T_2(s)^2$.
There could be a mismatch between the zeroes of $T_2(s)$ and those of $T(s)$ in which case the IAM generates a pathological behavior
in some narrow energy regions \cite{oller.181115.1,oller.181115.1b,gomez.181116.1}, as discussed above. 
%In the subsequent we neglect this term.

The ChPT expansion of $G(s)$ up NLO is invoked to fix the constants $G(0)$, $G'(0)$ and $G''(0)$ with the result
\begin{align}
\label{181116.6}
G(s)&=\frac{T_2(s)^2}{T_2(s)+T_4(s)+{\cal O}(p^6)}=T_2(s)-T_4(s)+{\cal O}(p^6)~,\\
G(0)&=T_2(0)-T_4(0)~,\nn\\
G'(0)&=T_2'(0)-T_4'(0)~,\nn\\
G''(0)&=T_2''(0)-T_4''(0)~.\nn
\end{align}
By the same reason $LC(G)$ is approximated by the crossed-channel cut  $T_4(s)$, $LC(T_4)$. 
Furthermore, the dispersive integral along the RHC of  Eq.~\eqref{181116.5} is minus the one in $T_4(s)$,
which imaginary part  along the RHC is $\Im T_4(s)=T_2(s)\rho(s)T_2(s)$ because of perturbative unitarity.
Thus, we conclude that $G(s)=T_2(s)-T_4(s)$  and  $T(s)$ is given by  Eq.~\eqref{181116.3}. 
The IAM for the two-loop ChPT PWAs is deduced in Ref.~\cite{arriola.181116.1}.

The IAM has  been applied to many processes in a large number of studies. 
Just to quote a few of them  we refer to meson-meson scattering \cite{pelaez.181115.1,oller.181115.1,oller.181115.1b,nicola.181116.2},
quark-mass dependence of masses and decay constants \cite{nebreda.181116.1}, 
$W_LW_L$ scattering \cite{dobado.181111.1,pelaez.181111.1}, 
$\pi N$ scattering \cite{nicola.181116.1}, etc.

%%%%%%%%%%%%%%%%%%%%%%%%%%%%%%%%%%%%%%%%%%%%%%%%%%%%%%%%%%%%%%%%%%%%%%%
\subsection{Solving the $N/D$ equations with $\Delta(p^2)$ calculated perturbatively}
\label{sec.181112.1}

Now we take the $N/D$ IEs and solve them for a given discontinuity of the PWA along the LHC, which is  
denoted by $\Delta(p^2)$. This function is supposed to be calculated perturbatively in a given EFT. 
We review here  the presentation in Ref.~\cite{oller.181112.1}, which employs the $N/D$ method to study 
 the low-energy  $NN$ interactions, with $\Delta(p^2)$ calculated from the ChPT Lagrangians with 0, 1 and 2 baryons.
The PWAs are denoted by employing  the $LSJ$ basis and the mixing between the triplet states 
($S=1$)  with $\ell=J\pm 1$ occurs (except for the $^3P_0$). 
The $NN$ PWAs are analytic functions of $p^2$ with a RHC ($p^2>0$) and a  LHC ($p^2<-m_\pi^2/4=L$). 
This threshold is due to the one-pion exchange (OPE) in the $t$- and $u$-channels, the lightest among all the states exchanged.

We reintroduce the $N/D$ method in a different manner with respect to 
Sec.~\ref{sec.181104.1} because the LHC is explicitly included.
We gather here for convenience the discontinuity equations 
for the functions $N(p^2)$ and $D(p^2)$,
\begin{align}
\label{181112.5}
T(p^2)&=\frac{N(p^2)}{D(p^2)}~,\\
\Im D(p^2)&=-\rho(p^2) N(p^2)~,~p^2\geq 0~,\nn\\
\Im N(p^2)&=D(p^2)\Delta(p^2)~,~p^2\leq L~,\nn
\end{align}
being zero in any other interval of the real $s$ axis.  
We use the non-relativistic reduction of the phase-space factor 
\begin{align}
\label{181112.6}
\rho(p^2)&=\frac{m_N\sqrt{p^2}}{4\pi}~,
\end{align}
cf. Eq.~\eqref{180804.16} with $\mu=m_N/2$.  
The $N/D$ IEs written with $m$ subtractions in $N$ and $n$ subtractions in $D$ are indicated by ${\rm ND}_{mn}$.
They read, cf. Eq.~\eqref{181112.5},
\begin{align}
\label{181112.7}
D(p^2)&=1+\sum_{i=1}^{n-1}\delta_i (p^{2}-p_0^2)^i
-\frac{(p^2-p_0^2)^n}{\pi}\int_0^\infty\frac{\rho(q^2)N(q^2)dq^2}{(q^{2}-p_0^2)^n(q^2-p^2)}~,\\
\label{181112.7b}
N(p^2)&=\sum_{i=0}^{m-1} \nu_i (p^{2}-p_0^2)^i+\frac{(p^2-p_0^2)^m}{\pi}\int_{-\infty}^{L}
\frac{\Delta(q^2)D(q^2)dq^2}{(q^2-p_0^2)^m(q^2-q^2)}~.
\end{align}
We have fixed the normalization of $D(p^2)$ at threshold so that $D(0)=1$, 
because we can divide simultaneously the functions $N(p^2)$ and $D(p^2)$ 
by the same real constant without modifying $T(p^2)$. 
In Eqs.~\eqref{181112.7} and \eqref{181112.7b} we have always taken the same subtraction point $p_0^2$, 
but it is not necessary and one could use several of them. 
The IE for $D(p^2)$ is obtained after substituting the DR for $N(p^2)$ in the one of $D(p^2)$. It reads,
\begin{align}
\label{181113.1}
D(p^2)&=1+\sum_{i=1}^{n-1}\delta_i (p_i^{2}-p_0^2)^i
-\sum_{i=0}^{m-1}\nu_i\frac{(p^2-p_0^2)^n}{\pi}\int_{0}^{\infty}\frac{\rho(q^2)dq^2}{(q^2-p^2)(q^2-p_0^2)^{n-i}}\\
&+\frac{(p^2-p_0^2)^n}{\pi^2}\int_{-\infty}^{L} \frac{\Delta(q^2)D(q^2)}{(q^2-p_0^2)^m}
\int_0^\infty\frac{\rho(k^2)dk^2}{(k^2-p^2)(k^2-q^2)(k^2-p_0^2)^{n-m}}~.\nn
\end{align}
In order to end with  finite integrals along the RHC in the previous equation we require that $m\leq n$.
This equation is a linear IE of $D(p^2)$ along the LHC, $p^2<L$,  which we typically solve numerically by discretization of the 
integral. Once $D(p^2)$ is known on the LHC then the functions $D(p^2)$ and $N(p^2)$ can be calculated for any complex value of 
$p^2$ from Eqs.~\eqref{181113.1} and Eq.~\eqref{181112.7b}, respectively, and with them the PWA $T(p^2)$ itself.

 The RHC integrals  in Eq.~\eqref{181113.1} can be given algebraically for which we introduce the loop function $g(p^2,k^2)$ given by
\begin{align}
g(p^2,k^2)&=\frac{m}{4\pi^2}\int_0^\infty \frac{q dq^2}{(q^2-p^2)(q^2-k^2)}
=\frac{im/4\pi}{\sqrt{p^2}+\sqrt{k^2}}~.
\end{align}
When $p^2$ and $k^2$ have values along the LHC, one should make the replacement $p^2\to p^2+i\ve$ and $k^2\to k^2+i\ve$ in the 
previous equation. By the derivative of $g(p^2,k^2)$ with respect to any of its arguments 
we have an expression with more subtractions involved. 

In the works \cite{alba.181113.1}, \cite{oller.181112.1} and \cite{oller.181113.1} the LHC discontinuity $\Delta(p^2)$ is calculated 
consecutively up to higher orders in the chiral expansion. Namely up to orders  $p^0$, $p^2$ and $p^3$, respectively. 
 A rather accurate reproduction of the $NN$ phase shifts results at ${\cal O}(p^3)$.

By adding a higher-order CDD pole  to $D(p^2)$ in Eq.~\eqref{181113.1} as $\gamma_{\ell}/p^{2\ell}$, one can impose 
by construction  that a PWA with orbital angular momentum $\ell$ vanishes  as $p^{2\ell}$ when $p^2\to0 $. 
 Notice that, as it is clear from Eq.~\eqref{181113.1}, this way of proceeding does not imply any spurious dependence on the subtraction point if the latter is varied. 
Similarly, more terms could be added by adding   to $D(p^2)$ CDD poles of lower order, 
 $\sum_{i=1}^{\ell}\gamma_i/p^{2i}$.
Its number is limited by the aim of reproducing a given number of low-energy scattering observables, e.g. 
a few parameters in the effective range expansion (ERE).

An interesting application of the $N/D$ method, as developed in Ref.~\cite{oller.181113.1}, is the fact that it allows 
to calculate with great numerical precision the shape parameters in the ERE up to very high orders. 
This improvement results because the PWA can be evaluated for complex $p^2$, so that the Cauchy theorem for complex 
integration can be applied to calculate the derivatives of $T^{-1}(p^2)$. 
Let us note that when performing an integration numerically one can take a huge number of points in its discretization. 
The ERE is given by the expansion around threshold of  
\begin{align}
\label{181113.4}
p^{2\ell+1}\cot\delta_\ell&=-\frac{1}{a}+\frac{1}{2}r p^2+\sum_{i=2}^\infty v_i p^{2i}~,
\end{align}
up to the nearest singularity   of $T^{-1}(p^2)$ [without considering  the threshold branch point].
In $NN$ scattering this discontinuity  is $L$, but for other systems this singularity could correspond to a zero 
of  $T(p^2)$ (that is, a CDD pole).

Following Ref.~\cite{oller.181113.1}  the function $H(p^2)$ is introduced,
\begin{align}
\label{181113.5}
H(p^2)&=\frac{4\pi}{m} \frac{D(p^2)}{N(p^2)}p^{2\ell}+ip^{2\ell}\sqrt{p^2} =
p^{2\ell}\sqrt{p^2}\cot\delta~,
\end{align}
where $\sqrt{p^2}$ is defined in the first Riemann sheet. 
The derivatives of $H(p^2)$ at the origin correspond to the  different terms in the ERE. 
These derivatives can be calculated by applying the Cauchy integral theorem,
\begin{align}
\label{181113.5b}
\left. \frac{d^n}{d(p^2)^n}H(p^2)\right|_{p^2=0}&
=\frac{n!}{2\pi i}\oint_{\cal C} \frac{H(z)dz}{z^{n+1}}~,
\end{align}
with ${\cal C}$ any close contour inside the radius of convergence of the ERE expansion, $p^2<L$.  
 In this way, we can write
\begin{align}
\label{181113.6}
a^{-1}&=-\frac{1}{2\pi i}\oint_{{\cal C}} \frac{H(z)dz}{z}~,\\
r&=\frac{1}{\pi i}\oint_{{\cal C}}\frac{H(z)dz}{z^2}~,\nn\\
v_i&=\frac{1}{2\pi i}\oint_{{\cal C}}\frac{H(z)dz}{z^{i+1}}~.\nn
\end{align}
For coupled PWAs a possible way of proceeding is to diagonalize first the $S$ matrix in partial waves \cite{oller.181113.1,pavon.190202.1}.
As an example of the capability of this method, Ref.~\cite{oller.181113.1} calculates a 
 shape parameter of an order as high as 10, $v_{10}$, within a $1\%$ of  numerical precision.

By employing the Fredholm theory for linear IEs \cite{tricomi.181021.1} Ref.~\cite{oller.181112.1} concludes 
that if the function $\Delta(p^2)$ has the asymptotic behavior $p^{2\gamma}$ with $\gamma<-1/2$ for $p^2\to -\infty$,  
then any IE of the type ${\rm ND}_{mn}$ has solution. 
The perturbative $\Delta(p^2)$ calculated in ChPT typically  scales like a
power of $p^2$, and it usually diverges by increasing the order of the calculation because of the derivative-coupling nature 
of the EFT.
For the case of $NN$ scattering, we have that $\Delta(p^2)$ for $p^2\to -\infty$  at LO  vanishes at least  as $1/p^2$, 
at NLO it diverges at most like $p^2$ and at next-to-next-to-leading order (NNLO) it does as $p^3$.

Let us briefly review the main steps of the argumentation in Ref.\cite{oller.181112.1}. 
We start with the once-subtracted DRs in ${\rm ND}_{11}$, which give the IE
\begin{align}
\label{181113.7}
D(p^2)&=1+\nu_1\frac{m \sqrt{-p^2}}{4\pi}+\frac{mp^2}{4\pi^2}\int_{-\infty}^L
\frac{\Delta(q^2)D(q^2)dq^2}{q^2(\sqrt{-q^2}+\sqrt{-p^2})}~.
\end{align}
Next, Ref.~\cite{oller.181112.1} replaces $\Delta(p^2)$ by its asymptotic behavior $(-p^2)^\gamma$, introduces 
the new variables $x=L/q^2$ and $y=L/p^2$, and multiplies the function $D(p^2)$  by $y^{-\gamma/2}$. 
In this way the kernel of the resulting IE for the new function $\widetilde{D}=y^{-\gamma/2}D$ is symmetric in $x$ and $y$, and 
it reads
\begin{align}
\label{181113.8}
\widetilde{D}(y)&=y^{-\gamma/2}
+y^{-\frac{\gamma+1}{2}}\nu_1\frac{m(-L)^{\frac{1}{2}}}{4\pi}
+\frac{\lambda m}{4\pi^2}(-L)^{\gamma + \frac{1}{2}}
\int_0^1\frac{\widetilde{D}(x)dx}{(xy)^{\frac{\gamma+1}{2}}\left({x}^\frac{1}{2}+{y}^\frac{1}{2}\right)}~.
\end{align}
The associated kernel and the independent term of this IE are,  
\begin{align}
\label{181113.9}
K(y,x)&=\frac{1}{(xy)^{\frac{\gamma+1}{2}}\left(x^\frac{1}{2}+y^\frac{1}{2}\right)}~,\\
f(y)&=y^{-\frac{\gamma}{2}}+y^{-\frac{\gamma+1}{2}}\nu_1\frac{m(-L)^\frac{1}{2}}{4\pi}~,\nn
\end{align}
respectively.  
Both  $K(x,y,)$ and $f(y)$ are quadratically integrable\footnote{By this it is understood that 
	\begin{align}
	\label{181113.10}
	\int_0^1dy \int_0^1 dx\, K(x,y)^2<\infty~,\\
	\int_0^1 dy\,  f(y)^2 <\infty~.\nn
	\end{align}} for $\gamma<-1/2$
and then, because of the Fredholm theorem,  the IE of Eq.~\eqref{181113.8} has a unique
solution for $\gamma<-1/2$.\footnote{The only exception would occur if 
$\frac{\lambda m}{4\pi^2}(-L)^{\gamma + \frac{1}{2}}$, which multiplies the integral in Eq.~\eqref{181113.8}, 
coincided with an eigenvalue of $K(y,x)$. However, given that $\lambda$ is continuous, and the eigenvalues of a kernel 
are discrete, one could invoke continuity to obtain the solution in the  unlikely case of such a coincidence.} 
This result can be easily extended to any IE of the type ${\rm ND}_{mn}$ from the analysis of the case ${\rm ND}_{nn}$ 
(the IEs with $m<n$ are more convergent).  
The argument of Ref.~\cite{oller.181112.1} is the following. 
If the number of subtractions in $N$ and $D$ is increased by one then there is a one more power of $p^2$
multiplying the integral in Eq.~\eqref{181113.7}. Regarding the integration along the RHC
a one more factor of $k^2$ results after $N$ is inserted 
in the expression for $D$ and another extra factor of $1/k^2$ happens from the RHC integral of $D$,
cf. Eqs.~\eqref{181112.7} and \eqref{181112.7b}.
The latter two factors cancel each other. 
There is also an extra factor $1/q^2$ in the LHC integral of Eq.~\eqref{181113.7} because of the extra subtraction in $N$.
The additional powers of $p^2$ and $1/q^2$ are removed once $D(y)$ is multiplied by an extra factor of $y$ so as
to end with a symmetric kernel, which becomes the same symmetric kernel $K(y,x)$ as given in  Eq.~\eqref{181113.9}.
Regarding the independent term it does not become more singular because the extra power of
$1/y$ is also canceled when the IE is multiplied by $y$.  
By iterating this process step by step, the necessary number of subtractions can be reached at the same time as
 the kernel and the independent term remain as   square integrable functions.

%%%%%%%%%%%%%%%%%%%%%%%%%%%%%%%%%%%%%%%%%%%%%%%%%%%%%%%%%%%%%%%%%%%%%
\subsection{The first-iterated solution of the $N/D$ method}
\label{sec.181116.1}

Given $\Delta(p^2)$  the first-iterated $N/D$ method consists of iterating once along the RHC, by
 substituting $D(q^2)\to 1$ in the expression for $N(p^2)$, Eq.~\eqref{181112.7b}, which is calculated by performing 
 the resulting integration. Afterwards, one can also calculate $D(p^2)$ by evaluating the integral in Eq.~\eqref{181112.7}. 
 In summary, we have
\begin{align}
\label{181116.7b}
D(p^2)&=1+\sum_{i=1}^{n-1}\delta_i (p^{2}-p_0^2)^i-\frac{(p^2-p_0^2)^n}{\pi}\int_0^\infty \frac{\rho(q^2)V(q^2)dq^2}{(q^2-p_0^2)^n(q^2-p^2)}~,\\
\label{181116.7}
N(p^2)&=\sum_{i=0}^{m-1}\nu_i (p^{2}-p_0^2)^i+\frac{(p^2-p_0^2)^m}{\pi}\int_{-\infty}^{L}
\frac{\Delta(q^2)dq^2}{(q^2-p_0^2)^m(q^2-q^2)}~.
\end{align}
The applications for which this method can be more straightforwardly implemented happen when $\Delta(p^2)$ 
 is obtained from  tree-level amplitudes. Then, one can directly identify $N(p^2)$ in Eq.~\eqref{181116.7} with this 
 tree-level calculation, denoted in the following by $V(p^2)$.\footnote{A possible $s$-channel exchange of a bare particle 
in $V(p^2)$ can be easily accommodated by including the corresponding pole in the DR of Eq.~\eqref{181116.7}.}  
 In this situation, the only unknown parameters in Eq.~\eqref{181116.7b}  are the subtraction constants $\delta_i$, 
 whose minimum number is fixed by  the possible  diverging degree in $p^2$ of $V(p^2)$ for  $p^2\to \infty$. 
More subtractions could also be added and their effects explored.

We are going to review on the application of this approach to the study of vector-vector interactions, first on the uncoupled case 
of $\rho\rho$ scattering \cite{gulmez.181101.2}, and afterwards its generalization for the coupled case \cite{gulmez.181116.1}. 
In the rest of this section we use again as argument the Mandelstam variable $s$, since all the crossed-channel contributions 
are already accounted for by the tree-level calculation of $V(s)$ and only RHC dispersive integrals are involved for calculating 
$D(s)$. 

\begin{figure}
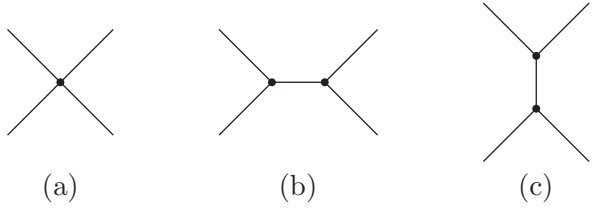

\begin{center}
\begin{axopicture}(240,100)
\Vertex(20,50){1.5}
\Line(0,30)(20,50)
\Line(0,70)(20,50)
\Line(20,50)(40,30)
\Line(20,50)(40,70)
\Text(20,10){{\small (a)}}%end contact interactions
\Vertex(100,50){1.5}
\Vertex(120,50){1.5}
\Line(100,50)(120,50)
\Line(80,30)(100,50)
\Line(80,70)(100,50)
\Line(120,50)(140,30)
\Line(120,50)(140,70)
\Text(110,10){{\small (b)}}%end s-exchange diagram 
\Vertex(200,40){1.5}
\Vertex(200,60){1.5}
\Line(200,40)(200,60)
\Line(180,80)(200,60)
\Line(180,20)(200,40)
\Line(220,80)(200,60)
\Line(220,20)(200,40)
\Text(200,10){{\small (c)}}%end t,u-exchange diagram 
\end{axopicture}
		\caption{{\small Vector-vector scattering diagrams resulting from the Lagrangian in Eq.~\eqref{181116.8}. 
The diagram (a) corresponds to contact interactions. 
The diagrams (b) and (c) represent the $s$-channel and $t$- and $u$-channel exchanges of a vector resonance, in this order. }
			\label{fig.190202.1}}
\end{center}
\end{figure}
The dynamical input of Ref.~\cite{gulmez.181101.2} consists of the tree-level amplitudes resulting 
 from the gauge-boson part of the hidden-gauge Lagrangian \cite{bando.181110.1,bando.181110.2}, which reads
\begin{align}
\label{181116.8}
{\cal L}'&=-\frac{1}{4}{\rm Tr}F_{\mu\nu}F^{\mu\nu}~,\\
F_{\mu\nu}&=\partial_\mu V_\nu-\partial_\nu V_\mu -ig[V_\mu,V_\nu]~,\nn\\
V_\mu&=\left(\begin{array}{ll}
\frac{1}{\sqrt{2}}\rho^0 & \rho^+ \\
\rho^- & -\frac{1}{\sqrt{2}}\rho^0  
\end{array}\right)~,\nn
\end{align}
and the resulting vertices involve three or four  vectors. 
The later vertices produce contact interactions and the former ones give rise to 
 the exchange of one vector resonance, which generates the LHC in $V(p^2)$ when the exchange takes place in the crossed channels, see  Fig.~\ref{fig.190202.1}. 
 The unitarization of tree-level amplitudes obtained from the hidden-gauge Lagrangian of Eq.~\eqref{181116.8} 
 was first explored in  Ref.~\cite{molina.181116.1}. 
 However, this reference only considers the extreme non-relativistic limit of the $\rho$ propagator, 
 so that the exchange diagrams in Figs.~\ref{fig.190202.1}(b) and (c) are reduced to contact interactions without LHC. 
 Nonetheless, the approach of Ref.~\cite{molina.181116.1} for the $\rho$ propagators is still sensible near the  $\rho\rho$ threshold $s_{\rm th}$,  and its unitarization is achieved by employing  Eq.~\eqref{181111.4} with $\cN(s)=\widetilde{V}(s)$ (a result that stems 
 from the application of the formalism of Sec.~\ref{sec.190203.1}),
\begin{align}
\label{181116.9}
\widetilde{T}(s)&=\frac{\widetilde{V}(s)}{1+\widetilde{V}(s)g(s)}~.
\end{align}
Since $V(s)\to \widetilde{V}(s)$ near threshold we also consider Eq.~\eqref{181116.9} with $\widetilde{V}(s)$ 
replaced by $V(s)$,
\begin{align}
\label{181116.9b}
\widetilde{A}(s)&=\frac{V(s)}{1+V(s)g(s)}~.
\end{align}
Three subtraction constants are taken in Eq.~\eqref{181116.7b} because $V(s)$ diverges at most like $s^2$ for $s\to\infty$ \cite{gulmez.181101.2}. The Eq.~\eqref{181116.7b} is rewritten with $p_0=0$, 
because the matching with Eq.~\eqref{181116.9b} is around threshold. Therefore, for $D(s)$ we use the expression 
\begin{align}
\label{181116.10}
D(s)&=\gamma_0+\gamma_1(s-s_{\rm th})+\frac{1}{2}\gamma_2(s-s_{\rm th})^2
-\frac{(s-s_{\rm th})s^2}{\pi}\int_{s_{\rm th}}^\infty \frac{\rho(s')V(s')ds'}{(s'-s_{\rm th})(s')^2(s'-s)}~.
\end{align}
The matching with the denominator of Eq.~\eqref{181116.9b} implies that 
\begin{align}
\label{181116.11}
\gamma_0+\gamma_1(s-_{\rm th})+\frac{1}{2}\gamma_2(s-s_{\rm th})^2&=1+V(s)g(s) \\
&+\frac{(s-s_{\rm th})s^2}{\pi}\int_{s_{\rm th}}^\infty \frac{\rho(s')V(s')ds'}{(s'-s_{\rm th})(s')^2(s'-s)}~.\nn
\end{align}
Denoting by $\omega(s)$ the rhs of the previous equation, the constants $\gamma_i$  are given by 
\begin{align}
\label{181116.12}
\gamma_0&=1+V(s_{\rm th})g(s_{\rm th})~,\\
\gamma_1&=\omega'(s_{\rm th})~,\nn\\
\gamma_2&=\omega''(s_{\rm th})~,\nn
\end{align}
where a prime corresponds to the derivative of $\omega(s)$ with respect to $s$. 
The subtraction constant in $g(s)$, on which $\gamma_0$ depends, is taken according to its natural value, cf. Eq.~\eqref{181106.8} (which, as expected, gives rise to very similar results as obtained by employing a three-momentum cutoff $\Lambda$, 
as discussed in Ref.~\cite{gulmez.181101.2}).
Within this scheme of work Ref.~\cite{gulmez.181101.2} confirms  a bound-state 
pole with $I=J=0$ below but close to the threshold of $\rho\rho$, as already obtained in Ref.~\cite{molina.181116.1}.  
This pole could be associated tentatively with the  resonance $f_0(1370)$. 
However, the deep pole for $I=0$ and $J=2$ found in Ref.~\cite{molina.181116.1} is not confirmed by 
Ref.~\cite{gulmez.181101.2}, because of the particularly strong effects for these quantum numbers of the branch-point singularity in $V(s)$ at $s=3M_\rho^2$. By employing the extremely non-relativistic approximation for the $\rho$ propagator 
(which is not justified for the large three-momenta involved in the region of this pole),
 Ref.~\cite{molina.181116.1} neglects altogether these important effects. 

The extension of the formalism of Ref.~\cite{gulmez.181101.2} for coupled scattering  is given in Ref.~\cite{gulmez.181116.1}, which 
studies the $SU(3)$  vector-vector scattering. The basic equations are completely analogous to the ones already reviewed here, 
and the extension is accomplished by making use of a matrix language.  
The resulting $T$ matrix of coupled PWAs is written as
\begin{align}
\label{181116.13}
T&=D(s)^{-1}N(s)~,
\end{align}
where the matrix elements of $N(s)$ and $D(s)$ are 
\begin{align}
\label{181116.14}
N_{ij}(s)&=V_{ij}(s)~,\\
D_{ij}(s)&=\gamma_{0;ij}+\gamma_{1;ij}(s-s_{{\rm th};j})+\frac{1}{2}\gamma_{2;ij}(s-s_{{\rm th};j})^2\\
&-\frac{(s-s_{{\rm th};j})s^2}{\pi}\int_{{s_{\rm th};j}}^\infty \frac{V_{ij}(s')\rho_j(s')ds'}{(s'-s_{{\rm th};j})(s')^2(s'-s)}~.\nn
\end{align}
We indicate with $s_{{\rm th};j}$ the threshold for the $j_{\rm th}$ channel. 
The analogous procedure explained above is followed and the matrix $D_{ij}(s)$ is matched with $I+V g$ around threshold, 
such that 
\begin{align}
\label{181116.15}
\omega_{ij}(s)&=
\delta_{ij}+V_{ij}(s)g_j(s)
+\frac{(s-s_{{\rm th};j})s^2}{\pi}\int_{s_{{\rm th};j}}^\infty \frac{V_{ij}(s')\rho_j(s')ds'}{(s'-s_{{\rm th};j})(s')^2(s'-s)}~.
\end{align} 
The Taylor expansion of   $V_{ij}(s)$ (present in $\omega_{ij}(s)$) around $s_{{\rm th};j}$ is meaningful because, as noticed in Ref.~\cite{gulmez.181116.1}, the LHC for $V_{ij}(s)$ runs below the thresholds $s_{{\rm th};i}$
and $s_{{\rm th};j}$.
The constants $\gamma_{k;ij}$ are given by analogous expressions to Eq.~\eqref{181116.12}, 
but now with matrix indices. Therefore, 
\begin{align}
\label{181116.16}
\gamma_{0;ij}&=1+V_{ij}(s_{{\rm th};j})g_j(s_{{\rm th};j})~,\\
\gamma_{1;ij}&=\omega_{ij}'(s_{{\rm th};j})~,\nn\\
\gamma_{2;ij}&=\omega_{ij}''(s_{{\rm th};j})~.\nn
\end{align}
A set of poles is found by Ref.~\cite{gulmez.181116.1} for  $J=0,~1$ with positive parity, since only  $S$-wave scattering is considered. 
In this respect, pole positions close to the $f_0(1370)$ and $f_0(1710)$ resonances are reported.
The coupled-channel study of Ref.~\cite{gulmez.181116.1} does not find any tensor pole,  contrarily to 
the results of Refs.~\cite{molina.181116.1,geng.181116.1}, but in agreement with the previous results 
of Ref.~\cite{gulmez.181101.2} for uncoupled $\rho\rho$ scattering.  
Similarly as in Ref.~\cite{molina.181116.1} the later study of \cite{geng.181116.1} also freezes the vector propagators 
in the tree-level amplitudes so that they become pure contact interactions.

Another strategy for determining the possible subtraction  constants
that could appear in the functions $N(s)$ and $D(s)$ is, if appropriate, to match  in some energy range with the 
perturbative  $T(s)$,  calculated from an EFT. 
One could also determine the power counting of the subtraction constants by varying the subtraction point 
and studying the subsequent change of the subtraction constants, as e.g. done in
Ref.~\cite{oller.181112.1} for the case of  $NN$ scattering with ChPT as EFT. 
In this way the matching could also be performed algebraically and not only numerically. 

The first-iterated solution of the $N/D$ method has the advantage of avoiding the generation of spurious LHC discontinuities 
which could occur when coupling channels with different thresholds if using a perturbative approximation for the matrix $\cN(s)$,
Eq.~\eqref{181110.11}, as well as the IAM \cite{weise.181116.1,oller.181116.1,pelaez.181109.1}.
% This is important so as to avoid the generation of spurious poles produced by such artifacts, which correspond to zeros of ${\rm det} D(s)$ in the appropriate RS.  
These possible artifacts are also avoided by fully solving the $N/D$  method for a given $\Delta(p^2)$, 
similarly as in Sec.~\ref{sec.181112.1}, for the case of coupled channels. 
This is done in Refs.~\cite{alba.181113.1,oller.181112.1,oller.181113.1,oller.181101.1} for studying $NN$ scattering.

%%%%%%%%%%%%%%%%%%%%%%%%%%%%%%%%%%%%%%%%%%%%%%%%%%%%%%%%%%%%%%%%%
%%%%%%%%%%%%%%%%%%%%%%%%%%%%%%%%%%%%%%%%%%%%%%%%%%%%%%%%%%%%%%%%%
\section[Final(Initial)-state interactions]{Final(Initial)-state interactions}
\label{sec.181117.1}
\setcounter{equation}{0}   

To test the dynamics and structure of a system it is customary to study its interactions with probes that only 
interact feebly with the system. 
Nonetheless, this relatively weaker interactions may trigger strong FSI\footnote{This acronym was already included above in Sec.~\ref{sec.190224.1}, it means final-state interactions.} that could also involve different final states. 
 By invoking $CPT$ invariance of QFT, there could also be strongly interacting processes driving to 
 weaker interacting probes in the initial state, in this case we would refer to initial-state interactions. 
Nonetheless, to fix ideas, we take the case of  FSI in the following, although the formalism developed could be applied 
also to the initial-state interactions. 

The full $T$ and $S$ matrices, related by Eq.~\eqref{181101.4},
 include all the processes involved, both the weaker and stronger ones. 
 E.g. the $S$ matrix for $\gamma\gamma\to \pi\pi, K\bar{K}$
(the states $|\pi\pi\rangle$ and $|K\bar{K}\rangle$ are taking with definite isospin $I$, since the strong interactions 
conserve it), may be represented by a $3\times 3$ $S$ matrix (without further specification of helicity and momentum arguments because a definite partial-wave projection is taken),  
\begin{align}
\label{181117.2}
S&=\left(
\begin{array}{lll}
S_{\gamma\gamma\to \pi\pi} & S_{\pi\pi\to \pi\pi} & S_{K\bar{K}\to \pi\pi} \\
S_{\gamma\gamma\to K\bar{K}} & S_{\pi\pi\to K\bar{K}} & S_{K\bar{K} \to K\bar{K}} \\
S_{\gamma\gamma \to \gamma\gamma} & S_{\pi\pi\to \gamma\gamma} &   S_{K\bar{K}\to \gamma\gamma}
\end{array}
\right)~.
\end{align}
The matrix element for the weakest interactions to produce the channel $i$ is
 indicated by $F_i$ and is denoted generically as a  form factor. 
For clarity, let us indicate that we always keep this name even if the $F_i$ had LHC.
Maintaining only terms linear in the weakest interactions the unitarity of the $S$ matrix, 
cf. Eqs.~\eqref{181101.5} and \eqref{181101.5b}, implies that 
\begin{align}
\label{181118.1}
F_i-F_i^\dagger&=i\sum_j \int dQ_j \theta_{j}(s) F_j T^\dagger_{ji}~,
\end{align}
where $\theta_j(s)=\theta(s-s_{th;j})$ and only the states that are open contribute in the sum. 
 The validity of this relation  above the lightest threshold rests on extended unitarity \cite{olive.181102.1}, 
 even if the final state $|i\rangle $ is closed.

As usual, the sum over intermediate states in the unitarity relation is simpler when the $T$ matrix is decomposed in PWAs. 
For a simple example, let us take the pion vector form factor $F_{\pi\pi}(s)$
which is associated to the transition $\langle \gamma|T|\pi^+\pi^-\rangle$ as
\begin{align}
\label{181118.2}
\langle \gamma(q)|T|\pi^+(p)\pi^-(p')\rangle&=e \ve(q)_\mu(p-p')^\mu F_{\pi\pi}(s)~,
\end{align}
where use has been of Lorentz covariance and $\ve_\mu(q)$ is the polarization vector of the photon. 
In the FSI affecting $F_{\pi\pi}(s)$ the two pions are in $P$ wave ($I=1$).

We restrict ourselves to  two-body interactions in the final state 
and assume that the PWAs driving the FSI are symmetric in the coupled case because of time-reversal symmetry. 
As a result, the rhs of the unitarity relation of Eq.~\eqref{181118.1}  expressed in terms of PWAs in the $LSJ$ basis 
reads\footnote{Depending on the initial state it could be necessary to express it in PWAs. 
For instance, for $\gamma\gamma\to$meson-meson the $\gamma\gamma$ state is decomposed in helicity amplitudes \cite{goldberger.181119.1,morgan.181119.1}.}
\begin{align}
\label{181118.3}
\Im F_i(s)&=\sum_j F_j(s)\rho_j(s)\theta_j T_{ij}(s)^*\\
&=\sum_j F_j(s)^*\rho_j(s) \theta_{j} T_{ij}(s)~.\nn
\end{align}

For the uncoupled case  Eq.~\eqref{181118.3} simplifies as 
\begin{align}
\label{181119.1}
\Im F_1(s)&=F_1(s)T_{11}(s)^* \rho_1(s) \theta_{1}~. 
\end{align}
Since the lhs is real it follows then that the phase of $F_1(s)$ is the same (modulo $\pi$) 
as the one of $T_{11}(s)$ along the RHC. This result is an important one and is known as 
 the Watson final-state theorem. 

We can think of two coupled-channel versions of the Watson final-state theorem. 
The first one is based on the diagonalization of the strong $T$ matrix along physical values of $s$ above the thresholds of the final 
states  $i$ considered. Let us notice that since the $S$ matrix is a unitary operator it can be written as $S=e^{i O}$, with $O$ a Hermitian operator, and then it can be diagonalized by a unitary matrix. 
Since $S=1+i\rho^{1/2}T\rho^{1/2}$, cf. Eq.~\eqref{051016.13}, 
the same can be said for $\rho^{1/2}T\rho^{1/2}$. Now, we multiply  the unitarity relation of 
Eq.~\eqref{181118.3} by $\rho^{1/2}$  and rewrite it in matrix notation as
\begin{align}
\label{190103.1}
\Im \left(\rho^{1/2} F(s) \right)&=\rho^{1/2}T(s)^*\rho^{1/2}\rho^{1/2}F(s)~.
\end{align}
The rhs turns out diagonal in the eigenbasis of $\rho^{1/2}T(s)\rho^{1/2}$ and the phase of every 
element of the vector-column $\rho^{1/2} F(s)$  in this basis 
 fulfills then the Watson final state theorem. The phase entering in this version of the theorem corresponds to the 
eigenphase of $\rho^{1/2}T(s)\rho^{1/2}$. 

The other way of proceeding with the generalization of the Watson final-state theorem in coupled channels is 
to employ Eq.~\eqref{181109.6b}, in terms of the matrices $\cN(s)$ and $g(s)$ there introduced.
Next, we  rewrite the lhs of Eq.~\eqref{181118.3} as $(F_i(s)-F_i(s)^*)/2i$ and put all the resulting $F_i(s)^*$
on the rhs. In this way,
\begin{align}
\label{181119.2}
F_i(s)&=\sum_j\left[\delta_{ij}+2i\rho_j(s) \theta_{j} T_{ij}(s)\right]F_j(s)^*~,
\end{align}
We also write this expression in matrix notation, extracting $T(s)=(\cN^{-1}+g)^{-1}$ as a common factor at the beginning of the rhs,
\begin{align}
\label{181119.2b}
F&=\left(\cN^{-1}+g\right)^{-1}\left(\cN^{-1}+g+2i\rho(s)\theta \right)F^*~.
\end{align}
Notice also that  $\rho\theta=\rho$. 
Since $\Im g(s)=-\rho(s)$ it follows  that $g(s)+2i\rho(s)\theta(s)=g(s)^*$,
and then from Eq.~\eqref{181119.2b} one can write  along the RHC that   
\begin{align}
\label{181119.3}
\left(\cN^{-1}+g\right)F&=\left(\cN^{-1}+g^* \right)F^*~.
\end{align}
Canceling $\cN$ from both sides, it follows the equality
\begin{align}
\label{181119.4}
\left[I+\cN(s) g(s)\right]F(s)&=\left[I+\cN(s) g(s)^* \right]F(s)^*~.
\end{align}
This is an expression that generalizes the Watson final-state theorem for the uncoupled case to coupled channels \cite{oller.180804.1,palomar.190203.1}.

Importantly, it follows from Eq.~\eqref{181119.4} that
\begin{align}
\label{181119.5}
\left[I+\cN(s) g(s)\right]F(s)
\end{align}
has no RHC \cite{basdevant.181119.1}, because this product of matrices is the same as its complex conjugate above the lightest threshold. 
Then, we can express  $F(s)$ as
\begin{align}
\label{181119.6}
F(s)&= \left[I+\cN(s) g(s)\right]^{-1}L(s)~,
\end{align}
with $L(s)$  an $n\times 1$ column vector with no RHC, being $n$ the number of PWAs coupled.

Analogous equations to Eqs.~\eqref{181119.4} and \eqref{181119.6} result by employing the $N/D$ method, so that 
$T(s)=D^{-1}(s)N(s)$ is used in  Eq.~\eqref{181119.2}. Following the  steps above to arrive to Eq.~\eqref{181119.2b}, 
we  have now the relation
\begin{align}
\label{181120.1}
F(s)=&D^{-1}(s)\left[D+ N 2i\rho \theta\right]F(s)^*=D^{-1}(s)D(s)^*F(s)^*~,
\end{align}
where we have taken into account that $\Im D(s)=- N(s)\rho$. Therefore, 
\begin{align}
\label{181120.1b}
D(s)F(s)=&D(s)^*F(s)^*~,  
\end{align}
so that $F(s)$ can be written too as
\begin{align}
\label{181120.2}
F(s)&=D(s)^{-1}L(s)~,
\end{align}
and $L(s)$ has only LHC (if any).  
This expression has the advantage over  Eq.~\eqref{181119.6} that it allows to write $F(s)$ as the product of 
two matrices, one of them having only RHC, $D(s)^{-1}$, and the other with only LHC, $L(s)$. 
Let us notice that $I+\cN(s) g(s)$ involves both types of cuts (since $\cN(s)$ has only LHC and $g(s)$ has a single RHC).
 For instance, if the pion form factor of  Eq.~\eqref{181118.2} is expressed according to Eq.~\eqref{181120.2}, 
$L(s)$ has no LHC because  $F(s)$ has only RHC.
The latter statement is clear if  we take into account that the only Lorentz invariant that can be built 
out of  two on-shell pion momenta is the Mandelstam variable $s$.

For the two coupled-channel case the two unitarity relations that result from Eq.~\eqref{181118.3} with $i=1,2$ allow us to 
express one form factor in terms of the other above the higher threshold \cite{basdevant.181119.1,jamin.181120.1}, 
e.g. $F_2(s)$ as a function $F_1(s)$.
In a similar way, we  expect quite generally that in a problem with $2n$ or $2n+1$ 
coupled PWAs we can make use of the unitarity relations to give $n$ of the form factors in terms of the other $n$ or $n+1$ ones.
For two channels the $S$ matrix in PWAs can be written according to the following parameterization which makes manifest its symmetric and unitary character,
\begin{align}
\label{181120.3}
S(s)&=\left(
\begin{array}{ll}
\eta e^{2i\delta_1} & i\sqrt{1-\eta^2}e^{i(\delta_1+\delta_2)} \\
i\sqrt{1-\eta^2}e^{i(\delta_1+\delta_2)} & \eta e^{2i\delta_2}
\end{array}
\right)~.
\end{align}
In the following we denote by $\cos 2\alpha=\eta$ ($\sin2\alpha=\sqrt{1-\eta^2}$) and express the $T$ matrix 
in terms of $\delta_i$ and $\eta$. In turn, the form factor is written as
\begin{align}
\label{181120.4}
F_i(s)&=f_i(s)e^{i(\delta_i(s)+\phi_i(s))}~,
\end{align}
where $f_i$ and $\phi_i$ are real functions and $f_i\geq 0$. 
Next, take the real and imaginary parts  in Eq.~\eqref{181119.2}, which is equivalent to the unitarity requirements.  
 One then has the following two relations, 
\begin{align}
\label{181120.5}
(1-\cos2\alpha)\cos\phi_1 \,f_1 & = \sqrt{\frac{\rho_2}{\rho_1}}\sin2\alpha\sin\phi_2 f_2~,\\
(1+\cos2\alpha)\sin\phi_1 \,f_1 & = \sqrt{\frac{\rho_2}{\rho_1}}\sin2\alpha\cos\phi_2 f_2\nn~.
\end{align}
 Dividing the first equality by the second one, it results
\begin{align}
\label{181120.6}
\tan\phi_1\tan\phi_2&=\tan^2 \alpha~.
\end{align}
Adding them squared,  one has
\begin{align}
\frac{\rho_2 f_2^2}{\rho_1 f_1^2}&=\tan^2\alpha+\frac{4\cos2\alpha}{\sin^2 2\alpha}\sin^2\phi_1~.
\label{181120.7}
\end{align}
The Eqs.~\eqref{181120.6} and \eqref{181120.7} allows one to express $f_2$ and $\phi_2$
in terms of $f_1$ and $\phi_1$, or vice versa.

%%%%%%%%%%%%%%%%%%%%%%%%%%%%%%%%%%%
%%%%%%%%%%%%%%%%%%%%%%%%%%%%%%%%%%%%%%%%%%%%%%%%%
\subsection[The Omn\`es solution]{The Omn\`es solution}
\label{sec.181117.2}

In this section we restrict ourselves to the uncoupled case (either because it is exact or just a good approximation), take the one-channel unitarity of Eq.~\eqref{181119.1} as valid and consider that the strong PWA is known.
 The continuous phase of the form factor $F(s)$,  because of the Watson final-state theorem, 
 is the same as the one of the PWA  $T(s)$ [both phases are denoted by the same function $\varphi(s)$].\footnote{If there is a difference between these two phases of $\pi$ then just take $-F(s)$ to make them equal.}
In the strict elastic region $\varphi(s)$ and the phase shifts $\delta(s)$ coincide but,  when inelasticity is important, 
 it might be that the phase of the form factor is still close to the one of the PWA, 
 while the latter differs markedly from $\delta(s)$.
The reason is that the form factor mostly couples  
to a given eigenchannel that diagonalizes the $S$ matrix, for which the elastic treatment holds \cite{yndu.181123.1,yndu.181123.2}.

The solution for an analytical function in the cut complex $s$ plane having a RHC which starts at threshold, ${s_{\rm th}}$, 
can be expressed in terms of the so-called Omn\`es function. The point is to make a DR for the logarithm of the function 
$f(s)=F(s)Q(s)/P(s)$, where $P(s)$ ad $Q(s)$ are the polynomials made out  of the possible
zeroes and poles of $F(s)$, respectively, so that the previous combination is free of them.  We assume that $F(s)$ has a finite 
 number of poles and zeroes. 
The discontinuity of the $\log f(s)$ along the RHC is given by the discontinuity of its imaginary 
part which is precisely $2i\varphi(s)$. Thus, it is pertinent to write down the following DR for $\omega(s)\equiv \log f(s)$,
\begin{align}
\label{181120.8}
\omega(s)&=\sum_{i=0}^{n-1}a_i s^i+\frac{s^n}{\pi}\int_{s_{\rm th}}^\infty \frac{\varphi(s')ds'}{(s')^n(s'-s)}~.
\end{align}
We have taken $n$ subtractions by assuming that $\vh(s)$ does not diverge stronger than $s^{n-1}$ for $s\to\infty$, with $n$ a finite integer. The Omn\`es function, $\Omega(s)$, is defined  as
\begin{align}
\label{181120.9}
\Omega(s)=\exp{\omega(s)}~.
\end{align}
We can always take the Omn\`es function such that $\Omega(0)=1$, and this fixes $a_0=1$. 
It is also clear that the ratio
\begin{align}
\label{181120.10}
R(s)&=\frac{F(s)}{\Omega(s)}~,
\end{align}
is a  meromorphic function of $s$ in the first RS of the complex $s$ plane.

Let us assume  that $\omega(s)$ is finite along the RHC, so that $0<|\Omega(s)|<\infty$,\footnote{Later we give an example for 
which this is not the case.} and $F(s)$ has no poles (i.e. there are no bound states).
A well-known theorem in complex analysis is that any function which is analytic in the whole complex $s$ plane is a constant or 
unbounded.
Under the present circumstances, we can apply this theorem to $R(s)$ in Eq.~\eqref{181120.10}, 
and then it must be either a constant or an analytical function unbounded at infinity. 
By writing $F(s)$ as  
\begin{align}
\label{181120.11}
F(s)&=R(s)\Omega(s)~,
\end{align}
it follows that $F(s)$ diverges as much as or stronger than $\Omega(s)$ for $s\to \infty$.
We can expect severe divergences in $\omega(s)$ if its DR requires for convergence more than one subtraction.  
This  conclusion can be reached by a similar analysis 
as the one undertaken in between Eqs.~\eqref{181012.5} and \eqref{181012.7}.
In this way, if $\varphi(s)/s^{n-1}$ ($n \geq 2$) does not vanish for $s\to \infty$, 
there would appear logarithmic divergences like $s^{n-1} \log s$. 
These terms cannot  be canceled by the $a_n s^{n-1}$ term in the subtractive polynomial.
Therefore, $R(s)$ would be an
exponential function so as to guarantee that $F(s)$  is susceptible to a DR.
In connection with this,  a hadronic form factor is expected to vanish for $s\to \infty$
because of the finite value of the non-perturbative scale of QCD, $\Lambda_{QCD}$. 
This fact is also suggested by the quark counting rules \cite{brodsky.181121.1,matveev.181121.1,llanes.181121.1}. 
By similar arguments, the phase of the form factor is also expected to tend to a constant limit for $s\to \infty$.  
However, for the case of singular interactions at the origin these expectations may fail \cite{frank.180502.1,oller.181101.1}.

If the conditions are met for a DR  of $\log F(s)Q(s)/P(s)$, cf. Eq.~\eqref{181120.8}, 
then  $R(s)=Q(s)/P(s)$ is a rational function. Therefore, from the previous analysis, we conclude that 
the DR of $\omega(s)$ in Eq.~\eqref{181120.8} involves only one subtraction. 
 It is then necessary that   $|\varphi(s)/s|<s^{-\gamma}$ for some $\gamma>0$ in the limit $s\to \infty$, 
 because otherwise we would end in this limit with the divergences of the type $s^{n-1}\log s$,
 as just discussed.\footnote{In potential scattering
the Levinson theorem \cite{levinson.181121.1,weinberg.181021.1} implies	that $\delta(0)-\delta(\infty)=(n+q/2)\pi$, 
where $n$ is the number of bound states and $q$ is the number of zero energy $S$-wave resonances. 
For more details, see Eq.~(95) of	Ref.~\cite{weinberg.181021.1}.\label{foot.181127.1}}
Thus, we arrive to the following representation for $F(s)$, 
\begin{align}
\label{181121.1}
F(s)&=\frac{P(s)}{Q(s)}\Omega(s)~,\\
\label{181121.2}
\Omega(s)&=\exp\omega(s)~,\\
\label{181121.3}
\omega(s)&=\frac{s}{\pi}\int_{s_{\rm th}}^\infty \frac{\vh(s')ds'}{s'(s'-s)}~.
\end{align}
In particular, let us notice that $P(s)$ would reabsorb the necessary normalization constant 
to guarantee that $\Omega(0)=1$ without loss of generality.
Let us calculate the asymptotic  behavior of $\Omega(s)$ in the limit $s\to \infty$ for $\vh(\infty)<\infty$.  
 We follow the procedure in Eq.~\eqref{181012.5} and re-express  $\omega(s)$ in
Eq.~\eqref{181121.3} as
\begin{align}
\label{181122.1}
\omega(s)&=\vh(\infty)\frac{s}{\pi}\int_{s_{\rm th}}^\infty \frac{ds'}{s'(s'-s)}
+\frac{s}{\pi}\int_{s_{\rm th}}^\infty \frac{\vh(s')-\vh(\infty)}{s'(s'-s)}ds'~.
\end{align}
We then have for  $s\to \infty$  
\begin{align}
\label{181122.2}
\omega(s+i\ve)&\xrightarrow[s\to\infty]{} -\frac{\vh(\infty)}{\pi}\log \frac{s}{s_{\rm th}} + i\vh(\infty)
-\frac{1}{\pi}\int_{s_{\rm th}}^\infty\frac{\vh(s')-\vh(\infty)}{s'}ds'~,
\end{align}
and the logarithmic divergence  dominates for $s\to \infty$, with the other two terms in this equation being 
constants.  It follows from here the limit behavior  
\begin{align}
\label{181122.3}
\Omega(s)&\xrightarrow[s\to \infty]{} {\cal C}_{\Omega}\, e^{i\vh(\infty)}\times
\left(\frac{s_{\rm th}}{s}\right)^\frac{\vh(\infty)}{\pi}~.
\end{align}
To determine the related asymptotic behavior for  $F(s)$ we employ  Eq.~\eqref{181121.1} with the result
\begin{align}
\label{181122.4}
F(s)&\xrightarrow[s\to\infty]{}{\cal C}_{F}\, e^{i\vh(\infty)}\times s^{p-q-\frac{\vh(\infty)}{\pi}}~,  
\end{align}
where $C_{\Omega}$ and $C_{F}$ are constants, and $p$ and $q$ are the degrees of $P(s)$ and $Q(s)$, in this order (or equivalently, 
the number of zeroes and poles of $F(s)$, respectively).  
 We can deduce from Eq.~\eqref{181122.4} the following interesting corollaries:

\noindent
i) If the actual asymptotic  high-energy behavior of $F(s)$ is of the form $s^\nu$, we have
%Eq.~\eqref{181122.4} that
\begin{align}
\label{181122.5}
p-q-\frac{\vh(\infty)}{\pi}=\nu~,
\end{align}
which one is tempted to interpret as a relativistic version of the Levinson theorem for a form factor. 

\noindent
ii) When modeling interactions, so that  some partial control
on the PWA and form factor is achieved, one should keep constant Eq.~\eqref{181122.5} under changes of the parameters 
in the approach. As $\nu$ is taken fixed then it follows that
\begin{align}
\label{181122.6}
p-q-\frac{\vh(\infty)}{\pi}={\rm fixed}~.
\end{align}
For instance,  if $\vh(\infty)/\pi$ decreases by one  and there are no bound states
 then an extra zero should be introduced in the form factor to satisfy Eq.~\eqref{181122.6}. 
A similar procedure would be applied for other scenarios.

iii) Let us emphasize that by using Eq.~\eqref{181121.1} one can satisfy Eq.~\eqref{181122.6} and compensate 
for variations in the modelling of strong interactions, while this is not the case for $\Omega(s)$.
 This function could  drive into an unstable behavior under changes of the parameters, e.g. in a fit to data, and 
 therefore it should be used with caution.

The pion scalar form factor is an important practical example in hadronic physics in relation to the points i)-iii) above. 
This form factor is associated with the light-quark scalar source, $\bar{u}u+\bar{d}d$, and is defined as
\begin{align}
\label{181122.7}
F(s)&=\hat{m}\int d^4 x e^{i(p+p')x}\langle0|\bar{u}(x)u(x)+\bar{d}(x)d(x)|0\rangle~,
\end{align}
where $u$ and $d$ are the up and down quarks,  $\hat{m}$ is the average of their current masses, 
and $s=(p+p')^2$. 

Because of the quantum numbers of the non-strange scalar source, the FSI 
occur in the isoscalar scalar meson-meson scattering. 
These interactions at enough low energy only involve the $\pi\pi$ channel and were discussed 
in Sec.~\ref{sec.181105.1}. We paid special attention to the appearance of an Adler zero, the pole of the 
$\sigma$ or $f_0(500)$ resonance, and to unitarity and analyticity, which link the mentioned zero and pole. 
At higher energies, another important aspect of these interactions is the $K\bar{K}$ channel with its threshold 
at around  991.4~MeV \cite{pdg.181106.1}. This energy is almost coincident with the appearance of the  $f_0(980)$ resonance,
 which is relatively narrow \cite{pdg.181106.1} and it gives rise to a rapid increase of the $\pi\pi$
  isoscalar scalar phase shifts at around the two-kaon threshold. In turn the elasticity parameter $\eta_{00}$ experiences a 
drastic reduction as soon as the $K\bar{K}$ opens because the $f_0(980)$ couples much more strongly 
to $K\bar{K}$ than to $\pi\pi$ \cite{guo.181123.1}, which causes an active conversion of pionic flux into kaonic one.
  These feature can be clearly seen in Fig.~\ref{fig.190130.1}.
   
The indicated rapid rise of the isoscalar scalar $\pi\pi$ phase shifts, $\delta_{00}$, 
also implies the corresponding rise of the phase of the isoscalar scalar PWA $T(s)$, $\vh(s)$, because they coincide 
below the $K\bar{K}$ threshold, i.e. for $\sqrt{s}<2m_K$. However, above this energy the rise of $\vh(s)$ is 
 interrupted  abruptly if $\delta_{00}(s_K)<\pi$, with $s_K=(2m_K)^2$, while in the opposite case $\vh(s)$ keeps rising. 
 One can go from one situation to the other under slight  changes in  the
hadronic model, being both compatible with the experimental phase shifts at around the $f_0(980)$.
This implies two drastically different types of behavior of the Omn\`es function for tiny differences in the parameters of the model, 
 according to whether $\delta(s_K)$ is larger or smaller than $\pi$.
In the former case $\Omega(s)$ becomes huge at the energy where $\delta(s)=\pi$ and for the latter one 
$\Omega(s)$ is nearly zero close to but below the $K\bar{K}$ threshold.
This problematic situation is discussed in detail by Ref.~\cite{oller.181123.1}.

In order to exemplify the rise of this phenomenon let us perform an explicit calculation by identifying 
$\varphi(s)$ with the phase of the  PWA $T(s)$ along the RHC. 
For the discussions that follow we rewrite the DR for $\omega(s)$, Eq.~\eqref{181121.3}, by   
isolating the contribution that involves the pole in the denominator of the integrand as
\begin{align}
\label{181123.1}
\omega(s)&=\vh(s)\frac{s}{\pi}\int_{s_{\rm th}}^\infty \frac{ds'}{s'(s'-s)}+
\frac{s}{\pi}\int_{s_{\rm th}}^\infty \frac{\vh(s')-\vh(s)}{s'(s'-s)}ds'~,
\end{align}
and the first integral can be evaluated algebraically.

\begin{figure}
	\begin{center}
		\scalebox{0.8}{
			\begin{tabular}{ll}
				\includegraphics[width=.5\textwidth]{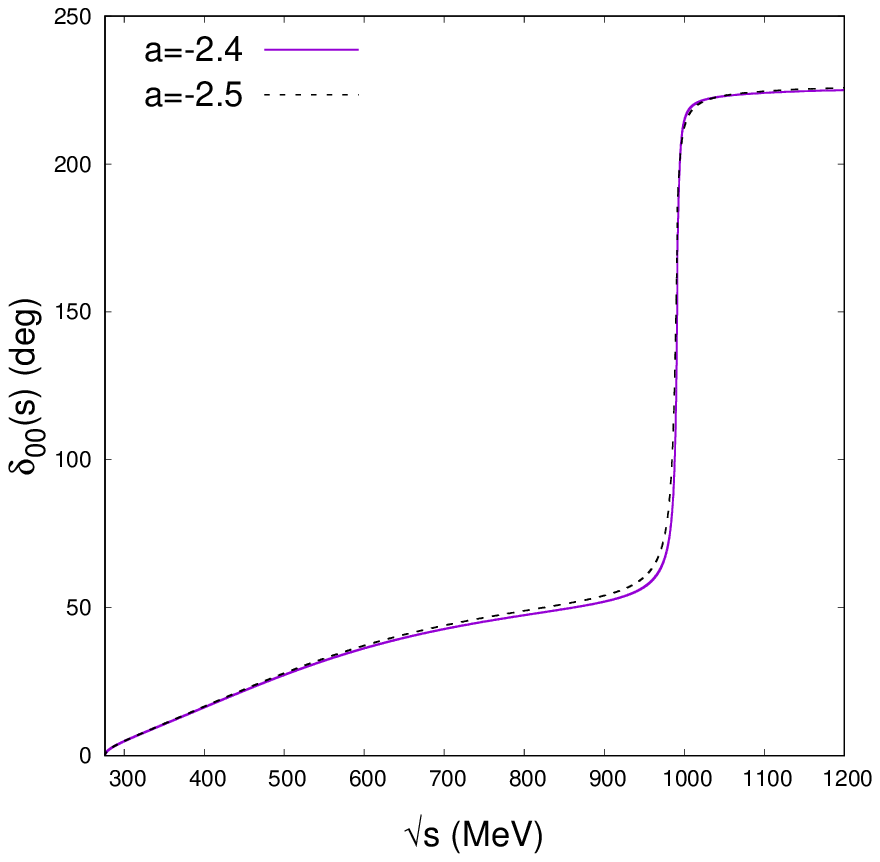} &
				\includegraphics[width=.5\textwidth]{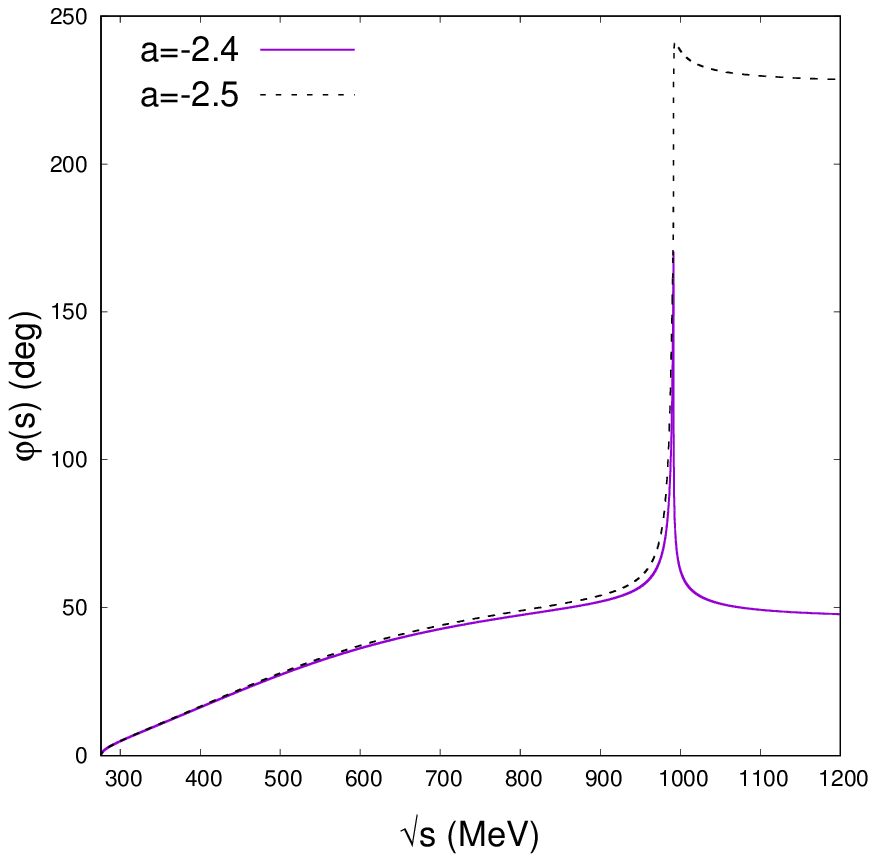} \\  
				&\includegraphics[width=.5\textwidth]{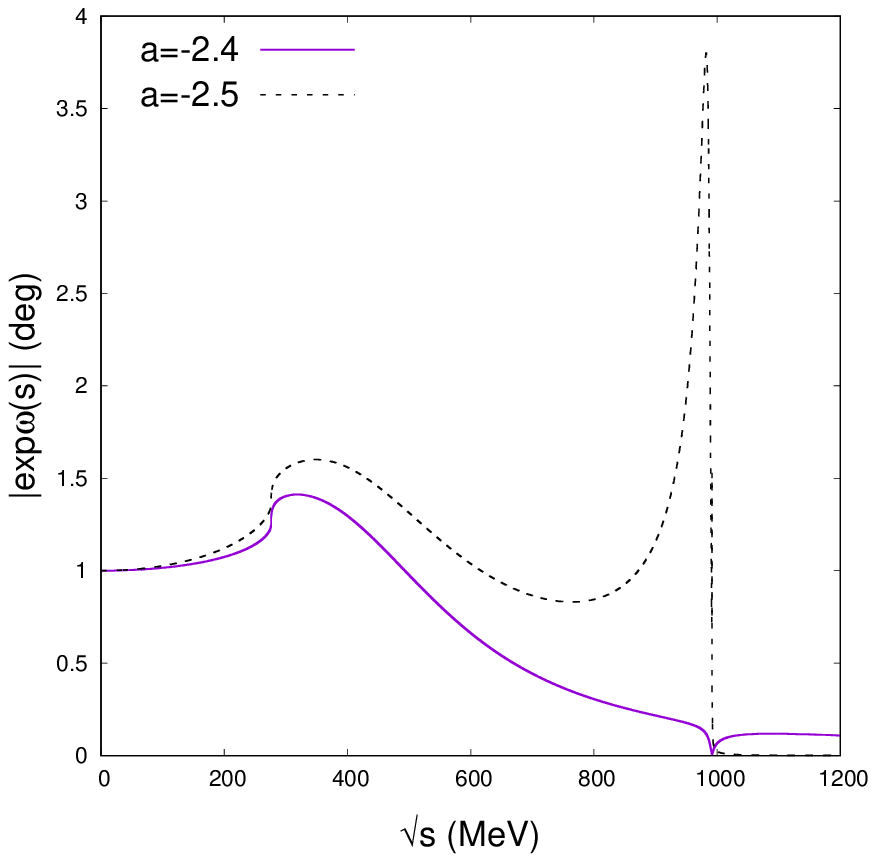} 
			\end{tabular}
		}
	\end{center}
	\caption{{\small From left to right and top to bottom: Phase shifts, $\delta_{00}(s)$,
			phase of $T(s)$, $\varphi(s)$, and modulus of the Omn\`es function, $|\exp\omega(s)|$. 
 For the solid lines we use the subtraction constant $a=-2.4$, and for the dashed lines $a=-2.5$. 
 More details are given in the text.}
		\label{fig.181124.1}
	}
\end{figure}

We make us of Eqs.~\eqref{181125.4} and \eqref{181125.5} for evaluating the isoscalar scalar 
PWAs involving the $\pi\pi$ and $K\bar{K}$. 
The limit $\delta_{00}(s_K)\to \pi$, which occurs with a subtraction constant around $-2.45$,  
gives rise to a singularity in the Omn\`es function $\Omega(s)$ because $\vh(s)$ turns out discontinuous at the 
two kaon threshold. The phase shifts $\delta_{00}(s)$ and $\vh(s)$ are represented in the left and right top panels 
of Fig.~\ref{fig.181124.1}, respectively.  One can understand in simple terms the discontinuous behavior of the phase 
$\vh(s)$ from the expression of $T(s)$ in terms of $\delta_{00}$ and $\eta_{00}(s)$, 
\begin{align}
\label{181123.2}
T(s)&=|T(s)|e^{i\varphi(s)}=\frac{1}{2\rho}\left[\eta_{00} \sin 2\delta_{00}+i(1-\eta_{00} \cos2\delta_{00}(s))\right]~.
\end{align}
A clear fact is that the imaginary part of $T(s)$ is positive for $s\gtrsim s_K$, since $\eta_{00}\leq 1$ and it experiences a strong decrease. Therefore, for $\delta_{00}(s_K)<\pi$ the real part of $T(s)$ changes sign above $s_K$ because $\delta_{00}(s)$ still raises, and the phase of $T(s)$  swiftly decreases from near $\pi$ below $s_K$ to  values in the interval $[0,\pi/2]$, see the solid line in the left top panel 
of Fig.~\ref{fig.181124.1}. This transition of $\vh(s)$ from $\pi$ to values  below $\pi/2$ finally becomes a discontinuity 
at $s=s_K$ in the limit $\delta_{00}(s_K)\to \pi^-$. 
On the other hand, for $\delta_{00}(s_K)>\pi$ the function
$\varphi(s)$ continues to grow because $\Re T(s)$ does not change sign and
$\Im T(s)>\Re T(s)$ due to the smallness of $ \eta_{00}$.
Thus, $\varphi(s)>\pi$ for $s>s_K$, which is represented by the dashed line on  
the right top panel of Fig.~\ref{fig.181124.1}. The discontinuity in $\varphi(s)$  at $s_K$
by an amount of $\pi/2$ implies a singularity in $\omega(s)$ at $s=s_K$ of the  end-point type because of the pole 
in the denominator of the integrand, with $\varphi(s_K-\ve)-\varphi(s_K+\ve)=\pm \pi/2$. 
 On the rhs of the latter equality, the plus sign applies for $\delta(s_K-\ve)\to\pi^-$ and the minus sign 
 for $\delta(s_K-\ve)\to \pi^+$. The resulting divergence is explicitly 
\begin{align}
\label{181123.3}
&\frac{1}{\pi}\left[\int^{s_K-\Delta}\frac{\varphi(s_K-\ve)ds'}{s'-s_K}
+\int_{s_K+\Delta}\frac{\varphi(s_K+\ve)ds'}{s'-s_K}\right]\to \frac{1}{\pi}\left[\varphi(s_K-\ve)-\varphi(s_K+\ve)\right]\log\Delta
=\pm \frac{1}{2} \log\Delta~,
\end{align}
with $\Delta\to 0^+$ and $\delta(s_K-\ve)\to \pi^{\mp}$, respectively.
After exponentiating $\omega(s)$ to obtain $\Omega(s)$, a factor $(\sqrt{\Delta})^{\pm 1}$ arises, 
which makes $\Omega(s_K)$ to become infinite 
when $\delta_{00}(s_K)\to \pi^+$ and  zero when $\delta_{00}(s_K)\to \pi^-$, 
as shown in the right bottom panel of Fig.~\ref{fig.181124.1}.\footnote{Of course, here we have plotted $\exp\omega(s)$ 
for values of $a$ that are very close to the value that exactly gives rise to a zero or an infinity in this function.}

It is interesting to remark that the transition $\delta_{00}(s_K-\ve)\to \pi^\pm$  implies a jump by one in Eq.~\eqref{181122.6}. 
In order to keep this relation fixed, there should be one more zero when passing from $\delta_{00}(s_K)<\pi$ to $\delta_{00}(s_K)>\pi$  ($p$ increases by one). 
Had we required the continuity from $\delta_{00}(s_K)>\pi$ to $\delta_{00}(s_K)<\pi$ then an extra pole (in the first RS) 
should be added for the latter case, so that $q$ would increase by one unit. 
This latter scenario can be safely ruled out in pion physics.
We then conclude that an Omn\`es representation of the isoscalar scalar $\pi\pi$ PWA in
the case $\delta_{00}(s_K)>\pi$ requires that it have a zero at the point at
which $\Im T(s)=0$ for $s<s_K$.
 This a lengthy derivation (but quite illustrative) of the zero of $T(s)$ in the elastic case when 
 $\delta_{00}(s_K)>\pi$ by applying that $T(s)=e^{i\delta_{00}(s)}\sin(\delta_{00}(s))/\rho$.

A similar reasoning was developed in Ref.~\cite{oller.181123.1} for the pion scalar form factor
$F(s)$, cf. Eq.~\eqref{181122.7}, because in good approximation its phase is very close to the phase of the isoscalar scalar $\pi\pi$ 
 PWA $T(s)$, even for $s\gtrsim s_K$. 
 This has been shown to be the case by calculating explicitly $F(s)$  in other approaches \cite{gasser.181123.1,oller.180804.1,moussallam.181123.1}.
Indeed, as mentioned above, the $f_0(980)$ resonance, which dominates the behavior of the isoscalar scalar meson-meson 
scattering around 1~GeV, couples much more strongly to kaons than to pions. 
E.g. Ref.~\cite{guo.181123.1} finds that its coupling to kaons is larger by around a factor 3.\footnote{Collections of miscellaneous properties, e.g. couplings, of the $f_0(980)$ and $a_0(980)$ as resulting from different models and fit to data 
can be found in Refs.~\cite{wu.190208.1,baru.181208.1}.}
As a result, the admixture between the pion and kaon form factors is suppressed and each of them 
follows its own eigenchannel of the isoscalar scalar meson-meson PWAs.
The Refs.~\cite{oller.181123.1,yndu.181123.1} provide the explicit expressions for
the eigenchannels and eigenphases.

One can deduce  the point $s_1$  at which the pion scalar form factor has a zero for $\delta(s_K)>\pi$. 
As explained in Ref.~\cite{oller.181123.1} we can  make use of the unitarity relation in the elastic region 
$|\Im F(s)|=|F(s)\sin  \delta_{00}(s)|/\rho(s)$ and the fact that there is only one zero. Therefore, it must be at the point 
$s_1<s_K$ at which $\delta(s_1)=\pi$. Notice that the $\delta_{00}(s)$ is a raising function in $s$ from $s_\pi$ up to energies
 above $s_K$ and, thus, for $s<s_K$ the $\Im F(s)$ (invoking the Watson final-state theorem) 
 can only be zero when $\delta_{00}(s)$ is an integer multiple of $\pi$.  
In this way one can fix  the first order polynomial  multiplying 
$\Omega(s)$ so that a continuous transition for $\delta(s_K)$ greater or smaller than $\pi$ originates. 
As a result, the Omn\`es representation of the pion scalar form factor involves the following modified Omn\`es function
\begin{align}
\label{181123.5}
\Omega(s)&=\left\{
\begin{array}{ll}
\exp\omega(s)& ~,~\delta(s_K)<\pi~,\\
\frac{s_1-s}{s_1}\exp\omega(s) & ~,~\delta(s_K)>\pi~. 
\end{array}
\right.
\end{align}
The context makes clear when the same symbol $\Omega(s)$ actually refers to its definition in Eq.~\eqref{181120.9} or the new one in 
Eq.~\eqref{181123.5}.

 The Ref.~\cite{oller.181123.1} also notices that one can write down a twice-subtracted DR for the pion scalar form factor,
\begin{align}
\label{181123.4}
F(s)&=F(0)+\frac{1}{6}\langle r^2\rangle_s^\pi s+\frac{s^2}{\pi}\int_{s_\pi}^\infty
\frac{\Im F(s') ds'}{(s')^2(s'-s)}~,
\end{align}
where  $\langle r^2\rangle_s^\pi$ is the quadratic scalar radius of the pion. 
Two subtractions are expected to be more than enough to guarantee the convergence of the previous DR because 
 from asymptotic QCD \cite{yndu.181123.2}  $F(s)$ is expected to vanish at infinity. 
 To include more subtractions than necessary in a DR is an appropriate method for emphasizing the information from the 
low-energy region.

A clear lesson from the discussion given is that one should use judiciously  an Omn\`es function 
  when performing fits to data. 
Possible troubles could occur if in the process of varying the free parameters in the fits an unstable behavior arises because of induced abrupt changes in the phase of the form factor which is integrated in $\omega(s)$. 
These regions of strongly different  behavior in $\exp\omega(s)$ are separated by the occurrence of a discontinuity of $\vh(s)$ in the parameter space. 
 In consequence, the fulfillment of the requirement in Eq.~\eqref{181122.6} should be  pursuit. 
 In particular, for the isoscalar scalar $\pi\pi$ PWA the function in Eq.~\eqref{181123.5} should be used, instead of a pure
Omn\`es function, $\exp\omega(s)$, cf. Eq.~\eqref{181121.2}.

Let us come back to Eq.~\eqref{181120.10}, but now we also consider that the  form factor $F(s)$ has 
LHC too,\footnote{Maybe for some readers it is unusual to employ the terminology of from factor for referring a production amplitude 
having also LHC. Nonetheless, in order to seek a more united notation, we follow the one introduced in Sec.~\ref{sec.181117.1}.} e.g. this 
the case for $\gamma\gamma\to \pi\pi$. Proceeding in the same manner now the resulting $R(s)$ also has  LHC, 
and then we denote it by $L(s)$, as in Eq.~\eqref{181120.2}. 
This way of proceeding allows a clear splitting between the RHC and LHC contributions, which is also 
exploited in the literature \cite{morgan.181119.1,pennington.190208.1,pennington.181125.1,danilkin.190208.1}. 
A  DR for $L(s)$ along the LHC is typically written,
\begin{align}
\label{181125.6}
L(s)&=\sum_{i=1}^{n-1}a_i s^i+\frac{s^n}{\pi}\int_{-\infty}^{s_L}\frac{\Im L(s')ds'}{(s')^n(s'-s)}~,\\
F(s)&=\Omega(s)L(s)~,\nn\\
\Im L(s)&=\Omega(s)^{-1} \Im F(s)~,~s<s_L~,\nn
\end{align}
where $s_L$ is the upper limit for the extension of the LHC. 
Once  the Omn\`es function is implemented from the knowledge  of the strong PWAs along the RHC, 
the required input of the previous DR is  $\Im F(s)$ along the LHC. 
In the uncoupled case the phase used for the Omn\`es function
is the phase shift of the strong PWA $T(s)$ and one could apply the approach introduced in this section.

Up to the end of this subsection let us concern ourselves with the $S$-wave of the reaction 
 $\gamma\gamma\to \pi^0\pi^0$, following the Refs.~\cite{pennington.181125.1,oller.181125.1}. 
 These references use a somewhat different non-perturbative procedure than the one of  Eq.~\eqref{181125.6} in order to calculate the low-energy cross section for $\gamma\gamma\to \pi^0\pi^0$.
The $\pi^0\pi^0$ system is considered at $S$ wave (there is no $P$ wave because of Bose-Einstein symmetry) and $I=0$ or 2. 
First,  a function ${\cal F}_I(s)$ with only RHC is built out of the  $F_I(s)$ by subtracting to it  a function $\widetilde{L}_I(s)$ suitably 
constructed so that it contains the LHC of $F_I(s)$,
 \begin{align}
 \label{181125.7}
 {\cal F}_I(s)&=\frac{F_I(s)-\widetilde{L}_I(s)}{\Omega_I(s)}~.
 \end{align}
 Next, a twice subtracted DR is implemented for this function, 
 \begin{align}
 \label{181125.8}
 F_I(s)&=\widetilde{L}_I(s)+a_I \Omega_I(s)+c_I s \Omega_I(s)
 + \Omega_I(s) \frac{s^2}{\pi} \int_{4m_\pi^2}^\infty \frac{\widetilde{L}_I(s')\sin\varphi_I(s')ds'}{(s')^2(s'-s)|\Omega_I(s')|}~.
 \end{align}
   One of the subtraction constants in Eq.~\eqref{181125.8} is fixed by taking into account the Low's theorem \cite{low.190208.1}, 
so that $F_I(s)$ in the limit $s\to 0$ tends to its renormalized Born term contribution.
The other subtraction constant is determined by matching with the NLO  ChPT calculation of Refs.~\cite{cornet.181125.1,donoghue.181125.1}.
The $\Im F(s)$  along the LHC is approximated by the contributions from the Born terms and the crossed exchanges of
the $J^{PC}$ resonance multiplets $1^{--}$ and $1^{++}$ as in Ref.~\cite{oller.181125.1}, where explicit
formulas for the resonance-exchange tree-level amplitudes can be found.
 In the actual calculations  of Refs.~\cite{pennington.181125.1,oller.181125.1}, $\widetilde{L}_I(s)$ 
is calculated by the tree-level amplitudes made out of the Born terms and the exchange of the lightest $1^{--}$ and
$1^{++}$ multiplets of vector and axial resonances, respectively. 
In this way, because of the Low's theorem 
\begin{align}
\label{181125.9}
\lim_{s\to 0}\left[F_I(s)-\widetilde{L}_I(s)\right]&={\cal O}(s)~,
\end{align}
so that   $a_I=0$ in Eq.~\eqref{181125.8}.
The contributions of the $1^{++}$ axial resonances appear one order lower in the chiral expansion than those from the 
exchange of the vector resonances. This is a clear indication that the former are more important than the latter at lower energies. 
Nonetheless,  the explicit axial exchanges are not included in Ref.~\cite{pennington.181125.1}, 
while Ref.~\cite{oller.181125.1} includes them in its results. 
This study confirms  the phenomenological relevance of the $1^{++}$ exchanges and calculates that their
 contributions amount around a 30\% of the complete result. 
A major step forward of Ref.~\cite{oller.181125.1} in comparison with Ref.~\cite{pennington.181125.1} is the use of the 
stable $\Omega_0(s)$ function defined in Eq.~\eqref{181123.5}, instead of just the pure Omn\`es function $\exp\omega(s)$.
In this way, the results at low energies are  much more stable and robust  under changes of the parameterizations 
for the PWAs used to account for  the isoscalar scalar $\pi\pi$ phase shifts in the delicate region of the $f_0(980)$ resonance 
around the $K\bar{K}$ threshold. 
A reduction of around a factor of 2 in the uncertainty of the cross section for $\gamma\gamma\to \pi^0\pi^0$ 
is achieved in this way for $\sqrt{s}\simeq M_\rho$, and  a reduction around 25\%  at $\sqrt{s}=500$~MeV.
Let us remark that  for $\delta_{00}(s_K)>\pi$ one has a zero in the denominator of Eq.~\eqref{181125.8}
for $I=0$ because $\Omega_0(s_1)=0$, cf. Eq.~\eqref{181123.5}.
However,  $\sin\varphi_0(s')/|\Omega_0(s')|$, the combination that appears in the integrand of Eq.~\eqref{181125.8},
is  finite because the zero of $\Omega_0(s')$ is at  $s_1$, where   $\varphi_0(s_1)=\pi$,
cf. Eq.~\eqref{181123.4}.

%%%%%%%%%%%%%%%%%%%%%%%%%%%%%%%%%%%%%%%%%%%%%%%%%%%%%%%%%%%%%%%%%%%%%%
\subsection[The Muskhelishvili-Omn\`es problem]{The Muskhelishvili-Omn\`es problem in coupled-channel form factors}
\label{sec.181117.3}

The basic problem under consideration is to find the solutions for a set of form factors $\{F_i(s), i=1,\ldots, n \}$, 
ordered from smaller to larger values of their thresholds $s_{{\rm th};i}$.
 The  $F_i(s)$ have LHC for $s<s_L$ and RHC for $s>s_{\rm th}$, with
$s_{\rm th}$  the lightest threshold and  $s_L$ as defined above.

The imaginary part of $F_i(s)$ for $s>s_{\rm th}$ is given by Eq.~\eqref{181118.3} because of extended unitarity.
Since the form factors are real for   $s_L<s<s_{\rm th}$ they fulfill the Schwarz reflection principle, 
and their discontinuities along the cuts obey,
\begin{align}
\label{181125.11}
\Im F_i(s+i\ve)-\Im F_i(s-i\ve)&=2i \Im F_i(s+i\ve)~.
\end{align}
As in Eq.~\eqref{181125.6} we assume that the discontinuity of these functions along the LHC is given, 
being denoted as $\Delta_L F_i(s)$. Namely,  
\begin{align}
\label{181125.12}
F_i(s+i\ve)-F_i(s-i\ve)&=\Delta_L F_i(s)~,~s<s_L~.
\end{align}
It is shown in  Eq.~\eqref{181120.2} that the  column vector $F(s)$ of matrix elements
$F_i(s)$ can be written  as the product of the %error 
inverse of the matrix $D(s)$  times $L(s)$. The former only has RHC and the latter LHC (if any). 
The possible bound-state poles of $F(s)$ would correspond to zeroes in the determinant of $D(s)$,  ${\rm det}D(s)$. 

For the study of this problem it is  convenient to consider the matrix ${\cal S}(s)$, already defined in Eq.~\eqref{181125.13}. 
%\begin{align}
%\label{181125.13}
%{\cal S}(s)&=I+T(s) 2i\rho(s)~.
%\end{align}
Let us remark that  in general $\cS(s)$ is not symmetric, even though $T(s)$ is. 
Recalling also Eq.~\eqref{181125.14}, it follows then that  $\cS(s)$ is not a symmetric or unitary matrix for $n>1$.
%It also follows from the unitarity relation satisfied by $T(s)$, Eq.~\eqref{051016.12a}, that
%\begin{align}
%\label{181125.14}
%\cS(s)\cS(s)^*&=\cS(s)^*\cS(s)=I~,
%\end{align}
%where the asterisk refers to complex conjugation and not Hermitian conjugation. 
%Therefore, $\cS(s)$ is not a symmetric nor unitary matrix for $n>1$.

Taking the $N/D$ method expression for $T(s)$,  $T(s)=D(s)^{-1}N(s)$,  the matrix $\cS(s)$ can be written as
\begin{align}
\label{181125.15}
\cS(s)&=I+2i D(s)^{-1}N(s)\rho(s)=
D(s)^{-1}\left[D(s)+2iN(s)\rho(s)\right]=D(s)^{-1}D(s)^*~,
\end{align}
an expression valid in the whole complex $s$ plane.
From this equation and taking into account that $D(s)^*=D(s^*)$ it also follows the complex-conjugation  relation
\begin{align}
\label{181126.1}
D(s)^{-1}&=\cS(s)D(s^*)^{-1}~.
\end{align}
Multiplying both sides by $L(s)$, cf. Eq.~\eqref{181120.2}, and recalling that $L(s)$ has no RHC,
an analogous relation for the form factors results
\begin{align}
\label{181126.1b}
F(s)&=\cS(s)F(s^*)~.
\end{align}

The discontinuity of  $L(s)=D(s)F(s)$ along the  LHC is 
\begin{align}
\label{181126.2}
\Delta_L L(s)&=D(s)\Delta_L F(s)~.
\end{align}
Considering  that $L(s)$ diverges  at infinity less strongly than $s^m$ for some integer $m\geq 0$,\footnote{We 
should say more rigorously that the divergence of $L(s)$ at infinity is not stronger than $s^{m-1}$, $m\geq 1$, to avoid just a logarithmic vanishing of $L(s)/s^m$. 
	However, we have in mind a power-like vanishing, $|L(s)/s^m|<|s|^{-\gamma}$, $\gamma>0$, 
	when $s\to\infty$.} we propose the following $m$-times subtracted DR
\begin{align}
\label{181126.3}
L(s)&=\sum_{i=0}^{m-1}a_i s^i+\frac{s^m}{\pi}\int_{-\infty}^{s_L}\frac{D(s')\Delta_LF(s') ds'}{(s')^m(s'-s)}~.
\end{align}
This equation in turns implies for $F(s)$ that
\begin{align}
\label{181126.4}
F(s)&=D(s)^{-1}\sum_{i=0}^{m-1}a_i s^i+\frac{s^m}{\pi}\int_{-\infty}^{s_L}\frac{D(s)^{-1}D(s')\Delta_LF(s') ds'}{(s')^m(s'-s)}~.
\end{align}

For a given matrix of PWAs $T(s)$, the problem of finding an $n\times n$ matrix $D(s)$  with only RHC 
which fulfills  $\cS(s)=D(s)^{-1}D(s)^*$, Eq.~\eqref{181125.15}, is known as the Hilbert problem.
It follows from Eq.~\eqref{181126.1} that every column of $D(s)^{-1}$ can be considered as a set of coupled form factors 
because it satisfies Eq.~\eqref{181126.1b} along the RHC. Furthermore, the form factors $D^{-1}(s)_{ij}$ with $j$ fixed and 
$i=1,\ldots,n$ have only RHC. In this way we can interpret Eq.~\eqref{181126.4} by considering that the physical 
form factors $F_i(s)$ are obtained by taking a linear combination of the column vectors from $D(s)^{-1}$, being 
the functions $L_i(s)$ the coefficients in this linear superposition.

An important property to notice is that the determinants  of $S(s)$ and $\cS(s)$ are equal, 
\begin{align}
\label{181126.5}
{\rm det}S(s)&={\rm det}\cS(s)~.
\end{align}
A few manipulations lead to this conclusion since 
\begin{align}
\label{181127.1}
{\rm det} S&={\rm det}\left(I+2i\rho^{\frac{1}{2}}T\rho^{\frac{1}{2}}\right)
={\rm det}\left(\rho^{\frac{1}{2}}\left[\rho^{-\frac{1}{2}}+2iT\rho^{\frac{1}{2}}\right]\right)
={\rm det}\left(I+2iT\rho\right)={\rm det}\cS~.
\end{align}
After diagonalizing the $S$ matrix, ${\rm det}S$ is given by the sum of the eigenphase shifts,
\begin{align}
\label{181127.2}
{\rm det}S&=\exp2i\sum_{i=1}^n\varphi_i(s)~.%={\rm det}\cS~.
\end{align}
For the two-coupled channel case the sum of the eigenphase shifts is equal to the sum of the phase shifts, 
cf. Eq.~\eqref{181120.3}.

Making use of $\cS(s)=D(s)^{-1}D(s)^*$, Eq.~\eqref{181125.15}, we can write down an Omn\`es representation for ${\rm det}D^{-1}(s)$.
As $D^{-1}(s)$ itself, the function  ${\rm det}D^{-1}(s)$ has only RHC, and
from Eq.~\eqref{181125.15} we can obtain the phase of ${\rm det}D^{-1}$ along this cut, which is half 
of the phase of ${\rm det}S$. In the following we denote the latter by $\Phi(s)$, 
namely, $\Phi(s)=2\sum_{i}\varphi_i(s)$.\footnote{The number of open channels changes with $s$. However, $\Phi(s)$ 
	is  continuous along the RHC.} 
One also has to take into account for the Omn\`es representation the possible zeros and poles of ${\rm det}D^{-1}$, 
being the former  the generalization of the CDD poles for the coupled-channel case. 
As in Eq.~\eqref{181121.1} we introduce the polynomials $P(s)$ and $Q(s)$ in factorized form, with their roots 
being the zeros and poles of ${\rm det}D(s)$, respectively. 
To simplify the notation we further introduce the symbols
\begin{align}
\label{181127.3}
\Delta(s)&={\rm det}D(s)^{-1}~,\\
s_R&=s_{{\rm th};1}~.\nn  
\end{align}
For writing down the Omn\`es representation we consider the function $e^{-i\frac{\Phi(s_R)}{2}}Q(s)\Delta(s)/P(s)$, so that 
it is equal to 1 at $s=s_R$. We then have,  
\begin{align}
\label{181127.4}
\Delta(s)&=\frac{P(s)}{Q(s)}\exp\omega(s)~,\\
\label{181127.4b}
\omega(s)&=\frac{\Phi(s_R)}{2}+\frac{s-s_R}{2\pi}\int_{s_R}^\infty \frac{\Phi(s')-\Phi(s_R)}{(s'-s_R)(s'-s)}ds'~.
\end{align}
In this form, the integral in the DR stays finite in the limit $s\to s_R$. 

By employing Eq.~\eqref{181122.4} we then have  the following limit behavior for $\Delta(s)$,
\begin{align}
\label{181127.5}
\Delta(s)&\xrightarrow[s\to\infty]{} s^{p-q-\frac{\Phi(\infty)-\Phi(s_R)}{2\pi}}~.
\end{align}

An interesting point of Eq.~\eqref{181127.5} is that it establishes a relation between  the asymptotic behavior of
$\Delta(s)$ and the leading power behavior in $s$ of the columns of $D(s)^{-1}$
\cite{musk.181126.1,warnock.181127.1,basdevant.181119.1}. 
Let us denote by  $\phi_i(s)$ the $i_{\rm th}$ column of $D(s)^{-1}$ which satisfies,
as follows from Eq.~\eqref{181126.1}, that
\begin{align}
\label{181127.6}
\cS(s)\phi_i(s)^*&=\phi_i(s)~,~s>s_R~.
\end{align}
Assuming  that $\cS(s)\to I$ at infinite as in Refs.~\cite{warnock.181127.1,basdevant.181119.1}, it follows that 
 the leading behavior of $\phi_i(s)$ should be integer-power like in this limit 
 (as no cut remains for $s\to \infty$ and it is taken for granted that  these functions can be expressed as DRs). 
 By suitable linear combinations we can always choose these $\phi_i(s)$ such that 
\begin{align}
\label{181127.7}
\Delta(s)\xrightarrow[s\to\infty]{}s^{\chi_1+\chi_2+\ldots+\chi_n}~,
\end{align}
where  $\chi_i$ is the leading degree in $s$ of $\phi_i(s)$ in this limit [which is defined by the highest 
 degree in $s$ of all the components of $\phi_i(s)$]. 
 
To reach this result let us first consider the simpler two-coupled channel case and, for definiteness, let us take that  $\chi_2\geq \chi_1$. If the leading behavior for $s\to\infty$ of  $\phi_1$ and $\phi_2$  are linearly independent vectors, then Eq.~\eqref{181127.7} is clear.     However, if the vectors made out of the leading components of $\phi_1$ and $\phi_2$ are linearly dependent,
then remove to $\phi_2$ the vector $\phi_1$ multiplied by $s^{\chi_2-\chi_1}$ times a constant, which is a new
$\phi_2$. By iterating this procedure if necessary, one ends up with a vector $\phi_2$,  whose leading behavior is linearly independent 
of $\phi_1$ and Eq.~\eqref{181127.7} holds then. 
 Of course, for $\chi_1>\chi_2$ one would proceed analogously by exchanging $1\leftrightarrow 2$.
It is quite clear that this process can be applied iteratively to treat the case with $n$ coupled PWAs (e.g. take $\chi_i\geq \chi_{i+1}$ and proceed from lower to higher values of $i$). 
Thus, Eq.~\eqref{181127.7} is fulfilled for an appropriately built matrix $D(s)^{-1}$.
By matching the rhs of Eq.~\eqref{181127.7} with the one of Eq.~\eqref{181127.5}, one has
\begin{align}
\label{181127.7b}
\chi_1+\chi_2+\ldots+\chi_n=p-q-\frac{\Phi(\infty)-\Phi(s_R)}{2\pi}~. 
\end{align}
This formalism is applied in Ref.~\cite{jamin.181120.1} to describe the strangeness-changing scalar
form factors for $K\pi$(1), $K\eta$(2) and $K\eta'$(3). 
This set up is also used previously for the study of the  $\pi\pi$ and $K\bar{K}$ isoscalar scalar form factors \cite{gasser.181123.1}, a 
problem also addressed more recently in Ref.~\cite{moussallam.181123.1} within a similar approach. 
The strangeness-changing or $\Delta S=1$ scalar  form factors are defined  by
\begin{align}
\label{181127.8}
\langle 0| \partial^\mu(\bar{s}\gamma_\mu u)(0)|K\phi_k\rangle&=-i\sqrt{\frac{3}{2}}\Delta_{K\pi}F_k(s)~,\\
\Delta_{K\pi}=m_K^2-m_\pi^2~.\nn
\end{align}
The state $|K\pi\rangle$ has definite $I=1/2$ and its form factor is  $ \sqrt{3}$ times the one of $|K^+\pi^0\rangle$, 
with $|0\rangle$ the vacuum state.

The $I=1/2$ scalar PWAs in coupled channels involving the  $K\pi$, $K\eta$ and  $K\eta'$ were studied in depth in  Ref.~\cite{jamin.181127.1}. These amplitudes were used in Ref.~\cite{jamin.181120.1} as the source for the  FSI. 
The Ref.~\cite{jamin.181120.1} also concludes that the results just change barely if the $K\eta$ channel is included or 
removed. In this way, the finest results obtained in this reference are worked out with two coupled channels, the $K\pi$ and $K\eta'$ ones. 
Because of this reason, we also take this two-coupled channel scenario in the following. 
The Ref.~\cite{jamin.181120.1} assumes that the $I=1/2$ scalar form factors vanish for $s\to \infty$
 because the hadrons are composite objects. This aspect is in agreement with expectations from QCD
counting rules \cite{brodsky.181121.1,matveev.181121.1,llanes.181121.1}.
Thus, unsubtracted DRs are applied in Ref.~\cite{jamin.181120.1} for the study of the hadronic form factors $F_1(s)$ and $F_3(s)$,
\begin{align}
\label{181127.9}
F_1(s)&=\frac{1}{\pi}\int_{s_{{\rm th};1}}^\infty\frac{\rho_1(s')F_1(s')T_{11}(s')^*}{s'-s}
+\frac{1}{\pi}\int_{s_{{\rm th};3}}^\infty\frac{\rho_3(s')F_3(s')T_{13}(s')^*}{s'-s}~,  \\
F_3(s)&=\frac{1}{\pi}\int_{s_{{\rm th};1}}^\infty\frac{\rho_1(s')F_1(s')T_{13}(s')^*}{s'-s}
+\frac{1}{\pi}\int_{s_{{\rm th};3}}^\infty\frac{\rho_3(s')F_3(s')T_{33}(s')^*}{s'-s}~.\nn
\end{align}
These coupled linear IEs are solved numerically in Ref.~\cite{jamin.181127.1} by iteration, employing an original 
method which is  summarized in the appendix C of \cite{jamin.190111.1} (which also gives a Fortran code for the Gauss 
routine in its appendix D).

The PWAs considered in Ref.~\cite{jamin.181127.1} fulfill that $\Phi(s_R)=0$, 
$q=0$ (no bound states), and $p=0$. The latter property implies the absence of CDD poles in $D(s)$ 
[they are reabsorbed in the matrix $N(s)$]. In these circumstances, the rhs of Eq.~\eqref{181127.5} simplifies to
\begin{align}
\label{181127.10}
\Delta(s)&\xrightarrow[s\to\infty]{} s^{-\frac{\Phi(\infty)}{2\pi}}~.
\end{align}
The first set of $T$ matrices discussed in Ref.~\cite{jamin.181127.1}, originally derived in Ref.~\cite{jamin.181120.1},
produces $\Phi(\infty)=2\pi$ ($\delta_1(\infty)=\pi$ and $\delta_3(\infty)=0$). 
The Eq.~\eqref{181127.7b} then implies that
\begin{align}
\label{181127.11}
\chi_1+\chi_2&=-1~.
\end{align}
Since  it is not possible that $\chi_1$ and $\chi_2$  are negative integers at the same time, then 
there is  only one linearly independent solution vanishing for $s\to\infty$ with $\chi_1=-1$. 
This is the solution that results by solving Eq.~\eqref{181127.9} with the PWAs that stem from the fits (6.10) and
(6.11) of Ref.~\cite{jamin.181120.1}.
As starting input to implement the iterative method for solving the IEs of Eq.~\eqref{181127.9} the
 Ref.~\cite{jamin.181127.1} takes for $F_{1}(s)$  an Omn\`es representation with constant $P(s)$ and $Q(s)$,
 cf. Eq.~\eqref{181121.1}, and $F_3(s)$ is set to zero in the first round.
The normalization factor for $F_1(s)$ is the value of $F_{K\pi}(0)$ according to  NLO ChPT
\cite{CT.181127.1,CT.181127.2}.

Another set of PWAs are also used in Ref.~\cite{jamin.181127.1} which is obtained by matching smoothly the unitarized
ChPT PWAs of Ref.~\cite{jamin.181120.1} with a $K$-matrix parameterization  at $\sqrt{s}$  around $1.75$~GeV.
This step is considered so as to improve the reproduction of the experimental data  on $K\pi$ scattering \cite{aston.181127.1} 
for energies above 1.9~GeV. 
The new PWAs allow to study the transition from $\Phi(\infty)=2\pi$ to $\Phi(\infty)=4\pi$ by varying some parameters
of the $K$-matrix parameterizations, at the same time that the experimental data of Ref.~\cite{aston.181127.1} are 
well reproduced in the whole energy range in which data exists (up to $\sqrt{s}=2.5$~GeV). 
 For the case $\Phi(\infty)=4\pi$ it follows from  from Eq.~\eqref{181127.7b} that (let us recall again that $q=p=\Phi(s_R)=0$)
\begin{align}
\label{181127.12}
\chi_1+\chi_2&=-2~.
\end{align}
We can then have simultaneously that  $\chi_1=\chi_2=-1$, so that there are two linearly independent solutions 
that vanish at infinity.
Now two constants are needed for fixing the linear superposition of the two linearly-independent solutions. 
One of them is the same as above, namely the value of $F_{1}(0)$ as given by NLO ChPT \cite{CT.181127.2}. 
 For the other parameter,   Ref.~\cite{jamin.181127.1} considers the value of the $K\pi$ form factor at $s=\Delta_{K\pi}$, 
 the Callan-Treiman point. The reason is the existence at this point of an accurate relation with the ratio of the weak decay constants of 
  kaons ($f_K$) and pions ($f_\pi$), which reads
\begin{align}
\label{181127.13}
F_{K\pi}(\Delta_{K\pi})&=\frac{f_K}{f_\pi}+\Delta_{CT}~. 
\end{align}
The Reference \cite{CT.181127.2} gives a very small value for  $\Delta_{CT}$, which is estimated as $-3\times 10^{-3}$. 
The ratio  $f_K/f_\pi=1.22\pm 0.01$, as taken in Ref.~\cite{jamin.181127.1} accordingly to the phenomenological information 
available at its time. 
It is worth emphasizing that Ref.~\cite{jamin.181120.1} could vary between $\Phi(\infty)=2\pi$
(one linearly-independent solution) and $4\pi$ (two linearly-independent solutions) by choosing adequately the $K$-matrix ans\"atze,
and for all the cases the value of $F_{K\pi}(\Delta_{K\pi})$ is  compatible. 
This clearly shows the great stability of the results. 
More specifically, for $\Phi(\infty)=2\pi$ (one linearly-independent solution), it results that 
$F_{K\pi}(\Delta_{K\pi})=1.219-1.22$, in profound agreement with Eq.~\eqref{181127.13}.

Finally, let us come back to Eq.~\eqref{181119.6} in order to express the form factors  by using the matrix of functions
$\left[I+\cN(s)g(s)\right]^{-1}$. 
When $\cN(s)$ is modelled without LHCs, cf. Sec.~\ref{sec.181104.1}, 
 explicit expressions for $\Omega(s)$  and $D(s)$  can be found. 
 For the uncoupled case one can write
\begin{align}
\label{181127.14}
\Omega(s)&=\frac{\displaystyle{\prod_{i=1}^q(s-s_{P;i})}}{\displaystyle{\prod_{j=1}^p(s-s_{Z;j})}}\frac{1}{1+\cN(s)g(s)}~,
\end{align}
where the zeroes ($s_{Z;i}$) and poles ($s_{P;i}$)  of $1/[1+\cN(s)g(s)]$ are explicitly removed. 
In the coupled-channel case, the matrix $D(s)$ in Eq.~\eqref{181126.1} can be taken as 
\begin{align}
\label{181127.15}
D(s)&=\left[I+\cN(s)g(s)\right]~.
%  D(s)&=\left[I+\cN(s)g(s)\right]^{-1}~. % error
\end{align}

Coming back to a general matrix $\cN(s)$, we  
can also introduce the analogous of $\Omega(s)$ in coupled channels, similarly as in Ref.~\cite{warnock.181127.1}. 
This matrix is denoted by $\cD^{-1}(s)$, such that $\cD^{-1}(s)$ fulfills Eq.~\eqref{181126.1}, 
and it is holomorphic in the cut complex $s$ plane. 
Given the matrices  $D(s)$ and $\cD(s)$ one can conclude from Eq.~\eqref{181126.1} that the product $\cD(s)D(s)^{-1}$ has no cuts    because
\begin{align}
\label{181127.16}
&\cD(s+i\ve)D(s+i\ve)^{-1}-\cD(s-i\ve)D(s-i\ve)^{-1}\\
=&\cD(s-i\ve)\cS(s-i\ve)\cS(s+i\ve)D(s-i\ve)^{-1}-
%D(s-i\ve)\cD(s-i\ve)^{-1}=0~,\nn %error
\cD(s-i\ve) D(s-i\ve)^{-1}=0~.\nn
\end{align}
Here we have also used Eq.~\eqref{181125.14} and the property $\cD(s^*)=\cD(s)^*$, which also holds for $D(s)\cD(s)^{-1}$. 
It follows then that the product $\cD(s) D(s)^{-1}$ is a matrix of rational functions $R(s)$, which allows to 
write  $D(s)$ in terms of $\cD(s)$ as \cite{warnock.181127.1}
\begin{align}
\label{181127.17}
D(s)^{-1}&=\cD(s)^{-1}R(s)~.
\end{align}
Of course, this result applies to any possible matrix of functions $D(s)$ satisfying Eq.~\eqref{181126.1},
independently of the modelling of $\cN(s)$.

%%%%%%%%%%%%%%%%%%%%%%%%%%%%%%%%%%%%%%%%%%%%%%%%%%%%%%%%%%%%%%%%%%%%%%%%%%%%%%%%%%%%%%%%%%%%%%
%%%%%%%%%%%%%%%%%%%%%%%%%%%%%%%%%%%%%%%%%%%%%%%%%%%%%%%%%%%%%%%%%%%%%%%%%%%%%%%%%%%%%%%%%%%%%%
\subsection{The extended Khuri-Treiman formalism for $\eta\to 3\pi$ decays}
\label{sec.1903291.1}

Here we review on the recent approach developed in Refs.~\cite{alba.190329.1,alba.190329.2}, in which
the Khuri-Treiman (KT) formalism  for studying the decays $\eta\to 3\pi$ is extended to include
explicitly coupled channels in the equations.
It is an interesting problem from the point of view of coupled-channel dynamics
since both FSI and initial-state interactions interplay. 
The standard Khuri-Treiman formalism, originally derived for $K\to 3\pi$ in Ref.~\cite{khuri.190329.1},
implements elastic $\pi\pi$ unitarity and an undermined crossing symmetry.
This approach was further analyzed by  other authors in the subsequent 
Refs.~\cite{gr.2,gr.3,gr.4,gr.5,gr.6,gr.7,gr.8,gr.9,gr.10,neveu.190329.1}.
In more recent times  it has been perfected and often applied  to
$\eta\to 3\pi$ decays \cite{kambor.190329.1,anisovich.190329.1,jpac.190329.1,alba.190329.1,colangelo.190329.1,gasser.190329.1}.\footnote{For
its application to other three-body decays we quote \cite{kubis.190330.1,kubis.190330.2}, and references therein.}
The main aim of Refs~\cite{alba.190329.1,alba.190329.2} is to disentangle the influence of the coupled channels  
$\pi\pi$, $K\bar{K}$  and $\pi\eta$ through FSI and initial-state interactions.
The $K\bar{K}$ contributes to both of them, while  $\pi\pi$ and $\eta\pi$  give rise only to FSI and
initial-state interactions, respectively.
The $K\bar{K}$ channel couples strongly to the $f_0(980)$ and $a_0(980)$ scalar resonances, and the $\pi\eta$
channel does so to the $a_0(980)$. In this way Ref.~\cite{alba.190329.1} can
 keep track explicitly of the imprint of these resonances in these $\eta$ decays, apart from the $f_0(500)$ and $\rho(770)$
 already accounted for by the elastic-$\pi\pi$ KT studies.
 In addition, the $\eta\pi$ threshold is close to the physical region of the $\eta\to 3\pi$ decays, which implies that
 it is more accurate to keep it explicitly (as in the coupled-channel KT approach)
 than to proceed with a Taylor expansion in the Mandelstam variables, as done in the
 elastic case (which is discussed in more detail below).   
The KT  formalism has been applied   recently to the study of $\pi\pi$ scattering in Ref.~\cite{alba.190329.4}.

There is an ongoing experimental effort that has allowed to know with great
precision the squared of the $\eta\to \pi^+\pi^-\pi^0$
decay amplitude across the entire Dalitz plot \cite{13.190329.1,14.190329.1,15.190329.1,16.190329.1}.
The improvement in the experimental determination of the analogous function for $\eta\to 3\pi^0$ has also been remarkable
\cite{8.190329.1,9.190329.1,10.190329.1,mami2.190329.1,11.190329.1,12.190329.1}. To accomplish an equally precise
theoretical description of the $\eta\to 3\pi$ decays from first principles is an interesting prospect
in low-energy hadron physics, as 
several crucial aspects of the standard model enter. 

The $\eta\to 3\pi$ decays violate isospin, as it can be easily seen from the fact that the $G$-parity of the $\eta$
is $+1$ and the one of a $\pi$ is $-1$.
In pure QCD this implies that the $\eta\to 3\pi$ decay amplitudes are proportional
to the difference of the lightest current quark masses, $m_u-m_d$.
This difference is conveniently encoded within the parameter $Q^{-2}$
\begin{eqnarray}
  Q^{-2}&=&\frac{m_d^2-m_u^2}{m_s^2-\hat{m}^2}~,
  \end{eqnarray}
as introduced in Ref.~\cite{CT.181127.1}, with $\hat{m}=(m_u+m_d)/2$  [already defined after Eq.~\eqref{181122.7}].
Isospin symmetry is also violated by electromagnetic interactions, whose precise evaluation cannot be pinned
down in ChPT due to the proliferation of low-energy counterterms \cite{4.190329.1}.
Nonetheless, their contribution to $\eta\to 3\pi$ are
suppressed because they vanish in the $SU(2)$ chiral limit \cite{5.190329.1}.
The Refs.~\cite{6.190329.1,7.190329.1} evaluate explicitly the contributions
of order $e^2 m_u$ and $e^2 m_d$. The Ref.~\cite{7.190329.1} calculates also the contributions
of order $e^2(m_u-m_d)$ that were neglected in \cite{6.190329.1}, and finds no extra small parameter
suppressing them apart from the ratio $(m_u-m_d)/(m_d+m_d)\approx 1/3$. The Ref.~\cite{7.190329.1}
corroborates that these electromagnetic corrections remain typically very small (at the percent level)
 throughout the physical region for the $\eta\to 3\pi$ decays.

The application of ChPT to describe the $\eta\to 3\pi$ decays is not without troubles.
From the early calculations within current algebra  the  resulting $\eta\to\pi^+\pi^-\pi^0$ width is
around 65~eV \cite{weinberg.190329.1,CT.181127.3}, which is  much smaller than the experimental result
$\Gamma(\eta\to \pi^+\pi^+\pi^0)=(300 \pm 12)$~eV \cite{pdg.181106.1}.
Part of this discrepancy is accounted for by the NLO calculation in ChPT of Gasser and Leutwyler \cite{CT.181127.3},
who obtain a value of $(160\pm 50)~$eV for this decay width.
This is an increase by a factor $2.4\pm 0.8$ as compared to the LO
result because of the infrared singularities in ChPT at NLO.
Despite this improvement the decay width is still too small at NLO, by around a factor of 2, as compared with
its experimental value. 
 Indeed, a few years earlier, Roiesnel and Truong \cite{truong.190329.1}
stressed that  a non-perturbative calculation handling the isoscalar-scalar
$\pi\pi$ interactions increases the current algebra result for the decay width up to around 200~eV.
In addition the parameter $ \alpha$ used for the parameterization of the Dalitz plot for
$\eta\to 3\pi^0$ turns out with positive  sign in NLO ChPT \cite{CT.181127.3},
while experimentally it is $\alpha=-0.0318\pm 0.0015$ \cite{pdg.181106.1}. 
These disagreements indeed stimulated the calculations based on
dispersive methods (as referred above within the KT formalism),
or within unitarized ChPT \cite{truong.190329.1,borasoy.190329.1,borasoy.190329.2},
so as to improve upon the results from one-loop ChPT.  
 The calculation of the $\eta\to 3\pi$ decays at NNLO  ChPT is undertaken in Ref.\cite{bijnens.190329.1}. 
However, the prediction is not sharp because of the proliferation of the NNLO  ChPT counterterms.
If one assumes resonance saturation then the Daliz plot parameters are not well reproduced.
This clearly indicates that the $\eta\to 3\pi$ decay amplitudes are sensitive to the detailed values
 of the ${\cal O}(p^6)$ counterterms, being not enough to estimate them on the basis of  gross features.

Let us revise firstly the formalism for the so-called elastic case, in which only the $\pi\pi$
states are included in the unitarity relations. Next, we discuss its extension to coupled channels,
as developed in Ref.~\cite{alba.190329.3}, but employing a much more compact notation.

%%%%%%%%%%%%%%%%%%%%%%%%%%%%%%%%%%%%%%%%%%%%%%%%%%%%%%%%%%%%%%%%%%%%%%%%%%%%%%%%%%%%%%%%%%%%
\subsubsection{The elastic case: $\pi\pi$ final-state interactions}
\label{sec.190330.1}

Because of crossing symmetry the same amplitudes that describe $\eta(p_0)\to \pi(p_1)\pi(p_2)\pi(p_3)$
also correspond to the $\eta(p_0)\pi\to \pi\pi$ scattering amplitudes. We define the Mandelstam variables $s$, $t$ and $u$ as
\begin{align}
\label{190330.1}
  s&=(p_0-p_3)^2=(p_1+p_2)^2~,\\
  t&=(p_0-p_1)^2=(p_2+p_3)^2~,\nn\\
  u&=(p_0-p_2)^2=(p_1+p_3)^2~.\nn
\end{align}

In particular, the amplitude for $\eta\to \pi^+\pi^-\pi^0$ is $A(s,t,u)$ and the one for $\eta\to 3\pi^0$ is $B(s,t,u)$
[the latter can be expressed in terms of the former by an isospin analysis, cf. Eq.~\eqref{190403.7}].
 In this way, the different set of scattering reactions that are listed next are given by the decay amplitudes as
 \begin{align}
   \label{190331.1}
T(\eta\pi^0\to \pi^+\pi^-)&=A(s,t,u)~,\\
T(\eta \pi^- \to \pi^-\pi^0)&=A(t,s,u)~,\nn\\
T(\eta \pi^+\to \pi^+\pi^0)&=A(u,t,s)~,\nn\\
%T(\eta\to \pi^+\pi^-\pi^0)&=A(s,t,u)~,\nn\\
T(\eta\pi^0\to \pi^0\pi^0)&=B(s,t,u)~.\nn
 \end{align}
 For the last reaction it does not matter which is the $\pi^0$  crossed from the
 final to the initial state because of the Bose-Einstein symmetry [so that $B(s,t,u)$ is symmetric under
any permutation of its arguments]. 

%\begin{figure}[tb]
%\begin{center}
%\includegraphics{./Pics/MandelstamPlane.pdf}
%\caption{Mandelstam plane with the scattering and decay physical regions for $\eta\to 3\pi$ and $\eta\pi\to \pi\pi$, respectively. 
%\label{fig.190330.1}}
%\end{center}
%\end{figure}

\begin{figure}[tb]
\begin{center}
\epsfig{file=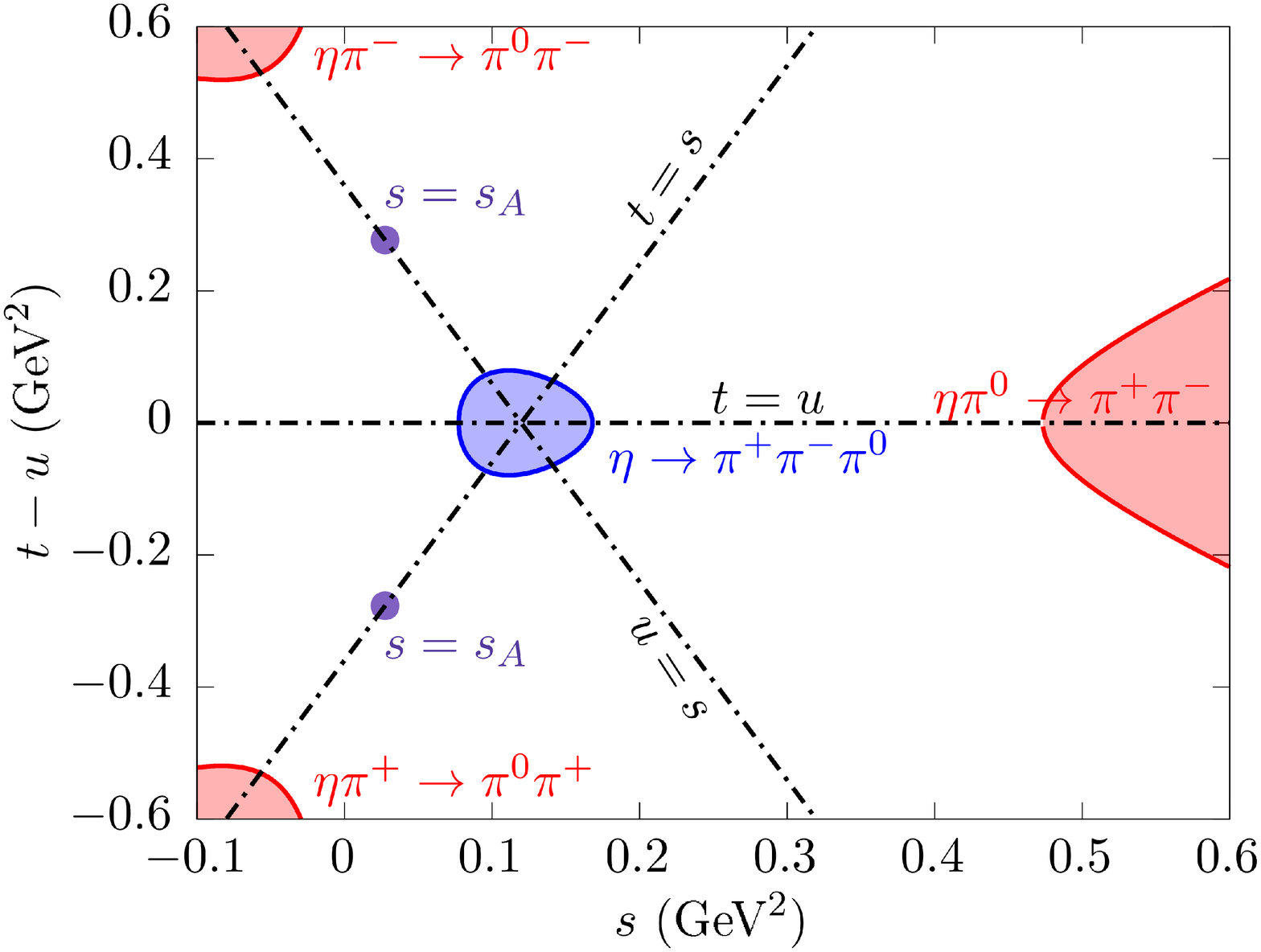,scale=0.35,angle=-90}
\caption{{\small Mandelstam plane with the scattering and decay physical regions for $\eta\to 3\pi$ and $\eta\pi\to \pi\pi$, respectively. In the figure, $s_A$ refers to the Adler zero of the amplitude $A(s,t,u)$, which at leading order is given by
  $s_A=4m_\pi^2/3$ \cite{CT.181127.3,colangelo.190329.1}.}
\label{fig.190330.1}}
\end{center}
\end{figure}

The different physical regions for $\eta\to \pi^+\pi^-\pi^0$ are indicated in Fig.~\ref{fig.190330.1}, with 
the variables $s$ and $t-u$ used as the $x$ and $y$ axes, respectively.
For the scattering processes, the $s$-channel corresponds to $\eta\pi^0\to \pi^+\pi^-$ , the $t$-channel to
$\eta\pi^-\to\pi^-\pi^0$ and the $u$-channel to $\eta\pi^+\to\pi^+\pi^0$.
In terms of $s$ and the scattering angle in the $s$-channel, $\theta$, it is straightforward
to find the following expression for $t$ and $u$, 
\begin{align}
\label{190330.2}
%t,\,u(s,\cos\theta)&=\frac{1}{2}\left(m_\eta^2+3m_\pi^2-s\pm \cos\theta \kappa(s)\right)~,
t,\,u(s,\cos\theta)&=\frac{1}{2}\left(m_\eta^2+3m_\pi^2-s\pm \cos\theta \sqrt{\lambda(s)\sigma(s)}\right)~,
\end{align}
respectively, where
\begin{align}
\label{190330.3}
\lambda(s)&=\lambda(s,m_1^2,m_2^2)=s^2+m_\eta^4+m_\pi^4-2s(m_\pi^2+m_\eta^2)-2m_\pi^2m_\eta^2~,\\
\sigma(s)&=1-\frac{4m_\pi^2}{s}~.\nn
\end{align}
and for later convenience we also introduce the function $\kappa(s)=\sqrt{\sigma(s)\lambda(s)}$.  
It is convenient to rewrite here the K\"allen function  as in the first line of the previous equation
because below we also add a small positive imaginary part to $m_\eta^2$.
The variable $t-u$ is from Eq.~\eqref{190330.2} 
\begin{align}
\label{190330.4}
t-u&=\cos\theta \sqrt{\lambda(s)\sigma(s)}~.
\end{align}

The  $s$-channel physical region ($s\geq (m_\eta+m_\pi)^2$) in the Mandelstam plane is determined
from Eq.~\eqref{190330.2} by varying $\cos\theta\in [-1,1]$. Notice that the variable $t-u$ is then
symmetric around the $s$ axis, as it is clear from Eq.~\eqref{190330.4} and the plot in Fig.~\ref{fig.190330.1}.  
The boundary of the $s$-channel physical  region is the set of points $(s,t-u)$ given by
\begin{align}
\label{190331.2}
\partial D_{s{\rm-channel}}=\left\{(s,\pm \sqrt{\lambda(s)\sigma(s)})~,~s\geq m_{\eta\pi}^2\right\}~.
\end{align}
In the following we denote by $m^2_{\eta\pi}=(m_\eta+m_\pi)^2$. 
A similar analysis also holds for the decay channel. In this case $(m_\eta-m_\pi)^2\geq s\geq 4m_\pi^2$,
and these limits are  determined by requiring that $t$ and $u$ be real in Eq.~\eqref{190330.2}.
Regarding $t-u$, its possible values follow by varying $\cos\theta\in[-1,1]$
in the allowed interval of values of $s$, and they are symmetric around the real $s$ axis.
Thus, the physical region for the decay channel is limited by the boundary
\begin{align}
\label{190401.1}
\partial D_{{\rm decay-channel}}=\left\{(s,\pm \sqrt{\lambda(s)\sigma(s)})~,~(m_\eta-m_\pi)^2\geq s\geq 4m_\pi^2\right\}~.
\end{align}
One can determine the absolute maximum of $|t-u|$ by looking for the zero of $\partial (t-u)(s,\cos\theta=1)/\partial s$.
This occurs at $s=0.110$~GeV$^2$ where the absolute value for $|t-u|$ is  $0.080$~GeV$^2$.
A characteristic feature of the decay channel, with important consequence in the analytical properties of the
PWAs, is that all the Mandelstam variables are larger than or equal to
$4m_\pi^2$, as it is clear from Eq.~\eqref{190330.1} since all the pions are physical.
Indeed, given the symmetric kinematics under the exchange of $s$, $t$ and $u$, when varying $s$ and $\cos\theta$
within the decay-channel physical
region all these variables finally sweep the whole interval $[4m_\pi^2,(m_\eta-m_\pi)^2]$.

The determination of the physical regions for the $t$-  and  $u$-channels, $t\geq m^2_{\pi\eta}$
and  $u\geq m_{\pi\eta}^2$, respectively, is more involved. It is discussed in detail in
the Appendix \ref{app.190331.1}, so that the familiar reader with these techniques can skip it.

Let us discuss now the isospin decomposition of the decay amplitudes $A(s,t,u)$ and $B(s,t,u)$ in the $s$-channel.
For this respect, one can use equivalently the scattering or decay processes, although we choose the former to settle the
formalism since the notation required in the different steps is simpler (which also applies to the PWAs below).\footnote{In this way we can use the same language employed along this work for two-body scattering.} 
For the isospin decompositions it is important to keep in mind that 
the operator in the QCD Lagrangian that gives rise to isospin breaking,
 $-1/2(m_u-m_d)(\bar{u}u-\bar{d}d)$, is isovector ($I=1$) with $t_3=0$.
This is also the case for the ${\cal O}(e^2)$ isospin-breaking operator resulting from the photon exchange
between the lightest quarks given the form of the charge matrix squared of the lightest quarks,
${\rm diag}(4/9,2/9)=5\,I/18+{\rm diag}(1,-1)/6$.
Thus, at first order in the isospin-breaking contributions,
we apply in the following  the Wigner-Eckart theorem \cite{rose.021016.2}
treating the scattering operator for the isospin-breaking transition as having $I=1$, $t_3=0$.

The different isospin states $|It_3, \pi\pi\rangle$ and $|It_3,\eta\pi\rangle$
are\footnote{We follow here the sign convention of Ref.~\cite{oller.181101.2}, so that 
$|\pi^+\rangle=-|1+1\rangle$ and
  $K^+=-|1/2-1/2\rangle$, where the state on the rhs of the equality belongs to the isospin basis $|It_3\rangle$.
  One can then apply the standard convention for the Clebsch-Gordan coefficients, as listed e.g. in Ref.~\cite{pdg.181106.1}.
  We always use that $C|\pi^\pm\rangle=|\pi^\mp\rangle$, $C|K^\pm\rangle=|K^\mp\rangle$ and that
  $C|K^0\rangle=|\bar{K}^0\rangle$.
\label{foot.190402.1}}
\begin{eqnarray}
  \label{190402.1}
|\pi^+\pi^-\rangle&=&-\frac{1}{\sqrt{3}}|00,\pi\pi\rangle-\frac{1}{\sqrt{2}}|10,\pi\pi\rangle-\frac{1}{\sqrt{6}}|20,\pi\pi\rangle~,\\
|\pi^+\pi^0\rangle&=&-\frac{1}{\sqrt{2}}|2+1,\pi\pi\rangle-\frac{1}{\sqrt{2}}|1+1,\pi\pi\rangle~,\nn\\
|\pi^0\pi^-\rangle&=&+\frac{1}{\sqrt{2}}|2-1,\pi\pi\rangle+\frac{1}{\sqrt{2}}|1-1,\pi\pi\rangle~,\nn\\
|\pi^0\pi^0\rangle&=&\sqrt{\frac{2}{3}}|20,\pi\pi\rangle-\frac{1}{\sqrt{3}}|00,\pi\pi\rangle~,\nn\\
|\eta\pi^0\rangle&=&+|10,\eta\pi\rangle~,\nn\\
|\eta\pi^+\rangle&=&-|1+1,\eta\pi\rangle~,\nn\\
|\eta\pi^-\rangle&=&+|1-1,\eta\pi\rangle~.\nn
\end{eqnarray}

In this way, the isospin amplitudes $M^I(s,t,u)$ results by combining in isospin the initial state
with the $|10\rangle$ isospin-breaking operator to produce the final state. All the necessary Clebsch-Gordan coefficients
  can be read form Eq.~\eqref{190402.1}.  We then have the following relations:
\begin{eqnarray}
  \label{190403.1}
A(\eta\pi^0\to \pi^+\pi^-)=A(s,t,u)&=&-\frac{1}{3}M^2(s,t,u)+\frac{1}{3}M^0(s,t,u)~,\\
% \label{190403.2}
A(\eta\pi^+\to \pi^+\pi^0)=A(u,t,s)&=&+\frac{1}{2}M^2(s,t,u)+\frac{1}{2}M^1(s,t,u)~,\nn\\
% \label{190403.3}
A(\eta\pi^-\to \pi^0\pi^-)=A(t,s,u)&=&+\frac{1}{2}M^2(s,t,u)-\frac{1}{2}M^1(s,t,u)~.\nn
\end{eqnarray}
From these equations we can express the different isospin amplitudes $M^I(s,t,u)$ in terms of the basic amplitude $A(s,t,u)$
as\footnote{The relation between our isospin amplitudes and those of Ref.~\cite{alba.190329.1} is $M^0=-\sqrt{3}M^{0,0}$,
  $M^1=\sqrt{2}M^{1,1}$ and $M^2=\sqrt{2}M^{2,0}$. 
There is an errata in the first line of Eq.~(19) of Ref.~\cite{alba.190329.1} so that instead of $1/3$ it should be $-1/\sqrt{3}$.}
\begin{align}
  \label{190403.4}
  M^0(s,t,u)&=3A(s,t,u)+A(u,t,s)+A(t,s,u)~,\\
%  \label{190403.5}
  M^1(s,t,u)&=A(u,t,s)-A(t,s,u)~,\nn\\
%  \label{190403.6}
  M^2(s,t,u)&=A(u,t,s)+A(t,s,u)~.\nn
\end{align}

The isospin decomposition for the $\eta\pi^0\to \pi^0\pi^0$ amplitude reads
 \begin{eqnarray}
 \label{190403.7}
B(s,t,u)&=&\frac{2}{3}M^2(s,t,u)+\frac{1}{3}M^0(s,t,u)~,\\
        &=&A(s,t,u)+A(t,s,u)+A(u,t,s)~.\nn
 \end{eqnarray}

The next step is the PWA decomposition of the isospin amplitudes $M^I(s,t,u)$ by applying Eq.~\eqref{051016.6}.
Since now the particles have spin 0 this formula simplifies  to
\begin{align}
\label{190403.8}
M^{(IJ)}(s)=\frac{1}{2}\int_{-1}^{+1} d\cos\theta  P_J(\cos\theta) M^I(s,t,u)~.
\end{align}
In the calculation of the angular integration the variables $t$ and $u$ are expressed as a function of
$s$ and $\cos\theta$, as given in Eq.~\eqref{190330.2}. The inversion of the previous equation can be
accomplished easily by employing Eq.~\eqref{290916.3}, where the two-body state $|\vp\rangle$ is decomposed in the spherical basis.
Therefore,
\begin{align}
\label{190403.9}
M^I(s,t,u)&= %\langle \vp'|T^I||\vp|\hat{\vz}\rangle=4\pi\sum_{J =0}^\infty
4\pi\sum_{J=0}^\infty \sum_{M=-J}^J Y_J^M(\hat{\vp}')Y_J^M(\hat{\vz})M^{(IJ)}(s)
=\sum_{J=0}^\infty (2J+1)P_J(\cos\theta)M^{(IJ)}(s)~,
\end{align}
where we have made use of the fact that $Y_J^M(\hat{\vz})\neq 0$ only for $M=0$ and of the well-known relation between
$Y^0_J(\vp')$ and $P_J(\cos\theta)$ \cite{rose.021016.2}. 

The PWAs $M^{IJ}(s)$ satisfy unitarity constraints which stem from Eq.~\eqref{181101.5b} evaluated between states with well-defined
total angular momentum. We apply this equation by analytical continuation in the $\eta$ mass.
For $m_\eta<3m_\pi$ it implies the standard unitarity relation for form factors in PWAs given in Eq.~\eqref{181118.3}, where
the isospin-violating $\eta\pi\to\pi\pi$ is considered only at first order.
Thus, for $m_\eta<3m_\pi$ and $s\geq 4m_\pi^2$ we have that 
\begin{align}
\label{190403.10}
M^{(IJ)}(s+i\vep)-M^{(IJ)}(s-i\vep) =2i \Im  M^{(IJ)}(s+i\vep)
&=i\frac{\sigma(s)^{1/2}}{8\pi} M^{(IJ)}(s+i\vep) T^{(IJ)}(s+i\vep)^*\\
  &=2i e^{-i\delta^{(IJ)}(s)}\sin\delta^{(IJ)}(s) M^{(IJ)}(s+i\vep)~,\nn
\end{align}
where $T^{(IJ)}(s)$ is the $\pi\pi$ PWA with isospin $I$ and angular momentum $J$ and
$\delta^{(IJ)}(s)$ is the corresponding phase-shift [cf. Eqs.~\eqref{051016.13} and \eqref{181109.4}]. 

For the transition to the physical situation
in which $m_\eta>3m_\pi$ one gives a vanishing positive imaginary part to $m_\eta^2$ so that $m_\eta^2\to m_\eta^2 +i\ep$
for the external legs in the scattering process \cite{mandelstam.190403.1}.
Indeed, this analytical extrapolation in $m_\eta^2$ for the relationship
between $t$, $u$ and $s$ in
Eq.~\eqref{190330.2} is necessary even in order to settle the PWA in Eq.~\eqref{190403.8}. 
The reason is because for $\ep=0$  we have that for $s=(m_\eta^2-m_\pi^2)/2$ and $\cos\theta=\mp 1$ 
the Mandelstam variables $t$ and $u$ become $4m_\pi^2$, respectively.
As a result, the threshold branch-point singularity in the crossed channels prevents the PWA expansion
at $s=(m_\eta^2-m_\pi^2)/2$, because the Lehmann ellipse \cite{lehmann.190429.1,martin.290916.1} 
collapses then into a line and no analytical extrapolation
in $\cos\theta$ is possible.\footnote{Let us recall that the region of convergence of a Legendre polynomial
  expansion in the complex $\cos\theta$ plane is the largest ellipse which can be drawn
  with foci at $\pm 1$ so that the function
  represented is analytic inside the ellipse. This is the so-called Lehmann ellipse \cite{lehmann.190429.1}.}
This singularity is an example of an anomalous threshold, because in the $s$ channel there is no threshold
at $(m_\eta^2-m_\pi^2)/2$, being in the interior of the allowed region for $s$. However, 
 the threshold branch-point singularity in the crossed channels
at $t,\,u=4m_\pi^2$ for $\cos\theta=\mp 1$ is  reached, producing a cusp effect.  
The analytical extrapolation in the external resonance masses is
used in Refs.~\cite{alvarez.190409.1,alvarez.190409.2} to study the scattering
involving the scalar resonances $f_0(980)$ and $a_0(980)$  with the $\phi(1020)$.
 
As a result,  Eq.~\eqref{181101.5b} instead of giving the imaginary part of  $M^{IJ}(s)$
(which does not have a proper analytical continuation)  gives its discontinuity across
the real $s$ axis for $s\geq 4m_\pi^2$, that is, across the RHC.
The imaginary part of the PWA has now two sources, one due to the physical particles across the RHC and
another because of the infinitesimal proximity of the crossed cuts. 
We denote the discontinuity of the PWA across the RHC by $\Delta M^{IJ}(s)$, which corresponds explicitly to 
\begin{align}
\label{190403.11}
\Delta M^{(IJ)}(s)&=M^{(IJ)}(s+i\vep)-M^{(IJ)}(s-i\vep)~,~s\geq 4m_\pi^2~,\nn\\
\Delta M^{(IJ)}(s)&=
%\frac{\sigma(s)^{1/2}}{16\pi} M^{IJ}(s+i\vep) T^{IJ}(s+i\vep)^*~,~m_\eta<3m_\pi~ {\rm and}~s\geq 4m_\pi^2\\
2i e^{-i\delta^{(IJ)}(s)}\sin\delta^{(IJ)}(s) M^{(IJ)}(s+i\vep)~.
\end{align}
On the other hand, the imaginary part of $M^{(IJ)}(s)$ can be obtained by a DR representation of the function.
In the region $s\in [4m_\pi^2,(m_\eta-m_\pi)^2]$ the pions in all the three-channels are physical.
However, by using the complex $\eta$ mass squared prescription ($m_\eta^2+i\ep$) one can avoid the overlap of the branch-point singularities at $t,\,u=4m_\pi^2$ % and $(m_\eta-m_\pi)^2$
with the unitarity cut for $s$ along these points.
This is enough to settle the KT approach as we show below \cite{anisovich.190329.1}. 

The next crucial step in the KT formalism is to write down the amplitude $A(s,t,u)$ as the sum of
three functions of only one Mandelstam variable, $M_0(s)$, $M_1(s)$ and $M_2(t)$ as \cite{alba.190329.1,colangelo.190329.1}
\begin{align}
\label{190407.1}
A(s,t,u)&=M_0(s)-\frac{2}{3}M_2(s)+(s-u)M_1(t)+(s-t)M_1(u)+M_2(t)+M_2(u)~.
\end{align}
Notice that this equation is invariant under the exchange $t\leftrightarrow u$.
This is a consequence of invariance under charge conjugation, which then implies that the amplitude
is symmetric under the exchange $\pi^+\leftrightarrow \pi^-$ and, therefore, of $t$ and $u$.

The representation of $A(s,t,u)$ in Eq.~\eqref{190407.1} only holds in ChPT up to ${\cal O}(p^8)$
because then the $\pi\pi$ $D$-wave
PWAs start to contribute in the discontinuity relations \cite{bijnens.190329.1,colangelo.190329.1}.
The structure of this equation can be understood by taking into account the isospin decomposition
for the process $\eta\pi^0\to \pi^+\pi^-$ and the crossed-channel ones, and by neglecting PWAs with $J\geq 2$ in the
intermediate states for the corresponding unitarity relations \cite{colangelo.190329.1}.
In this way,   the $I=0$ amplitude only occurs for the $s$-channel with $J=0$ (the $M_0(s)$ piece).
In the $t$- and $u$-channel we have $I=1$ and 2 because it involves $\pi^\mp \pi^0$, respectively.
For $I=1$ we only have $P$-wave, which is proportional to $\cos\theta_t$, with $\theta_t$ the scattering angle in the $t$ channel.
The contribution $(s-u)M_1(t)$ emerges in Eq.~\eqref{190407.1} because  $\cos\theta_t$ is proportional to $s-u$, cf. Eq.~\eqref{190331.13}.
Next, this term is symmetrized in $t$ and $u$.
Regarding the $I=2$ $S$-wave contribution, associated to the function $M_2(s)$, it can appear in all three channels, as shown explicitly in Eq.~\eqref{190403.1}.
Taking into account the weights in front of $M^2(s,t,u)$, the structure $-2 M_2(s)/3+M_2(t)+M_2(u)$
results in Eq.~\eqref{190407.1} after summing over the $s$-, $t$- and $u$-channels. 
The Ref.~\cite{colangelo.190329.1} derives Eq.~\eqref{190407.1} by considering only the discontinuity structure all over the different channels, neglecting $J\geq 2$,  
and assuming that the scattering amplitude does not grow faster than $\alpha^2$ if $s$ and $t-u$ are simultaneously multiplied by this constant. 

As a consequence of Eqs.~\eqref{190403.7} and \eqref{190407.1}, the $\eta\to \pi^0\pi^0\pi^0$ amplitude reads
\begin{align}
\label{190408.1}
B(s,t,u)&=\frac{1}{3}\left[M_n(s)+M_n(t)+M_n(u)\right]~,\\
M_n(s)&=M_0(s)+4M_2(s)~.\nn
\end{align}

The isospin amplitudes $M^{I}(s,t,u)$ in Eq.~\eqref{190403.4} can be rewritten in terms of the $M_I(s)$ functions
in Eq.~\eqref{190407.1} as 
\begin{align}
\label{190408.2}
M^0(s,t,u)&=3M_0(s)+M_0(t)+M_0(u)+\frac{10}{3}\big[M_2(t)+M_2(u)\big]+2(s-u)M_1(t)+2(s-t)M_1(u)~,\nn\\
M^1(s,t,u)&=2(u-t)M_1(s)+(u-s)M_1(t)-(t-s)M_1(u)+M_0(u)-M_0(t)+\frac{5}{3}\big[M_2(t)-M_2(u)\big]~,\nn\\
M^2(s,t,u)&=2M_2(s)+\frac{1}{3}\big[M_2(t)+M_2(u)\big]+M_0(t)+M_0(u)-(s-u)M_1(t)-(s-t)M_1(u)~.
\end{align}

Next, we project in $S$ wave the $I=0$, 2 amplitudes and in $P$ wave the isovector one, employing Eq.~\eqref{190403.8}.
As usual \cite{anisovich.190329.1}, the angular projections of the functions with arguments $t$ and $u$ are denoted by the
symbol $\langle z^n M_I\rangle(s)$, which is given by
\begin{align}
\label{190408.3}
\langle z^n M_I\rangle(s)&=\frac{1}{2}\int_{-1}^{+1}dz z^n M^I(s,t(s,z),u(s,z))~. 
\end{align}
In this expression $t(s,z)$ and $u(s,z)$ are calculated by using Eqs.~\eqref{190330.2}
 and \eqref{190330.3} with $m_\eta^2+i\ep$.
As a result we can write that
 \begin{align}
\label{190408.4}
M^{00}(s)&\equiv 3\big[M_0(s)+\hat{M}_0(s)\big]\\
%&=3\big[M_0(s)+\frac{2}{3}\langle M_0\rangle+\frac{20}{9}\langle M_2\rangle+2(s-s_0)\langle M_1\rangle
%+\frac{2}{3}\kappa \langle M_1\rangle \big]~,\nn\\
M^{11}(s)&\equiv -\frac{2}{3}\kappa\big[ M_1(s)+\hat{M}_1(s)\big] \nn\\
%&=-\frac{2}{3}\big[\kappa M_1(s)+\frac{9}{2}(s-s_0)\langle zM_1\rangle +\frac{3\kappa}{2}\langle z^2M_1\rangle
%  +3\langle zM_0\rangle-5\langle z M_2\rangle \big]~,\nn\\
M^{20}(s)&\equiv 2\big[M_2(s)+\hat{M}_2(s)\big]\nn
%&=2\big[M_2(s)+ \frac{1}{3}\langle M_2\rangle+ \langle M_0\rangle-\frac{3}{2}(s-s_0)\langle M_1\rangle
%-\frac{\kappa}{2}\langle zM_1\rangle\big]~,\nn
\end{align}
where
\begin{align}
\label{190408.5b}
\hat{M}_0(s)&=%M_0(s)+
\frac{2}{3}\langle M_0\rangle+\frac{20}{9}\langle M_2\rangle+2(s-s_0)\langle M_1\rangle
+\frac{2}{3}\kappa \langle M_1\rangle~,\\
\kappa(s)\hat{M}_1(s)&=%\kappa M_1(s)+
\frac{9}{2}(s-s_0)\langle zM_1\rangle +\frac{3\kappa}{2}\langle z^2M_1\rangle
  +3\langle zM_0\rangle-5\langle z M_2\rangle ~,\nn\\
\hat{M}_2(s)&=%M_2(s)+
  \frac{1}{3}\langle M_2\rangle+ \langle M_0\rangle-\frac{3}{2}(s-s_0)\langle M_1\rangle
-\frac{\kappa}{2}\langle zM_1\rangle~,\nn
\end{align}
  and
\begin{align}
\label{190408.5}
3s_0&=m_\eta^2+3m_\pi^2~.
\end{align}

We can combine these equations with Eq.~\eqref{190403.11} so as to determine the discontinuities of the functions $M_I(s)$ along the
RHC. It then results that
\begin{align}
\label{190408.6}
\Delta M_0(s)&=2i e^{-i\delta^{(00)}(s)}\sin\delta^{(00)}(s) \big[M_0(s)+\hat{M}_0(s)\big]~,\\
\Delta M_1(s)&=2i e^{-i\delta^{(11)}(s)}\sin\delta^{(11)}(s) \big[M_1(s)+\hat{M}_1(s)\big]~,\nn\\
\Delta M_2(s)&=2i e^{-i\delta^{(20)}(s)}\sin\delta^{(20)}(s) \big[M_2(s)+\hat{M}_2(s)\big]~.\nn
\end{align}
These discontinuities can be used to write down  DRs for the $M_I(s)$ functions once a high-energy behavior is
assumed for each of them.
 In this regard, Ref.~\cite{alba.190329.1} invokes Regge theory and assumes that the functions $M_0(s)$ and $M_2(s)$ are
 not expected to grow faster than $s$ for $s\to \infty$,
 and $M_1(s)$ should be bounded by a constant in the same limit
 (notice that in the amplitudes it appears multiplied by $\kappa$, cf. Eq.~\eqref{190408.4}).
 In this way, $A(s,t,u)$ grows at most linearly in all directions $s,t,u\to \infty$, as originally required in
 Ref.~\cite{anisovich.190329.1}. 
 Once this high-energy behavior is taken for granted there is still a possible redefinition of the $M_I$ functions
 in Eq.~\eqref{190407.1} that leaves invariant $A(s,t,u)$. 
In this way, by shifting $M_1(s)$ and $M_2(s)$ accordingly with their assumed asymptotic behavior,
\begin{align}
\label{190409.1}
M_1(s)&\to M_1(s)+a_1~,\\
M_2(s)&\to M_2(s)+a_2+b_2 s~,\nn
\end{align}
one then has the following shift in $M_0(s)$ to guarantee the invariance of $A(s,t,u)$,
\begin{align}
\label{190409.2}
M_0(s)&\to M_0(s)+a_0+b_0 s~,
\end{align}
where
\begin{align}
\label{190409.3}
a_0&=-\frac{4}{3}a_2+3s_0(a_1-b_1)~,\\
b_0&=-3a_1+\frac{5}{3}b_2~, \nn
\end{align}
and used has been made of the relation $t+s+u=3s_0$, cf. Eq.~\eqref{181102.2b}.
This three free-parameter ambiguity in the definition of the $M_I(s)$ is used to impose the conditions
\begin{align}
\label{190409.4}
M_1(0)&=0~,\\
M_2(0)&=0~,\nn\\
M'_2(0)&=0~,\nn 
\end{align}
where the prime indicates the derivative with respect to its argument [a notation already used,
  e.g. in Eq.~\eqref{181116.12}]. 
As a result of the discontinuities in Eq.~\eqref{190408.6},
the conditions in Eq.~\eqref{190409.4} and the asymptotic behavior assumed,
one can then write the following DRs for the
functions $M_I(s)$,
\begin{align}
\label{190409.5}
M_0(s)&=\widetilde{\alpha}_0+\widetilde{\beta}_0 s+\frac{s^2}{\pi}\int_{4m_\pi^2}^\infty ds'\frac{\Delta M_0(s')}{(s')^2(s'-s)}~,\\
M_1(s)&=\frac{s}{\pi}\int_{4m_\pi^2}^\infty ds'\frac{\Delta M_1(s')}{(s')(s'-s)}~,\nn\\
M_2(s)&=\frac{s^2}{\pi}\int_{4m_\pi^2}^\infty ds'\frac{\Delta M_2(s')}{(s')^2(s'-s)}~.\nn
\end{align}
In virtue of the study of the asymptotic behavior of DRs in Sec.~\ref{sec.190124.1},
cf. Eqs.~\eqref{181012.5}--\eqref{181012.7}, if $\Delta M_{I}(s)/s^{2-J}$
vanishes for $s\to \infty$ at least as fast as $s^{-\nu}$, with $\nu>0$,
then terms involving divergent logarithms, e.g. of the form  
$s^{1+I(I-2)}\log s$, are avoided. In this way, the asymptotic behavior of the $M_I(s)$ is truly power-like,
as explicitly stated in the standard assumptions, that we have indicated here and in the literature
\cite{anisovich.190329.1,alba.190329.4,colangelo.190329.1}.

 As already mentioned above,  Ref.~\cite{colangelo.190329.1} assumes an asymptotic quadratic behavior for $A(s,t,u)$.  
The resulting redefinition in the $M_I(s)$ is used to impose convenient conditions on these functions, reducing its 
 degree of divergence in $s$ to a linear one for $M_1(s)$ and $M_2(s)$.

The set of  Eqs.~\eqref{190408.6}, \eqref{190408.5b},  and \eqref{190409.5} constitute a system of coupled linear IEs
for the function $M_0(s)$, $M_1(s)$ and $M_2(s)$ for $s\geq 4m_\pi^2$. 
Its solution solves the KT problem posed for given values of the subtraction constants
$\widetilde{\alpha}_0$, $\widetilde{\beta}_0$ and  
the input $\pi\pi$ phase shifts  $\delta^{(IJ)}(s)$.
To accomplish it we proceed similarly as above for obtaining Eq.~\eqref{181126.1b},
so that the we can recast the content of the discontinuity in Eqs.~\eqref{190408.6} as
\begin{align}
\label{190410.1}
M_I(s+i\vep)&=M_I(s-i\vep)e^{2i\delta^{(IJ)}}+2i\hat{M}_I(s)e^{i\delta^{IJ}}\sin\delta^{(IJ)}~,~s\geq 4m_\pi^2~.
\end{align}
In this relation it is important to notice that the crossed cuts in $\hat{M}_I(s)$ has been separated from the RHC
by employing the complex $m_\eta^2+i\ep$.\footnote{In the Fig.~4 of Ref.~\cite{descotes.190329.1}
  the position of these cuts in the complex $s$ plane are explicitly drawn.}
Next, we employ an Omn\`es function for each $\pi\pi$ PWA involved, $\Omega^{(IJ)}(s)$,
cf. Eqs.~\eqref{181121.2} and \eqref{181121.3},
and divide Eq.~\eqref{190410.1} by $\Omega^{(IJ)}(s+i\vep)$, so that it reads
\begin{align}
\label{190410.2}
\frac{M_I(s+i\vep)}{\Omega^{(IJ)}(s+i\vep)}&=\frac{M_I(s-i\vep)}{\Omega^{(IJ)}(s-i\vep)}+2i
\frac{\hat{M}_I(s) \sin\delta^{(IJ)}(s)}{|\Omega^{(IJ)}(s)|}~,~s\geq 4m_\pi^2~,
\end{align}
where we have taken into account that
\begin{align}
\label{190410.3}
\Omega^{(IJ)}(s+i\vep)e^{-2i\delta^{(IJ)}(s)}&=\Omega^{(IJ)}(s-i\vep)~,~s\geq 4m_\pi^2~,\\  
\Omega^{(IJ)}(s+i\vep)e^{-i\delta^{(IJ)}(s)}&=|\Omega^{(IJ)}(s)|~,~s\geq 4m_\pi^2~,  \nn
\end{align}
as follows from Eqs.~\eqref{181121.2} and \eqref{181121.3}.
The Eq.~\eqref{190410.2} provides us with the discontinuity along the RHC
of $M_I(s+i\vep)/\Omega^{(IJ)}(s+i\vep)$.
This function has only RHC, so that we can readily write down a DR for it in the form
\begin{align}
\label{190410.4}
M_I(s)&=\Omega^{(IJ)}(s)\Big[
P_I^{(m)}(s)+\frac{s^n}{\pi}\int_{4m_\pi^2}^{\infty}ds'\frac{\hat{M}_I(s') \sin\delta^{(IJ)}(s')}{|\Omega^{(IJ)}(s')|(s')^n(s'-s)}\Big]~,
\end{align}
where $n$ is the number of subtractions and $P_I^{(m)}(s)$ is a polynomial of degree $m$ with $m\geq n-1$.

The number of subtractions needed to be taken in Eq.~\eqref{190410.4} depends on the asymptotic behavior of
$\Omega^{(IJ)}(s)$, which in
turn is given by $\delta^{(IJ)}(\infty)$ according to Eq.~\eqref{181122.3}.
In this regard, Ref.~\cite{alba.190329.1} takes that  asymptotically  the  $\pi\pi$ phases shifts satisfy the limits
\begin{align}
\label{190410.5}
\delta^{(00)}(\infty)=\delta^{(11)}(\infty)=\pi~,~\delta^{(20)}(\infty)=0~.
\end{align}
These limits are phenomenologically meaningful, but not necessarily without controversy \cite{oller.181123.1,yndu.181123.1}, as also discussed in Sec.~\ref{sec.181117.2}.
Nonetheless, the Omn\`es functions that result are well behaved according to the discussion 
with respect to the constraint of Eq.~\eqref{181122.6} in Sec.~\ref{sec.181117.2}.
 Finally, the implementation of the thresholds constraints of Eq.~\eqref{190409.4},
allows us to write  the following DRs for $M_I(s)$,
\begin{align}
\label{190410.6}  
M_0(s)&=\Omega^{(00)}(s)\Big[
\alpha_0+\beta_0 s +\gamma_0 s^2+\frac{s^2}{\pi}\int_{4m_\pi^2}^{\infty}ds'\frac{\hat{M}_0(s') \sin\delta^{(00)}(s')}{|\Omega^{(00)}(s')|(s')^2(s'-s)}\Big]~,\\
%\label{190410.6b}  
M_1(s)&=\Omega^{(11)}(s)\Big[\beta_1 s +
  \frac{s}{\pi}\int_{4m_\pi^2}^{\infty}ds'\frac{\hat{M}_1(s') \sin\delta^{(11)}(s')}{|\Omega^{(11)}(s')|(s')(s'-s)}\Big]~,\nn\\
%\label{190410.6c}  
M_2(s)&=\Omega^{(20)}(s)\Big[
\frac{s^2}{\pi}\int_{4m_\pi^2}^{\infty}ds'\frac{\hat{M}_2(s') \sin\delta^{(20)}(s')}{|\Omega^{(20)}(s')|(s')^2(s'-s)}\Big]~.\nn
\end{align}
There are four constants $\gamma_0$, $\beta_0$, $\gamma_0$ and $\beta_1$ that should be fixed.
Notice that the convergence of the DRs, in terms of the assumed asymptotic behaviors of $\delta^{(IJ)}(s)$ and $M_I(s)$,  
requires only two subtractions constants ($\beta_0$ and $\gamma_0)$. 
Two more constants are added, $\gamma_0$ and $\beta_1$, because of the  high-energy behavior assumed for the $M_I(s)$,
on account of the Sugawara-Kanazawa theorem introduced in  Sec.~\ref{sec.190124.1}.  
In this regard, one should notice Eq.~\eqref{190410.4} which implies that $\Omega^{(IJ)}(s)$
tends to $1/s^{1-I(I-1)/2}$ for $s\to \infty$. 
%Since in all the cases $\sin \delta^{(IJ)}(\infty)=0$, $|\Delta (M_I/\Omega^{(IJ)})|$, cf. Eq.~\eqref{190410.2}, 
%has the same type of asymptotic behavior in $s$ as the one of the corresponding $M_I(s)$
%[which was already discussed above after Eq.~\eqref{190408.6}].
 %In this way, o
One can also match simultaneously with the one-loop calculation of $A(s,t,u)$
in ChPT at  ${\cal O}(p^4)$ \cite{CT.181127.3},
including as well the electromagnetic contributions of order $e^2(m_u+m_d)$ \cite{6.190329.1}.
As explained below this matching implies to fulfill 4 equations, cf. Eq.~\eqref{190410.9}.
%This matching implies to keep at most ${\cal O}(s^2)$ contributions in the $P_I^{(m)}(s)$, as reflected in Eq.~\eqref{190410.6}. 

Let us denote by ${\cal I}_I$ the integrals appearing in Eq.~\eqref{190410.6}, so that,
\begin{align}
\label{190410.8}
{\cal I}_I(s)&
=\frac{1}{\pi}\int_{4m_\pi^2}^{\infty}ds'\frac{\hat{M}_I(s') \sin\delta^{(IJ)}(s')}{|\Omega^{(IJ)}(s')|(s')^{2-2J}(s'-s)}~.  
\end{align}

The matching process at NLO in the chiral counting  is direct by taking into account that $\Delta M_I(s)$ is already
${\cal O}(p^4)$, since it implies at least one loop. 
This statement is not modified by the integration in the DRs, because there are the same number of $s$ factors
$[s\sim s'\sim {\cal O}(p^2)]$ in the numerator and denominator \cite{oller.181113.1}.
As a result, any difference with respect to the NLO ChPT result can be accounted for by the subtractive polynomials. 
 At the practical level, one first performs the expansion of the $M_I(s)$
in powers of $s$ up to ${\cal O}(s^2)$ for $I=0,2$ and up to ${\cal O}(s)$ for $M_1(s)$.
These expansions are performed both for the $M_I(s)$ given by NLO ChPT (in which case they are barred, $\bar{M}_I(s)$)
and for the ones in Eq.~\eqref{190410.6}. For instance, for $M_0(s)$ and $\bar{M}_0(s)$ one has,
\begin{align}
\label{190410.7}
M_0(s)&=\alpha_0+s\Big[\beta_0+\alpha_0 {\Omega^{(00)}}'(0)\Big]
+s^2\Big[\gamma_0+{\cal I}_0(0)+\beta_0{\Omega^{(00)}}'(0)+\frac{1}{2}\alpha_0{\Omega^{(00)}}''(0)\Big]+{\cal O}(s^3)~,\\
\bar{M}_0(s)&=\bar{M}_0(0)+s \bar{M}_0'(0)+\frac{1}{2}s^2\bar{M}_0''(0)+{\cal O}(s^3)~,\nn
\end{align}
and similarly for the other functions. 
The resulting expansions are substituted in $A(s,t,u)$, with $u=3s_0-s-t$, which corresponds to a polynomial with two independent variables which
coefficients multiply $s$, $t$, $st$, $s^2$ and $t^2$, plus the independent term.
Then, one imposes that the resulting polynomials must be the same, which implies six relations that at the
end give rise to only four linearly independent  equations in the free parameters.
The latter can be easily solved with the solution \cite{alba.190329.1},
\begin{align}
\label{190410.9}
\alpha_0&=\bar{M}_0(0)+\frac{4}{3}\bar{M}_2(0)+3s_0\Big[\bar{M}_2'(0)-\bar{M}_1(0)\Big]+9s_0^2\bar{M}_2^r~,\\
\beta_0&=\bar{M}_0'(0)+3\bar{M}_1(0)-\frac{5}{3}\bar{M}_2'(0)-9s_0 \bar{M}_2^r-{\Omega^{(00)}}'(0)\alpha_0~,\nn\\
\beta_1&=\bar{M}_1'(0)-{\cal I}_1(0)+\bar{M}_2^r~,\nn\\
\gamma_0&=\frac{1}{2}\bar{M}_0''(0)-{\cal I}_0(0)+\frac{4}{3}\bar{M}_2^r-\frac{{\Omega^{(00)}}''(0)}{2}\alpha_0
-{\Omega^{(00)}}'(0)\beta_0~,\nn
\end{align}
with $\bar{M}_2^r=\bar{M}_2''(0)/2-{\cal I}_2(0)$. Notice that these constants also enter linearly on the
rhs of the previous equations. 

%Still there would be room for an extra free parameter for $M_2(s)$ in Eq.~\eqref{190410.6}, giving rise to a term of the form $\Omega^{(20)}(s)\gamma_2 s^2$, which is compatible with both the high-energy behavior and the constraints in Eq.~\eqref{190409.4}. This is not included in Ref.~\cite{alba.190329.1} so as to fix unambiguously the free parameters in Eq.~\eqref{190410.6}, because as just shown only four equations are possible at ${\cal O}(p^4)$. We consider that this is a sensible procedure, since the FSI in the non-resonant $I=2$ $\pi\pi$ PWA are expected to be significantly smaller than in the other isospin channels. 

The numerical procedure followed in Ref.~\cite{alba.190329.1} to solve the IEs expressed in Eqs.~\eqref{190410.6} and \eqref{190410.9}
is by iteration, taking as starting point the case with all the ${\cal I}_I(s)=0$ in these equations. One first solve for the constants
$\alpha_0$, $\beta_I$ and $\gamma_0$, substitute them in Eq.~\eqref{190410.6} to obtain the zeroth order result for the $M_I(s)$
and, in terms of them, one can perform the angular integrations of Eq.~\eqref{190408.3}  to determine the first iterated
$\hat{M}_I(s)$ functions and ${\cal I}_I(s)$.
The latter are employed in  Eq.~\eqref{190410.9} to obtain the first iterated constants,
which  are substituted in Eq.~\eqref{190410.6} and we have then the first iterated $M_I(s)$.
These are used again in the same manner to calculate the second iterated $\hat{M}_I(s)$ and ${\cal I}_I(s)$ and the
process is repeated until numerical convergence in reached.
The Ref.~\cite{alba.190329.1} reports a fairly fast convergence with the number of iterations,
reaching a numerical precision after seven iterations with a numerical maximum relative error of around $4\times 10^{-5}$.

\begin{figure}[tb]
\begin{center}
\begin{minipage}[t]{8 cm}
\epsfig{file=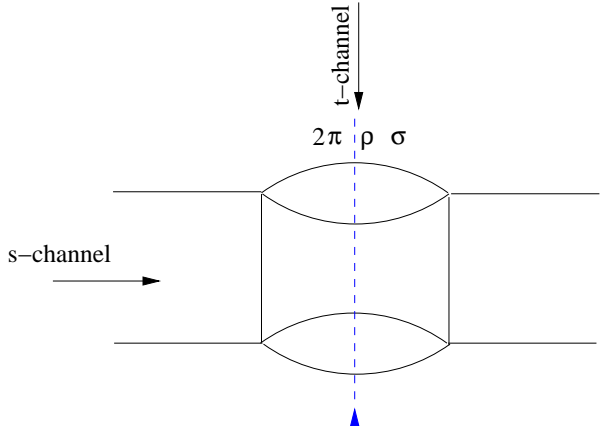,scale=0.8}
\end{minipage}
\begin{minipage}[t]{16.5 cm}
  \caption{{\small Feynman diagram involving the $s$-channel singularity at $s=16 m_\pi^2$ (indicated by the dashed line) 
    that cannot be accounted for by the KT formalism. See the discussion in the text. }
\label{fig.190501.1}}
\end{minipage}
\end{center}
\end{figure}

Before ending this section, we show in Fig.~\ref{fig.190501.1} a Feynman diagram that cannot be accounted
for by the KT formalism,
both for $\eta\to 3\pi$ decays and  $\pi\pi$ scattering \cite{alba.190329.4}.
For the former we can take e.g. the line in the bottom left as corresponding to the $\eta$, and the rest of external
lines to pions, while for $\pi\pi$ scattering all the external legs are pions.
Regarding the internal lines they are pions.\footnote{The two pion lines in the ovals could be replaced by
  the exchange of $\pi\pi$ resonances, like the $\sigma$ or $\rho$ within a phenomenologically driven approach.}
We notice that the $s$-channel singularity associated with the $4\pi$ threshold in Fig.~\ref{fig.190501.1}
cannot be accounted for by the $s$-channel $M_I(s)$ functions in Eq.~\eqref{190407.1},
because they are generated by the KT approach that only takes into account two-body unitarity.
It cannot either be accounted for by the $t$- and $u-$channel functions $M_I(t)$ and $M_I(u)$, respectively,
since the Legendre expansion in any of the crossed channels cannot give rise to a singularity as it converges only
within the Lehmann ellipse.
Inside this ellipse the Legendre expansion generates an analytical function in $\cos\theta_t$ and $\cos\theta_u$, with 
$\theta_t$ and $\theta_u$ being the scattering angles in the corresponding crossed channels.
This limitation associated with the $4\pi$ threshold
is a source of systematic uncertainty in the KT formalism, which is possibly enhanced
because in order to calculate the functions $M_I(s)$ one has to evaluate  
the DRs in Eq.~\eqref{190410.6}. In this way, an integration from the $\pi\pi$ threshold up to $\infty$ has to be
performed involving values of $s$ for which the KT formalism is controversial.

%%%%%%%%%%%%%%%%%%%%%%%%%%%%%%%%%%%%%%%%%%%%%%%%%%%%%%%%%%%%%%%%%%%%%%%%%%%%%%%%%%%%%%%%%%%%
\subsubsection{The coupled case: $\pi\eta$ initial- and $\pi\pi$, $K\bar{K}$
  final-state interactions}
\label{sec.190410.1}

The Ref.~\cite{alba.190329.1} extended the traditional one-channel KT formalism to include the effects of coupled channels.
In this way, the coupled-channel initial-state interactions by the $\eta\pi$ and $K\bar{K}$ channels (with $I=1$, $J=0$)
and the FSI of the $K\bar{K}$ and $\pi\pi$ ones ($J=I=0,~$1) can be taken into account, while the $I=2$ $\pi\pi$ FSI remain
elastic.
One of the main interests of Ref.~\cite{alba.190329.1} for including coupled-channel dynamics is to take into account
explicitly in the results the influence of the resonances $f_0(980)$ and $a_0(980)$,
which give rise to marked structures in the isoscalar and isovector $S$-wave meson-meson scattering amplitudes around
$1$~GeV, respectively. We add here that another point of interest of this formalism is its capability to account for
the $\pi\eta$ threshold in the initial-state interactions, which is not far away from the end-point $(m_\eta-m_\pi)^2$
of the physical values of the Mandelstam variables in the $\eta\to 3\pi$ decays.

%As initial states, denoted by  $i=1,2,3$, we have the $J=0$ $\eta\pi$, $K^+K^-$ and $K^0\bar{K}^0$ states,  while for $J=1$ there is only the $\eta\pi$ state. The reason is because the  $\eta\pi^0$ and $K\bar{K}$ states  have opposite charge conjugation  in $P$ wave (for other charge states they have different $G$ parity). As final states, designated by $f=1,\ldots,5$,  we consider the sates $\pi^0\pi^0$, $\pi^+\pi^-$, %$\pi^+(\vp)\pi^-(-\vp)$, $\pi^-(\vp)\pi^+(-\vp)$,  ,$K^+K^-$ and $K^0\bar{K}^0$, which are ordered according to their increasing thresholds.

Let us study the unitarity relations of Eq.~\eqref{051016.11} for the coupled-channel processes.
As initial states,  we have the $I=1$, $J=0$ $\eta\pi$, $K\bar{K}$  states, denoted by  $i=1,2$, respectively,
and for $J=1$ there is only the $\eta\pi$ state ($i=1$).
The reason is because the  $\eta\pi^0$ and $K\bar{K}$ states 
have opposite charge conjugation  in $P$ wave (for other charge states they have different $G$ parity).
As final states, denoted by $f=1,2$,  we consider the $I=J=0,\,1$ states
$\pi\pi$ and $K\bar{K}$, respectively, which are ordered according to their increasing thresholds.
For $I=2$ and $J=0$ we only have the $\pi\pi$ state ($i=1$).
Therefore, we can write in general from Eq.~\eqref{051016.11} that
\begin{align}
\label{190411.1}
\Im T^{(J)}_{fi}&=\sum_{h}\frac{|\vq_h|}{8\pi\sqrt{s}}T^{(J)}_{fh}(s)^* T^{(J)}_{hi}(s)~,
\end{align}
where the sum is over all the states indicated, initial and final ones.
We neglect in the following the $P$-wave $\eta\pi$ initial-state interactions as in Ref.~\cite{alba.190329.1}. 
These interactions have exotic quantum numbers, $J^{PC}=1^{-+}$, though some structure around 1.4-1.5~GeV is observed
\cite{39.190410.1,rodas.190414.1}.  
 The $T$ matrix elements in Eq.~\eqref{190411.1} contain isospin conserving and
also first-order isospin-breaking contributions with $I=1$, $t_3=0$, as discussed above.
The former ones are denoted with the subscript $S$ and the latter by the subscript $W$.

An important source for isospin mixing between the $f_0(980)$ and $a_0(980)$ resonances is the mass difference between the
charged and neutral kaons \cite{achasov.190410.1,hanhart.190410.1}.
As a result, when the isospin of the final state is null, we keep the mass difference between the thresholds
of the $K^+K^-$ and $K^0\bar{K}^0$ states in Eq.~\eqref{190411.1},
since then the unitarity relation mixes strong amplitudes with $I=0,1$ in $S$-wave.
Similarly as in Ref.~\cite{alba.190329.1} we then introduce the diagonal matrix
$\Sigma_K$, which is given by the difference between the phase space of the $K^+K^-$ and $K^0\bar{K}^0$ states.
Namely,
\begin{align}
  \Sigma_K=\left(
  \begin{array}{ll}
0 & 0  \\
   0 & \frac{\sigma_{K^+}^{1/2}(s)-\sigma_{K^0}^{1/2}(s)}{16\pi\sqrt{s}}  \\
  \end{array}\right)~,\\
    \sigma_P(s)=s/4-m_P^2~.\nn
\end{align}
In all the other instances we  keep the average kaon mass and $\sigma_K(s)=s/4-m_K^2$.
We take the same value for all the pion thresholds as in the elastic case. 
The needed Clebsch-Gordan coefficients for the isospin decomposition of the $K\bar{K}$ states can be read from 
\begin{align}
\label{190411.3}
|K^+K^-\rangle&=-\frac{1}{\sqrt{2}}\left(|00,K\bar{K}\rangle+|10,K\bar{K}\rangle\right)~,\\
|K^0\bar{K}^0\rangle &=-\frac{1}{\sqrt{2}}\left(|00,K\bar{K}\rangle-|10,K\bar{K}\rangle\right)~,\nn
\end{align}
where we have followed the sign convention in the footnote \ref{foot.190402.1}.

Therefore, by keeping terms linear in isospin breaking and indicating the isospin $I$ of the final states, we can rewrite
Eq.~\eqref{190411.1} as
\begin{align}
\label{190411.2}
%\Im T^{(J)}_{V;f'i}&=\sum_{f}\frac{|\vq_f|}{8\pi\sqrt{s}}\Big[T^{(J)}_{S;f'f}(s)^* T^{(J)}_{W;fi}(s)
%  +T^{(J)}_{W;f'f}(s)^* T^{(J)}_{S;fi}(s)
%  +T^{(J)}_{S;f'f}(s)^*\Delta_K T^{(J)}_{S;fi}\Big]~.
\Im T^{(IJ)}_{W;fi}&=\sum_{h}\frac{|\vq_h|}{8\pi\sqrt{s}}\Big[T^{(IJ)}_{S;fh}(s)^* T^{(IJ)}_{W;hi}(s)
  +\delta_{J0}T^{(IJ)}_{W;fh}(s)^* T^{(1J)}_{S;hi}(s)\Big]
  +\delta_{I0}T^{(00)}_{S;f2}(s)^*\Sigma_K|_{22} T^{(10)}_{S;2i}~.
\end{align}

The $P$-wave isovector $K\bar{K}$ FSI are finally neglected in Ref.~\cite{alba.190329.1} 
because the $\rho(770)$ signal can be well reproduced with only the $\pi\pi$ channel,
since the inclusion of the $K\bar{K}$ channel gives small effects  \cite{guerrero.190411.1}.
There are coupled-channel initial-state interactions in $I=1$, $J=0$ and coupled FSI for $I=J=0$, while
 elastic FSI occur for $I=2$, $J=0$ and $I=J=1$.
Thus, in order to apply Eq.~\eqref{190411.2} we have:
\vs
\begin{tabular}{rll}
  $T^{(00)}_{W;fi}~$:& $f=1(\pi\pi)$, $2(K\bar{K})$ &  $i=1(\eta\pi)$, $2(K\bar{K})$. \\
& &\\$T^{(20)}_{W;fi}~$, $T^{(11)}_{W;fi}~$:& $f=1(\pi\pi)$ & $i=1(\eta\pi)$, $2(K\bar{K})$ [only for $J=0$].\\ 
  &&\\
  $T^{(00)}_{S;fh}~$:& $f,h= 1(\pi\pi)$, $ 2(K\bar{K})$. & \\
  &&\\
  $T^{(20)}_{S;fh}$, $T^{(11)}_{S,fh}~$:& $f=h=1(\pi\pi)$.& \\
  &&\\
%  $T^{(20)}_{S;fh}$, $T^{(11)}_{S,fh}~$:& $f=h=1(\pi\pi)$.& \\
%  &&\\
  $T^{(10)}_{S;hi}~$:& $h,i= 1(\eta\pi)$, $2(K\bar{K})$. & \\
\end{tabular}
\vs
Notice that $T^{(20)}_{W;fi}$ and $T^{(11)}_{W;fi}$ are $1\times 2 $ matrices.
The next original step in Ref.~\cite{alba.190329.1} is to proceed by analogy with the elastic case and
split  every isospin-breaking amplitude as a sum of a function with only RHC and another with only
crossed-channel cuts,  cf. Eq.~\eqref{190408.4}. Due to the overlapping of the $s$-channel with the $t$- and $u$-channels
for some values of $s$, it is necessary to proceed with the analytical continuation in the masses squared
of the $\eta$ (as in the elastic case), and now also of the kaons ($m_K^2+i\ep$).
For the former mass extrapolation, we already
discussed that this is due to the $\eta\to 3\pi$ physical decay region,
where all the three pions are physical  (indeed this overlapping between cuts happens up to $s=m_\eta^2-5m_\pi^2$).
For the later, the $K\bar{K}\to K\bar{K}$ amplitude has a LHC that extends for $s\in (-\infty,4m_K^2-4m_\pi^2]$,
which overlaps with the unitarity cut that extends for $s\geq 4m_\pi^2$.
Therefore, the $T^{(IJ)}_{W;fi}(s)$ amplitudes are expressed as
\begin{align}
\label{190412.1}
T^{(IJ)}_{W;fi}(s)&=M^{(IJ)}_{fi}(s)+\hat{M}^{(IJ)}_{fi}(s)~.
\end{align}

From the unitarity relation in Eq.~\eqref{190411.2} we can then write the  equations for the
discontinuity of the functions $M^{(IJ)}_{fi}(s)$ [denoted by $\Delta M^{(IJ)}_{fi}(s)$],
with the slightly complex masses squared of the $\eta$ and $K$, as
\begin{align}
\label{190412.2}
  \Delta M^{(IJ)}_{fi}&=\sum_{h}\frac{|\vq_h|}{8\pi\sqrt{s}}\Big(
T^{(IJ)}_{S;fh}(s)^* \big[ M^{(IJ)}_{hi}(s+i\vep)+\hat{M}^{(IJ)}_{hi}(s)\big]
  +\delta_{J0}\big[M^{(IJ)}_{fh}(s-i\vep)+\hat{M}^{(IJ)}_{fh}(s)\big] T^{(1J)}_{S;hi}(s)\Big)\\
&  +\delta_{I0}T^{(00)}_{S;f2}(s)^*\Sigma_K|_{22} T^{(10)}_{S;2i}~.\nn
\end{align}

In order to solve these IEs it is interesting to isolate the contribution from the crossed-channels by
employing the generalization of Eq.~\eqref{190410.2} in the elastic FSI case to the coupled-channel
FSI and initial-state interactions.
One now has to solve the Muskhelishvili-Omn\`es problem, as discussed in
Sec.~\ref{sec.181117.3}, for the $S$-wave $I=0$ and $I=1$ quantum numbers, each of which involves two coupled-channels.
We then assume that this has been done in terms of the input $T^{(IJ)}_{S;fh}$ PWAs, 
and that the matrices ${\cal D}^{(00)}(s)$, ${\cal D}^{(10)}(s)$,
as well as the one-channel functions ${\cal D}^{(20)}(s)$ and ${\cal D}^{(11)}(s)$, are at our disposal.
In this way, we then consider the matrix of functions
\begin{align}
\label{190412.3}
X^{(IJ)}(s)&={\cal D}^{(IJ)}(s) M^{(IJ)}(s)\,{\cal D}^{(1J)}_0(s)^T~.  
\end{align}
The first factor from left to right on the rhs of the previous equation takes care of the FSI
and the last one of the initial-state interactions.
In Eq.~\eqref{190412.3}  the superscript $T$ indicates the transpose of the matrix and
the subscript $0$ refers to the initial-state interactions, so that
$\cD^{(10)}_0(s)=\cD^{(10)}(s)$ and $\cD^{(11)}_0(s)=1$, to avoid confusing the latter with $\cD^{(11)}(s)$ (which accounts 
for the $\pi\pi$ $I=J=1$ FSI). 
All these matrices of functions ${\cal D}^{(IJ)}(s)$ are calculated in Ref.~\cite{alba.190329.1}, and details can be
found there and in Refs.~\cite{alba.190329.3,martin.190412.1,moussallam.190412.1}.
Notice that the $T$ matrices  used in Ref.~\cite{alba.190329.1} to evaluate the ${\cal D}^{(IJ)}(s)$
do not incorporate any LHC. As a result, there is no
need to calculate them with complex $m_\eta^2+i\ep$ and $m_K^2+i\ep$. Indeed, this is  already used to settle
the way in which the different PWAs $T^{(IJ)}_{S}(s)$ enter in Eq.~\eqref{190412.2}
(they are calculated with $\ep=0$). This is also the case for the ${\cal D}^{(IJ)}(s)$ employed.

Let us rewrite  the discontinuity relation of Eq.~\eqref{190412.2} in matrix form as
\begin{align}
\label{190412.5}
\big[I-2iT^{(IJ)}_{S}(s)^*\rho\big] M^{(IJ)}(s+i\ep)&=
M^{(IJ)}(s-i\ep)\big[I+\delta_{J0}2i\rho T^{(1J)}_S(s)\big]+2i T^{(IJ)}_S(s)^* \rho \hat{M}^{(IJ)}(s) \\
&+\delta_{J0}2i \hat{M}^{(IJ)}\rho T^{(1J)}_S(s)
  +\delta_{I0}T^{(00)}_{S}(s)^*\Sigma_K T^{(10)}_{S}(s)~.\nn
\end{align}
Rewriting  $M^{(IJ)}(s)$ in terms of $X^{(IJ)}(s)$ by inverting the matrices $\cD^{(IJ)}(s)$ and $\cD^{(10)}(s)$
in Eq.~\eqref{190412.3}, and taking into account that  Eq.~\eqref{181126.1} is satisfied by the ${\cal D}^{(IJ)\,-1}$,
one concludes that
\begin{align}
\label{190412.6}
\Delta X^{(IJ)}(s)&=X^{(IJ)}(s+i\ep)-X^{(IJ)}(s-i\ep)\\
&={\cD}^{(IJ)}(s)^* T^{(IJ)}_S(s)^*\rho\hat{M}^{(IJ)}(s){\cD}^{(1J)}_0(s)^T
+\delta_{J0}{\cD}^{(IJ)}(s)^*\hat{M}^{(IJ)}(s)\rho T^{(1J)}(s){\cD}^{(1J)}_0(s)^T\nn\\
&  +\delta_{I0}{\cD}^{(00)}(s)^*T^{(00)}_{S}(s)^*\Sigma_K T^{(10)}_{S}(s){\cD}^{(10)}(s)^T~.\nn
%&={\cD}^{(IJ)}(s-i\ep)T^{(IJ)}_S(s)^*\rho\hat{M}^{(IJ)}(s){\cD}^{(1J)}(s+i\ep)^T
%+{\cD}^{(IJ)}(s-i\ep)\hat{M}^{(IJ)}(s)\rho T^{(1J)}(s){\cD}^{(1J)}(s+i\ep)\nn\\
%&  +\delta_{I0}{\cD}^{(00)}(s)^*T^{(00)}_{S}(s)^*\Sigma_K T^{(10)}_{S}(s){\cD}^{(10)}(s)^T~.\nn
\end{align}  
Let us rewrite Eq.~\eqref{181126.1} as
\begin{align}
\label{190412.7}
{\cD}^{(IJ)}(s)^*&={\cD}^{(IJ)}(s) \cS^{(IJ)}(s)~,  
\end{align}
which also implies that
\begin{align}
\label{190412.8}
\Im {\cD}^{(IJ)}(s)=-{\cD}^{(IJ)}(s)T^{(IJ)}_S(s)\rho~.
\end{align}
This result allows us to rewrite Eq.~\eqref{190412.6} as
\begin{align}
\label{190412.9}
\Delta X^{(IJ)}(s)&=-2i\Im{\cD}^{(IJ)}(s)\hat{M}^{(IJ)}(s){\cD}^{(1J)}_0(s)^T
-\delta_{J0}2i{\cD}^{(IJ)}(s)^*\hat{M}^{(IJ)}(s) \Im{\cD}^{(1J)}_0(s)^T\\
&  +\delta_{I0}{\cD}^{(00)}(s)^*T^{(00)}_{S}(s)^*\Sigma_K T^{(10)}_{S}(s){\cD}^{(10)}_0(s)^T\nn\\
&=-\Delta\left[\cD^{(IJ)}(s)\hat{M}^{(IJ)}(s)\cD^{(1J)}_0(s)^T\right]
+\delta_{I0}{\cD}^{(00)}(s)^* T^{(00)}_{S}(s)^*\Sigma_K T^{(10)}_{S}(s){\cD}^{(10)}_0(s)^T~.\nn
%\Delta X^{(IJ)}(s)&=-\Im{\cD}^{(IJ)}(s)\hat{M}^{(IJ)}(s){\cD}^{(1J)}(s+i\ep)^T
%-{\cD}^{(IJ)}(s-i\ep)\hat{M}^{(IJ)}(s) \Im{\cD}^{(IJ)}(s)^T\\
%&  +\delta_{I0}{\cD}^{(00)}(s)^*T^{(00)}_{S}(s)^*\Sigma_K T^{(10)}_{S}(s){\cD}^{(10)}(s)^T\nn\\
%&=-\Delta\left[\cD^{(IJ)}(s)\hat{M}^{(IJ)}(s)\cD^{(1J)}(s)^T\right]
%+\delta_{I0}{\cD}^{(00)}(s)^* T^{(00)}_{S}(s)^*\Sigma_K T^{(10)}_{S}(s){\cD}^{(10)}(s)^T~.\nn
\end{align}
where the last step is an immediate consequence of the first line in the same equation.

At this stage we are ready to write down the DR representations for the different $M^{(IJ)}(s)$. Notice that
the expression for $M^{(11)}(s)/\kappa(s)$ is the same as in the elastic case given in Eq.~\eqref{190410.6}.
The Ref.~\cite{alba.190329.1} writes down DRs with at most two subtractions and keeping the same form of the
subtraction polynomial for $M^{(00)}_{11}$ and $M^{(20)}_{11}$ as in Eq.~\eqref{190410.6}. %$\hat{M}^{(00)}_{11}$ and $\hat{M}^{(20)}_{11}$
 Therefore, we write  that
\begin{align}
\label{190412.9b}
{M}^{(IJ)}(s)&=\kappa(s)^J\,{\cD}^{(IJ)}(s)^{-1}\left[P^{(IJ)}(s)+s^{2-J} \cI^{(IJ)}(s)\right]{\cD}^{(1J)}_0(s)^{-1\,T}~,\\
\cI^{(IJ)}(s)&=\frac{1}{\pi}\int_{4m_\pi^2}^\infty ds'\frac{\Delta X^{(IJ)}(s')}{(s')^{2-J}(s'-s)}~.\nn
\end{align}
The matrix $P^{(00)}(s)$ is a $2\times 2 $ matrix of second-degree polynomials,
 $P^{(11)}(s)$ is a linear function in $s$,
and $P^{(20)}(s)$ is $1\times 2$ row vector, with
$P^{(20)}_{11}=0$ and $P^{(20)}_{12}$ a polynomial of second degree.
The Omn\`es functions $\cD^{(11)}(s)^{-1}$ and $\cD^{(20)}(s)^{-1}$ are the same as in the elastic case.
However, the high-energy behavior for the matrix elements of
$\cD^{(00)}(s)^{-1}$ and $\cD^{(10)}(s)^{-1}$ follows the discussion in Sec.~\ref{sec.181117.3} and
vanish as $1/s$ (for the $T$ matrices taken in Ref.~\cite{alba.190329.1}).
 As a result, we have that the high-energy behavior for $M^{(11)}(s)$ and $M^{(20)}(s)$ is the same as in the
 elastic case, while $M^{(00)}(s)$ now tends to constant asymptotically and, thus,
 changes compared with the elastic case.\footnote{This fact is properly taken into account by the DR written
   for $X^{(00)}(s)$  in Eq.~\eqref{190412.9b}, in virtue of the application of the Sugawara-Kanazawa theorem.}
 This fact implies that in the reshuffling transformation of Eq.~\eqref{190409.1}
 the parameters $a_1$ and $b_2$ are not  independent any more. 
They must fulfill that $a_1=5/9 b_2$, so that the linear term in Eq.~\eqref{190409.2} vanishes.

There is a total of 16 subtraction constants in the polynomial matrices $P^{(IJ)}(s)$.
In addition, one also has the  matrix elements $\hat{M}^{(IJ)}_{2i}(s)$ and
$\hat{M}^{(IJ)}_{f2}(s)$,  which involve the $K\bar{K}$ states, with $i,f=1$ and/or $2$
(accordingly to the given $IJ$ quantum numbers).
Their full determination, by ending with a closed set of IEs,
would also imply to work out the different crossed channels involving kaons.
Thus, extra one-variable functions would be required.
The Ref.~\cite{alba.190329.1} altogether neglects these crossed-channel
$\hat{M}^{(IJ)}_{i2}(s)$ and $\hat{M}^{(IJ)}_{2f}(s)$ contributions,
arguing that their main interest is focused on the $\eta\pi \to\pi\pi$ amplitudes.
Its procedure is expected to be enough for a realistic calculation
of the effects of incorporating explicitly the $a_0(980)$ and $f_0(980)$ through coupled-channel dynamics. 

The 16 subtractions constants are fixed by matching to the chiral expansion of the amplitudes. For the
$\eta\to 3\pi$ amplitudes one employes the same procedure as in the elastic case by matching $A(s,t,u)$ with the
NLO ChPT calculation. 
Nonetheless, now $M^{(20)}_{11}(0)'\neq 0$ because of the coupled channel effects due to the initial-state interactions,
 and its value is given by the solution of the equations.
For the isospin-violating amplitudes with $K\bar{K}$ asymptotic states, Ref.~\cite{alba.190329.1}
matches them  with the LO ChPT ones up to and including the second derivatives in $s$ (which
are zero because the leading ChPT amplitudes are polynomial of first degree  in $s$).

Once the subtraction constants are given in terms of the ChPT amplitudes,
the numerical solutions of the IEs from the DRs in Eq.~\eqref{190412.9b} are found by applying
an analogous iterative method as described for the elastic case at the end of
Sec.~\ref{sec.190330.1}.

We do not dwell on a detailed comparison with phenomenology and refer directly to Ref.~\cite{alba.190329.1}, since
our emphasis in this review is on the methodology. Nonetheless, it is remarkable the improvement achieved for
the $\alpha$ parameter in the $\eta\to 3\pi^0$ decay by including coupled-channel effects at the level of around a
$20\%$. The resulting figure for $\alpha$
is then perfectly compatible with the experimental value \cite{pdg.181106.1}.
For the other Dalitz plot parameters in the charged and neutral three-pion decay modes of the $\eta$,
the changes are also in the right direction to improve the agreement with experiment. 

%%%%%%%%%%%%%%%%%%%%%%%%%%%%%%%%%%%%%%%%%%%%%%%%%%%%
\section{Summary and outlook}
\setcounter{equation}{0}   
\label{sec.190516.1}

We have revisited here some techniques in coupled-channel dynamics of relevance in current research,
as it is attested by the large number of applications that have used them.
As remark in the Introduction we did not pretend to offer a complete list of standard techniques, but rather
those that we have found of more use in our own research or have captured our attention.
The techniques here discussed rest either on unitarity and analyticity of partial-wave amplitudes or
in the use of the Lippmann-Schwinger equation in terms of a potential.
The review has then proceeded logically from these basics starting points towards more concrete aspects,
giving a unified presentation of all the contents here offered.

Other interesting additions to the discussions included in the manuscript would have been addressing the connection of
coupled-channel dynamics and resonances in more depth, e.g. by considering matrix elements of resonances,
a broad aspect that would have required a full review by itself.
Other topics of interest, not discussed here, are recent advances in scattering theory that allows one to
solve the scattering problem by just invoking analyticity and unitarity, in terms of the discontinuity along the
left-hand cut of a given potential.
Of course, it is also of great interest the adaptation of the methods discussed along the review to finite volume.
This is a field that has experienced a great growth in the last years so as
to interpret the results obtained from lattice Quantum Chromodynamics.

The coupled-channel  techniques have shown very useful in phenomenology, so as 
to reproduce experimental data and extract from them interesting conclusions regarding other processes or
spectroscopy in the strong-interacting regime.
Unfortunately, they lack to offer a closed set of equations that  allow us to
calculate the $S$ matrix just from some data.
Nonetheless, its application in conjunction with modern techniques from effective field theories
is a powerful tool whose scope and capabilities are worth exploiting to their maximum extent.

%%%%%%%%%%%%%%%%%%%%%%%%%%%%%%%%%%%%%%%%%%%%%%%%%%%%%%%%%
\subsection*{Acknowledgments}
I would like to thank Zhi-Hui Guo for reading the manuscript and giving me useful remarks.  
 I also acknowledge Bachir Moussallam for his readiness to compare with his results,
and Miguel Albaladejo for providing me the Fig.~\ref{fig.190330.1}. 
This work is partially supported  by the MINECO (Spain) and FEDER (EU) grant FPA2016-77313-P. 

\appendix

%%%%%%%%%%%%%%%%%%%%%%%%%%%%%%%%%%%%%%%%%%%%%%%%%%%%%%%%%%%%%%%%%%%%%%%%%%%%%%%%%%%%%%%%%
\section{Determination of the physical regions for the $t$-, $u$- and decay-channels of $\eta\to 3\pi$}
\setcounter{equation}{0}
\label{app.190331.1}
\def\theequation{\Alph{section}.\arabic{equation}}

Let us discuss the $u$-channel physical region, $u\geq m_{\eta\pi}^2$, with $m_{\eta\pi}=(m_\eta+m_\pi)^2$, as defined above 
after Eq.~\eqref{190331.2}. In terms of $u$ and the
scattering angle in the $u$-channel, $\theta_u$, we can express the variables $s$ and $t$ as
\begin{align}
  \label{190331.4}
  t,\,s&=\frac{1}{2}\left(m_\eta^2+3m_\pi^2-u\pm \cos\theta_u \sqrt{\lambda(u)\sigma(u)}\right)~,
\end{align}
respectively. This equation is analogous to Eq.~\eqref{190330.2}, where $t$ and $u$ are expressed
in terms of $s$ and $\theta$ (the scattering angle in the $s$-channel).
 We can deduce Eq.~\eqref{190331.4} by taking into account Eq.~\eqref{190330.1} and making use of the following
kinematical identities in the $\pi(p_1)\pi(p_3)$ rest frame:
\begin{align}
  m_\eta^2&=P^2=(p_1+p_2+p_3)^2=u+m_\pi^2+2p_2^0\sqrt{u}~,
\end{align}
so that
\begin{align}
  \label{190331.5}
  p_2^0&=\frac{m_\eta^2-m_\pi^2-u}{2\sqrt{u}}~.
\end{align}
In the same reference frame it is clear that $p_1^0=p_3^0=\sqrt{u}/2$.
 It follows that the associated three-momenta $|\vp|_1=\sqrt{{p_1^0}^2-m_\pi^2}$ and $|\vp|_2=\sqrt{{p_2^0}^2-m_\pi^2}$ are
\begin{align}
  \label{190331.6}
|\vp|_2&=\sqrt{\frac{\lambda(u)}{4u}}~,\\
|\vp|_1&=\sigma(u)^{1/2}~.\nn
\end{align}
As a result, the Eq.~\eqref{190331.4} is obtained  by substituting the expressions for $p_{1,2}^0$ and $|\vp|_{1,2}$
in the formulas,  
\begin{align}
  \label{190331.7}
  s=(p_1+p_2)^2&=2m_\pi^2+2 p_1^0 p_2^0-2\cos\theta_u |\vp_1||\vp_2|~,\\
  t=(p_2+p_3)^2&=2m_\pi^2+2 p_1^0 p_2^0+2\cos\theta_u |\vp_1||\vp_2|~.\nn
\end{align}

In order to disentangle the physical $u$-channel region depicted in Fig.~\ref{fig.190330.1} we make use of 
\begin{align}
  \label{190331.8}
t-u&=\frac{1}{2}\left(m_\eta^2+3m_\pi^2-3u+ \cos\theta_u \sqrt{\lambda(u)\sigma(u)}\right)~,\\
s&=\frac{1}{2}\left(m_\eta^2+3m_\pi^2-u- \cos\theta_u \sqrt{\lambda(u)\sigma(u)}\right)~,\nn
\end{align}
as it follows from Eq.~\eqref{190331.4}. For a given $u\geq m_{\eta\pi}^2$, the variable $s$ lies within the
interval
\begin{align}
  \label{190331.9}
  s&\in [s_-(u),s_{+}(u)]=\big[\frac{1}{2}\left(m_\eta^2+3m_\pi^2-u- \sqrt{\lambda(u)\sigma(u)}\right),\frac{1}{2}\left(m_\eta^2+3m_\pi^2-u+ \sqrt{\lambda(u)\sigma(u)}\right)\big]~.
\end{align}
The lower limit decreases as $u$ increases without bound. However, the upper limit is bounded and it increases with $u$, having
the asymptotic value
\begin{align}
\label{190331.10}
\lim_{u\to \infty}\frac{1}{2}\left(m_\eta^2+3m_\pi^2-u+ \sqrt{\lambda(u)\sigma(u)}\right)=0~.
\end{align}
The fact that $s_+(u)$ keeps growing with increasing $u$
can be seen by calculating its derivative which does not vanishes for $u\geq m_{\eta\pi}^2$.
As a result, the range of values for $s$, $s_+(u)-s_-(u)$, increases progressively with $u$, being zero
at the threshold $u=m_{\eta\pi}^2$.

Regarding $t-u$ we have from Eq.~\eqref{190331.8} that it lies within the interval of values
\begin{align}
\label{190331.11}
t-u\in [tu_-(u),tu_+(u)]= \big[\frac{1}{2}\left(m_\eta^2+3m_\pi^2-3u- \sqrt{\lambda(u)\sigma(u)}\right),\frac{1}{2}
  \left(m_\eta^2+3m_\pi^2-3u
 +\sqrt{\lambda(u)\sigma(u)}\right)\big]~.
\end{align}
As in the case of $s(u)$, the lower limit $tu_-(u)$ decreases with $u$ without bound, and
the upper limit $tu_+(u)$ is bounded. Its position can be
calculated by the point in which the derivative of $tu_+(u)$ with respect to $u$  vanishes.
This can be worked out numerically, and the absolute maximum of $t-u$ in the $u$-channel
is at $u=-0.478$~GeV$^2$, where $tu_+=-0.516$~GeV$^2$.
Notice that $tu_-(u)$ happens when $s=s_+(u)$ and $tu_+(u)$ does for  $s=s_-(u)$, because of the different sign
in front of the $\cos\theta_u$ in Eq.~\eqref{190331.8}. Thus, $s$ moves in the interval $[s_-(u),s_+(u)]$
for increasing $u\geq m_{\eta\pi}^2$  (with every time a  lower $s_-(u)$ and a  higher $s_+(u)$), and then
$t-u$ sweeps the values $tu_+(u)\to tu_-(u)$. In this way the $u$-physical region is filled in Fig.~\ref{fig.190330.1}.
The boundary is the set of points $(s,t-u)$ given by
\begin{align}
\label{190331.12}
\partial D_{u{\rm-channel}}=\left\{(\frac{1}{2}\left(m_\eta^2+3m_\pi^2-u\mp\sqrt{\lambda(u)\sigma(u)}\right),\frac{1}{2}\left(m_\eta^2+3m_\pi^2-3u\pm \sqrt{\lambda(u)\sigma(u)}\right))~,~u\geq m_{\eta\pi}^2\right\}~.
\end{align}

The analysis for the $t$-channel physical region can be done analogously as for the $u$-channel case, by expressing
$s$ and $u$ in terms of $t$ and $\theta_t$, the $t$-channel scattering angle. In this way, instead of
Eq.~\eqref{190331.8}, we have now
\begin{align}
\label{190331.13}
t-u&=\frac{1}{2}\left(3t-m_\eta^2-3m_\pi^2- \cos\theta_t \sqrt{\lambda(t)\sigma(t)}\right)~,\\
s&=\frac{1}{2}\left(m_\eta^2+3m_\pi^2-t+ \cos\theta_t \sqrt{\lambda(t)\sigma(t)}\right)~.\nn
\end{align}
Thus, the only difference is a change of sign in the rhs of $t-u$ in the analogy between the $t$- and $u$-channel physical
regions, which implies that $t-u>0$  for the physical region of the $t$-channel.
Therefore, we have now that for $t\geq m_{\eta\pi}^2$ the variable $s$ lies within the interval $s\in[s_-(t),s_+(t)]$, and
$t-u\in[-tu_+(t),-tu_-(t)]$. As a result, the $t$-channel physical region is
the mirror image by the $s$ axis of the $u$-channel one, as depicted in Fig.~\ref{fig.190330.1}. 

\newpage
%%%%%%%%%%%%%%%%%%%%%%%%%%%%%%%%%%%%%%%%%%%%%%%%%%%%%%%%%%%%%%%%

\end{document}